\documentclass[a4paper,11pt]{book}

\usepackage[T1]{fontenc}
\usepackage[utf8]{inputenc}
\usepackage{doi}
\usepackage[autostyle]{csquotes}
\usepackage[
    backend=biber,
    style=numeric-comp,
    date=year,
    giveninits=true,
    sorting=none,
    url=false, 
    doi=true,
    eprint=true,
]{biblatex}

\usepackage{biblatex}
\usepackage[french,german,english]{babel}

\usepackage{lmodern} % use this to fix blurry typewriter text font
\usepackage{setspace} % increase interline spacing slightly

\usepackage{graphicx}
\usepackage{xcolor}
\graphicspath{{images/}}

\usepackage{subfig}
\usepackage{booktabs}
\usepackage{lipsum}
\usepackage[final]{microtype}
\usepackage{url}

\usepackage{fancyhdr}

\fancypagestyle{plain}{
	\fancyhf{}

	\fancyfoot[EL,OR]{\thepage}}
\fancypagestyle{addpagenumbersforpdfimports}{
	\fancyhead{}
	
	\fancyfoot{}
	\fancyfoot[RO,LE]{\thepage}
}

\usepackage{listings}
\usepackage{hyperref}
\hypersetup{pdfborder={0 0 0},
	colorlinks=true,
	linkcolor=teal,
	citecolor=teal,
	urlcolor=black}
\ifpdf
\usepackage[final]{pdfpages}
\else
\usepackage{calc}
\usepackage{breakurl}
\usepackage[nlwarning=false]{hypdvips}
\usepackage{backref}
\renewcommand*{\bacref}[1]{}
\fi
\usepackage{bookmark}

\makeatletter
\renewcommand\@pnumwidth{20pt}
\makeatother

\makeatletter
\def\cleardoublepage{\clearpage\if@twoside \ifodd\c@page\else
    \hbox{}
    \thispagestyle{empty}
    \newpage
    \if@twocolumn\hbox{}\newpage\fi\fi\fi}
\makeatother \clearpage{\pagestyle{plain}\cleardoublepage}

\usepackage{color}
\usepackage{tikz}
\usepackage[explicit]{titlesec}
\newcommand*\chapterlabel{}
\titleformat{\chapter}[display]  % type (section,chapter,etc...) to vary,  shape (eg display-type)
	{\normalfont\bfseries\Huge} % format of the chapter
	{\gdef\chapterlabel{\thechapter\ }}     % the label 
 	{0pt} % separation between label and chapter-title
 	  {\begin{tikzpicture}[remember picture,overlay]
    \node[yshift=-8cm] at (current page.north west)
      {\begin{tikzpicture}[remember picture, overlay]
        \draw[fill=teal] (0,0) rectangle(35.5mm,15mm);
        \node[anchor=north east,yshift=-7.2cm,xshift=34mm,minimum height=30mm,inner sep=0mm] at (current page.north west)
        {\parbox[top][30mm][t]{15mm}{\raggedleft \rule{0cm}{0.6cm}\color{white}\chapterlabel}};  %the empty rule is just to get better base-line alignment
        \node[anchor=north west,yshift=-7.2cm,xshift=37mm,text width=\textwidth,minimum height=30mm,inner sep=0mm] at (current page.north west)
              {\parbox[top][30mm][t]{\textwidth}{\rule{0cm}{0.6cm}\color{teal}#1}};
       \end{tikzpicture}
      };
   \end{tikzpicture}
   \gdef\chapterlabel{}
  } % code before the title body
\titlespacing*{name=\chapter,numberless}{-3.7cm}{83.2pt-\parskip}{-3.2pt+\parskip}
\titlespacing*{\chapter}{-3.7cm}{50pt-\parskip-\parskip}{30pt+\parskip+\parskip}
\titlespacing*{\section}{0pt}{13.2pt}{1em-\parskip}  % 13.2pt is line spacing for a text with 11pt font size
\titlespacing*{\subsection}{0pt}{13.2pt}{1em-\parskip}
\titlespacing*{\subsubsection}{0pt}{13.2pt}{1em-\parskip}
\titlespacing*{\paragraph}{0pt}{13.2pt}{1em-\parskip}

\newcounter{myparts}
\newcommand*\partlabel{}
\titleformat{\part}[display]  % type (section,chapter,etc...) to vary,  shape (eg display-type)
	{\normalfont\bfseries\Huge} % format of the part
	{\gdef\partlabel{\thepart\ }}     % the label 
 	{0pt} % separation between label and part-title
 	  {\ifpdf\setlength{\unitlength}{20mm}\else\setlength{\unitlength}{0mm}\fi
	  \addtocounter{myparts}{1}
	  \begin{tikzpicture}[remember picture,overlay]
    \node[anchor=north west,xshift=-65mm,yshift=-6.9cm-\value{myparts}*20mm] at (current page.north east) % for unknown reasons: 3mm missing -> 65 instead of 62
      {\begin{tikzpicture}[remember picture, overlay]
        \draw[fill=black] (0,0) rectangle(62mm,20mm);   % -\value{myparts}\unitlength
        \node[anchor=north west,yshift=-6.1cm-\value{myparts}*\unitlength,xshift=-60.5mm,minimum height=30mm,inner sep=0mm] at (current page.north east)
        {\parbox[top][30mm][t]{55mm}{\raggedright \color{white}Part \partlabel \rule{0cm}{0.6cm}}};  %the empty rule is just to get better base-line alignment
        \node[anchor=north east,yshift=-6.1cm-\value{myparts}*\unitlength,xshift=-63.5mm,text width=\textwidth,minimum height=30mm,inner sep=0mm] at (current page.north east)
              {\parbox[top][30mm][t]{\textwidth}{\raggedleft \rule{0cm}{0.6cm}\color{black}#1}};
       \end{tikzpicture}
      };
   \end{tikzpicture}
   \gdef\partlabel{}
  } % code before the title body
\titlespacing*{\part}{11.06cm}{26.4pt-\parskip-\parskip}{0pt}

\usepackage{amsmath}
\usepackage{amsfonts}
\usepackage{amssymb}
\usepackage{mathtools}
\usepackage{array}
\usepackage{graphicx}

\makeatletter
\def\resetMathstrut@{%
  \setbox\z@\hbox{%
    \mathchardef\@tempa\mathcode`\(\relax
      \def\@tempb##1"##2##3{\the\textfont"##3\char"}%
      \expandafter\@tempb\meaning\@tempa \relax
  }%
  \ht\Mathstrutbox@1.2\ht\z@ \dp\Mathstrutbox@1.2\dp\z@
}

\DeclareFontFamily{OMX}{MnSymbolE}{}
\DeclareSymbolFont{MnLargeSymbols}{OMX}{MnSymbolE}{m}{n}
\SetSymbolFont{MnLargeSymbols}{bold}{OMX}{MnSymbolE}{b}{n}
\DeclareFontShape{OMX}{MnSymbolE}{m}{n}{
    <-6>  MnSymbolE5
   <6-7>  MnSymbolE6
   <7-8>  MnSymbolE7
   <8-9>  MnSymbolE8
   <9-10> MnSymbolE9
  <10-12> MnSymbolE10
  <12->   MnSymbolE12
}{}
\DeclareFontShape{OMX}{MnSymbolE}{b}{n}{
    <-6>  MnSymbolE-Bold5
   <6-7>  MnSymbolE-Bold6
   <7-8>  MnSymbolE-Bold7
   <8-9>  MnSymbolE-Bold8
   <9-10> MnSymbolE-Bold9
  <10-12> MnSymbolE-Bold10
  <12->   MnSymbolE-Bold12
}{}

\let\llangle\@undefined
\let\rrangle\@undefined
\DeclareMathDelimiter{\llangle}{\mathopen}%
                     {MnLargeSymbols}{'164}{MnLargeSymbols}{'164}
\DeclareMathDelimiter{\rrangle}{\mathclose}%
                     {MnLargeSymbols}{'171}{MnLargeSymbols}{'171}

\makeatother

\DeclarePairedDelimiter{\bbra}{\llangle}{\rvert}
\DeclarePairedDelimiter{\kket}{\lvert}{\rrangle}
\usepackage{braket}
\usepackage{bm}
\usepackage{qcircuit}
\usepackage{algorithm}
\usepackage{algorithmic}
\usepackage{pdflscape}
\usepackage[framemethod=TikZ]{mdframed}
\usepackage{makecell}
\usepackage{bbm}
\usepackage{algorithm}
\usepackage{algorithmic}
\usepackage{tabularx}
\usepackage{makecell}
\usepackage{booktabs}
\usepackage{upgreek}
\usepackage{mathtools}
\usepackage{physics}
\usepackage{tabularray}
\usepackage{tcolorbox}
\usepackage{acronym}  
\usepackage{mathrsfs}
\usepackage{multirow}
\usepackage{graphicx}

\DeclareFieldFormat
  [article,inbook,incollection,inproceedings,patent,thesis,
   unpublished,techreport,misc,book]
  {title}{\mkbibquote{#1}}

\mdfdefinestyle{summarybox}{
    roundcorner=10pt,
    innertopmargin=\topskip,
    frametitle={Summary},
    frametitlerule=true,
    frametitlebackgroundcolor=gray!20,
}
\mdfdefinestyle{paperbox}{
    roundcorner=10pt,
    innertopmargin=\topskip,
    frametitle={Note},
    frametitlerule=true,
    frametitlebackgroundcolor=gray!20,
}
\usepackage{adjustbox}
\usepackage{dashbox}
\newcommand\dboxed[2]{\setlength{\fboxsep}{#2pt} \dbox{\ensuremath{#1}}}

\newcommand{\summary}[1]{
\begin{mdframed}[style=summarybox]
#1
\end{mdframed}
}

\newcommand{\printpublication}[1]{\AtNextCite{\defcounter{maxnames}{99}}\fullcite{#1}}
\newcommand{\fidelity}[0]{\overline{\mathcal{F}}}

\renewcommand{\tr}{\mathrm{Tr}}

\renewcommand{\ketbra}[2]{|#1 \rangle\langle #2|}

\newcommand{\HS}{\mathcal{H}}
\newcommand{\K}{{\mathcal{K}}}
\newcommand{\F}{\mathcal{F}}
\DeclareMathOperator{\EE}{\mathbb{E}}
\newcommand\E[1]{\ensuremath\EE[#1]}
\newcommand{\mc}{\mathcal}

\newcommand{\mr}{\mathrm}
\newcommand{\mf}{\mathfrak}

\newcommand{\w}{\omega}
\newcommand{\herm}{{\textrm{Herm}(\HS)}}

\newcommand{\inprod}[2]{\langle #1, #2 \rangle}
\newcommand{\shortrightarrow}{\mathrel{\mkern-3mu\rightarrow}}
\newcommand{\bbrakket}[2]{ \llangle{#1}\lvert{#2}\rrangle}

\DeclareMathAlphabet{\mathcal}{OMS}{cmsy}{m}{n}
\DeclarePairedDelimiter{\expect}{\langle}{\rangle}

\DeclareSourcemap{
  \maps[datatype=bibtex]{
     \map{
        \step[fieldsource=doi,final]
        \step[fieldset=isbn,null]
        }
      }
}
\DeclareSourcemap{
  \maps[datatype=bibtex]{
     \map{
        \step[fieldsource=doi,final]
        \step[fieldset=issn,null]
        }
      }
}
\DeclareSourcemap{
  \maps[datatype=bibtex]{
    \map{
      \pertype{incollection}
      \step[fieldset=address, null]
      \step[fieldset=publisher, null]
    }
  }
}
\DeclareSourcemap{
  \maps[datatype=bibtex]{
    \map{
      \pertype{inproceedings}
      \step[fieldset=address, null]
      \step[fieldset=publisher, null]
      \step[fieldset=location, null]
      \step[fieldset=series, null]
    }
  }
}

\definecolor{problemcolor}{RGB}{50, 136, 140}

\newtcolorbox[auto counter]{pbbox}[2][]{fonttitle=\bfseries,
title=Theory Box~\thetcbcounter: #2,#1,colframe=gray}
\newtcolorbox[use counter from=pbbox]{problembox}[2][]{
floatplacement=h,float,
colback=problemcolor!5!white,colframe=problemcolor!75!black,title=Box~\thetcbcounter: #2,#1}

\AtEveryBibitem{\clearfield{pages}} 

\usepackage{chngcntr}
\counterwithout{footnote}{chapter}

\begin{document}

\frontmatter
\begin{titlepage}
\begin{otherlanguage}{french}

\sffamily

\begin{flushleft}
\parbox{0.3\textwidth}{\includegraphics[width=4cm]{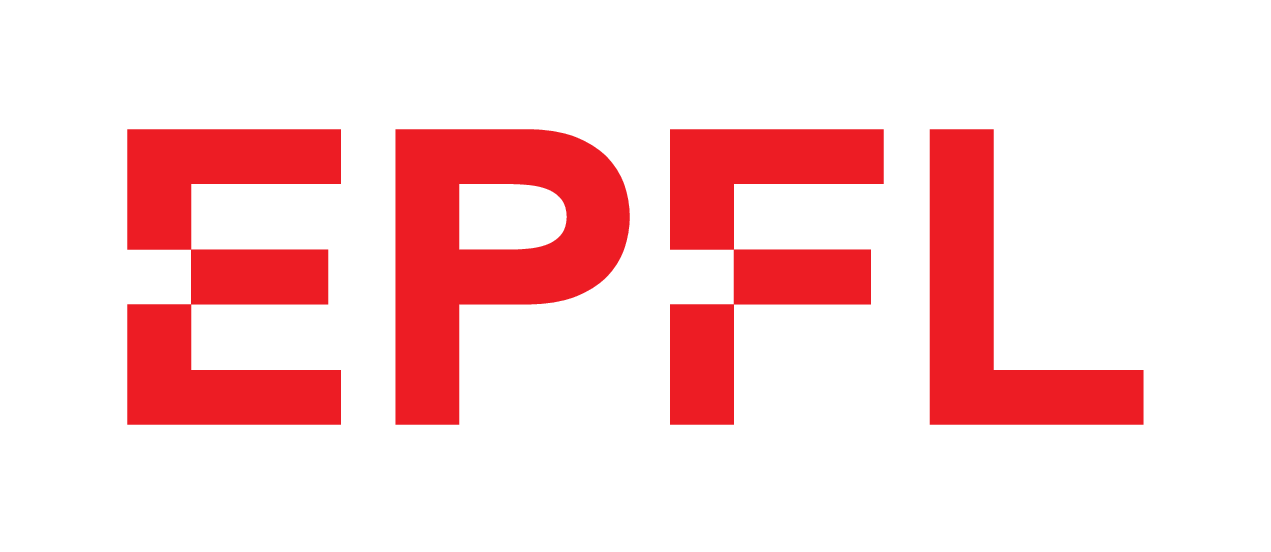}}
\end{flushleft}

\begin{flushright}
Thèse n.~11705
\end{flushright}

\null\vspace{2cm}

\begin{minipage}{4cm}
    % nothing
\end{minipage}
    \hfill
\begin{minipage}{11cm}
% {\Large Scalable Quantum Algorithms for Noisy Quantum \\ [8pt] Computers} \\
{\Large Enabling large-scale digital quantum simulations with superconducting qubits} \\

\vspace{2cm}

Présentée le 28 octobre 2025\\[8pt]
Faculté des sciences et techniques de l'ingénieur\\
Laboratoire de théorie et simulation des matériaux \\
Programme doctoral en science et génie des matériaux\\

pour l'obtention du grade de Docteur ès Sciences\\[8pt]
par\\ [12pt]
{\Large \textbf{Laurin Elias Fischer}}\\[9pt]

Acceptée sur proposition du jury\\[4pt]
     Prof. Harald Brune, président du jury\\
     Prof. Nicola Marzari, Dr. Ivano Tavernelli, directeurs de thèse\\
     Prof. Zo\"{e} Holmes, rapporteuse\\
     Prof. Zoltán Zimborás, rapporteur\\
     Prof. Frank Wilhelm-Mauch, rapporteur\\
\end{minipage}
\vspace{2cm}
\begin{flushright}
     2025
\end{flushright}

\end{otherlanguage}
\end{titlepage}

\cleardoublepage
\thispagestyle{empty}

\vspace*{1cm}

\begin{raggedright}
	\textit{``my silly work\\
	...\\
	everything is still unclear to me;\\
	perhaps the basic ideas are right after all\\
	...\\
    maybe those more capable than I \\ may yet make something sensible of it''}\\
      \textcolor{white}{.}\\
    Werner Heisenberg, \\
    summer 1925
\end{raggedright}

\vspace{4cm}

\begin{center}
    {To my grandfather}
\end{center}

\setcounter{page}{0}
\cleardoublepage
\chapter*{Abstract}
\markboth{Abstract}{Abstract}
\addcontentsline{toc}{chapter}{Abstract (English/Deutsch)} % adds an entry to the table of 

Quantum computing promises to revolutionize several scientific and technological domains through fundamentally new ways of processing information. Among its most compelling applications is digital quantum simulation, where quantum computers are used to replicate the behavior of other quantum systems. This could enable the study of problems that are otherwise intractable on classical computers, transforming fields such as quantum chemistry, condensed matter physics, and materials science. 

Despite this potential, realizations of practical quantum advantage for relevant problems are hindered by imperfections of current devices. This also affects quantum hardware based on superconducting circuits which is among the most advanced and scalable platforms.The envisaged long-term solution of fault-tolerant quantum computers that correct their own errors remains out of reach mainly due to the associated qubit number overhead. As a result, the field has developed strategies that combine quantum and classical resources, exploit hardware-native operations, and employ error mitigation techniques to extract meaningful results from noisy data. This thesis contributes to this broader effort by exploring methods for advancing quantum simulation across the full computational stack, including hardware-level innovations, refined techniques for noise modeling and error mitigation, and algorithmic improvements enabled by efficient measurement processing. 

On the hardware side, we develop a framework for full qudit-based quantum computation using superconducting transmon circuits. This makes optimal use of the available quantum resources by exploiting higher energy levels of transmons beyond the usually employed two-dimensional qubit subspace. We demonstrate how to implement a universal qudit gate set, enabling more efficient circuit synthesis and implementations of informationally-complete (IC) measurements. On the algorithmic and software side, we present techniques for the efficient classical processing of quantum data using IC measurements. We introduce a method that reduces the variance of observable estimators in post-processing, thereby improving the efficiency of quantum algorithms without additional quantum resources. On the error mitigation side, we employ IC measurements to perform accurate quantum computations that mitigate noise in classical post-processing. We demonstrate two such error mitigation techniques on IBM Quantum hardware at scales that lie beyond the brute-force simulation capabilities of classical computers. Firstly, we use IC measurements to parallelize the subspace expansion algorithm for ground state estimation. Secondly, we leverage a tensor-network pipeline that systematically compensates for the effects of noise to simulate many-body quantum dynamics. The latter method relies on accurate models of the device's noise channels, which were in the past challenged by experimental ambiguities that led to biased noise models. We demonstrate that a novel noise learning scheme can remove these inconsistencies, improving both the bias and variance of noise-model-based error mitigation.

The results of this thesis showcase how digital quantum simulation can be advanced despite the presence of noise, pushing the frontiers of quantum computing with superconducting hardware on several fronts. This both extends the reach of error mitigation and informs future strategies towards fault-tolerant quantum computing.

\begin{otherlanguage}{german}
\cleardoublepage
\chapter*{Zusammenfassung}
\markboth{Zusammenfassung}{Zusammenfassung}

Quantencomputing verspricht mehrere wissenschaftliche und technologische Bereiche durch grundlegend neue Arten der Informationsverarbeitung zu revolutionieren. 
Eine der wichtigsten Anwendungen ist die digitale Quantensimulation, bei der Quantencomputer verwendet werden, um das Verhalten anderer Quantensysteme nachzubilden. 
Dies könnte neue Ansätze für Fragestellungen ermöglichen, die auf klassischen Computern unlösbar sind, und damit Bereiche wie Quantenchemie, Festkörperphysik und Materialwissenschaften transformieren.

Trotz dieses Potenzials wird die Realisierung eines Vorteils durch Quantencomputer für relevante Probleme durch einige Unzulänglichkeiten erschwert.
Dies betrifft auch Geräte auf Basis supraleitender Schaltkreise, die zu den fortgeschrittensten und skalierbarsten Hardware-Plattformen zählen. 
Die langfristig angestrebte Lösung von fehlertoleranten Quantencomputern, die ihre eigenen Fehler korrigieren, ist aktuell vor allem aufgrund des damit verbundenen Mehrbedarfs an Qubits außer Reichweite. 
Es werden daher Strategien entwickelt, die Quanten- und klassische Ressourcen kombinieren, und Fehler mitigieren, um aus verrauschten Daten sinnvolle Ergebnisse zu extrahieren.
Diese Dissertation trägt zu diesen Bestrebungen bei, indem vielfältige Methoden zur Weiterentwicklung der Quantensimulation untersucht werden – einschließlich Innovationen auf Hardwareebene, verbesserter Techniken zur Rauschmodellierung und Fehlermitigation, sowie Verbesserungen von Algorithmen durch effiziente Verarbeitung von Messergebnissen.

Auf der Hardwareseite entwickeln wir ein Konzept für qudit-basierte Quantenberechnungen mit supraleitenden Transmon-Schaltkreisen. 
Dieses nutzt die verfügbaren Quantenzustände optimal aus, da höhere Anregungen von Transmons zusätzlich zum üblicherweise verwendeten zweidimensionalen Qubit-Unterraum verwendet werden. 
Wir zeigen, wie universelle Quantengatter implementiert werden können, was eine effizientere Synthese von Schaltkreisen und informationsvollständige (IV) Messungen ermöglicht.
Auf der algorithmischen und softwareseitigen Ebene entwickeln wir Techniken zur effizienten klassischen Verarbeitung von Quantendaten mithilfe von IV-Messungen. Wir präsentieren eine Methode, die die Varianz von Observablen-Schätzern reduziert und so die Effizienz von Quantenalgorithmen ohne zusätzliche Quantenressourcen verbessert. 
Zur Fehlermitigation verwenden wir IV-Messungen, um präzise Quantenberechnungen durchzuführen, bei denen das Rauschen in der klassischen Nachverarbeitung kompensiert wird. 
Wir demonstrieren zwei solcher Fehlermitigationstechniken auf IBM Quantum Hardware auf einer Größenskala, die über die Möglichkeiten exakter klassischer Simulationen hinausgeht. 
Erstens nutzen wir IV-Messungen zur Parallelisierung des \emph{subspace-expansion}-Algorithmus zur Schätzung von Grundzustandsenergien. 
Zweitens verwenden wir eine \emph{tensor-network}-Methode, die systematisch die Auswirkungen von Rauschen kompensiert, um Vielteilchen-Quantendynamik zu simulieren. 
Letztere Methode basiert auf präzisen Modellen der Fehlerkanäle des Quantencomputers, was in der Vergangenheit durch ungerechtfertigte Annahmen zu Ungenauigkeiten geführt hat. 
Wir zeigen, dass ein neues Lernverfahren für Fehlerkanäle diese Inkonsistenzen beseitigen, und sowohl die Verzerrung als auch die Varianz der modellbasierten Fehlermitigation verbessern kann.

Die Ergebnisse dieser Arbeit zeigen, wie digitale Quantensimulation trotz Rauschens weiterentwickelt werden kann, was die Grenzen des Quantencomputings mit supraleitender Hardware auf mehreren Ebenen erweitert. 
Dies vergrößert die Reichweite von Fehlermitigationsmethoden und dient als Orientierung für zukünftige Strategien hin zu fehlertolerantem Quantencomputing.

\end{otherlanguage}

\cleardoublepage
\chapter*{Publications}
\markboth{Publications}{Publications}
\addcontentsline{toc}{chapter}{Publications} % adds an entry to the table of contents

The following publications are covered in this thesis:
\begin{itemize}
    \item \printpublication{fischer2022ancillafree} 
    \item \printpublication{fischer2023universal}
    \item \printpublication{fischer2024dual}
    \item \printpublication{fischer2024dynamical} 
    \item \printpublication{chen2025disambiguating}
    \item \printpublication{fischer2025large}
\end{itemize}

\clearpage
\noindent The following research has been conducted during this doctorate but is not explicitly covered:
\begin{itemize}
    \item \printpublication{miller2024hardwaretailored}
    \item \printpublication{egger2023pulse}
    \item \printpublication{hope2025quantum}
    \item \printpublication{haupt2025statepreparation}
\end{itemize}
\chapter*{Acknowledgements}
\markboth{Acknowledgements}{Acknowledgements}
\addcontentsline{toc}{chapter}{Acknowledgements}

As I reflect upon the past years, I am honored and moved by how many people have supported me along the way.
Their guidance, encouragement, and belief in me have been invaluable, and I am sincerely thankful to each of them.

First I'd like to thank Ivano Tavernelli, my primary supervisor at IBM Quantum. 
You have always had my back and have inspired me with your incredibly broad knowledge across various fields in physics and chemistry. 
Moreover, I am grateful to my academic supervisor Prof. Nicola Marzari for the trust put in me and welcoming me as a doctoral student at EPFL. 
Thank you to Francesco Tacchino for always finding the time and some space on the white board to discuss the nitty-gritty details and for your steady mentorship in all matters big and small. 
Also, a special thanks to Daniel Egger for introducing me to the world of superconducting qubits and patiently showing me how to write clean code and bibliographies.  
I am grateful to Zo{\"e} Holmes, Zolt{\'a}n Zimbor{\'a}s and Frank Wilhelm-Mauch for serving as members of the committee for this thesis. 

I could not have imagined a better environment for my studies than the exceptional team at IBM Research -- Zürich managed by Stefan Wörner who is leading by example. 
The great spirit of collegiality and an open-door culture have continuously nurtured my passion for quantum science and technology. 
Thank you to Alberto, Alex, Almudena, Anthony, Bence, Christa, Conrad, Daniel M., David, Elena, Elisa, Igor, Julian, Julien, Lukas, Marc, Max, Panos, Pauline, Sabina, Samuele, Timothée, and so many others who came and went throughout the years, for the many wonderful hours spent away from the desk, be it at the ping-pong table, hiking in the mountains, riding bikes, hanging out by the lake, or even road-tripping through California. 

Over the years, I have had the privilege of working with many brilliant scientists from across the globe. 
I am grateful to my IBM colleagues from the United States including E. Chen, A. Seif, A. Eddins, A. Kandala, and Y. Kim for involving me in some of the most exciting projects of my PhD and infecting me with their relentless enthusiasm. 
Another highlight of my journey has been the workshops and discussions with the team from Algorithmiq including (among many others) S. Filippov, G. Garc{\'i}a-P{\'e}rez, S. Maniscalco, M. Rossi, M. Leahy, and J. Goold.
Thank you also to S. Chen and L. Jiang from the University of Chicago and A. Chiesa and S. Carretta from the University of Parma for enriching collaborations. 

Moreover, I want to acknowledge the Marie Skłodowska-Curie training network \emph{Molecular Quantum Simulations} through which this research was funded under the European Union Horizon 2020 research and innovation program (grant agreement No. 955479).
Thank you to the coordinators Juliane Sauer and Guido Pupillo and all the other fellows for creating this unique and enriching program. 
My special thanks go also to Prof. Oriol Vendrell for hosting my visit at Heidelberg University where I got a glimpse into the theoretical chemistry community and Daniel Bultrini with whom I had the pleasure to bring more ``quantum'' into this ``classical'' world. 

Finally, I am deeply grateful to my friends and family who have supported me along the way and keep reminding me that there are more important things to life than physics. 
To my parents and my sisters -- thank you for always providing safe havens to return home to and putting up with me when I get a little smarty-pants.
Thank you also to my grandfather and uncle for fostering my interest in the natural sciences from an early age. 
Finally, my sincerest gratitude goes to my amazing partner Jana for filling my life with kindness, love, and joy. 

\bigskip

\hfill Laurin Fischer,

\hfill summer 2025

\cleardoublepage
\pdfbookmark{\contentsname}{toc}
\tableofcontents
 
\chapter*{List of Abbreviations}
\begin{acronym}	

 \acro{CPTP}{Completely-positive Trace-preserving}
 
 \acro{DRAG}{Derivative Removal by Adiabatic Gate}
 
 \acro{DSE}{Direct Sum Extension}

 \acro{DFT}{Density Functional Theory}
 
 \acro{(E)CR}{(Echoed) Cross-Resonance}
 
 \acro{FTQC}{Fault-tolerant Quantum Computation} 
 
 \acro{GCX}{Generalized Controlled X} 
 
 \acro{GST}{Gate Set Tomography}
 
 \acro{IC}{Informationally Complete}
 
 \acro{LBCS}{Locally-Biased Classical Shadows}
 
 \acro{ML}{Maximum Likelihood} 
 
 \acro{MPO}{Matrix Product Operator}
  
 \acro{MPS}{Matrix Product State} 
 
 \acro{MSE}{Mean Square Error} 
 
 \acro{MUB}{Mutually Unbiased Bases}
 
 \acro{OD}{Operational Distance}
 
 \acro{PBC}{Periodic Boundary Conditions}

 \acro{PEA}{Probabilistic Error Amplification}

 \acro{PEC}{Probabilistic Error Cancellation}

 \acro{PM}{Projective Measurement}

 \acro{POVM}{Positive Operator-valued Measure} 
 
 \acro{PSE}{Parallelized Subspace Expansion} 

 \acro{PTM}{Pauli Transfer Matrix}
 
 \acro{QAE}{Quantum Amplitude Estimation}
 
 \acro{QDT}{Quantum Detector Tomography}
 
 \acro{QEC}{Quantum Error Correction}
 
 \acro{QED}{Quantum Electrodynamics}

 \acro{QPE}{Quantum Phase Estimation}
 
 \acro{QPT}{Quantum Process Tomography}
 
 \acro{QSD}{Quantum Shannon Decomposition}

 \acro{QST}{Quantum State Tomography}
 
 \acro{SIC}{Symmetric Informationally Complete}
 
 \acro{SPT}{Symmetry-Protected Topological}
 
 \acro{SPAM}{State Preparation and Measurement} 
 
 \acro{SPL}{Sparse Pauli-Lindblad}
 
 \acro{SSV}{Single-Shot Variance}
  
 \acro{TEM}{Tensor-Network Error Mitigation}
 
 \acro{TPE}{Tensor product extension}
 
 \acro{TREX}{Twirled Readout Error Extinction}
 
 \acro{VQE}{Variational Quantum Eigensolver}

 \acro{VQA}{Variational Quantum Algorithms}

 \acro{ZNE}{Zero-noise Extrapolation}
\end{acronym} 
 
\cleardoublepage
\phantomsection
\addcontentsline{toc}{chapter}{List of Figures} 
\listoffigures

\cleardoublepage
\phantomsection
\addcontentsline{toc}{chapter}{List of Tables} 
\listoftables

 \cleardoublepage
 \phantomsection
 \addcontentsline{toc}{chapter}{Symbols} % adds an entry to 
 \chapter*{Notation\markboth{Notation}{}}

\begin{itemize}
  \setlength\itemsep{0.25em}
	\item[$N$] Number of qubits.
	\item[$\mathcal{H}$] A Hilbert space of a quantum system. 
	\item[$\text{Lin}(\mathcal{H})$] The space of (bounded) linear operators on a Hilbert space $\mathcal{H}$.
	\item[$\mathcal{S}(\mathcal{H})$] The space of quantum states of a Hilbert space $\mathcal{H}$.
	\item[$\text{Herm}(\mathcal{H})$] The set of Hermitian operators on a Hilbert space $\mathcal{H}$.
	\item[$\ket{\psi}$] Dirac notation for a pure state. 
    \item[$\rho$] A general or mixed quantum state. 
	\item[$H$] The Hamiltonian of a quantum system. Also sometimes the Hadamard gate. 
	\item[$O$] Usually a Hermitian operator $O \in \text{Herm}(\mathcal{H})$, i.e., an observable.
	\item[$\kket{O}$] Vectorized ``double-ket'' notation of an operator $O \in \text{Lin}(\mathcal{H})$.
    \item[$\mathcal{O}$] Order of magnitude.	
    \item[$\mathbbm{1}$] The identity operation, with dimension depending on the context.
    \item[$X, Y, Z, I$] Single-qubit Pauli matrices.
    \item[$P_i$] A generic Pauli operator.
%    	\item[$f_i$] The Pauli fidelity of a Pauli operator $P_i$.
	\item[$\mathbbm{P}_N$] The $N$-qubit Pauli basis.
	\item[$\mathcal{P}_N$] The $N$-qubit Pauli group.

	\item[$U$] A unitary operation.     
%	\item[$\mathcal{U}$] Depending on context, calligraphic letters are used to denote qudit-space unitaries.
	\item[$\Lambda$] A potentially non-unitary quantum channel.
	\item[$E_k$] Kraus operators of a quantum channel.
   
    \item[$\vb{M}$] A POVM with elements $\{ M_k\}_{k=\{1, \dots, n\}}$.
    \item[$S$] Usually the number of measurement shots sampled from a POVM. 
	\item[$n$] Usually the number of measurement outcomes of a POVM.
	\item[$\vb{D}$] A dual frame to a POVM with elements $\{ D_k\}_{k=\{1, \dots, n\}}$.
	\item[$\omega_k$] Coefficients of an observable decomposed into POVM operators. 
	\item[$\mathcal{F}$] The frame superoperator that translates between a POVM and its dual frame.

    \item[$\hat o$] Hats usually denote estimators of random variables, except for Sec.~\ref{sec:transmon_qubits}, where hats denote operators to distinguish them from their classical counterparts. 
	\item[$\E$] The expectation value of a random variable.

	\item[$\chi$] Usually the bond dimension of a tensor network.
	
	\item[$\mathcal{T}$] Time ordering symbol.
    \item[$\coloneqq$] This symbol is used to highlight the introduction of additional notation. 
\end{itemize}

%%%%%%%%%%%%%%%%%%%%%%%%%%%%%%%%%%%%%%%%%%%%%%
%%%%% MAIN: The chapters of the thesis
%%%%%%%%%%%%%%%%%%%%%%%%%%%%%%%%%%%%%%%%%%%%%%
\mainmatter
\chapter{Introduction}
\label{chap:introduction}

\section{The case for quantum computing}

Computation has played a pivotal role in advancing science and shaping modern society over the last decades.
The use of mathematics, algorithms, and computational modeling has become a cornerstone of almost every scientific domain and is commonly referred to as ``computational science''.
This has led to breakthroughs in fields as varied as genomics, climate modeling, theoretical chemistry, particle physics, 
and artificial intelligence. 
Advances in computational capabilities have been driven, one the one hand, by novel algorithms and advances in software and, on the other hand, by decades-long exponentially increasing hardware performance fueled by Moore's law. 
However, the trend of ever more powerful processors is coming to an end as physical limits of transistor down-scaling are approached~\cite{moores_law}.
Moreover, there are complex problems for which it is widely accepted that classical computers cannot offer efficient algorithms, both in terms of time and memory requirements~\cite{vaezi2023quantum}.   
This calls for novel computational paradigms that process information in a fundamentally different way in order to overcome these limitations of so-called \emph{classical computing}.  
Such efforts have gathered growing attention in the last years and include the fields of neuromorphic computing~\cite{markovic2020physics}, reservoir computing~\cite{yan2024emerging}, and, notably, quantum computing. 

Information processing is inevitably done with physical degrees of freedom that obey the laws of nature, intrinsically connecting computation and physics. 
This is revealed by Landauer's principle which links the erasure of information to the dissipation of heat and has motivated early works on the concept of reversible computers that would dissipate less energy~\cite{bennett1982thermodynamics}. 
About 100 years after its inception~\cite{born1925quantenmechanik}, quantum mechanics has become physics' most complete theory of nature. 
From a modern point of view, it then seems natural to consider devices that process information in ways that are governed by the laws of quantum mechanics.
We call such devices \emph{quantum computers}. 

Historically, in the early 1980s, first ideas were formulated that a quantum automaton which leverages phenomena unique to quantum physics (such as superposition and entanglement) could be used for computation~\cite{manin1980computable}. 
Since these concepts have no classical analogue, they might be able to perform certain information processing tasks more efficiently than classical computers.
This idea was further developed and famously popularized by Richard Feynman~\cite{feynman1982simulating}.
He was motivated by the question of how to efficiently simulate quantum systems.
The state of a quantum system is described by a complex-numbered vector called the wavefunction. 
Classical computers face a fundamental bottleneck as the dimension of this vector scales exponentially with the number of constituent particles of the system, such as spins on a lattice in condensed matter physics or molecular orbitals in theoretical chemistry. 
This limits exact solutions to small systems.
However, if the computing device itself is a quantum system, we can hope to efficiently encode the wavefunction of the system of interest into that of the quantum computer. 
Moreover, if we can manipulate the quantum state of the computer in a sufficiently general way, we can simulate, for example, the time evolution of the studied system~\cite{lloyd1996universal}. 
This approach is known as \emph{quantum simulation}.

Quantum simulation is at the heart of many of the most pressing questions in physics, theoretical chemistry and material science, rendering it one of quantum computing's most prominent applications\footnote{Some authors use ``quantum simulation'' to exclusively refer to simulating \emph{quantum dynamics}. We use this term more broadly for any simulation of a quantum system's properties, such as spectral calculations.}.  
This is of particular importance for \emph{ab-initio} computational studies of molecular systems and materials~\cite{bauer2020quantum, mcardle2020quantum}. 
In the context of chemistry, target problems include the prediction of reaction rates and pathways through ground-state~\cite{tilly2022variational} and excited-state properties~\cite{ollitrault2020quantum}, molecular dynamics~\cite{miessen2022quantum, ollitraultmolecularquantumdynamics2021}, as well as electronic dynamics or molecular vibrations~\cite{motta2021emerging}.
The long-term vision for material science foresees the use of quantum computers to design new materials with specific properties, such as superconductors, magnets, or semiconductors, batteries, photovoltaics, and catalysts for chemical processes~\cite{working_group_material_science}.
In the realm of high-energy physics, quantum computers promise the simulation of classically difficult models such as lattice field theories, and are also explored to alleviate the data processing bottleneck of large collider experiments in particle physics~\cite{working_group_HEP}.

While quantum simulation is a natural use of a quantum computer, it was a priori unclear if ``classical'' problems can benefit from quantum computing. 
Indeed, several such quantum algorithms that give a provable speedup over the best known classical algorithms ones were found shortly after Feynman's proposal by Deutsch/Jozsa~\cite{deutsch1992rapid} and Bernstein/Vazirani~\cite{bernstein1993quantum}.
However, they solve artificial problems of little practical interest.
This changed in 1994 with the famous Shor's algorithm that can solve integer factorization with an exponential speedup~\cite{shor1994algorithms} which sparked lasting interest in the field due to the potential of breaking conventional encryption methods.
Since then, finding novel quantum algorithms for various problem domains has become an active field of research. % rephrase
While Shor's algorithm perhaps remains unique in providing a rigorously understood exponential speedup, 
quantum algorithms have been brought forth for various application domains such as optimization problems~\cite{working_group_optimization}, machine learning~\cite{schuld2021machine}, finance~\cite{herman2023quantum}, health care~\cite{working_group_life_sciences}, and drug design~\cite{santagati2024drug}.

\section{Quantum computing in a nutshell} 
\label{sec:how_to_build_QC}
\begin{figure}[t]
    \centering
    \includegraphics[width=1\textwidth]{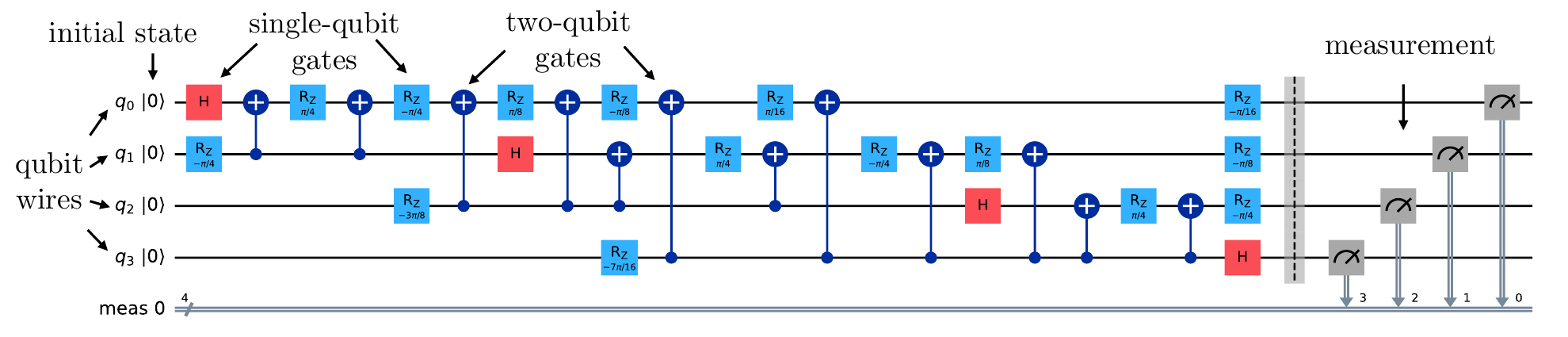}
    \caption[A quantum circuit schematic]{\small Schematic of a four-qubit quantum circuit. Qubits are initialized in the all-zero state $\ket{0\dots 0}$. A series of single- and two-qubit gates apply a unitary transformation to the state. 
Finally, a measurement yields a classical bitstring sampled according to the Born rule from the final state. 
}
\label{fig:quantum_circ}
\end{figure}

Constructing a physical quantum information processor requires balancing a fundamental trade-off between isolation, controllability, and scalability. 
The quantum components need to be well isolated to prevent decoherence from environmental noise, yet also controllable for logical operations and readout, while supporting the scaling to large system sizes. 
A plethora of different quantum computation paradigms have been proposed which navigate these challenges differently.
For example, analog quantum simulation uses continuous-time evolution of engineered quantum systems to study specific Hamiltonians~\cite{hangleiter2022analogue}.
This has been scaled to large isolated systems but lacks the full universal controllability of more general approaches.
By contrast, gate-based or ``digital'' quantum computing applies a sequence of logical gates to a register of low-dimensional quantum information carriers called \emph{qubits} (for two-dimensional systems with a basis $\{\ket{0}, \ket{1}\}$) or \emph{qudits} (for higher-dimensional systems), see Fig.~\ref{fig:quantum_circ}. 
Other quantum computing paradigms include measurement-based computation, boson sampling, adiabatic quantum computing, and topological quantum computing models. 
In this thesis we focus on digital, gate-based quantum computation, which is currently the most advanced and versatile framework. 

A gate-based quantum computation can be visualized as a quantum circuit, see Fig.~\ref{fig:quantum_circ}. 
Each of the $N$ qubits in the register is shown as a wire. 
The quantum state of a qubit register can be represented by a $2^N$-dimensional wavefunction known as the \emph{state vector} $\ket{\psi} = \sum_{j=0}^{2^N-1} c_j \ket{\vec{j}}$
with $c_j \in \mathbb{C}$ and $\sum_{j=0}^{2^N-1} |c_j|^2 = 1$, where the states $\{\ket{\vec{j}}\} = \{ \ket{0\dots 0}, \ket{0 \dots 1}, \dots, \ket{1 \dots 1} \}$ form the \emph{computational basis}.  
For gate-based quantum computing, we need the ability to initialize our qubits in a well-defined state, usually taken to be $\ket{0\dots 0}$, i.e., all qubits are in their $\ket{0}$ state. 
Transformations of our qubits are described by unitary transformations $U$ called gates that act on the vector $(c_0, \dots, c_{2^N-1})^T$ as matrices that conserve the norm $\sum_{j=0}^{2^N-1} |c_j|^2$. 
The circuit picture in Fig.~\ref{fig:quantum_circ} indicates the order (from left to right) in which the gates are applied to the initial states, and which qubits the gate acts on. 
A finite set of gates that can efficiently realize any unitary $U$ is known as a \emph{universal gate set}. 
This is typically realized by single-qubit gates and one particular entangling two-qubit gate, such as the CNOT gate shown in Fig.~\ref{fig:quantum_circ}.
A quantum circuit usually terminates with a measurement that yields the outcome $\vec{j}$ with probability $|c_j|^2$ from the final state prior to measurement. 

Several physical platforms are being pursued to encode qubit (or qudit) degrees of freedom.
These include, e.g., hyperfine states in trapped ions~\cite{bruzewicz_trapped-ion_2019}, electronic states in neutral atoms with Rydberg interactions~\cite{rydberg_atoms_review}, the polarization or path information of photons~\cite{couteau_quantum_2025}, electron spin states in semiconductors~\cite{spin_qubits_review} or exotic excitations in topological systems~\cite{RevModPhys.80.1083}. 
Besides the aforementioned considerations around qubit quality and scalability, an important aspect is the speed at which quantum gates can be physically realized~\cite{wack2021quality}. 
In this regard, superconducting architectures that exploit quantized current or charge states in superconducting circuits have emerged as a prominent platform to implement not only high-quality but also fast quantum operations. 
This thesis explores superconducting quantum computers with a circuit design known as \emph{transmons}~\cite{koch2007chargeinsensitive}. 

\section{Current state of the field}

Its transformative technological potential has turned quantum computing into a field with one foot in academia and one foot in an expanding industrial ecosystem.
This includes both global tech giants such as Google, IBM, and AWS, as well as a rich landscape of start-ups across both hardware and software backed by large amounts of private and governmental funding~\cite{seskir_landscape_2022}.
A recent report of the Boston Consulting Group estimates that quantum computing will sustain a market of around $\$100$ billion by 2040~\cite{bobier2024long}. 
However, as of today, quantum computers have yet to demonstrate an industrially relevant advantage over conventional methods. 
The current climate is thus characterized by a race to find an industrially relevant problem where current quantum devices can provide value over classical computers, a goal referred to as \emph{quantum advantage}.

\subsection{The quest for a quantum advantage}

We typically think of a quantum advantage as a scenario where a quantum computation yields a ``better'' result than any classical computation in terms of accuracy.
While this notion is well-defined in some cases, e.g., for classical optimization or variational problems, it may be more subtle in general. 
A quantum advantage could also manifest as a more efficient solution, e.g., in terms of energy consumption or a faster time to solution. 
Several experiments have reported a quantum device outperforming classical computers at the task of sampling from the distribution of measurement outcomes of some (noisy) quantum state~\cite{arute_quantum_2019, zhong2020quantum}. 
However, over time, classical methods to simulate these experiments have been refined, leading to refutations of the earlier advantage claims~\cite{larose2024brief}. 
While these sampling problems are of little practical value, this illustrates two main themes in the race to quantum advantage. 
Firstly, there will likely be a period of back-and-forth between quantum and classical methods surpassing each other. 
Secondly, quantum advantage claims may break down if the noise present in the quantum system can be exploited to classically simulate the quantum computation to a good approximation. 

In their race to advantage, quantum computers have to compete with conventional computational methods that have been refined for decades and are often highly specialized to certain problem classes. 
For example, in quantum chemistry, famously successful methods include Density Functional Theory (DFT)~\cite{engel2011density} and {Coupled Cluster}~\cite{bishop1991overview} techniques for electronic structure calculations, or {Car-Parinello molecular dynamics}~\cite{CPMD} and path integral methods~\cite{schulman2012techniques} for dynamics, to name just a few.
The core question is thus which specific problems offer the most potential for an imminent quantum advantage.
In this context, resource estimations in terms of the required number of qubits and quantum gates guide the search for promising applications~\cite{beverland2022assessing, dalzell2023quantum}. 
From these surveys, quantum simulation for physics and chemistry has emerged as the most promising candidate for pursuing a quantum advantage with comparatively moderate quantum resources~\cite{scholten2024assessing}. 
Moreover, classical and quantum computing communities are more and more collaborating to identify use cases across different fields~\cite{working_group_optimization, working_group_HEP, working_group_life_sciences, working_group_material_science}. 
A crucial idea that is also central to this thesis is to combine classical high-performance computing and quantum resources in a way that leverages their respective strengths~\cite{jie2024quntum-centric, pascuzzi2024quantum}.
For example, in chemistry, embedding schemes may treat only the strongly correlated part of a molecule as an active space on the quantum processor and employ classical methods like Hartree-Fock or DFT on the remaining degrees of freedom~\cite{rossmannek2021quantum, ma2020quantum}.

\subsection{Challenges and roadblocks}

Over the last years, superconducting quantum computers have evolved from chips of a few individual qubits to now supporting up to $\sim 150$ physical qubits~\cite{abughanem2025ibm}. 
The central challenge is that these qubits are subject to noise.
All operations like qubit initializations, quantum gates, and measurements come with small unavoidable errors.
This limits the depth of quantum computations that can reliably be executed. 
Despite continuous improvements to the error rates~\cite{jurcevic2021demonstration, abughanem2025ibm}, many established quantum algorithms require significantly larger circuit depths than those that are currently affordable~\cite{scholten2024assessing, blunt2022perspective}.
This class of algorithms is commonly referred to as ``fault-tolerant quantum algorithms'' (see Sec.~\ref{sec:fault-tolerant_QC_perspective}).
Besides noise, superconducting qubits suffer from a limited connectivity, i.e., they can usually only be coupled to their nearest neighbors in a sparse 2D graph layout.
This poses significant constraints on the implementable circuits.

The above limitations have led to the development of quantum algorithms that only require shallow, hardware-native circuits, also known as ``near-term quantum algorithms''.
For example, {variational quantum algorithms} (VQAs) have become a popular choice for noisy quantum devices~\cite{cerezo2021variational}. 
They prepare a parametrized trial state $\ket{\psi(\vec{\theta})}$ with hardware-friendly circuits. 
An expectation value of an observable $\langle O \rangle(\vec{\theta})$ is then classically optimized by tuning the parameters $\vec{\theta}$.
This has enabled successful demonstrations of ground state calculations for small molecules on noisy hardware~\cite{kandala2017hardwareefficient}.
However, due to the probabilistic nature of quantum mechanics, each evaluation of $\langle O \rangle(\vec{\theta})$ 
during the optimization takes many state preparations and measurements of $\ket{\psi(\vec{\theta})}$.
As applications are scaled to industrially relevant system sizes, these requirements on the number of measurements become a significant bottleneck~\cite{gonthier2022measurements}, which we refer to as the \emph{measurement overhead problem}. 
With traditional measurement techniques, run times of some variational algorithms would reach weeks or even months~\cite{miessen2021spin-boson}. 

Variational algorithms exemplify that quantum algorithms often come with a trade-off between circuit depth and an increased overhead of the required measurements. 
While this has been successful for small proof-of-principle experiments, it poses challenges for scaling to meaningful system sizes.
This calls for the development of methods to process measurement outcomes more efficiently, which forms a central theme of this thesis. 
Furthermore, scaling quantum applications requires strategies to deal with noisy quantum circuit executions, which are discussed in the following.

\subsection{Fault-tolerance vs. error mitigation}
\label{sec:fault-tolerant_QC_perspective}

It is a fundamental question whether reliable quantum computation in the presence of unavoidable experimental imperfections -- a goal known as \emph{fault-tolerant quantum computation} (FTQC) -- is feasible even in principle.
Remarkably, it is indeed possible to preserve quantum information by actively correcting errors that occur during computation, which is known as \emph{Quantum Error Correction} (QEC), as first proposed in 1995~\cite{shor1995scheme}. 
The idea is that fragile entangled quantum states can be protected from decoherence through the use of entanglement itself. 
That is to say, QEC codes encode logical qubit states into highly entangled states of multiple physical qubits.
This distributes information such that errors can be detected and corrected without directly measuring (and thus collapsing) the logical information. 
A key parameter is the \emph{code distance} $d$, which is related to the maximal number of simultaneous physical qubit errors that can be detected and corrected. 
The seminal \emph{threshold theorem} guarantees that logical error rates are suppressed exponentially with increasing code distance, provided the physical error rate is below some code-specific threshold~\cite{knill1998resilient}.  
This in theory enables (almost) arbitrarily long quantum computations to be performed reliably. 
However, increasing $d$ comes at the cost of increased qubit overhead, characterized by the number of physical qubits $n$ required to encode $k$ logical qubits. 
The challenge of designing QEC codes that achieve a good tradeoff for the values $[[n, k, d]]$, while requiring realistic error thresholds has led to myriad of error correction approaches and is an active field of research~\cite{ErrorCorrectionZoo}.

First demonstrations of code distance scaling below the error threshold have been reported recently for a single logical qubit in $[[25, 1, 5]]$ and $[[49, 1, 7]]$ surface codes~\cite{acharya_quantum_2025}. 
More modern codes have aimed to reduce the qubit overhead. 
For example, 12 logical qubits can be encoded into 288 physical qubits in the so-called ``Gross code'', assuming a physical error rate of $0.1\%$~\cite{bravyi_high-threshold_2024}. 
Beyond realizing a quantum memory, implementing a universal set of logical gates usually requires subroutines such as magic state distillation that lead to additional physical qubit overhead~\cite{gupta_encoding_2024}. 
Thus, while fundamental building blocks of fault-tolerant quantum computers have been demonstrated in proof-of-principle experiments~\cite{acharya_quantum_2025, gupta_encoding_2024, stean_code_innsbruck}, scaling hardware to support FTQC remains an elusive goal that is out of reach for today's devices. 
IBM's current long-term projection is for full-scale FTQC to be realized in the next decade~\cite{ibm_roadmap}. 

\begin{figure}
  \begin{minipage}[c]{0.5\textwidth}
    \includegraphics[width=\textwidth]{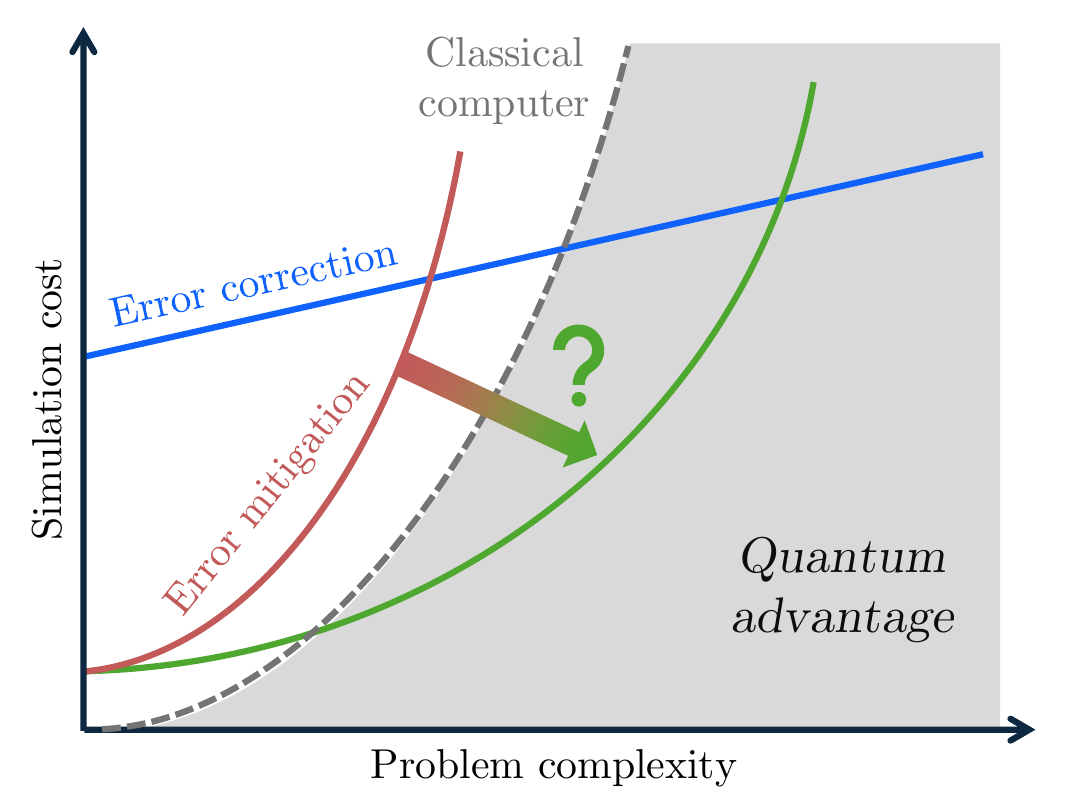}
  \end{minipage}\hfill
  \begin{minipage}[c]{0.48\textwidth}
    \caption[Relationship between error mitigation, error correction, and quantum advantage]{\small 
    Classical simulations of quantum circuits scale exponentially with the circuit size (dashed line).
    For error-corrected quantum computers (blue) this is polynomial, but comes at the cost of a large resource overhead, represented as a y-axis offset. 
    Error-mitigated quantum computers (red/green) also scale exponentially in general. It is an open question whether the basis of this exponential can be reduced to a point where this beats classical methods to provide a quantum advantage (shaded region). 
    } \label{fig:schema_error_mitigation_advantage}
  \end{minipage}
\end{figure}

Before the onset of fault-tolerance, so-called \emph{error mitigation} techniques offer a pragmatic alternative. 
Unlike FTQC, which involves extracting measurements from effectively perfect logical states, error mitigation combines noisy measurements taken from uncorrected quantum states to infer expectation values that approximate ideal outcomes. 
Remarkably, when the noise channels acting in the device are known, error mitigation can be used to yield in principle noise-free estimators for expectation values~\cite{temme2017error}.
However, this comes at a significant cost. 
The mitigated estimator will have a variance that is increased by a factor of $\gamma^d$, where $d$ is related to the number of gates in the circuit and $\gamma$ is a measure for the amount of noise in the device. 
That is, error mitigation in general comes with an exponential overhead in measurements, further contributing to existing measurement bottlenecks of (near-term) algorithms. 

An abundance of different error mitigation techniques has been proposed that range from more heuristic and noise-agnostic to more rigorous, see Ref.~\cite{cai2023quantum} for a comprehensive review.
However, the in principle exponential scaling of resources required to perform mitigation cannot be circumvented, as attested by several works on theoretical lower bounds on the measurement overhead~\cite{takagi2022fundamental, tsubouchi2023universal, quek_exponentially_2024}.
These no-go theorems have sparked an ongoing discussion around the central question: \textit{Can error-mitigated noisy quantum computers provide relevant quantum advantages?}
This essentially boils down to the question whether the exponential overhead of error mitigation can outperform the generally exponential cost of classical simulations~\cite{zimboras2025myths}, see Fig.~\ref{fig:schema_error_mitigation_advantage} for a schematic representation.
A more subtle issue is that classical methods may circumvent exponential scaling when we allow for them to make approximations. 
The more noise there is in the quantum system, the better these classical approximations become, as has recently been exploited by various techniques to simulate noisy quantum circuits efficiently~\cite{schuster2024polynomial, angrisani2025simulating}.
The importance to reduce quantum noise levels is thus two-fold; It exponentially improves the overhead of error mitigation while simultaneously making classical approximations more difficult. 

Fig.~\ref{fig:schema_error_mitigation_advantage} may indicate that the importance of error mitigation will fade when FTQC reaches a break-even point. 
However, even fault-tolerant machines will have a non-zero logical error rate which will ultimately limit their affordable circuit volumes. 
Error mitigation is thus projected to remain a useful tool to extend the reach of (early) fault-tolerant devices when applied to logical errors~\cite{zimboras2025myths, aharonov2025importance}.

\subsection{Previous experimental demonstrations}

Let us now review previous experimental demonstrations of digital quantum simulations. 
In the context of chemistry, the largest variational ground state calculation to date has been performed on twenty superconducting qubits~\cite{obrien2022purificationbased}, while more recent works have performed quantum-informed selected configuration interaction (CI) experiments on systems exceeding 50 qubits~\cite{robledomoreno2025, SQD_2}.
We note that both of these approaches are heuristic in nature and that their viability is currently under debate~\cite{cerezo2023does, reinholdt2025fundamental}.
Scaling these algorithms further is also complicated by the high number of terms of electronic molecular Hamiltonians.
For $M$ orbitals, this typically scales as $\mathcal{O}(M^4)$ when mapped to a quantum computer, further exacerbating the measurement overhead problem.
Therefore, recent attempts to scale to larger qubit numbers have focused more on the quantum simulation of condensed matter~\cite{rosenberg2024dynamics} or high-energy physics~\cite{physrevd.109.114510}, where interesting lattice model Hamiltonians with $N$ sites only have $\mathcal{O}(N)$ terms.
With the use of error mitigation, several experiments have successfully simulated Ising models in a regime where an brute-force classical simulation is not possible using 56~\cite{haghshenas2025digital} or up to over 100 qubits~\cite{kim2023evidence}, whereas exact classical simulations break down beyond $\sim 40$ qubits. 
At these scales, the results of the quantum computation need to be benchmarked against \emph{approximate} classical methods and it is a priori unclear which methods achieve the best convergence to the ground truth. 
This era where quantum computers are starting to compete with classical simulations has been coined \emph{quantum utility}~\cite{kim2023evidence}.

\section{Thesis contributions overview and outline}

\begin{problembox}[label={box:path_foward}]{Avenues to advance near-term quantum computers}
\begin{enumerate}
    \item \textbf{Navigate hardware constraints:} Current quantum computers are noisy and have limited qubit numbers and connectivity. Hence, it is paramount to co-design applications with device constraints in mind and make optimal use of the available quantum resources. 
    
    \item \textbf{Efficient measurement processes:} Near-term quantum simulation algorithms call for statistical methods that process measurement outcomes as efficiently as possible.
    This originates from the ubiquitous subroutine of expectation value estimations which require many individual measurements \textit{and} additional measurement overheads arising from error mitigation. 
        
    \item \textbf{Leverage quantum and classical:} It is essential to develop methods that combine high-performance classical and quantum resources.   
    Quantum computers should only address the relevant part of the problem that is classically intractable while offloading as much of the remaining processing to classical computers. 

    \item \textbf{How to verify outcomes?} Recently, digital quantum simulations are reaching ``utility'' scale where the outcome can no longer be simulated with exact classical methods. 
    Without the rigor of full error correction, we need suitable benchmarks to verify the output of large-scale noisy quantum computations and build trust in the underlying error mitigation techniques.  
    
    \item \textbf{Noise characterization:} We need scalable techniques to accurately characterize the noise of the hardware. 
    Not only is this a prerequisite to perform the most powerful error mitigation techniques but it is also required to understand the resource overhead that comes with error mitigation and eventually error correction. 
\end{enumerate}
\end{problembox}

Based on the discussions above, we summarize different aspects that will contribute to advance quantum simulation with today's error-mitigated superconducting quantum computers towards its goal of achieving quantum advantage in Box~\ref{box:path_foward}. 
In this thesis we contribute to these goals from different perspectives that range across the entire stack of a superconducting quantum computer.
This thesis is written in a self-contained way that is suitable for readers with a physics or general STEM background with knowledge of quantum mechanics. 
Chapters~\ref{chap:theory_of_qc}--\ref{chap:algorithms_and_subroutines} provide background information and review existing work relevant for this thesis. 
Chapter~\ref{chap:theory_of_qc} provides the necessary theoretical background from the field of quantum information, while Chapter~\ref{chap:superconducting_qubits} introduces the hardware platform of superconducting qubits.  
Chapter~\ref{chap:algorithms_and_subroutines} summarizes relevant quantum algorithms and error mitigation techniques on which we build in this thesis.

Novel contributions of this thesis are covered in Chapters~\ref{chap:qudit_processing}--\ref{chap:gauge_learning}. 
On the hardware side, we explore a novel paradigm of information processing with superconducting circuits in Chapter~\ref{chap:qudit_processing}. 
Qubits are usually encoded into the two energetically lowest states of a transmon circuit. 
However, additional higher-excited states can be accessed and controlled with high fidelity. 
We develop a blueprint to harness these additional energy levels and use transmons for universal, qudit-based computation. 
We identify the available control pulses for single-qudit and entangling multi-qudit operations and present a gate decomposition algorithm that enables the implementation of arbitrary multi-qudit unitaries. 
Since the control of qudit levels comes with additional error sources, we identify optimal transmon parameter regimes for qudit operation through pulse-level simulations.

On the quantum algorithms side, a central theme of this work is to explore how generalized measurements described by informationally-complete (IC) positive, operator-valued measures (POVMs) can offer advantages over standard computational basis measurements. 
IC POVMs offer great flexibility as they can be used to efficiently estimate any sufficiently local observable. 
The most general class of POVMs, however, has previously suffered from an overhead in ancilla qubits which are not readily available under typical device connectivity constraints. 
We demonstrate a new scheme to implement such POVMs using our newly developed qudit processing toolbox (Section~\ref{sec:ancilla-free-POVMs}).
A more restricted class of IC-POVMs, also known as \emph{Classical Shadows}~\cite{huang2020predicting}, can be implemented without ancilla qubits. 
We establish a new method to process outcomes of such shadow measurements more efficiently by leveraging previously neglected degrees of freedom in the formalism, which is presented in Chapter~\ref{chap:duals_optimization}.

Furthermore, we consider two complementary error mitigation approaches that combine classical and quantum resources using POVM measurements and efficient classical post-processing.
We demonstrate these techniques in quantum simulation experiments at scales beyond brute-force classical simulability.

Firstly, we use POVM measurements to parallelize the heavy measurement workload required by subspace expansion techniques. 
We demonstrate that this approach can successfully mitigate errors of ground state energy evaluations at circuit sizes of up to 80 qubits, see Chapter~\ref{chap:subspace_expansion}.

Secondly, we present first experimental realizations of a noise-model-based error mitigation strategy that removes noise in post-processing by exploiting classical tensor networks, as proposed in Ref.~\cite{filippov2023scalable}, see Chapter~\ref{chap:dual_unitary_TEM}. 
We apply this method to simulate the infinite-temperature autocorrelation function of a kicked Ising model using up to 91 qubits. 
We leverage a special class of ``dual-unitary'' circuits in this model which offer exact analytical solutions despite featuring non-integrable dynamics. 
This serves as an elegant way to verify the outcomes of the computation.
We also probe dynamics beyond such exact verification and compare our results to classical approximations with tensor networks. 
These results cement error-mitigated digital quantum simulation on pre-fault-tolerant quantum processors as a trustworthy platform for the exploration and discovery of quantum many-body physics.

In Chapter~\ref{chap:gauge_learning}, we dive into the task of noise characterization and expose flaws in previous noise learning strategies.
We overcome these limitations by using a novel noise learning framework that includes different noise sources in a self-consistent way, as attested by a range of experiments. 
Finally, we conclude in Chapter~\ref{chap:conclusion} with a summary and an outlook into future research directions. 

\chapter{Theory of quantum computation}
\label{chap:theory_of_qc}

\summary{
This chapter summarizes fundamental concepts in the field of quantum computing and quantum information that form the mathematical toolbox we will use throughout the thesis.
The presented material is standard textbook knowledge based mainly on Refs.~\cite{nielsen_chuang_2010, hayashi2006quantum, watrous2018theory} and may be skipped by the experienced reader. 

The idealized quantum circuits introduced in Sec.~\ref{sec:how_to_build_QC} consist of state vectors, unitary transformations, and projective measurements in the computational basis.
In a realistic setting, these building blocks are all subject to various imperfections. 
Here, we introduce the mathematical framework to describe quantum systems, their manipulations and classical simulations thereof in a more general context. 
}

\section{Quantum states}
Any quantum mechanical system is described by a \emph{density operator} $\rho$ which is an element of the \emph{state space} $\mathcal{S}(\mathcal{H})$. 
The state space is defined over a \emph{Hilbert space} $\mathcal{H}$, i.e., a complex vector space with a Hermitian inner product. 
For all relevant examples in this thesis, $\mathcal{H}$ has a finite dimension denoted by $d$.
The state space is formally defined as 
\begin{equation}
    \label{eq:definition_state_space}
    \mathcal{S}(\mathcal{H}) \coloneq \{ \rho \in \text{Lin}(\mathcal{H}) | \Tr[\rho] = 1, \rho \geq 0 \}.
\end{equation}
That is, valid quantum states $\rho$ are normalized, positive-semidefinite operators from the set of (bounded) linear operators on $\mathcal{H}$, denoted $\text{Lin}(\mathcal{H})$. 
If a state $\rho$ can be written as $\rho = \ketbra{\psi}{\psi}$ for a unit vector $\ket{\psi} \in \mathcal{H}$, it is referred to as a \emph{pure state} (and usually simply written as $\ket{\psi}$). 
Such pure states are typically sufficient to describe fully isolated quantum systems, such as the idealized states prepared by a quantum circuit with unitary gates.

If a state is not pure, it is referred to as a \emph{mixed state}, and has a (not necessarily unique) representation $\rho = \sum_i^N p_i \ketbra{\psi_i}{\psi_i}$, where the coefficients $p_i$ form a probability distribution. 
General quantum states can thus be interpreted as probabilistic mixtures of pure states. 
Evidently, pure states satisfy $\rho^2 = \rho$ and are hence projectors. 
The \emph{purity} of a state can be quantified through $\Tr[\rho^2]$ with $1/d \leq \Tr[\rho^2] \leq 1$. 
The state $\rho = \mathbbm{1} / d$ that satisfies the lower bound on the purity is called the \emph{maximally-mixed state}. 

\paragraph{Single-qubit states}

\begin{figure}
  \begin{minipage}[c]{0.35\textwidth}
    \includegraphics[width=\textwidth]{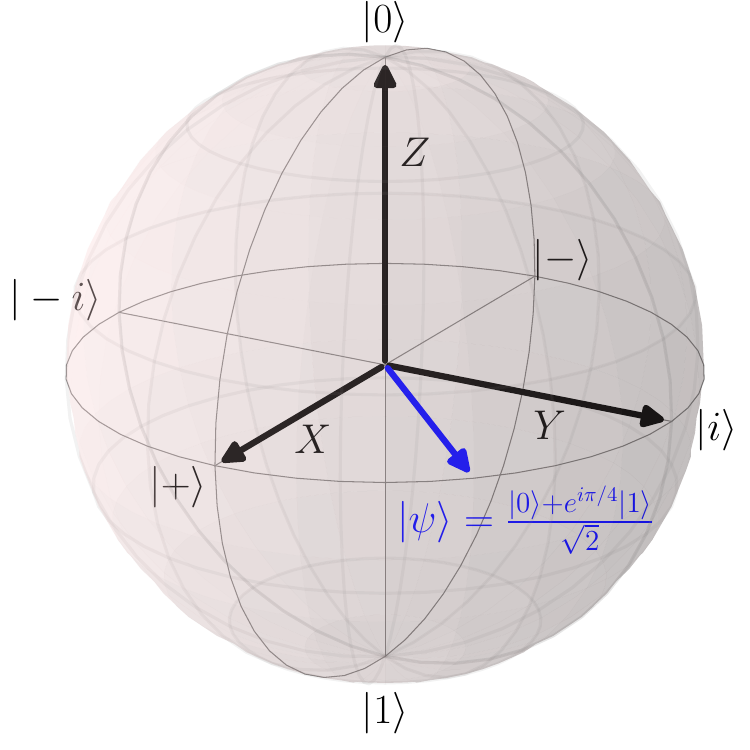}
  \end{minipage}\hfill
  \begin{minipage}[c]{0.63\textwidth}
    \caption[Bloch sphere representation of single-qubit states]{\small 
    The Bloch sphere represents all single-qubit quantum states through a vector $\vec{r}$ as introduced in Eq.~\eqref{eq:bloch_sphere_rep}. 
    The $+1$ ($-1$) Pauli eigenstates of the $Z$, $X$, and $Y$ operatos are labeled $\ket{0}$ ($\ket{1}$), $\ket{+}$ ($\ket{-}$), and $\ket{i}$ ($\ket{-i}$), respectively. 
    The blue arrow shows a pure state as an example. \\
    } \label{fig:bloch_sphere}
  \end{minipage}
\end{figure}

To work with quantum states numerically, they are represented in a particular basis. 
For single-qubit pure states, the standard basis for $\mathcal{H}$ is given by the \emph{computational basis} $\{ \ket{0}, \ket{1} \}$. 
For single-qubit mixed states, a convenient basis of $\mathcal{S}(\mathcal{H})$ is given by the four \emph{Pauli operators}, which are
\begin{equation}
\label{eq:paulis_def}
I = \begin{pmatrix} 1 & 0 \\ 0 & 1 \end{pmatrix}, \quad
X = \begin{pmatrix} 0 & 1 \\ 1 & 0 \end{pmatrix}, \quad
Y = \begin{pmatrix} 0 & -i \\ i & 0 \end{pmatrix}, \quad
Z = \begin{pmatrix} 1 & 0 \\ 0 & -1 \end{pmatrix}, \quad
\end{equation}
when written in the computational basis. 
Due to normalization, any single-qubit state can be written as 
\begin{equation}
\label{eq:bloch_sphere_rep}
\rho = \frac{1}{2} \left( I + r_x \sigma_x + r_y \sigma_y + r_z \sigma_z \right),
\end{equation}
 where $\vec{r} = (r_x, r_y, r_z)^T$ is a vector with $|\vec{r}| \leq 1$.
Thus, the space of single-qubit states can be represented as a unit sphere known as the \emph{Bloch sphere}, as shown in Fig.~\ref{fig:bloch_sphere}.
Pure states correspond to the surface of the sphere while mixed state lie inside the sphere, with the maximally mixed state at the origin. 				
The vector representation of a pure state for some basis is also called the ``state vector'', while the matrix representation of a general state $\rho \in \mathcal{S}(\mathcal{H})$ is called ``density matrix''. 

\paragraph{Composite systems and multi-qubit states}

Two subsystems described by Hilbert spaces $\mathcal{H}_A$ with dimension $d_A$ and $\mathcal{H}_B$ with dimension $d_B$ can be combined to form a composite system described by the tensor product Hilbert space $\mathcal{H}_A \otimes \mathcal{H}_B$ with dimension $d_A \cdot d_B$.
For an $N$-qubit system, the Hilbert space of the full  system thus has dimension $2^N$ with a computational basis written as $\{\ket{\boldsymbol{j}}\} = \{ \ket{0\dots 0}, \ket{0 \dots 1}, \dots, \ket{1 \dots 1} \}$, as introduced in Sec.~\ref{sec:how_to_build_QC}. 
The space of Hermitian operators on an $N$-qubit system is spanned by the $4^N$-dimensional multi-qubit Pauli basis
\begin{equation}
\label{eq:Pauli_basis}
\mathbbm{P}_N \coloneq \{ P_1 \otimes P_2 \otimes \dots \otimes P_N\,  | \, P_j \in \{I, X, Y, Z\} \}. 
\end{equation}
These operators are often referred to as Pauli strings and abbreviated to, e.g., $XZIX$ for a four-qubit term, where the number of non-identity terms is known as the \emph{weight} $k$ of the Pauli operator, e.g., $k=3$ for this example. 
Pauli operators are the primary basis used for observables $O$, which typically admit a representation $O = \sum_i c_i P_i$ with only polynomially many terms.
Moreover, they form a group under multiplication when accounting for additional phases originating, e.g., from $Y = iZX$. 
We formally define the $N$-qubit Pauli group as
\begin{equation}
\label{eq:Pauli_group}
 \mathcal{P}_N \coloneqq \{ i^k P  \, | \, P \in \mathbbm{P}_N, \, k \in \{0, 1, 2, 3\} \}. 
\end{equation}

\paragraph{Entanglement}
A pure state $\ket{\psi}$ of a bipartite system $\mathcal{H}_A \otimes \mathcal{H}_B$ is called \emph{separable} if it can be expressed as a product state such that $\ket{\psi} = \ket{\phi}_A \otimes \ket{\phi}_B$, else it is called \emph{entangled}. 
The amount of entanglement can be quantified by the \emph{von-Neumann entropy} $ S_{\ket{\psi}} \coloneqq-\Tr [ \rho_A \log(\rho_A)]$, where $\rho_A \coloneqq \Tr_B[\ketbra{\psi}{\psi}]$ is the reduced density matrix of the subsystem $A$. 
The maximum amount of entanglement for a two-qubit state is achieved, for example, by the state  $\ket{\text{Bell}} \coloneqq \left( \ket{00} + \ket{11} \right) / \sqrt{2}$ known as a \emph{Bell state}. 
The multi-qubit generalization of this which exhibits maximal multi-partite entanglement  is called a GHZ state and defined as $\ket{\text{GHZ}} \coloneqq \left(	\ket{0}^{\otimes N} + \ket{1}^{\otimes N} \right) / \sqrt{2}$.

Any mixed state $\rho \in \mathcal{S}(\mathcal{H}_S)$  can be obtained from a pure state in a larger space $\ket{\psi} \in  \mathcal{H}_S \otimes \mathcal{H}_E$ such that $\rho = \Tr_E[ \ketbra{\psi}{\psi}] $. 
The state $\ket{\psi}$ is not unique and is called a ``purification'' of $\rho$. 
The dimension of $\mathcal{H}_E$ needs to be at least equal to the rank of $\rho$, so at most match the dimension of $\mathcal{H}_S $. 
The purification picture suggests that any mixed state can be viewed as arising from the entanglement between a system $\mathcal{H}_S$ and its environment $\mathcal{H}_E$ and that the mixture of $\rho$ originates from discarding the degrees of freedom of the environment.

\paragraph{Haar randomness}
In the context of quantum information we are often interested in sampling pure states taken uniformly at random from the underlying Hilbert space.  
Since unitary operations turn pure states into pure states, the associated probability measure should be invariant under the unitary group action of $U(d)$. 
This is satisfied by the \emph{Haar measure} $d\mu_H(u), u \in U(n)$~\cite{mele2024introduction}. 
We can generate so-called Haar-random states by sampling unitaries distributed as $d\mu_H$ and applying them to an arbitrary reference state. 
The induced measure on the space of pure states is denoted as $d\psi$ and (with some abuse of terminology) also often referred to as the Haar measure.   

\section{Quantum channels}
\label{sec:quantum_channels_representations}

Having established the mathematical foundation of the quantum state space, we now classify the possible transformations of states.

\subsection{Quantum gates}
\label{sec:quantum_gates}

\begin{table}
\begin{center}
\includegraphics[width=\textwidth]{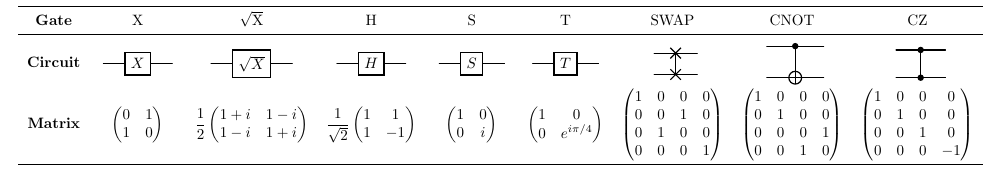}
\caption[Summary of common quantum gates]{
\small Quantum gates with circuit symbols and matrices in the computational basis. 
} 
\label{tab:quantum_gates_summary}
\end{center}
\end{table}
 
In Sec.~\ref{sec:how_to_build_QC}, we have introduced unitary evolutions, usually denoted $U$, which act on a state $\rho$ as $\rho \mapsto \rho^\prime = U \rho U^\dagger$ and map any pure state $\ket{\psi}$ to the pure state $U \ket{\psi}$. 
In circuit-based quantum computing, the unitary of large quantum circuits are built up from individual few-qubit unitaries called ``gates'', see Fig.~\ref{fig:quantum_circ}. 
We now summarize the quantum gates relevant to this thesis. 
We note that any unitary transformation $U$ is generated by a Hermitian operator, i.e., $U = e^{-i O}$ for some $O \in \herm$. 
A \emph{Pauli rotation} is a unitary that is generated as $R_P(\theta) \coloneqq e^{-i P \theta /2}$ where $P$ is a Pauli string and $\theta$ is the rotation angle of the gate.   
For example, a single-qubit rotation $R_Z(\theta) = e^{-i Z \theta /2}$ can be interpreted geometrically as a rotation with angle $\theta$ around the $Z$ axis on the Bloch sphere (similarly for $X$ and $Y$). 

Common single-qubit phase gates are the $S$ gate $S \coloneqq R_Z(\pi / 2)$ and the $T$ gate $T \coloneqq \sqrt{S} = R_Z(\pi / 4)$.
The \emph{Hadamard} gate $H$ performs a basis change between the $X$ eigenbasis and the $Z$ eigenbasis and is given by the unitary $H \coloneqq \exp(-i \pi /2 (X + Z)/\sqrt{2})$. 
Furthermore, note that the Pauli matrices $\{X, Y, Z\}$ themselves are unitary and correspond to $\pi$ rotations along the respective axis such that $R_P(\pi)= -i P$ (equality up to an irrelevant global phase). For example, for a bit-flip gate $R_X(\pi)$ we thus simply write $X$. 
We create entanglement between qubits with two-qubit gates.
In IBM quantum devices, this is commonly done with the controlled-NOT (CNOT) gate which flips the state of the ``target'' qubit based on the state of the state of the ``control'' qubit, i.e., $U_{\text{CNOT}} \coloneqq \ketbra{0}{0}_c \otimes I_t + \ketbra{1}{1}_c \otimes X_t$. 
A single CNOT gate is sufficient to create a Bell pair between two unentangled qubits due to $\ket{\text{Bell}} = U_{\text{CNOT}}  (H \otimes I) \ket{00}$. 
Hence the CNOT gate is a ``maximally-entangling'' two-qubit gate. 
A non-entangling two-qubit gate is the SWAP gate, which acts as $\text{SWAP}(\ket{\psi} \otimes \ket{\phi}) = \ket{\phi} \otimes \ket{\psi}$, i.e., it exchanges the state of two qubits. 
A SWAP gate can be realized with three CNOT gates and is required to ``move around'' qubits on devices with limited qubit connectivity.  
For a summary of the matrix representation and circuit symbols used throughout this thesis see Tab.~\ref{tab:quantum_gates_summary}.

\subsection{Non-unitary channels}
\label{sec:channel_representations}

In order to describe non-unitary physical processes that arise from interactions with the environment it is convenient to introduce the formalism of \emph{quantum channels}, also known as quantum operations or quantum processes.
We define quantum channels as maps that take density operators to density operators.
Hence, we require that the normalization and positivity of the input states are preserved, resulting in \emph{completely-positive, trace-preserving (CPTP) maps}.
A CPTP quantum channel can always be expressed as\footnote{One can also define channels that map states from some Hilbert space $\HS_1$ to states from another space $\HS_2$. In this thesis, we mostly consider channels that act on $N$-qubit states where $\HS_1 = \HS_2$.}
\begin{equation}
    \label{eqn:def_channel_kraus_form}
    \Lambda: \,\mathcal{S}(\mathcal{H}) \rightarrow \mathcal{S}(\mathcal{H})\,\, , \,\, \rho \mapsto \sum_{k=1}^L E_k \rho E_k^\dagger,
\end{equation}
where $\{E_k\}_{k=1,\dots, L}$ are linear operators that satisfy $\sum_{k=1}^L E_k^\dagger E_k = \mathbbm{1}$.
Eq.~\eqref{eqn:def_channel_kraus_form} is known as the \emph{operator-sum} or \emph{Kraus representation} of quantum channels and $E_k$ are called the \emph{Kraus operators}. 
While they are not necessarily unique for a given channel, their number $L$ can always be assumed to be $\leq d^2$. 

The CPTP characterization of quantum channels covers all operations discussed in this thesis. 
However, more general definitions of quantum operations exist that allow for violations of the trace-preserving condition. 
This can occur for example when including dynamical circuit operations like measurement and classical feed-forward in the quantum channel. 
Moreover, some constructions allow for non-completely positive maps that arise in the context of correlations between a system and environment~\cite{pechukas1994reduced}.
Besides the Kraus representation of quantum channels there exist several other representations. 
Below, we introduce the ones that we will frequently make use of throughout this thesis. 

\paragraph{Liouvillian superoperators}
When applying two quantum channels $\Lambda_1$ and $\Lambda_2$ in sequence, it is often of interest to compute the composite channel $\Lambda_2 \circ \Lambda_1$. 
When composing two unitary channels the total unitary is obtained from simply multiplying the two unitary matrices. 
However, for non-unitary operations, the Kraus representation can be inconvenient for representing composite channels. 
This task is more naturally expressed in the \emph{Liouvillian superoperator} representation of quantum channels. 

Let $\rho$ be the $d\times d$ density matrix of a state with respect to some basis $\ket{i}_{i \in\{1,\dots, d\}}$.  
In the Liouvillian formalism, we represent $\rho$ not as a matrix but as a $d^2$-dimensional column vector, where the $d$ rows of $\rho$ are stacked on top of each other. 
This ``vectorized'' state is denoted with the double-ket notation $\kket{\rho}$, see Appendix~\ref{app:sec_double_ket_notation}.
A channel $\Lambda$ can now be represented as a $d^2 \times d^2$ matrix $\mathcal{L}^\Lambda$ known as the Liouvillian superoperator\footnote{The term ``superoperator'' refers to the fact that it acts on \emph{operators} on the underlying Hilbert space.}. 
This lets us compute the output state from a simple matrix-vector multiplication as $\kket{\Lambda(\rho)} = \mathcal{L}^\Lambda\kket{\rho}$. 
The composition of channels is now conveniently expressed as a matrix multiplication $\mathcal{L}^{\Lambda_2 \circ \Lambda_1} = \mathcal{L}^{\Lambda_2} \mathcal{L}^{\Lambda_1}$. 
The Liouvillian superoperator can be computed from the Kraus operators as $\mathcal{L}^\Lambda = \sum_k (E_k^\dagger)^T \otimes E_k$.

\paragraph{Pauli transfer matrix}
Instead of the column-vectorization leading to the Liouvillian superoperater introduced above, a matrix representation of quantum channels can be obtained based on vectorization with respect to any orthonormal basis of $\text{Lin}(\mathcal{H})$. 
For $N$-qubit systems, a particularly convenient choice is given by the $4^N$-dimensional Pauli basis introduced in Eq.~\eqref{eq:Pauli_basis}.
This leads to the \emph{Pauli transfer matrix} (PTM) representation of (multi)-qubit channels. 
The PTM is a $4^N \times 4^N$-dimensional matrix $T^\Lambda$ with entries
\begin{equation}
\label{eq:PTM_def}
T^\Lambda_{a,b} = \Tr[P_a \Lambda(P_b) ]  / 2^N
\end{equation}
Of particular importance are channels with a diagonal PTM which are referred to as \emph{Pauli channels}. 
This is equivalent to a Kraus representation where all Kraus operators are Pauli operators themselves (up to prefactors). 
The diagonal entries of a PTM 
\begin{equation}
\label{eq:Pauli_fidelity_def}
f_i \coloneqq \Tr[ P_i {\Lambda} \left( P_i \right) ] / 2^N.
\end{equation}
are called the \emph{Pauli fidelities} and indicate how much a given Pauli is corrupted by the channel.
This forms the basis of noise learning protocols used for error mitigation, see Sec.~\ref{sec:noise_learning_theory}. 

\paragraph{Stinespring dilation}
Similarly to how any mixed state can be purified on a larger Hilbert space, any channel can be represented by a unitary operation on a larger composite Hilbert space $\HS_S \otimes \HS_\text{E}$ of the system $\HS_S$ and some environment $\HS_\text{E}$. 
This is known as the \emph{Stinespring dilation} of quantum channels. 
The unitary evolution of the full channel reduces to those of the given channel acts when considering only the partial dynamics on the system $\HS_S$ .  
The dimension required for $\HS_\text{E}$ is at most that of the system itself. 
We can thus think of any non-unitary quantum channel as a unitary interaction with an environment. 
The price we pay for not explicitly dealing with the environment's degrees of freedom is the added complexity of CPTP maps over unitary channels.

\section{Time evolution}
\label{sec:time_evolution}
The fundamental equation of motion in (non-relativistic) quantum mechanics is given by the Schr{\"o}dinger equation which describes the time evolution of a pure state $\ket{\psi(t)}$. Changes of the state are generated by a Hamiltonian $H(t)$ as
\begin{equation}
    \label{eq:schrödinger_equation}
    i \hbar \frac{d}{dt} \ket{\psi(t)} = H(t) \ket{\psi(t)}.
\end{equation}
The formal solution to this equation can be expressed by a time evolution operator such that $\ket{\psi(t)} = U(t, t_0) \ket{\psi(t_0)}$ with\footnote{$\mathcal{T}$ is the time-ordering symbol which can be omitted for commuting Hamiltonians.}
\begin{equation}
    \label{eq:time_evolution_operator}
    U(t, t_0) = \mathcal{T} \exp \left( -\frac{i}{\hbar} \int_{t_0}^t H(t) d t\right).
\end{equation}

In the \emph{Schr{\"o}dinger picture} of quantum mechanics, observables -- described by Hermitian operators $O \in \herm$ --, are time-independent, while states evolve according to Eq.~\eqref{eq:schrödinger_equation}.
This is akin to the quantum circuit picture, where the initial state $\ket{\psi_0} = \ket{0}^{\otimes N}$ evolves into a final state $\ket{\psi_f} = U_\text{circ}\ket{\psi_0} $ by applying unitary gates, after which an expectation value $\bra{\psi_f} O \ket{\psi_f}$ is computed. 
In an equivalent framework, known as the \emph{Heisenberg picture}, the time dependence is moved from states to operators which evolve according to
\begin{equation}
    \label{eq:heisenberg_picture_evo}
    O(t) = U(t, t_0)^\dagger O(t_0) U(t, t_0).
\end{equation}
Hence, all expectation values remain unchanged from the Schr{\"o}dinger picture. 
This offers an alternative interpretation of a quantum circuit. 
Instead of applying gates to the qubit state, one can instead ``backpropagate'' the observable through the gates to obtain a final observable $O_f = U_\text{circ}^\dagger O U_\text{circ}$ and subsequently evaluate the expectation value over the (fixed) qubit initial state as $\bra{\psi_0} O_f \ket{\psi_0}$. 
While the Schr{\"o}dinger and Heisenberg pictures are formally equivalent, it can be insightful to  investigate a given quantum circuit from both perspectives. 
This can shed light on, e.g., how different errors propagate through the circuit, and how much individual gates affect a given observable. 
Moreover, the accuracy of classical simulations can heavily depend on the chosen picture, see Chapter~\ref{chap:dual_unitary_TEM}.
Alternatively, the time evolution can be split between states and observables in a way such that states are evolved by certain parts of the Hamiltonian while observables are evolved by the remaining Hamiltonian. 
This is known as the \emph{interaction picture}. 
In this thesis, we employ the interaction picture whenever we model the dynamics of superconducting qubit or qudit systems under the periodic driving of microwave pulses, see Chapters~\ref{chap:superconducting_qubits} and \ref{chap:qudit_processing}.

While the Schr{\"o}dinger equation describes the unitary time evolution of an isolated quantum system, a more general framework can describe a system's interaction with its environment. 
In the context of quantum computing the system is frequently taken to be the qubit register where the interaction with its environment leads to error terms. 
In this thesis, we model non-unitary dynamics under a Markovian noise approximation with a Lindblad master equation
\begin{align}
    \label{eq:lindblad_general}
    \dot{\rho}(t) = -\frac{i}{\hbar} \left[H(t),\rho(t) \right]  + \frac{1}{2} \sum_i  2 L_i \rho(t)  L_i^\dagger - \{  L_i^\dagger L_i, \rho(t) \}
\end{align}
where the first commutator term describes the unitary part of the dynamics and the set of Lindblad jump operators $\{L_i\}$ describe non-unitary, dissipative dynamics. 

\section{Measurements}

\subsection{Positive, operator-valued measures}
\label{sec:measurements_theory}

The probabilistic nature of quantum mechanics manifests in the act of measurements.
Formally, measurements are described by \emph{positive, operator-valued measures} (POVMs). 
Let $\rho_S$ be a state of a system Hilbert space $\HS_S$.
A measurement of $\rho_S$ with $n$ possible outcomes is modeled by an $n$-outcome POVM which is a set of $n$ positive semi-definite Hermitian operators $\vb{M} = \{M_k\}_{k \in \{1, \dots, n \}}$, also called \emph{POVM effects}, that sum to the identity, i.e., $\sum_{k=1}^n M_k = \mathbbm{1}$. 
The probability of observing outcome $k$ is given by Born's rule as $p_k = \Tr[\rho_S M_k]$.

Projective measurements (PMs) are a special case of POVMs, where each POVM operator is a projector such that $M_k = \ketbra{\phi_k}{\phi_k}$ for some set of pure states $\{ \phi_k \}$.
A POVM is said to be \emph{informationally complete} (IC) if it spans the space of Hermitian operators~\cite{d2004informationally}, i.e., any observable $O$ can be decomposed as $O = \sum_k \omega_k M_k$. 
The minimal number of linearly independent POVM elements for an IC-POVM is $n = d^2$. 
We call such POVMs \emph{minimally informationally complete}. 
In that case, the coefficients $\w_k$ are unique. 
However, for POVMs with $n>d^2$, this decomposition is not unique, and the POVM is called \emph{overcomplete}.
This redundancy is described by frame theory, as outlined in Ref.~\cite{innocenti2023shadow} and detailed in Appendix~\ref{app:chap_frame_theory}.

A quantum computer in the standard circuit picture from Fig.~\ref{fig:quantum_circ} a priori only allows for projective measurements in the computational basis. 
A central theme of this thesis is the question of how to implement more complex POVM measurements in such a setting and how to leverage them for quantum algorithms. 
Below, we give a summary of common approaches to realize more complex POVM measurements from underlying projective measurements. 

\paragraph{Dilation POVMs}
Any POVM on $\mathcal{H}_{\text{S}}$ can be implemented in an extended space $\mathcal{H}_{\text{ext}}$ by coupling to additional degrees of freedom either through a tensor product extension (TPE) $\mathcal{H}_{\text{ext}} = \mathcal{H}_{\text{S}} \otimes \mathcal{H}_{\text{A}}$ or a direct sum extension (DSE) $\mathcal{H}_{\text{ext}} = \mathcal{H}_{\text{S}} \oplus \mathcal{H}_{\text{A}}$~\cite{chen2007ancilla}.
To realize POVM measurements, a specific unitary $U$ is applied to $\mathcal{H}_{\text{ext}}$ such that the probability distribution of a subsequent $M$-outcome projective measurement on $\mathcal{H}_{\text{ext}}$ coincides with the POVM outcome distribution $\{p_m\}$ for the original state $\rho_\text{S}$. 
Before applying $U$, the initial state on $\mathcal{H}_{\text{ext}}$ is of the form $\rho = \rho_{\text{S}} \otimes \rho_{\text{A}}$ in a TPE
while in a DSE it has no support on $\mathcal{H}_{\text{A}}$.
In both cases, the existence of $U$ is guaranteed by Naimark's dilation theorem~\cite{gelfand1943imbedding}.
In Appendix~\ref{app:sec_naimark_single_qubit}, we show the explicit construction for the dilation of a four-outcome POVM on a single qubit. 
Similar constructions to these dilation procedures are possible when leveraging mid-circuit measurements and classical feed-forward~\cite{ivashkov2024high}.

\paragraph{PM-simulable POVMs}
While dilation POVMs require additional quantum resources, e.g., in the form of a higher-dimensional space, certain POVMs can be simulated only through projective measurements in the available measurement bases.
This is achieved through convex combinations of POVMs:
For two $n$-outcome POVMs $M^1$ and $M^2$ acting on the same space, their convex combination with elements $M_k = p M^1_{k} + (1-p) M^2_{k}$ for some $p \in [0,1]$ is also a valid POVM. 
This can be achieved in practice by a \emph{randomization of measurements} procedure, which simply consists of the following two steps for each measurement shot.
First, randomly pick $\vb{M}^1$ or $\vb{M}^2$ with probability $p$ or $1-p$, respectively, then perform the measurement associated with the chosen POVM.
We call POVMs that can be achieved by randomizations of projective measurements \emph{PM-simulable}.
We use the notation 
\begin{equation}
\label{eqn:multiset_sum}
 \biguplus_i q_i \vb{M}^i = \left\{ q_i {M^i_k} \right\}_{i,k} 
\end{equation}
to denote a convex combination of POVMs $\vb{M}^{i}$ acting on the same Hilbert space, such that the POVM $\vb{M}^{i}$ is implemented with probability $q_i$.
Importantly, PM-simulable informationally-complete POVMs are overcomplete~\cite{dariano_classical_2005}. 
The decomposition of observables is thus not unique.
In Chapter~\ref{chap:duals_optimization}, we leverage the additional degrees of freedom associated with the decomposition of observables to build better estimators.

\paragraph{Product POVMs}
%\label{sec:single_qubit_POVMs}
The POVM formalism is only practically useful if the POVM operators $M_k$ can be efficiently stored and processed. 
Therefore, most POVM-based quantum algorithms employ local POVMs, where every $M_k$ is a tensor product of operators that each act non-trivially only on a limited number of qubits. 
An important class of such POVMs are \emph{product POVMs} where the POVM effects factor into single-qubit $n$-outcome POVM effects. That is, each global effect can be written as
\begin{equation}
\label{eqn:povm_product_form}
    M_{{k}} =
    M_{k_1,k_2,\dots,k_N} =
    M^{(1)}_{k_1} \otimes M^{(2)}_{k_2} \otimes \cdots \otimes M^{(N)}_{k_N}  
\end{equation}
where $\vb{M}^{(i)} = \{ M^{(i)}_{k_i} \}_{k_i=1}^{n}$ is a $n$-outcome single-qubit POVM acting on qubit $i$. 
In Appendix~\ref{app:povm_classes}, we  summarize different classes of increasingly complex single-qubit IC POVMs that can be used to realize product POVMs. 

\paragraph{SIC POVMs}
A famous class of POVMs are symmetric, minimal IC (SIC) POVMs. 
They consist of $d^2$ rank-one operators $M_i = \frac{1}{d} \ket{\psi_i}\bra{\psi_i}$ with equal pairwise inner products, i.e., $\Tr[M_i, M_k] = |\langle \psi_i | \psi_j\rangle|^2 = \frac{1}{d+1}, \, \text{for all } i \ne j$.
For a single-qubit, it the four states point towards the corners of a regular tetrahedron and are given by $\ket{\psi_0} = \ket{0}$ and $\ket{\psi_m} = \left( \ket{0} + \sqrt{2} e^{2\pi i (m-1)/3} \ket{1} \right)/\sqrt{3}$ with $ m\in \{1, 2, 3 \}$.

\subsection{Quantum detector tomography}
\label{sec:detector_tomography}

In an experiment that performs a quantum measurement, the implemented POVM operators can be characterized through \emph{quantum detector tomography} (QDT) \cite{fiurasek2001maximumlikelihood, dariano2004quantum}.
Similarly to the better known quantum state tomography (QST) and quantum process tomography (QPT), QDT is an important tool to characterize a quantum experiment. 
In QST an unknown state $\rho$ is estimated from measurements of a known set of reference POVM operators $\{M_k^{\text{ref}}\}$. By contrast, in QDT the unknown POVM operators $\vb{M} = \{M_k \}$ are estimated from a known set of prepared reference quantum states $\{ \rho^{\text{ref}}_i \}$. 
As in QST, there is the concept of \emph{informational completeness}: For a full characterization of $\{M_k\}$ through QDT, the  reference states $\{\rho^{\text{ref}}_i\}$ need to span the operator space of $\text{Herm}(\mathcal{H})$~\cite{lundeen2009tomography}. 
One possible set of such states for single-qubit POVMs are the six Pauli eigenstates $\{ \ket{0}, \ket{1}, \ket{+}, \ket{-}, \ket{+i}, \ket{-i} \}$. 
The POVM measurement is carried out on each such reference state, sampling from the probability distributions $p^{(j)}_k = \tr[\rho^{\text{ref}}_j M_k] $.
Let $\mathcal{N}^{(j)}_k$ be the number of times outcome $k$ is recorded for initial state ${\rho_j}$. 
One way to obtain an estimator for the underlying single-qubit POVM operators is to directly invert the system of linear equations
\begin{equation}
\label{eq_theo:POVM_linear_inversion}
\tr[\rho^{\text{ref}}_j M_k] \sim \frac{\mathcal{N}^{(j)}_k}{\sum\limits_{k'} \mathcal{N}^{(j)}_{k'}}
\end{equation}
to obtain the entries of $M_km$. 
This approach suffers from the fact that the obtained POVM operators might be non-physical, as they are not necessarily positive. 
An analogous issue exists for QST through linear inversion of Eq.~\eqref{eq_theo:POVM_linear_inversion}~\cite{neugebauer2020neuralnetwork}. 
Positivity can be enforced with a maximum-likelihood (ML) estimation by maximizing the likelihood functional 
\begin{equation}
\label{eq_theo:detector_tomog_likelihood}
\mathcal{L}\left(\vb{M} \right) = \prod_{m, k} \left(\tr[\rho^{\text{ref}}_j M_k]\right)^{\mathcal{N}^{(j)}_m}
\end{equation}
under the constraint that the operators $M_k$ form a valid POVM~\cite{fiurasek2001maximumlikelihood}. 
As laid out in Ref.~\cite{hradil2004maximumlikelihood}, the optimization can be performed with an iterative algorithm that converges to the ML estimator. 
In this work, we make use of ML quantum detector tomography to reconstruct experimentally implemented POVM operators in Sec.~\ref{sec:ancilla-free-POVMs}.

\section{Distance measures}
\label{sec:distance_measures}
In quantum computing, we often seek a measure for how close an imperfectly prepared state, channel, or POVM is to its ideal theoretical counterpart.
Quantum information theory offers distance measures to evaluate this, which are introduced in the following. 

\paragraph{State fidelity}
The \emph{state fidelity} between two density operators is defined as\footnote{We adopt the definition with a square of the trace. This deviates from some standard sources~\cite{nielsen_chuang_2010}, but has the advantage of relating closely to distance measures of channels~\cite{gilchrist2005distance}.}
\begin{equation}
    \label{eq:state_fidelity}
    F(\rho_1, \rho_2) \coloneq \Tr\left[\sqrt{\sqrt{\rho_1}\rho_2\sqrt{\rho_1}}\right]^2.
\end{equation}
The fidelity is symmetric in its inputs and satisfies $0 \leq F(\rho_1, \rho_2) \leq 1$ with $F(\rho_1, \rho_2) = 1$ if and only if $\rho_1 = \rho_2$. 
Therefore, the \emph{infidelity} $1 - F(\rho_1, \rho_2)$ is a common measure to quantify the mismatch between two states. 
In many situations, one of the input states is a pure state, in which case the expression simplifies to $F(\rho, \ketbra{\psi}{\psi}) = \bra{\psi}\rho\ket{\psi}$. 

\paragraph{Gate fidelity}
The state fidelity can be built upon to define a natural distance measure between channels. 
One possible approach is to take two copies of the same input state and compare the output states of two channels, then averaging over Haar-random input states. 
This leads to the definition of an \emph{average channel fidelity} 
\begin{equation}
        \overline{F}(\Lambda_1, \Lambda_2) \coloneq \int d\psi \, F\left(\Lambda_1(\ketbra{\psi}{\psi}), \Lambda_2(\ketbra{\psi}{\psi})\right).
            % &= \int d\psi \langle\psi|U^\dagger
            %     \Lambda(|\psi\rangle\!\langle\psi|)U|\psi\rangle \\
            % &= \frac{d F_{\text{pro}}(\Lambda, U) + 1}{d + 1}
\end{equation}
If one of the channels is given by a unitary $U$, the definition simplifies to the \emph{average gate fidelity}
\begin{equation}
\label{eq:avg_gate_fidelity}
    \overline{F}(\Lambda, U) \coloneq \int \text{d}\psi \, \bra{\psi}U^\dagger \Lambda(\ketbra{\psi}{\psi})U\ket{\psi}.
        % &= \int d\psi \langle\psi|U^\dagger
        %     \Lambda(|\psi\rangle\!\langle\psi|)U|\psi\rangle \\
        % &= \frac{d F_{\text{pro}}(\Lambda, U) + 1}{d + 1}
\end{equation}
This is the quantity most commonly employed to benchmark the quality of a noisy quantum gate, and often simply referred to as the ``gate fidelity''. 
Notably, there are efficient experimental protocols known as \emph{randomized benchmarking} that measure the average gate fidelity independent of other error sources like state preparation and measurement errors~\cite{helsen2022general}.

\paragraph{Operational distance}
The \emph{operational distance} (OD) is a distance measure between POVM operators~\cite{maciejewski2020mitigation, puchala2018strategies}. 
For two $n$-outcome POVMs $\boldsymbol{\Pi}=\{\Pi_k \}$ and $\boldsymbol{\Sigma}=\{\Sigma_k\}$ the OD is defined as
\begin{align}
\label{eq_theo:Operational_distance_definition}
D_{\text{OD}}(\boldsymbol{\Pi}, \boldsymbol{\Sigma}) \coloneqq \max_{\rho}
\frac{1}{2} \sum_{k=1}^{n} \left| \Tr[\rho \Pi_k] - \Tr[\rho \Sigma_k] \right|.
\end{align}
The OD is thus the worst-case \emph{total variation} between the probability distribution of measurement outcomes obtained with the two POVMs. Importantly, ${0\leq D_{\text{OD}}(\boldsymbol{\Pi}, \boldsymbol{\Sigma}) \leq 1}$ where  $D_{\text{OD}}(\boldsymbol{\Pi}, \boldsymbol{\Sigma})=0$ if and only if the two POVMs coincide. 
The OD can be calculated directly from the POVM operators through
\begin{align}
\label{eq_theo:Operational_distance_calculation}
D_{\text{OD}}(\boldsymbol{\Pi}, \boldsymbol{\Sigma}) = \max_{I'  \subset I } 
\|\sum_{k \in I'} \Pi_k - \Sigma_k \|_{\infty}, 
\end{align}
where $I$ is the set of all outcomes $I={\{1, \dots, n\}}$.

\section{Classical simulations of quantum circuits}
\label{sec:classical_simulation_techniques}
Classical simulation techniques of quantum circuits have become an indispensable tool for exploring quantum algorithms at small scales. 
As quantum devices scale up, they remain important for benchmarking and understanding the limits of current quantum hardware. 
Countless methods have been brought forth which navigate a trade-off between scalability in terms of the circuit volume and the accuracy of the underlying approximations, see. Ref.~\cite{cicero2024simulation} for an overview. 
Here, we summarize the techniques used throughout this thesis. 

\subsection{Exact diagonalization}

The brute-force simulation of a quantum circuit where the full $2^N$-dimensional complex state vector is stored classically and unitary gates are applied as matrices is known as \emph{exact diagonalization}. 
This method is exact by construction but scales exponentially with the system size, which limits its application to about 30 to 40 qubits on typical classical hardware.
Hence, quantum computations beyond this scale are intractable by exact classical methods, a concept recently labelled as \emph{quantum utility}~\cite{kim2023evidence}.
However, there are some circuits where an exact classical simulation can be scaled efficiently. 
This is typically the case in the presence of certain symmetries. 
The most important class of such symmetries are Clifford gates, as detailed in the following.

\subsection{Clifford circuits}
\label{sec:clifford_circuit_theory}

A unitary quantum gate $U$ is called a \emph{Clifford gate} if it maps the Pauli group to itself under conjugation. 
That is, $U P U^\dagger \in  \mathcal{P}_N$ for any $P\in \mathcal{P}_N$. 
Clifford gates form a group under multiplication referred to as the {Clifford group} $\mathcal{C}_N$. 
Important Clifford gates are Pauli rotations with angles that are multiples of $\pi/2$ (such as the $S$ gate), the Hadamard gate $H$, and the CNOT gate. 
In fact, these three gates generate the entire Clifford group. 

According to the Gotteman-Knill theorem, a circuit that only consists of Clifford gates and measurements of Pauli observables can be simulated efficiently in the so-called \emph{stabilizer formalism}. 
A state $\ket{\psi}$ is said to be stabilized by a Pauli operator $P\in \mathcal{P}_N$ if $P\ket{\psi} = \ket{\psi}$. 
Stabilizer states are a special class of quantum states that are defined as the simultaneously stabilized states of an abelian subgroup of $\mathcal{P}_N$. 
This stabilizer group is generated by up to $N$ independent Pauli operators and thus at most $N$ operators are sufficient to uniquely specify a stabilizer state. 
For example, the initial all-zero state $\ket{0\dots 0}$ is specified by the stabilizer generators $\{Z_1, \dots, Z_N \}$. 

By definition, Clifford gates map stabilizer states to stabilizer states. 
The action of a Clifford gate can thus be simulated efficiently by simply conjugating the generators of the stabilizer group with the Clifford unitary, resulting in an updated set of (at most) $N$ generators. 
Finally, the expectation value of any Pauli operator $P \in \mathcal{P}_N$ is $+1$ ($-1$) if $P$ ($-P$) is in the stabilizer group of the final state, else it is $0$.
When including non-unitary channels, the simulation can still be carried out in the same manner if all channels are Pauli channels. 
In this case the expectation values can deviate from $\pm 1$ as they get multiplied by the fidelities $f_i$ of the given Paulis $P_i$ (see Eq.~\eqref{eq:Pauli_fidelity_def}).

The Gottesman-Knill theorem suggests that entanglement alone is not sufficient to achieve computational advantage with quantum circuits. 
 Interestingly, even quantum circuits with only a few non-Clifford gates or small deviations from Clifford unitaries often remain approximately tractable using recently popularized techniques like Clifford perturbation theory~\cite{beguvsic2025simulating} or Pauli path propagation methods~\cite{rudolph2025pauli}. 
These approaches exploit the fact that small amounts of non-Cliffordness, also known as "magic," can be treated as perturbations on an efficiently simulable Clifford backbone. 
In this context, magic refers to the non-stabilizer resources that lie outside the Clifford framework and are essential for universal quantum computation, such as non-stabilizer states or $T$ gates. 
The ability to classically simulate low-magic circuits implies that quantum advantage is not just a matter of circuit volume, but of accumulating sufficient magic to break classical simulability. 
Rigorous understanding and quantification of magic is still actively being researched~\cite{magic2022}.

\subsection{Tensor networks}
\label{sec:tensor_networks_intro}

 \begin{figure}[t]
     \centering
     \includegraphics[width=1\textwidth]{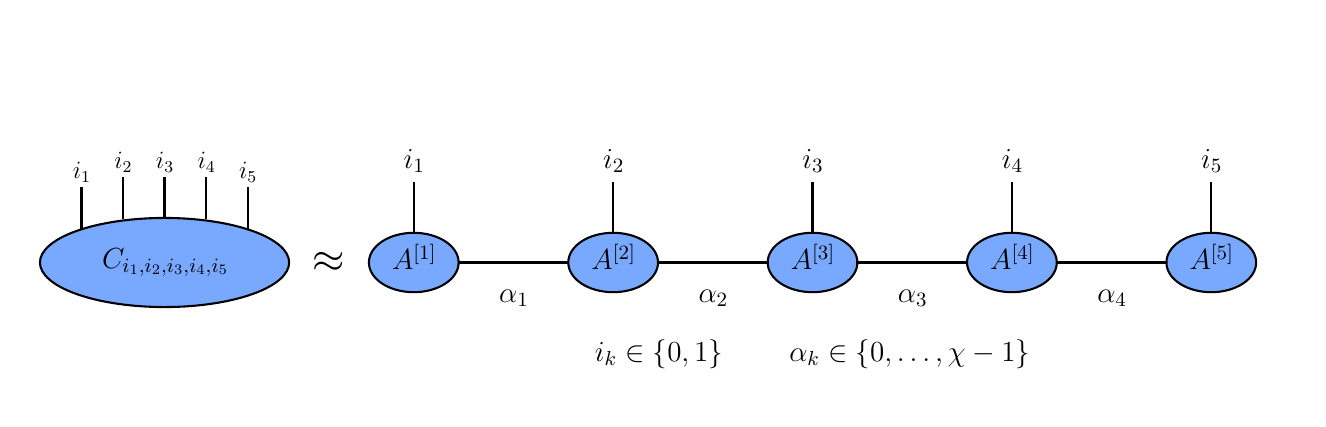}
    \caption[Schematic of matrix product states tensor networks]{\small 
    Schematic of a 5-qubit matrix product state (MPS) that approximates the state vector through a chain of tensors connected through bonds with dimension $\chi$. 
    } \label{fig:MPS_figure}
 \end{figure}

While Clifford-based simulation techniques exploit low magic in quantum circuits, low entanglement is exploited by tensor network techniques which have become a cornerstone of classical simulation in quantum physics, see Ref.~\cite{tensor_networks_review} for a modern review.
Tensor networks exploit the fact that many physically relevant quantum states exhibit limited entanglement and thus admit compressed representations. 
In this context, a tensor is an array of complex numbers of some rank, e.g., a rank-one tensor a vector, a rank-two tensor is a matrix, etc. 
A general $N$-qubit state vector $\ket{\psi} = \sum_{i_1, \dots, i_N = 0}^1 C_{i_1, \dots, i_N} \ket{i_1 \dots i_N}$ is thus represented by a single rank-$N$ tensor $C_{i_1, \dots, i_N}$ with $2^N$ entries. 

A tensor network is a set of tensors where some indices are summed over, or ``contracted'', according to a given pattern.
This can be used to simplify the representation of the state vector.
The most common example is the \emph{matrix product state} (MPS) which uses a one-dimensional chain where each tensor carries a physical qubit index and shares one contracted index with its neighbors, see Fig.~\ref{fig:MPS_figure}, such that
\begin{equation}
\label{eq:MPS_definition}
C_{i_1, \dots, i_N} =  \sum\limits_{\substack{i_1, \dots, i_N \\ = 0}}^1  \sum\limits_{\substack{\alpha_1, \dots, \alpha_{N-1} \\ = 0}}^{\chi - 1} 
A^{[1]}_{i_1, \alpha_1} A^{[2]}_{\alpha_1, i_2, \alpha_2} \cdots 
A^{[n-1]}_{\alpha_{n-2}, i_{n-1}, \alpha_{n-1}} A^{[n]}_{\alpha_{n-1}, i_n} 
|i_1 \cdots i_n\rangle.
\end{equation}
The dimension $\chi$ that the connecting legs $\alpha_i$ carry is known as the \emph{bond dimension}, here assumed to be uniform. 
This is the central quantity that determines the complexity of the MPS, as each tensor carries $2 \chi^2$ parameters (except the ones on the edge for open boundary conditions). 
The bond dimension determines the maximum amount of entanglement that the MPS can capture.
For quantum circuit simulations in the Schrödinger picture, the quantum register starts out in a product state with $\chi = 1$. 
As more and more entangling gates are applied, the bond dimension of the exact MPS representation can in general increase exponentially fast. 
This necessitates truncation of the bonds to some maximal bond dimension $\chi_\text{max}$. 
A common approach is to compute the singular value decomposition across a given bond and only keeping the $\chi_\text{max}$ largest singular values, discarding any remaining non-zero singular values. 

MPS representations can also be used to simulate a quantum circuit in the Heisenberg picture.
In this setting, we start from an observable $O$ at the end of the circuit and, step by step, evolve it by $\mathcal{U}^\dagger = U^\dagger \boldsymbol{\cdot} U$ corresponding to the unitary layers $U$, proceeding backwards in the circuit. 
In the PTM representation (see Sec.~\ref{sec:channel_representations}), the observable takes the form of an MPS with physical dimension 4 (labelling $I, X, Y, Z$) and the superoperators $\mathcal{U}^\dagger$ take the form of \emph{matrix product operators} (MPOs).
An MPO is defined analogously to an MPS, but with two physical legs per site corresponding to the input and output spaces.
The backwards circuit evolution therefore reduces to the sequential application of MPOs to MPS and compression of the resulting MPS if the bond dimension exceeds $\chi_\text{max}$ (see Ref.~\cite{dmrg_2007} for a suitable compression algorithm). 
Finally, the target expectation value is the overlap between the MPS of the evolved operator in the Heisenberg picture, and the PTM representation of the initial state $\ket{0}$, which can be computed from the appropriate tensor contractions.

Tensor network simulations with MPSs and MPOs in both the Schrödinger and Heisenberg pictures as summarized above form the basis of classical simulations of large-scale quantum dynamics simulations presented in Chap.~\ref{chap:dual_unitary_TEM}. 
\chapter{Superconducting quantum hardware}
\label{chap:superconducting_qubits}

\summary{This thesis is centered around the quantum computing architecture of superconducting transmon qubits. 
In this chapter, we review how to perform the fundamental information processing steps -- initialization, single-qubit and two-qubit gates, and measurements -- in transmons. 
Moreover, we discuss the most relevant noise sources present in these devices.
Finally, we provide background information on the devices used for hardware demonstrations in this thesis.
\\
Parts of this chapter are based on Refs.~\cite{fischer2022ancillafree} and~\cite{fischer2023universal}.
}
Superconducting qubits are among the leading platforms for quantum computation. 
They are realized with microelectronic circuits where superconductivity is achieved by cooling to millikelvin temperatures using dilution refrigerators.
Their popularity stems from an appealing combination of scalability, fast gate operations, and compatibility with well-established microfabrication techniques~\cite{krantz_quantum_2019}.

As superconducting circuits offer discrete energy spectra they are often referred to as ``artificial atoms''. 
By coupling them to microwave resonators, one can engineer coherent light-matter interactions, which forms a framework called \emph{circuit quantum electrodynamics} (cQED).
This is a solid-state analogy to the more traditional cavity QED which studies the behavior of individual atoms in optical cavities. 
The strong and controllable interactions with microwave photons and the possibility to perform high-fidelity readout with dispersive couplings renders circuit QED a versatile platform for high-fidelity quantum information processing~\cite{blais2021circuit}. 

There is a broad family of different types of superconducting qubits that  exploit different degrees of freedom in superconducting circuits, see Ref.~\cite{kjaergaard2020superconducting} for a review.
For example, flux qubits encode quantum information in the direction of circulating supercurrents within a loop, generating quantized magnetic flux states~\cite{chiorescu2003coherent}.
Alternatively, phase qubits utilize the phase difference across a current-biased Josephson junction as the computational degree of freedom~\cite{martinis2009superconducting}.
Finally, charge qubits store quantum information in the number of Cooper pairs on a superconducting island and were among the first to demonstrate coherent control~\cite{nakamura1999coherent}. 
However, their sensitivity to charge noise limited their practical use. 
The \emph{transmon qubit} is a variant of the charge qubit which mitigates this sensitivity by operating in a regime where the Josephson energy dominates the charging energy~\cite{koch2007chargeinsensitive}.
While research continues into alternative designs such as fluxonium~\cite{PhysRevLett.129.010502} (capacitively shunted flux qubits), the transmon remains the most mature and scalable architecture. 
In recent years, transmon systems with over 100 qubits have been made available through a cloud access by IBM~\cite{abughanem2025ibm}. 

Accordingly, this thesis focuses on the transmon qubit architecture which is introduced in Sec.~\ref{sec:transmon_qubits}.
Next, we describe the basic building blocks of quantum computation with transmons. 
Single-qubit gates are covered in Sec.~\ref{sec:single-qubit_gates}, while entangling gates are discussed in Sec.~\ref{sec:superconducting_two_qubit_gates} focussing on the cross-resonance gate. 
We further cover measurements in Sec.~\ref{sec:transmons_measurements} and common noise sources in Sec.~\ref{sec:common_noise_sources}. 
Finally, we provide details on the IBM Quantum devices used throughout this work in Sec.~\ref{sec:IBM_Quantum_devices}. 

\section{The transmon circuit}
\label{sec:transmon_qubits}

\begin{figure}
  \begin{minipage}[c]{0.64\textwidth}
    \includegraphics[width=\textwidth]{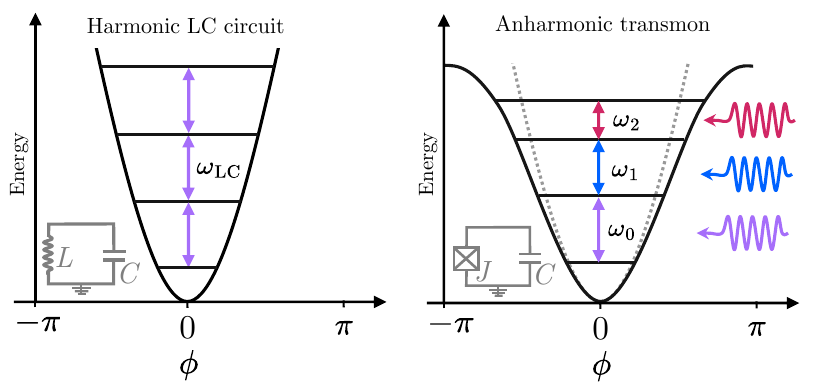}
  \end{minipage}\hfill
  \begin{minipage}[c]{0.34\textwidth}
    \caption[The transmon circuit as an anharmonic oscillator]{\small Schematic of low-lying energy spectra of a harmonic LC circuit (left) and the transmon circuit (right). The non-linear circuit element of the Josephson junction creates an anharmonic spectrum that allows for selected driving of transitions.  
    } \label{fig:transmon_schema}
  \end{minipage}
\end{figure}

The transmon is a superconducting circuit element that can be described as a weakly anharmonic oscillator. 
To understand the design principles behind it, let us briefly review the physics of a simple harmonic LC resonant circuit.
This well-known textbook circuit features an inductance $L$ with flux $\Phi$ and a capacitance $C$ that carries a charge $Q$, see Fig.~\ref{fig:transmon_schema}. 
The classical Hamiltonian of this system is simply given by the sum of the magnetic energy of the inductor and the electrical energy of the capacitor as $H = {Q^2}/{2C} + {\Phi^2}/{2L}$. 
We recognize a harmonic oscillator where, upon canonical quantization, the flux and charge form conjugate variables with a canonical commutation relation of $[ \hat{\Phi}, \hat{Q}] = i \hbar$.
In the context of superconducting circuits, it is convenient to work with dimensionless operators $\hat{n} \coloneqq {\hat{Q}}/{2e}$ and $\hat{\phi} \coloneqq {2 \pi \Phi}/{\Phi_0 }$ where $2e$ is the charge of a Cooper pair and $\Phi_0 = {h}/{2e}$ is the associated magnetic flux quantum. 
These denote the number of Cooper pairs on the capacitor and the superconducting phase, respectively.
With the charging energy $E_\text{C} \coloneqq {e^2}/{2C}$ and the the inductive energy $E_\text{J} \coloneqq {\Phi_0^2}/{4\pi^2 L}$, the Hamiltonian becomes $ H_\text{LC} = 4 E_C \hat{n}^2 + E_L \hat{\phi}^2/2$.
The low-energy spectrum of this Hamiltonian consists of evenly spaced states with energy difference $\hbar \omega_\text{LC}$ with $\omega_\text{LC} \coloneqq \sqrt{8 E_\text{L} E_\text{C}}$.

Encoding a qubit into the discrete spectrum of a superconducting circuit requires the ability to address individual states. 
However, in a harmonic spectrum, a coherent drive of frequency $\omega_\text{LC}$ would couple all states of the ladder, rendering the LC circuit impractical for information processing. 
Instead, we require an anharmonic spectrum to address transitions individually.
This can be achieved by replacing the linear inductance of the circuit with a non-linear Josephson junction.
The current through the junction is given by the DC Josephson effect as $I(\Phi) = I_0 \sin(\Phi / \Phi_0)$, where $I_0$ is the critical current of the junction. 
Computing the energy stored in the junction as $E(\Phi) = \int_0^\Phi I(\Phi) d\Phi$ we arrive at the new Hamiltonian
\begin{align}
\label{eq:transmon_hamiltonian}
\hat H_{\text{TM}} \coloneqq 4E_\text{C}\left( \hat{n} - n_g \right)^2 - E_\text{J} \cos(\hat{\phi}).
\end{align}
The charging energy $E_C$ now includes the capacitance from both the original ``shunt'' capacitor and the self-capacitance of the junction and $E_\text{J} = \frac{I_0 \Phi_0}{2\pi} $ is the Josephson energy.
We have also included a constant $n_g$ known as the \emph{offset charge} that results from capacitive coupling of undesired voltage sources due to imperfect isolation from the environment. 

\begin{figure}
  \begin{minipage}[c]{0.59\textwidth}
    \includegraphics[width=\textwidth]{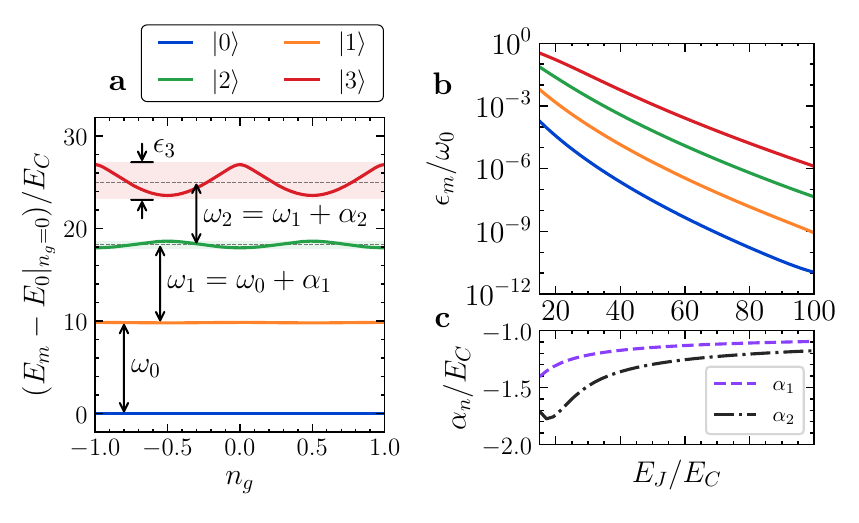}
  \end{minipage}\hfill
  \begin{minipage}[c]{0.39\textwidth}
    \caption[The transmon spectrum and charge noise]{\small Properties of the lowest energy eigenstates $\ket{m}$ of a transmon obtained from numerical diagonalization of the Hamiltonian in Eq.~\eqref{eq:transmon_hamiltonian}. 
\textbf{a)} Fluctuations of the eigenenergies with the offset charge $n_g$ at $E_\text{J}/E_\text{C} = 15$, which become exponentially stronger for higher levels. 
\textbf{b)} Reduction of the charge dispersion $\epsilon_n$ with increasing $E_\text{J}/E_\text{C}$.
{\textbf{c)} Anharmonicities} $\alpha_n$ as a function of $E_\text{J}/E_\text{C}$.
    } \label{fig:transmon_properties}
  \end{minipage}
\end{figure}

Let us now study the spectrum of the transmon Hamiltonian.
We denote the Hamiltonian in its eigenbasis by $\hat H_{\text{TM}} = \sum_{n} E_n \ket{n}\bra{n}$, such that $\ket{0}$ is the ground state.
Through the expansion $\cos{{\phi}} = 1 - \frac{{\phi}^2}{2} + \frac{{\phi}^4}{24} - \dots$, we see that, for small ${\phi}$, the transmon resembles a harmonic oscillator. 
However, the higher powers of ${\phi}$ create an anharmonic spectrum where the spacing of the eigenenergies is not equidistant, but decreases with higher levels, see Fig.~\ref{fig:transmon_schema}. 
This motivates us to encode a qubit into the two energetically lowest states $\ket{0}$ and $\ket{1}$ as the computational subspace.
With the excitation energies\footnote{For the remainder of this chapter we will set $\hbar = 1$ for ease of notation.} $\omega_n \coloneqq E_{n+1} - E_n$ between adjacent levels, we define the \emph{anharmonicities} $\alpha_n \coloneqq \omega_n - \omega_{n-1}, \,\, n \geq 1$, as the difference in adjacent transition frequencies. 

The characteristics of the Hamiltonian in Eq.~\eqref{eq:transmon_hamiltonian} are governed by the ratio of $E_\text{J}$ to $E_\text{C}$. 
First qubits were realized in a regime with $E_\text{C} \gg E_\text{J}$ in a design known as the Cooper-pair box~\cite{shnirman1997quantum, nakamura1999coherent}. 
In this regime, qubit eigenstates essentially correspond to states with fixed charge number. 
However, these eigenenergies $E_n$ have a strong periodic dependence on the offset charge $n_g$.
This is quantified by the \emph{charge dispersion}, i.e., the maximal difference in eigenenergies of
$\epsilon_n = \left|E_n(n_g=0) - E_n(n_g = 1/2) \right|$.
In a realistic experimental setting, $n_g$ is subject to fluctuations which thus leads to variations in the qubit transition frequency $\omega_0$, creating phase errors~\cite{schreier2008suppressing}.
Over time, qubit designs have therefore moved to values with larger $E_\text{J} \gg E_\text{C}$ where charge dispersion is suppressed. 
This is known as the \emph{transmon} regime. 
However, increasing $E_\text{J} / E_\text{C}$ comes with a trade-off; as the ratio increases, the absolute value of the anharmonicity $\alpha_0$ is reduced, which complicates driving the individual transitions due an increased risk of leakage into adjacent levels and phase errors. 

We illustrate the interplay of $E_\text{J} / E_\text{C}$, the charge dispersion $\epsilon_n$, and anharmonicities by numerically diagonalizing the spectrum of $H_{\text{TM}}$. 
For this, we chose the charge representation and expand the Hamiltonian with a Fourier series in the superconducting phase $\hat{\phi}$ following Ref.~\cite{gambetta2013quantum}. 
We truncate to 20 Fourier modes and diagonalize the resulting matrix to obtain the low-energy spectrum.   
Fig.~\ref{fig:transmon_properties}\textbf{a} shows the spectrum as a function of $n_g$ which show essentially flat bands for the lowest two energies but exponentially increasing charge dispersion for higher-excited states. 
The transmon relies on the fact that the charge dispersion decreases exponentially with $E_\text{J}/E_\text{C}$ (see Fig.~\ref{fig:transmon_properties}\textbf{b}) while the anharmonicity is only reduced with a weak power-law (see Fig.~\ref{fig:transmon_properties}\textbf{c}), making control at high $E_\text{J}/E_\text{C}$ favorable~\cite{koch2007chargeinsensitive}. 
The devices relevant to this thesis use frequencies around $\omega_0/(2\pi) \sim 5 \, \text{GHz}$, anharmonicities of $\alpha_1/(2\pi) \sim 300\,\text{MHz}$ and $E_{\mathrm{J}}/E_{\mathrm{C}} \sim 35\,\text{--}\,45$.

\section{Fundamental operations}

\subsection{Single-qubit gates}
\label{sec:single-qubit_gates}

We now discuss how to drive single-qubit unitaries in the subspace of the states $\ket{0}$ and $\ket{1}$.
Let $H_0^{\text{lf}} = \sum_{n=0}^{1} E_n \ketbra{n}{n} = {\omega_0}Z /2$ denote the qubit Hamiltonian in its eigenbasis in the laboratory frame (lf).
An external microwave drive 
with envelope $\Omega(t)$, drive frequency $\omega_D$, and phase $\delta$ leads to an interaction Hamiltonian (see, e.g., Ref.~\cite{gambetta2013quantum} and Chap.~\ref{chap:qudit_processing})
\begin{align}
\label{eq_theo:interaction_Ham_lab_frame}
  H_{\text{int}}^{\text{lf}}(t) \coloneqq \Omega(t)\cos\left(\omega_D t - \delta \right) X.
\end{align}
It is convenient to transform into the rotating frame (rf) of the drive, i.e., an interaction picture with the transformation term $R(t) = e^{- i \omega_D  t Z / 2}$, see Appendix~\ref{app:rotating_frame}.
We perform the full calculation in Chap.~\ref{chap:qudit_processing} for the more general case of a $d$-dimensional qudit. 
Here, we only quote the well-known results for the qubit case $d=2$ in the rotating wave approximation, where the qubit Hamiltonian becomes
\begin{align}
\label{eq:qubit_interaction_SQ}
    H^{\text{rf}} \coloneqq \frac{\omega_0 - \omega_D}{2} Z + \frac{\Omega(t)}{2} \left( \cos(\delta) X + \sin(\delta) Y \right)
\end{align}
Hence, resonant driving with $\omega_D = \omega_0$ results in rotations around an axis in the $x$/$y$ plane with a polar angle given by the driving phase $\delta$. 
The rotation angle when driving for a total time $T$ is given by $\int_0^T \Omega(t) dt$, which corresponds to the area under the pulse envelope. 
We further see that unintended, non-zero detuning $\Delta \coloneqq \omega_0 - \omega_D$ leads to phase errors. 

$Z$-rotations could be implemented by carefully engineering the detuning or by combining $X$- and $Y$-rotations.
However, there is a more elegant and accurate way of implementing effective $Z$-rotations known as \emph{virtual $Z$ gates}~\cite{mckay2017efficient}. 
Instead of applying physical pulses, $Z$-gates can be implemented by appropriately shifting the phase of all subsequent operations.
This is accomplished by simply adjusting the reference frame of the control electronics in software to account for the rotated angle of the $Z$-rotation. 
Due to their instantaneous nature, virtual $Z$ gates can be considered error-free (only limited by the numerical precision of the control electronics). 
We will also generalize the concept of virtual phase gates to qudits in Chap.~\ref{chap:qudit_processing}.

Let us now discuss how to implement arbitrary single-qubit unitaries  in practice.
Although the effective interaction Hamiltonian in Eq.~\eqref{eq:qubit_interaction_SQ} provides direct $X$-, $Y$-, and $Z$-couplings with tunable rotation angles, it is preferable to run physical pulses only for a small set of fixed angles to keep the calibration overhead minimal.
Hence, it is convenient to make use of a decomposition that moves any angular dependence into virtual $Z$ gates.
Any $U \in \text{SU(2)}$ can be represented (up to global phase) by three real parameters through Euler angles as 
$U(\theta_1, \theta_2, \theta_3)  = R_z(\theta_3) \times R_x(\theta_2) \times R_z(\theta_1)$.
The $x$-rotation is commonly further decomposed into fixed-angle $\sqrt{X} \coloneqq  R_x(\frac{\pi}{2})$ gates such that~\cite{mckay2017efficient}
\begin{align}
    \label{eq:decomp_RZ_SX}
    U(\theta_1, \theta_2, \theta_3) = R_z(\theta_3 - \frac{\pi}{2}) \times \sqrt{X} \times R_z(\pi - \theta_2) \times \sqrt{X}\times R_z({\theta_1 - \frac{\pi}{2}}).
\end{align}
With this decomposition, $\sqrt{X}$ gates are the only physical pulse that needs to be calibrated well to implement arbitrary high-fidelity single-qubit gates.  

\subsection{Two-qubit gates}
\label{sec:superconducting_two_qubit_gates}

Entagling two qubits requires two-qubit gates such as the CNOT gate, see Sec.~\ref{sec:quantum_gates}. 
In the devices most relevant to this thesis, CNOT gates are realized through a qubit-qubit interaction known as the cross-resonance (CR) effect~\cite{rigetti2010fully}. 
The CR effect occurs between two transmons with a weak capacitive coupling mediated by a common resonator. 
Here, one transmon, referred to as the \emph{control}, is driven at the $\ket{0}_t \leftrightarrow \ket{1}_t$ qubit frequency $\omega_t$ of the second transmon, referred to as the \emph{target}.
This leads to a rotation of the target qubit that depends on the state of the control qubit. 

More specifically, a cross-resonance tone entangles the two qubit systems through a complicated interaction dominated by a $Z_c\otimes X_t$ generator~\cite{magesan2020effective}. 
When tuning this rotation to $R_{ZX}(\pi/2) = \exp(-i\tfrac{\pi}{4} Z \otimes X )$, the CR gate is equivalent to a CNOT up to local Clifford gates. 
Analytical studies of CR tones based on perturbation theory show that the effective two-qubit interaction Hamiltonian contains various single-qubit terms ($I\otimes X$, $I\otimes Z$, $Z\otimes I$), as well as a weak $Z\otimes Z$ term~\cite{tripathi2019operation, malekakhlagh2020firstprinciples}. 
A popular approach to largely cancel these unwanted terms employs an echoed pulse sequence, see Fig.~\ref{fig:ECR_pulse_overview}\textbf{a}.
In the echoed cross-resonance (ECR) gate the effects of the $Z\otimes I$, $Z\otimes Z$, and $I \otimes X$ destructively interfere, thus isolating the desired $Z\otimes X$ generator~\cite{sheldon2016procedure}.
The echo sequence can further be improved with resonant rotary pulses on the target qubit~\cite{sundaresan2020reducing}.
We study the cross-resonance effect and its extension to higher-dimensional qudit subspaces in Sec.~\ref{sec:two_qudit_operations}.

The CR gate has become popular due to its high achievable fidelities and relative simplicity, requiring only microwave pulses and no additional flux-bias lines~\cite{rigetti2010fully, patterson2019calibration}. 
The qubits can be kept at fixed frequencies and the coupling can remain "always-on" if crosstalk is mitigated by ensuring that all relevant qubit frequencies are sufficiently detuned from one another.
However, additional functionality such as frequency-tunable qubits or tunable couplers enable other types of entangling gates. 
In architectures with frequency-tunable transmons, two-qubit gates are typically implemented by dynamically adjusting the qubit frequencies to activate a resonant or near-resonant interaction. 
For example, moving one qubit into resonance with another induces a conditional phase shift~\cite{dicarlo2009demonstration}. 
Alternatively, parametric modulation techniques can be used to activate iSWAP-like interactions by modulating the qubit or coupler at the frequency difference between the qubits~\cite{Google_2q_gates}.
In systems with tunable couplers (but fixed-frequency qubits) the interaction strength can be dynamically controlled by adjusting the coupler’s flux bias, allowing high-fidelity gates with minimal residual coupling when idle~\cite{IBM_tunable_coupler}. 

\subsection{Measurement}
\label{sec:transmons_measurements}

The state of a transmon is read out by using a dispersive coupling to a microwave resonator mode~\cite{wallraff2005approaching}. 
Every qubit is capacitively coupled to a resonator that hosts a mode with a frequency $\omega_r$ that is off-resonant to the qubit frequency. 
In IBM Quantum devices, typical values are $\omega_r/(2\pi) = 6 \,\text{--}\, 7 \, \text{GHz}$, compared to $\sim 5\,\text{GHz}$ qubit frequencies. 
In the dispersive coupling regime, the frequency of the resonator mode is shifted by $\pm \chi$ depending on the state of the qubit ($-\chi$ for $\ket{0}$ and $+\chi$ for $\ket{1}$). 
This can be thought of as weakly-entangled dressed states between the resonator mode and the transmon. 

\begin{figure}
  \begin{minipage}[c]{0.5\textwidth}
    \includegraphics[width=\textwidth]{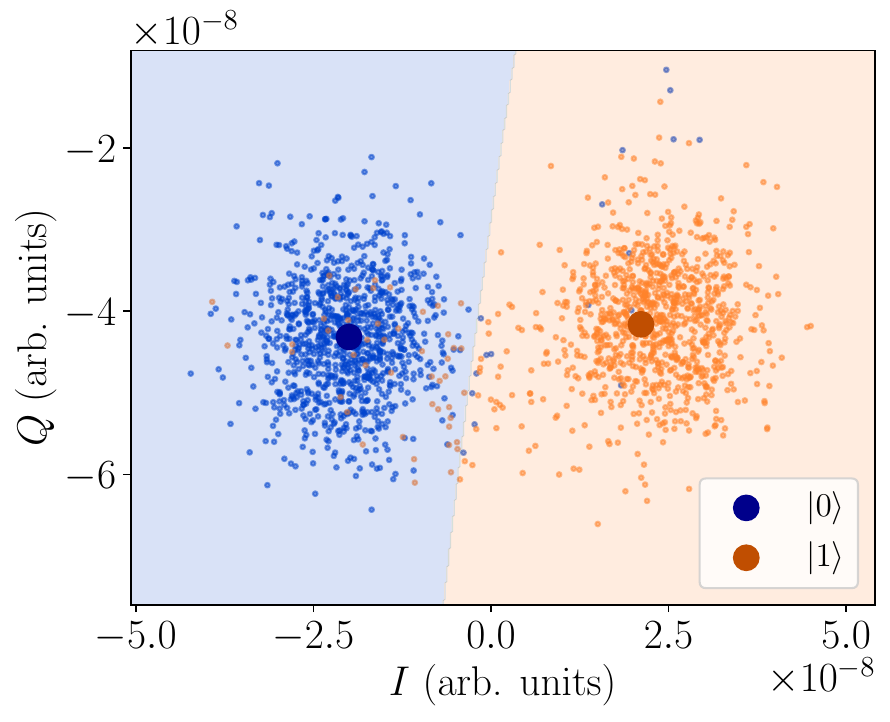}
  \end{minipage}\hfill
  \begin{minipage}[c]{0.48\textwidth}
    \caption[IQ-plane of a dispersive readout]{\small Measurement outcomes of a dispersive readout of a superconducting qubit. Blue dots represent measured signals from a prepare-$\ket{0}$ circuit, while orange dots are measured on prepare-$\ket{1}$ circuits. Large dots represent the mean of the respective distributions.
    } \label{fig:IQ_plane_readout_example}
  \end{minipage}
\end{figure}

In practice, a probe tone is sent into the resonator at a frequency around $\omega_r$, which leads to a state-dependent response. 
As the resonator's frequency is shifted by the qubit, the probe tone is off-resonant by $\pm \chi$. 
This causes a state-dependent phase shift in the reflected signal. 
This signal is downconverted and its in-phase and quadrature components $I$ and $Q$ are collected as integrated values over some acquisition time (typically of $\mathcal{O}(100\,\text{ns}) - \mathcal{O}(1\,\upmu\text{s})$), resulting in a single complex number for each measurement shot~\cite{bronn2017fast}. 
An example measurement of a qubit on the device \textit{ibmq\_manhattan} is shown in Fig.~\ref{fig:IQ_plane_readout_example}. 
The signal upon preparing the state in $\ket{0}$ clusters around one region of the IQ-plane, while the signals collected from the state $\ket{1}$ cluster in a well-separated region. 
This raw data is used to train a classifier which assigns the final value of 0 or 1. 

While the raw IQ value is usually discarded, this information can be used for more sophisticated measurement processing. 
For example, the confidence with which one assigns 0 and 1 can be used to inform better decoders of syndromes in error correction~\cite{hanisch2024soft}. 
Or, the presence of a third, distinct region can indicate leakage into higher-excited states (see also Chap.~\ref{chap:qudit_processing} for a discussion of dispersive readout in the qudit context). 

The last 20 years saw continuous further improvement in dispersive readout with the goals of decreasing readout misassignments and reducing the required acquisition times, see Ref.~\cite{dispersive_readout_review} for a recent overview.
For example, parametric amplifiers improve the collected signal-to-noise~\cite{macklin2015near} and Purcell filters~\cite{purcell_filter_ref} prevent qubit decay into the resonator mode. 
As the number of qubits on a chip increases, it becomes infeasible to operate a single readout transmission line for every qubit. 
Instead, multiple qubits can be read out simultaneously with a shared transmission line through multiplexed readout~\cite{multiplexed_readout}. 
This employs a multi-tone probe pulse that contains the different readout resonator frequencies (which must be sufficiently detuned), and is subsequently demodulated to extract the IQ signal for each qubit. 

\subsection{Qubit reset}

A quantum computation typically starts by initializing all qubits in the $\ket{0}$ state. 
One way to achieve this is to simply wait until the state thermalizes. 
This creates a thermal state where the fidelity of the reset operation depends on the residual population of the $\ket{1}$ state.
As coherence times of qubits improve, this becomes impractical due to long waiting times, see Sec.~\ref{sec:amplitude_damping}. 
As a workaround, specifically tailored pulse sequences may be used that transfer the population to a lossy resonator~\cite{undonditional_Reset_ZRL}. 
Alternatively, one may use a conditional reset, where a qubit is measured and an $X$ gate is applied in case the outcome was $\ket{1}$. 
While other ``passive'' reset techniques exist, the experiments reported throughout this thesis use multiple conditional resets in quick succession. 
This directly links readout fidelities to state preparation fidelities.
However, the connection between those two is more fundamental and further explored in Chap.~\ref{chap:gauge_learning}.

\section{Decoherence and errror sources}
\label{sec:common_noise_sources}

In practice, the performance of superconducting qubits is impacted by various noise sources that can limit coherence and gate fidelities. 
Here, we summarize different noise mechanism and their theoretical description. 

\subsection{Amplitude damping}
\label{sec:amplitude_damping}
Qubits prepared in the excited $\ket{1}$ states eventually decay into the ground state $\ket{0}$ due to energy dissipating into the environment. 
Formally, this can be modelled by a quantum channel known as \emph{(generalized) amplitude damping} with single-qubit Kraus operators (see Sec.~\ref{sec:channel_representations}) 
\begin{align}
\label{eqn:amplitude_damping}
E_1 &= \sqrt{p_0} \begin{pmatrix} 1 & 0 \\ 0 & \sqrt{e^{-t/ T_1 }} \end{pmatrix}, 
&&E_2 = \sqrt{p_0} \begin{pmatrix} 0 & \sqrt{1 - e^{-t/ T_1 }} \\ 0 & 0 \end{pmatrix}, \\
E_3 &= \sqrt{1 - p_0} \begin{pmatrix} \sqrt{e^{-t/ T_1 }} & 0 \\ 0 & 1 \end{pmatrix}, 
&&E_4 = \sqrt{1 - p_0} \begin{pmatrix} 0 & 0 \\ \sqrt{1 - e^{-t/ T_1 }} & 0 \end{pmatrix}.
\end{align}
Here, $t$ represents time and $T_1$ is the time constant of an exponential decay of the population of the $\ket{1}$ state. 
For $t\to\infty$, this channel produces a thermal state with a population given by the Maxwell-Boltzmann distribution as $p_0 = 1 / \left(1 + \exp(-\frac{\omega_0}{k_b T})\right)$. 
The $T_1$ time is also referred to as the ``lifetime'' of a qubit and has become a standard figure of merit in the benchmarking of quantum computers. 
In can be measured simply by preparing a qubit in the $\ket{1}$ state and fitting an exponential decay to that state's population in time. 

\subsection{Phase damping}
While $T_1$ characterizes the lifetime of an excitation to the $\ket{1}$ state, the $T_2$ time relates to the rate of decoherence of a relative phase between $\ket{0}$ and $\ket{1}$. 
Pure dephasing (dephasing without energy loss) is described by a time-dependent channel with Kraus operators 
\begin{align}
\label{eqn:phase_damping}
E_1 &= \sqrt{e^{-t/T_\phi}} I\quad, \quad E_2 = \sqrt{1 - e^{-t/T_\phi}} Z, 
\end{align}
which is equivalent to a Pauli error channel with $Z$ errors where the survival probability decays exponentially with a time constant $T_\phi$~\cite{nielsen2010quantum}.
The pure dephasing time $T_\phi$ is not directly accessible in experiments as $T_1$ decay also comes with an inherent loss of phase coherence. 
In practice, the dephasing time is measured, for example, with a spin echo sequence. 
Here, a qubit is put into an equal superposition $\ket{+} = (\ket{0} + \ket{1}) / \sqrt{2}$ (e.g. with a $H$ gate) and evolves for a time $\tau$, accumulating phase errors (while also subject to amplitude damping). 
Then, an $X$ gate is applied to flip the qubit state before another free evolution of duration $\tau$, where the direction of phase accumulation is now inverted. 
Finally, another $H$ gate shifts back to the computational basis and the population $p_0(t)$ is measured. 
Repeating this experiment with different evolution times leads to an exponential decay with a time constant $T_2$ that satisfies
\begin{equation}
\label{eq:T2_definition}
\frac{1}{T_2} = \frac{1}{2T_1} + \frac{1}{T_\phi}. 
\end{equation}
The dephasing time is thus upper bounded by $2 T_1$ in case there is no intrinsic, pure dephasing.

\subsection{Crosstalk} 
In a multi-qubit processor, unintended interactions between qubits or control lines can lead to crosstalk. 
In particular, during two-qubit gates such as the cross-resonance gate discussed above, microwave drives intended for one qubit may couple to neighboring qubits, inducing spurious coherent interactions~\cite{sheldon2016procedure}.
Crosstalk is exacerbated when there are so-called ``frequency collisions'' between neighboring qubits~\cite{qubit_frequency_design}. 
For example, this is the case when the base frequency of one qubit is close to the frequency of a (next-to) nearest-neighbor qubit.
Other types of frequency collisions can occur between qubits and readout resonators or different readout resonators (particularly in the case of multiplexed readout), or even between higher-level transitions of qubits.   
Crosstalk between qubits is particularly significant in architectures where the coupling between neighboring qubits is ``always-on'', i.e., not tunable.
Strategies to avoid crosstalk thus include careful frequency allocations during fabrication~\cite{qubit_frequency_design}, active cancellation pulses~\cite{vinay_crosstalk_DD}, and hardware-level isolation techniques~\cite{kono2020breaking}.

\subsection{Leakage}
Leakage refers to undesired population outside of the qubit subspace, most notably the $\ket{2}$ state. 
This presents a fundamental challenge for both error-mitigated and error-corrected quantum hardware, as many strategies to combat noise assume qubit-space noise channels such as Pauli channels~\cite{miao2022overcoming}. 
Several physical and control-related processes can lead to leakage.
First, leakage can be caused by off-resonant excitations from $0 \leftrightarrow 1$ pulses with a broad spectral range. 
Even though the $1 \leftrightarrow 2$ transition is detuned from the intended qubit transition, insufficient anharmonicity or short pulses (which are wider in frequency space) can cause residual leakage. 
This is commonly addressed by pulse shaping techniques such as \emph{Derivative Removal by Adiabatic Gate} (DRAG)~\cite{motzoi2009simple}, which can suppress leakage by orders of magnitude compared to naive pulses (e.g. Gaussian) of the same duration~\cite{gambetta2011analytic}. 
Nonetheless, this highlights a fundamental tradeoff in the design of transmons:  Increasing $E_J / E_C$ improves coherence but reduces anharmonicity, requiring slower gates to avoid leakage.
However, slower gates are subject to more dephasing and amplitude damping noise. 

Other potential sources of leakage include measurement-induced leakage, which  may occur due to measurement backaction, or breakdown of the dispersive approximation, caused by, e.g., too strong measurement pulses~\cite{readout_induced_leakage}. 
In many-qubit systems, a prominent source of leakage is given by frequency collisions between the $\ket{1}-\ket{2}$ transition and the $\ket{0}-\ket{1}$ transition of a neighboring qubit. 
Avoiding leakage thus calls for a careful design of qubit parameters and control pulses. 
If this is insufficient, active leakage reduction techniques can be used to depopulated excited levels~\cite{mcewen2021removing}.

\section{IBM Quantum devices and software stack}
\label{sec:IBM_Quantum_devices}

\begin{figure}
  \begin{minipage}[c]{0.56\textwidth}
    \includegraphics[width=\textwidth]{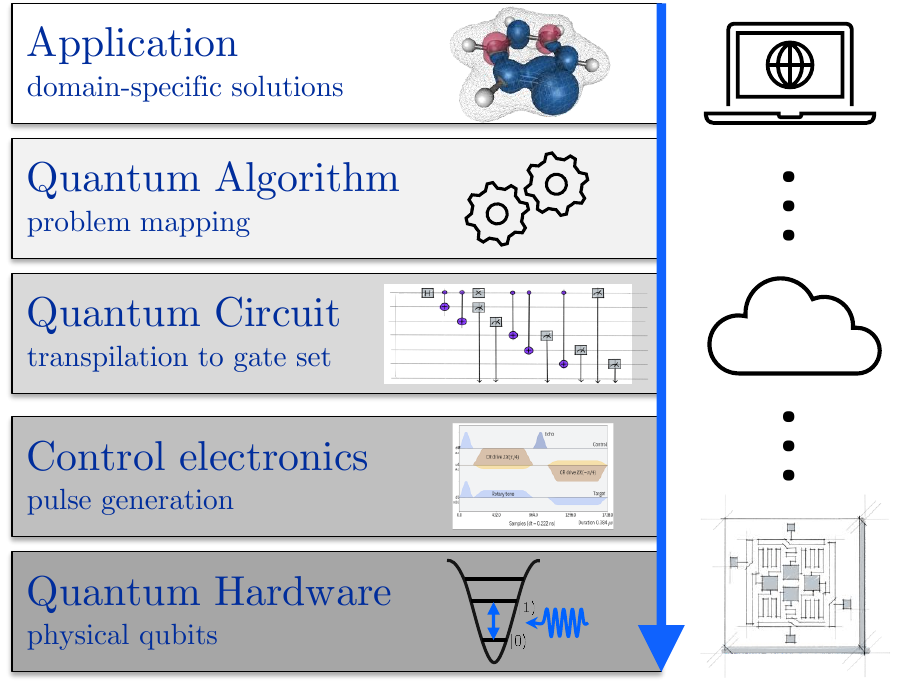}
  \end{minipage}\hfill
  \begin{minipage}[c]{0.42\textwidth}
    \caption[Abstraction layers of a quantum computer]{\small The quantum computing stack is formed from several abstraction layers that lie between the end user and the physical quantum hardware. IBM Quantum devices are accessible through a cloud service and may be programmed at different levels of the stack using the Qiskit software development kit~\cite{qiskit2024}. 
    } \label{fig:quantum_computing_stack}
  \end{minipage}
\end{figure}

IBM Quantum systems use fixed-frequency transmons. 
This design choice was made to maximise coherence times by avoiding additional complications of flux-tunable qubits such as flux noise and additional circuit and control line complexity~\cite{gambetta2017building}.
The qubits are arranged in a two-dimensional plane on a heavy-hexognal lattice, which has a unit cell of a hexagon of qubits with additional qubits on each edge (see inlays in Figs.~\ref{fig:PSE_MPS_results} and~\ref{fig:Fig2_DU_results}).
This layout offers a compromise between the desire of having highly-connected qubits and the need to avoid frequency collisions and crosstalk, as discussed above. 

The IBM Quantum devices are accessible to users in a cloud environment and can be interfaced with through a software development kit called \emph{Qiskit}~\cite{qiskit2024}.
Qiskit offers integrating with quantum computers on various layers of abstraction. 
These layers are referred to as the ``quantum computing stack'', in analogy to classical software stacks, see Fig.~\ref{fig:quantum_computing_stack}. 
On the most abstract layer, users may only define their domain-specific problem and use quantum solvers as black boxes. 
In the qiskit ecosystem, this is provided by application modules such as the ``qiskit-nature'' package~\cite{qiskit_nature_ref}. 
Such problems are then mapped to quantum algorithms, see Chap~\ref{chap:algorithms_and_subroutines} for an overview.
Quantum algorithms are built up from individual gates and measurement at the quantum circuit level of abstraction.
An important task at this level is the transpilation of a quantum circuit into a given set of basis gates supported by the hardware. 
Finally, to actually realize a circuit on the hardware, the gate sequence is translated into a scheduled sequence of control pulses.
We access this level through Qiskit's pulse module~\cite{alexander2020qiskit, mckay2018qiskit}. 

Throughout this thesis, we interface with IBM Quantum devices at different layers of the stack. 
In Chap.~\ref{chap:qudit_processing}, we explore the use of qudit levels in transmons through pulse-level experiments, while in Chap.~\ref{chap:subspace_expansion} --~\ref{chap:gauge_learning} we perform quantum simulation and error mitigation experiments at the quantum circuit level. 
In Tab.~\ref{tab:IBM_quantum_devices_specs}, we summarize details on the employed devices like median coherence times and gate fidelities. 
In general, single-qubit gate infidelities are of the order of $\mathcal{O}(10^{-4})$, two-qubit gate infidelities reach $\mathcal{O}(10^{-3})$ and readout assignment errors lie around $\mathcal{O}(10^{-2})$.
These numbers reflect the remarkable advancement of scale and performance that devices have undergone throughout the duration of this thesis.
While qubit numbers have increased from few-qubit devices to over a hundred qubits, gate fidelities and coherence times have improved significantly, see Ref.~\cite{abughanem2025ibm} for a comprehensive overview. 
Devices of the ``Falcon'' and ``Eagle'' design type employ ``always-on'' couplers and rely on cross-resonance gates as the fundamental two-qubit gate, while the more recent ``Heron'' devices use tunable couplers to implement a CZ gate. 
This enables lower crosstalk which leads to the improvement in two-qubit gate errors for \textit{ibm\_fez} at the expense of lower $T_2$ times. 
The imperfections of present-day quantum hardware calls for error mitigation techniques which we review in Sec.~\ref{sec:error_mitigation_section}.

\begin{table}
\begin{center}
\begin{tblr}{l ccc ccc c}
\hline \hline
{\\ device\\} & {\\type\\} & {\\qubit\\number}  &  {\\$T_1$  \\ $[\upmu \text{s}]$} & {\\$T_2$ \\ $[\upmu \text{s}]$}  & {1q gate \\ error \\$[\times 10^{-4}]$} &  {2q gate \\ error \\ $[\times 10^{-3}]$} & {readout \\ error \\ $[\times 10^{-2}]$} \\
\hline \hline
{\textit{ibmq\_lima}} 	    & Falcon r4    &  5   & 99  & 131 & $6.9$  &  $11.1 $ & $5.86$ \\
{\textit{ibm\_lagos}}        & Falcon r5.11 & 7    & 126 & 86  & $2.76$ &  $9.4 $  & $1.43$ \\
{\textit{ibm\_manhattan}}   & Falcon r4     & 65   & 60 & 78   & $<10$  &  $13 $  & $2$ \\
{\textit{ibm\_strasbourg}}   & Eagle r3     & 127  & 315 & 187 & $2.4$  &  $8.9 $  & $1.7$ \\
{\textit{ibm\_fez}}   	    & Heron r2     & 156  & 137 & 79  & $2.9$  &  $2.8 $  & $1.6$ \\
\hline \hline
\end{tblr}
\caption[Details on IBM Quantum superconducting hardware]{
\small Details on IBM Quantum superconducting processors used for experiments presented in this thesis, data taken from Refs.~\cite{abughanem2025ibm, manhattan_data_01, manhattan_data_02, fischer2024dynamical}.
The Falcon devices are used for qudit-space measurements presented in Chap.~\ref{chap:qudit_processing}, the \textit{ibm\_fez} device is used to explore subspace expansion in Chap.~\ref{chap:subspace_expansion}, while \textit{ibm\_strasbourg} is employed for quantum simulation and noise learning experiments presented in Chaps.~\ref{chap:dual_unitary_TEM} and Chap.~\ref{chap:gauge_learning}, respectively.
$T_1$ times, $T_2$ times, single-qubit (1q), two-qubit (2q) and readout errors refer to the median across all qubits of the chip.
Gate errors are infidelities measured with randomized benchmarking referring to $\sqrt{X}$ for single-qubit gates of all devices, CNOT gates two-qubit gates of Falcon devices, ECR gates for two-qubit gates on \textit{ibm\_strasbourg} and CZ gates for two-qubit gates on {\textit{ibm\_fez}}. 
}
\label{tab:IBM_quantum_devices_specs}
\end{center}
\end{table}
`
\chapter{Quantum algorithms and error mitigation}
\label{chap:algorithms_and_subroutines}
\summary {
In this chapter, we summarize quantum algorithms for applications in chemistry, physics, and material science. 
These broadly cover the simulation of quantum dynamics and the computation of the low-energy spectrum of eigenenergies.
We identify the estimation of expectation values from repeated measurements of a state as a core subroutine and discuss methods that address this problem. 
Finally, we summarize error mitigation techniques that help bring these algorithms onto real-world quantum devices in the presence of noise.\\
Parts of this chapter are based on Refs.~\cite{fischer2024dual} and ~\cite{fischer2024dynamical}.
}

\section{Quantum algorithms}

Since the advent of Shor's algorithm, the quest for finding novel quantum algorithms has become a mature field of research. 
To be considered both practical and impactful, a quantum algorithm should ideally satisfy three criteria.
First, it should be \emph{performant}. 
That is, it should come with provable performance guarantees that provide rigorous bounds on accuracy, runtime, and robustness under clearly defined assumptions.
Secondly, it should solve a \emph{hard} problem, i.e., the algorithm should exhibit a quantum advantage over classical methods, ideally demonstrated through formal complexity-theoretic separations or at the very least strong empirical evidence.
Thirdly, it should solve a \emph{relevant} problem that is not only theoretically interesting but also holds tangible value in the ``real world''. 

Throughout this thesis we focus primarily on digital quantum simulation as a central application of quantum computing. 
This entails encoding the Hamiltonian of a quantum system into a quantum computer and then using it to simulate the system’s time evolution or to calculate its spectral properties. 
This domain is generally considered \emph{hard} and \emph{relevant}, which has led to an influx of quantum algorithms that range from more rigorous to more heuristic.
In the following, we briefly discuss how to map various physical systems to qubits (Sec.~\ref{sec:mapping_to_quantum_comps}) and then review some of the most relevant quantum algorithms for spectral calculations (Sec.~\ref{sec:spectral_calculations}) and quantum dynamics (Sec.~\ref{sec:time_evolution}). 

\subsection{Mapping to quantum computers}
\label{sec:mapping_to_quantum_comps}

Digital quantum simulation requires the ability to map the quantum system of interest to the $N$-qubit system of the quantum computer. 
This is achieved by mapping the Hamiltonian $H^\text{sys}$ of the system to an isospectral $N$-qubit Hamiltonian $H^\text{q}$, i.e., they share dimension and eigenvalues.
Depending on the nature of the studied system, this is straightforward or may require more complicated mappings or approximations.
The resulting qubit Hamiltonian is expressed in the Pauli basis as 
\begin{equation}
\label{eq:qubit_hamiltonian}
H^\text{q} = \sum_{i=1}^M c_i P_{i_1} \otimes \dots \otimes P_{i_N}.
\end{equation}
To perform quantum simulation of $H^\text{q}$ efficiently, it is desirable to keep the total number qubits $N$, the number of terms $M$, as well as the Pauli weights (number of non-identity $P_{i_j}$ for some $i$) as low as possible. 

\paragraph{Spin systems}
Spin-$\frac{1}{2}$ models map most naturally to quantum computers as each spin can be identified with a specific qubit. 
The spin Hamiltonian can thus directly be interpreted as a qubit Hamiltonian. 
The complexity of quantum simulation mostly depends on how well the lattice of the spin model matches the layout of the qubit chip. 
Spin systems with only nearest-neighbor interactions such as the Ising or Heisenberg model have thus become popular model systems.
Our experiments in Chap.~\ref{chap:subspace_expansion} and~\ref{chap:dual_unitary_TEM} also focus on such systems. 

\paragraph{Fermionic systems}
Simulating fermionic systems with quantum computers requires mappings that respect the underlying anti-communtativity of fermions. 
For example, an electronic structure Hamiltonian in second quantization with $N$ spin-orbitals
${H}_\text{f} = \sum_{i,j = 1}^N h_{ij}  a^{\dagger}_i {a}_j + \frac{1}{2} \sum_{i,j,k,l = 1}^N h_{ijkl}  a^{\dagger}_i a^{\dagger}_j  a_k a_l $ can be mapped to an isospectral $N$-qubit Hamiltonian $H_\text{q}$ by mapping each $ a_i^\dagger$ (fermionic creators) and $ a_i$ (fermionic annihilators) to a qubit operator such that fermionic anti-commutation is preserved~\cite{seeley2012bravyikitaev}. 
The simplest mapping is given by the Jordan-Wigner transformation defined as
\begin{equation}
a_i = \left( \prod_{j=0}^{i-1} Z_j \right) \frac{X_i + iY_i}{2}, \quad a_i^\dagger = \left( \prod_{j=0}^{i-1} Z_j \right) \frac{X_i - iY_i}{2}.
\label{eq:JW-mapping}
\end{equation}
This has the advantage that the information of the occupation of a fermionic mode is encoded into a single-qubit operator as $a_i^\dagger a_i = \frac{1 - Z_i}{2}$. 
However, Eq.~\eqref{eq:JW-mapping} leads to Pauli strings of weight $\mathcal{O}(N)$ which complicates algorithms for spectral calculations and dynamics (see Secs.~\ref{sec:spectral_calculations}--\ref{sec:quantum_dynamics}).
More modern fermion-to-qubit mappings are designed to keep the weight of the resulting Pauli strings limited. 
For example, the Bravyi-Kitaev mapping achieves an encoding of fermionic operators with Pauli weights of $\mathcal{O}(\log N)$ while retaining the same $M$ and $N$ as the Jordan-Wigner mapping~\cite{seeley2012bravyikitaev}.
Finding efficient fermion-to-qubit mappings that optimize the qubit number, number of Pauli terms or their locality under certain hardware constraints is an active field of research~\cite{jiang2020optimal, kirby2022secondquantized}.

\paragraph{Bosonic and continuous-variable systems}
Digital simulations of bosonic systems typically require a truncation to a finite Fock basis due to their infinite-dimensional Hilbert spaces. 
These states are then mapped across multiple qubits. 
For example, a truncated harmonic oscillator with a cutoff at $n_\text{max}$ levels can be represented using $\lceil \log_2(n_\text{max}) \rceil$ qubits, with mappings tailored to preserve ladder operator algebra. 
Representing a system defined over continuous space requires discretization, e.g., by dividing space into a finite grid or lattice.
The resolution of the spatial grid determines how accurately the discretized model captures the original system and comes with a trade-off between physical accuracy and the number of required qubits.

\subsection{Spectral calculations}
\label{sec:spectral_calculations}

\paragraph{Phase estimation}
The flagship quantum algorithm for finding the ground state energy of the qubit Hamiltonian of Eq.~\eqref{eq:qubit_hamiltonian} is \emph{quantum phase estimation}~\cite{cleve1998quantum} (QPE), which is also at the heart of the famous Shor algorithm that breaks conventional RSA encryption~\cite{shor1999polynomial}. 
QPE estimates the eigenphase $\phi$ in the eigenvalue $e^{2\pi i\phi}$ of a unitary operator $U$, given an eigenstate $\ket{\psi}$ such that $U \ket{\psi} = e^{2\pi i\phi} \ket{\psi}$. 
The algorithm uses a register of ancilla qubits prepared in a superposition and applies controlled-$U^{2^k}$ operations, followed by an inverse quantum Fourier transform to extract a binary representation $\phi$~\cite{nielsen2010quantum}. 

If $U = e^{-iHt}$ for some Hamiltonian $H$ and $\ket{\psi_j}$ is an eigenstate of $H$ with energy $E_j$, the corresponding eigenphase obtained from QPE reveals the eigenenergy as $E_j = 2 \pi \phi / t$. 
If the initial state $\ket{\psi}$ is a superposition of energy eigenstates, the procedure yields eigenenergies with probabilities depending on the overlap of the corresponding eigenstate with $\ket{\psi}$. 
To successfully obtain the ground state energy, one generally requires the ability to prepare an approximation to the ground state that has an overlap of $1/\text{poly}(N)$. 
the need for controlled-$U^{2^k}$ gates leads to deep quantum circuits with high coherence requirements that are beyond the capabilities of near-term noisy quantum computers and are believed to require quantum error correction~\cite{wecker2015, blunt2022perspective, reiher2017elucidating}.

\paragraph{VQE}
Variational quantum algorithms (VQAs) have emerged as an avenue for quantum devices with limited qubit counts and coherence times~\cite{cerezo2021variational}, headlined by the \emph{variational quantum eigensolver} (VQE) for ground state search~\cite{peruzzo2014variational}. 
In VQAs, a parametrized quantum circuit with variational parameters $\vec{\theta}$ prepares a trial state $\ket{\psi(\vec{\theta})}$.
By updating the parameters $\vec{\theta}$ with a classical optimizer, the energy $E(\vec{\theta}) = \bra{\psi(\vec{\theta})} H \ket{\psi(\vec{\theta})}$ is minimized to approach the true ground state.
The \emph{variational principle} guarantees that the obtained energy forms an upper bound to the true ground state energy. 

The quantum computer merely prepares trial states with comparatively shallow circuits and samples the energies $E(\vec{\theta})$. 
This trades off the large circuit depths of QPE against the many measurements repetitions required to minimize $E(\vec{\theta})$. 
The three key areas of VQEs, namely ansatz expressibility~\cite{sokolov2020quantum}, energy measurements~\cite{crawford2021efficient}, and classical optimization or ``trainability'' of the circuit~\cite{holmes2022connecting, wang2021noiseinduced} have become rich research areas over the last years. 
It should be noted that VQEs are heuristic approaches which have so far evaded rigorous performance guarantees.
Their viability is thus still under debate~\cite{bittel2021training, lee2022there}.

\paragraph{Subspace expansion}

\begin{figure}
  \begin{minipage}[c]{0.4\textwidth}
    \includegraphics[width=\textwidth]{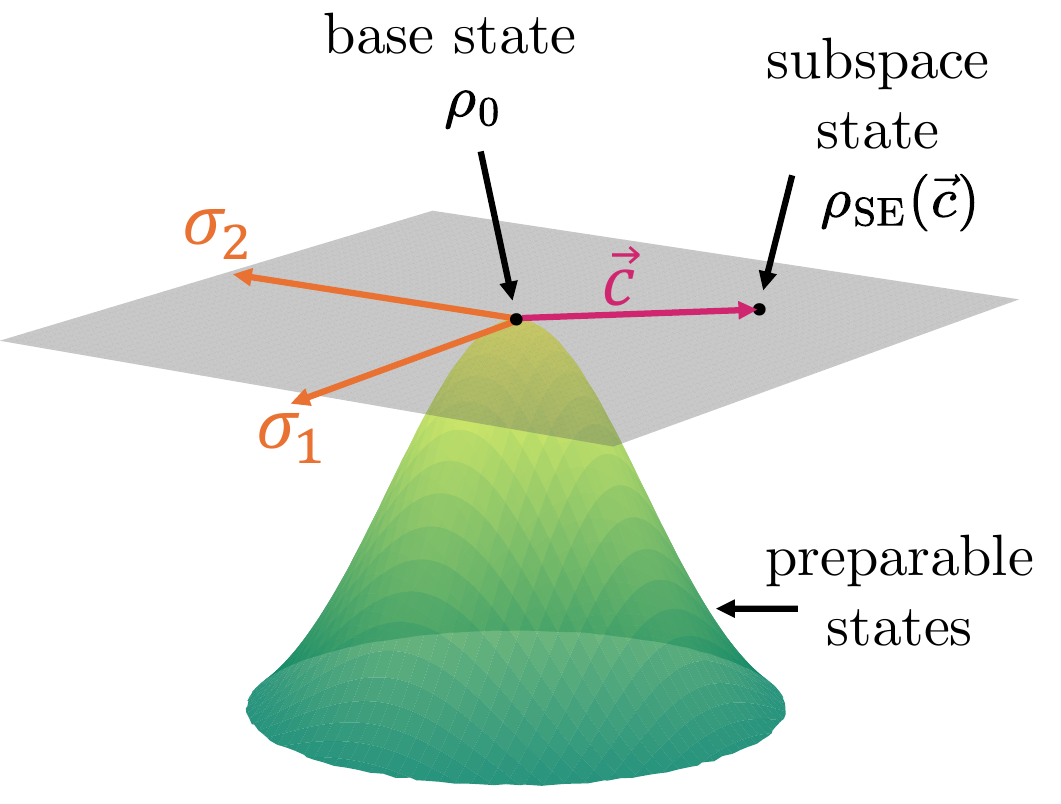}
  \end{minipage}\hfill
  \begin{minipage}[c]{0.58\textwidth}
    \caption[Schematic of quantum subspace expansion.]{\small 
    Schematic of quantum subspace expansion. A base state $\rho_0$ is prepared on a quantum computer. Expansion operators $\sigma_i$ span a subspace of states (gray shaded region) parametrized through coefficients $\vec{c}$. 
    } \label{fig:subspace_expansion_schematic}
  \end{minipage}
\end{figure}

A variety of algorithms for spectral calculations can be formulated in the \emph{subspace expansion} framework, see Ref.~\cite{Motta_2024} for a modern review.
Say our goal is to find the ground state $\ket{\psi}_0$ of some Hamiltonian $H$ and suppose that we can prepare a ``base state'' $\rho_0$ on a quantum computer which, for example, may be an approximation to $\ket{\psi}_0$. 
We can extend the reach of the quantum processor by using $L$ expansion operators $\sigma_i$ to span a space of states
\begin{align}
\label{eqn:rho_subspace_expanded}
\rho_\text{SE} (\vec{c}) \coloneqq \frac{W^\dagger \rho_0 W}{\Tr[W^\dagger \rho_0 W]}, \quad W \coloneqq \sum_{i=1}^L c_i \sigma_i .
\end{align}
parametrized by a coefficient vector $\vec{c} = \{c_1, \dots, c_L \} \in \mathbbm{C}^M$, as shown in a schematic in Fig.~\ref{fig:subspace_expansion_schematic}. 
The expectation value of an observable $O$ over a state in the subspace becomes 
\begin{align}
\label{eqn:exp_value_subspace}
\Tr[O \rho_\text{SE}(\vec{c}) ] = \frac{\sum_{i,j=1}^L c_i^* c_j \mathcal{O}_{ij}}{\sum_{i,j=1}^L c_i^* c_j\mathcal{S}_{ij} }
, \,\, \mathcal{O}_{ij} \coloneqq \Tr[\sigma_i^\dagger \rho_0 \sigma_j O ], \,\, \mathcal{S}_{ij} \coloneqq \Tr[\sigma_i^\dagger \rho_0 \sigma_j].
\end{align}
For variational ground state search, the goal is to find the minimal expectation value $\operatorname*{min}_{\vec{c}} \Tr[H \rho_\text{GSE}(\vec{c})]$. 
This problem is equivalent to finding the smallest pseudoeigenvalue $\lambda$ of the generalized eigenvalue problem 
\begin{align}
\label{eqn:GEVP}
\mathcal{O} \vec{c} = \lambda \mathcal{S} \vec{c}
\end{align}
where $\mathcal{O}$ is the subspace-projected observable and $\mathcal{S}$ is called the \emph{overlap matrix}.
For reasonably small values of $L$, this is tractable on a classical computer and generally solved by inverting the overlap matrix and computing the smallest eigenvalue of $\mathcal{S}^{-1} \mathcal{O}$.

The excitation operators $\sigma_i$ could be either unitary or Hermitian~\cite{yoshioka2022generalized}.
In the former case, further unitaries need to be added to the preparation circuit of $\rho_0$ to access the matrix elements $\mathcal{O}_{ij}$ and $\mathcal{S}_{ij}$. 
Here, we are more interested in the case where all expansion operators are Hermitian. 
In this case, the matrix elements $\mathcal{O}_{ij}$ and $\mathcal{S}_{ij}$ can be directly obtained from measurements of the state $\rho$. 
The measurement overhead of this method thus depends on the dimension of the subspace $L$ and the complexity of the expansion operators.

\paragraph{Krylov and Lanczos methods}
For ground state calculations, the powers of the Hamiltonian $H^p$ are ideal expansion operators, as the overlap of the (renormalized states) $H^p \rho {H^p}^\dagger$ with the true ground state is boosted exponentially with $p$, provided $\rho$ has non-zero overlap with $\ket{\psi}_0$. 
This forms the famous \emph{Krylov subspace}.
Is has long been used as the backbone of a classical numerical ``Lanczos method'' for finding extremal eigenvalues~\cite{cullum2002lanczos}.
In a quantum setting, the Krylov space can naturally be generated by applying time evolutions of the problem Hamiltonian in real~\cite{lanczos_real_time} or imaginary time~\cite{motta2020determining}, or with block encoding techniques~\cite{kirby2023exact}.
The intuition behind this is that the subspace spanned by $\{e^{-i\alpha \delta t  H}, \alpha \in {0, \dots, p-1}\}$ spans the Krylov subspace of oder $p$. 
This connects spectral calculations to quantum algorithms for time evolution discussed in Sec.~\ref{sec:time_evolution}.

\paragraph{Sample-based subspaces}
A recently popularized technique called \emph{Sample‑based Quantum Diagonalization} (SQD) spans a high-dimensional subspace based on individual measurement samples of an electronic wavefunction prepared with a Jordan-Wigner mapping~\cite{robledomoreno2025}. 
This can be undestood as a quantum-informed selcted configuration-interaction (CI) calculation that combines quantum and classical supercomputing resources. 
In our formalism, this corresponds to choosing a base state of $\rho_0 = \mathbbm{1}$ and applying expansion operators that are projectors onto the measured bitstrings $\sigma_i = \ketbra{i}{i}$. 
This had enabled scaling to system sizes beyond the capabilities of exact diagonalization methods~\cite{SQD_2}.
However, the suitability for non-concentrated ground states (in the measurement basis) and separation from classical methods remain debated.
In a more refined approach called \emph{Sample-based Krylov Quantum Diagonalization} (SKQD), the subspace is spanned with samples taken from a Krylov basis obtained from short-time evolution circuits~\cite{yu2025sample}.
This endows SKQD with performance guarantees similar to those of phase estimation, relying on a $1/\text{poly}(N)$ overlap of the reference state $\rho_0$ with the true ground state.

\paragraph{Extensions to higher-excited states}
Subspace expansion techniques are naturally suited to approximate eigenenergies of (low-lying) excited states by computing (the smallest) pseudo-eigenvalues of Eq.~\eqref{eqn:GEVP}. 
In this context, the expansion operators are chosen to represent the physically relevant excitations from the ground to excited states. 
For example, these could be $k$-body fermionic operators, e.g. single- and double-excitations or excitation operators inspired by the classical \emph{equation of motion} approach (qEOM)~\cite{ollitrault2020quantum}. 
As an alternative, a subspace spanned by parametrized states $\ket{\psi_1(\theta_1)}, \ket{\psi_2(\theta_2)}, \dots$ can be variationally optimized while enforcing orthogonality between them in a technique called \emph{subspace-search VQE}~\cite{nakanishi2019subspace}. 

\subsection{Quantum dynamics}
\label{sec:quantum_dynamics}

The second main domain of quantum simulation is to solve the dynamics of a system guided by time-dependent Schrödinger equation (or master equation for open systems), see Sec.~\ref{sec:time_evolution}.
This is at the heart of phenomena such as quantum transport, thermalization, entanglement growth and correlation functions.
In most cases, this boils down to applying the time evolution unitary $U = e^{-i H t}$ to some initial state and then measuring the expectation value of some observable. 
Here, we discuss how to implement the time evolution on a quantum computer while Sec.~\ref{sec:measuring_observables} deals with expectation value estimation. 

\paragraph{Product formulas}

Product formulas are the primary technique for implementing time evolution.
They approximate the time evolution operator by breaking the Hamiltonian into a sum of individual terms as $ H = \sum_j H_j$ such that each $H_j$ can be exponentiated efficiently.
For example, $H_j$ could be an individual Pauli or a set of commuting Paulis. 
In this case the exponentiation is accomplished by transforming into a basis in which the Pauli operator (or set of commuting Paulis) is diagonal using a Clifford circuit, applying a multi-qubit phase gate, and then undoing the basis transformation.
The full evolution can then be approximated with a first-order Trotter formula as 
\begin{equation}
\label{eq:trotter_first_order}
e^{-iHt} \approx \Bigl( \prod_j e^{-i H_j \Delta t} \Bigr)^n, \quad \text{where } t = n \Delta t
\end{equation}
which is accurate to \( \mathcal{O}(\Delta t^2) \).
Alternatively, higher-order product formulas improve the error scaling but require more gates. 
These approximations are especially effective when the $ H_j$ commute weakly or can be grouped into commuting sets. 
Recent research has improved error bounds and provided tighter resource estimates~\cite{morales2025selection}. 
Practical implementations must balance the Trotter step size $\Delta t$, circuit depth, and acceptable error, often guided by empirical benchmarks~\cite{miessen2024benchmarking}.

\paragraph{Floquet evolution}

Floquet theory describes the dynamics of quantum systems with periodic Hamiltonians $ H(t) = H(t + T)$. 
The corresponding evolution operator over one period defines the \textit{Floquet unitary}
$U_F = \mathcal{T} \exp\left(-i \int_0^T H(t)\, dt\right)$.
Although $H$ changes continuously in time, the full evolution over one period can be understood in terms of a ``stroboscopic'' evolution, i.e., $\ket{\psi(nT)} = U_F^T\ket{\psi(0)}$.
This can be leveraged to engineer effective Hamiltonians that are hard to implement directly. 
By discretizing the periodic drive, one constructs a sequence of unitary gates that realize \( U_F \), effectively simulating dynamics under a Hamiltonian \( H_{\text{eff}} \) such that \( U_F = e^{-i H_{\text{eff}} T} \).

The ability to simulate such discrete-time Floquet dynamics with specific repeated gate sequences makes them particularly suitable to explore with current, pre-fault-tolerant quantum computers. 
Moreover, they remain exact at any circuit depth, as they are by definition not subject to the algorithmic errors affecting Trotter decompositions.
Stroboscopic Floquet dynamics offer a wealth of new possibilities to probe unexplored universal and emergent phases~\cite{fischer2024dynamical}.
This forms the basis of Chap.~\ref{chap:dual_unitary_TEM}.

\paragraph{Variational time evolution}
Much as VQE was conceived as a hardware-efficient alternative to QPE for ground state search, variational algorithms also offer alternative ways to simulate quantum dynamics.
Instead of directly implementing $e^{-iHt}$, one evolves a parametrized ansatz state $|\psi(\vec{\theta}(t))\rangle$, where $ \vec{\theta}(t)$ is governed by an optimization rule derived from a time-dependent variational principle.
For example, The McLachlan variational principle leads to an evolution equation~\cite{gacon2024scalable}
% \sum_j \langle \partial_{\theta_i} \psi | \partial_{\theta_j} \psi \rangle \dot{\theta}_j 
\begin{align}
\label{eq:variational_time_evo}
\sum_k \mathrm{Re}\left( \langle\partial_j\psi(\vec\theta) | \partial_k\psi(\vec\theta)\rangle - \langle \partial_j \psi(\vec\theta)|\psi(\vec\theta)\rangle \langle\psi(\vec\theta)|\partial_k\psi(\vec\theta)\rangle \right) \dot{\theta}_k(t) \\
= \mathrm{Im}\left(\langle\partial_j\psi(\vec\theta)| H| \psi(\vec\theta) \rangle - \langle\partial_j\psi(\vec\theta)|\psi(\vec\theta)\rangle \langle\psi(\vec\theta)| H| \psi(\vec\theta) \rangle \right)
\end{align}
with $\partial_j \coloneqq \partial / \partial \theta_j$.
This can be solved classically to update the parameters at each time step. 
Quantum circuits are used to evaluate the necessary overlaps and expectation values of the variational state and its derivatives.
While this may by construction avoid the deep circuits needed for Trotterization, its accuracy is limited by the expressivity of the ansatz and numerical stability of the parameter update equations.
Moreover, the evaluation of expectation values that arise in Eq.~\eqref{eq:variational_time_evo} (and similar other time-dependent variational principles) pose a significant measurement overhead which is not present in direct product formula approaches. 

\paragraph{Advanced methods}
More advanced techniques for simulating time evolution are Linear Combination of Unitaries~\cite{childs2012hamiltonian} and Quantum Signal Processing~\cite{quantum_signal_prcoessing}. 
While these methods achieve better asymptotic error scaling compared to product formulas, they generally require higher circuit complexity as well as ancilla qubits. 
Hence they are firmly considered to be suitable only to fault-tolerant devices~\cite{miessen2023quantum}. 

\paragraph{Summary}
Quantum algorithms for quantum simulation often come with a trade-off. On the one hand, the more rigorous algorithms like phase estimation and product formula decompositions offer well-understood performance guarantees but come at the expense of relatively deep quantum circuits.
On the other hand, more heuristic methods such as variational algorithms have been proposed, which offer shorter and more hardware-tailored quantum circuits. 
However, this usually comes at the cost of an increased overhead in measurements that originates from the need to estimate many expectation values, e.g. for quantum-classical feedback loops. 
The runtime of such algorithms, and ultimately their feasibility, thus crucially relies on efficient protocols for extracting expectation values.

\section{Measuring observables}
\label{sec:measuring_observables}

The roundup of algorithms for quantum simulation presented above has shown that estimating expectation values of observables for some state $\rho$ prepared by a quantum circuit is a core subroutine in quantum computing.
Due to the collapse of the state upon measurement, this information is not accessible directly.
Instead, expectation values have to be statistically estimated from many individual samples obtained from running and measuring the state preparation circuit over and over again, see Box~\ref{box:expectation_values} for a formal problem statement.
Here, we discuss various methods for expectation value estimation. 
In Sec.~\ref{sec:pauli_groupings}, we review more traditional methods that are based on measuring the specific Pauli terms of a given observable.
Then, in Sec.~\ref{sec:estimators_povm_measurements}, we address this problem through the lens of (IC) POVM measurements, which are central to Chaps.~\ref{chap:duals_optimization},~\ref{chap:subspace_expansion}, and~\ref{chap:dual_unitary_TEM}.

\begin{problembox}[label={box:expectation_values}]{Expectation value estimation}
\textbf{Given:}
\begin{itemize}
\item A quantum circuit with unitary $U$ that prepares a state $\ket{\psi} = U\ket{0}$ 
\item An observable $O = \sum_{i}c_i P_i $ as a linear combination of Pauli strings $P_i$
\item A total ``measurement budget'' of $S$ repeated preparations of $\ket{\psi}$
\end{itemize}
\textbf{Task:}
Construct a statistical estimator $\hat{o}$ for $\langle O \rangle^\text{ideal} = \langle \psi |O|\psi \rangle$ based on $S$ individual samples taken with some measurement strategy on $\ket{\psi}$, such that:
\begin{itemize}
\item Ideally, the estimator is unbiased, i.e., $\mathbbm{E}[\hat{o}] = \langle O \rangle^\text{ideal}$
\item The variance of the estimator $\mathrm{Var}[\hat{o}] = \mathbbm{E}\left[(\hat{o} - \mathbbm{E}[\hat{o}])^2\right]$ is as low as possible
\end{itemize}
\end{problembox}

\subsection{Pauli groupings}
\label{sec:pauli_groupings}
Quantum computers generally offer measurements in the computational basis.
Estimating expectation values of observables that are not diagonal in this basis requires additional basis transformations prior to measurements. 
The ideal measurement would be given by a projective measurement in the eigenbasis of $O$, which is generally not known and thus unavailable even in principle.
Hence the observable from Eq.~\eqref{eq:qubit_hamiltonian} needs to be broken up into terms that can be diagonalized by appropriate basis transformations. 

A natural way would be to measure each individual Pauli operator, which can be diagonalized with single-qubit terms:
While $Z$ is already diagonal, $X$ is diagonalized by a single $H$ gate, while $Y$ is diagonalized by the sequence $H S^\dagger$. 
This naive approach can be easily improved upon by measuring commuting Pauli terms together, which is called \emph{Pauli grouping}. 
That is, we decompose $O = \sum_{i=1}^G O_i$ into $G$ groups $O_i  = \sum_{j=1}^{m_i} c_{i,j} P_{i,j}$ such that $[P_{i, j}, P_{i, j^\prime}] = 0$. 
If the Paulis in a group commute qubit-wise, the diagonalization circuits only require single-qubit gates. 
Else, the diagonalization circuits require in general up to $N(N-1)/2$ CNOT gates~\cite{gokhale_ON3_Measurement_2020}, which is considered prohibitively expensive for current quantum hardware which may face additional SWAP gate overheads. 

For $n_i$ measurements of $O_i$ (with $\sum_i n_i = N_S$), the resulting estimator has a variance~\cite{miller2024hardwaretailored}
\begin{align}
\label{eq:epsilon_with_covariances}
\epsilon^2 &= \sum\limits_{i=1}^N \frac{1}{n_i} \sum_{j, j'=1}^{m_i} c_{i,j} c_{i,j'} \text{Cov}\left[ P_{i,j},  P_{i,j'} \right].
\end{align}
This expression depends on the grouping itself, the shot allocations $n_i$, the Pauli coefficients, and implicitly the state $\ket{\psi}$ (through the covariances).
Optimizing the variance over each of these aspects has been extensively explored in the literature, see Ref.~\cite{patel2025quantum} for a recent review.
For example, coefficients stretching over several scales (such as for molecular Hamiltonians) can be exploited~\cite{crawford_efficient_quantum_2021}, or diagonalization circuits can be tailored to given hardware connectivities~\cite{miller2024hardwaretailored}.

While Pauli groupings are the go-to option for measuring observables with few individual commuting groups, they become inefficient for quantum algorithms that require estimating many different expectations values, such as subspace epxansions~\cite{choi2023measurementa}. 
Moreover, Pauli groupings measure specific bases, such that the observables of interest need to be known \emph{before} the measurement circuits themselves can be constructed and executed. 
These limitations are overcome by informationally complete measurement schemes discussed in the following. 

\subsection{Estimators from POVM samples}
\label{sec:estimators_povm_measurements}

%\paragraph{Frames and dual operators}
As an alternative to Pauli grouping methods, samples from POVM measurements can be used to build estimators of observable expectation values. 
Let $\rho$ be a state and $\vb{M} = \{M_k\}_{k \in \{1, \dots, n \}}$ be an informationally-complete POVM (see Sec.~\ref{sec:measurements_theory}). 
Any observable $O$ on the same space can then be expressed as 
\begin{equation}
\label{eqn:observable_POVM_decomp}
O = \sum_{k=1}^{n} \w_k M_k 
\end{equation}
for some $\w_k \in \mathbb{R}$. 
% If $\vb{M}$ is informationally complete such a decomposition exist for any , which is why we assume that the POVM is IC in the following. 
Given such a decomposition of $O$, the expectation value becomes
\begin{equation}
\label{eqn:expectation_value_decomp}
\expval{O}_\rho = \Tr[\rho O] = \sum_k \w_k \Tr[\rho M_k] = \mathbb{E}_{k \sim \{p_k\}}[\w_k].
\end{equation}
In other words, $\expval{O}_\rho$ can be expressed as the mean value of the random variable $\w_k$ over the probability distribution $\{p_k\}$.
Given a sample of $S$ measurement outcomes $\{ k^{(1)}, \dots, k^{(S)} \}$, we can thus construct an unbiased Monte-Carlo estimator of $\expval{O}_\rho$ as\footnote{Instead of using the sample mean to estimate the underlying mean of the distribution $\omega_k \sim p_k$, other standard estimators like median-of-means can be employed~\cite{huang_predicting_2020}. While this allows for tighter bounds on the resulting variance, it can perform worse in practice~\cite{stricker2022experimental}.}
\begin{equation}
\label{eqn:canonical_estimator}
    \hat{o} : \{k^{(1)},\dots, k^{(S)}\} \mapsto \frac{1}{S} \sum_{s=1}^{S} \w_{k^{(s)}}.
\end{equation}
We will now outline a formal approach to obtain the coefficients $\w_k$ in Eq.~\eqref{eqn:observable_POVM_decomp} for a given observable $O$ based on frame theory~\cite{d2004informationally} (see Appendix~\ref{app:chap_frame_theory}). 
First, if the POVM elements are linearly independent (such as for a minimally IC POVM) the coefficients $\w_k$ are unique. 
However, for linearly dependent POVMs, such as those that arise from overcomplete PM-simulable POVMs, the decomposition in Eq.~\eqref{eqn:observable_POVM_decomp} is not unique. 
The set of POVM operators $\vb{M}$ forms a \emph{frame} for the space of Hermitian operators.
For any frame, there exists at least one dual frame $\vb{D} = \{D_k\}_{k \in \{1, \dots, n \}}$, such that $O = \sum_{k=1}^n \Tr[OD_k] M_k$.
Therefore, the coefficients $\w_k$ can simply be obtained from the duals $\vb{D}$ as
\begin{equation} 
\label{eqn:coeffs_from_duals}
    \w_k = \Tr[OD_k].
\end{equation}
For a minimally IC POVM, only one dual frame exists.
It can be constructed from the POVM elements as 
\begin{equation}
\label{eqn:def_canonical_duals}
\kket{D_k} = \mathcal{F}^{-1} \kket{M_k} \, , \quad k =1,2,\dots,n
\end{equation}
with the \emph{canonical frame superoperator} 
\begin{equation}
\label{eqn:def_frame_superop}
\F \coloneqq \sum_{k=1}^n \kket{M_k}\bbra{M_k},
\end{equation}
where we have used the vectorized ``double-ket'' notation detailed in Appendix~\ref{app:sec_double_ket_notation}.
Thus, the frame superoperator can be used to transform between the POVM space and the dual space.  
For an overcomplete POVM, the canonical frame superoperator creates one of infinitely many possible dual frames, known as the ``canonical dual frame''. 

The theory of frames and duals enables a systematic approach to estimate observable expectation values from a given set $\{ k^{(1)}, \dots, k^{(S)}\}$ of POVM measurement outcomes:
First, one picks a valid dual frame $\vb{D}$ and construct the dual operators $\{D_{k^{(1)}}, \dots, D_{k^{(S)}}\}$ for the observed outcomes.
Then, one computes the corresponding operator coefficients $\{\w_{k^{(1)}}, \dots, \w_{k^{(S)}}\}$ through Eq.~\eqref{eqn:coeffs_from_duals}.
Finally, Eq.~\eqref{eqn:canonical_estimator} yields an estimate for $\expval{O}$.
The statistical variance of this estimator is given by the standard error on the mean 
\begin{equation}
\label{eqn:monte_carlo_variance}
    \mathrm{Var}[\hat{o}] = \frac{\mathrm{Var}[\w_k]}{S}\, .
\end{equation}
The numerator, also known as the \emph{single-shot variance} (SSV), depends explicitly on the POVM $\vb{M}$, the duals $\vb{D}$ (when they are not unique), the observable $O$ and the state $\rho$ as
\begin{equation} 
\label{eqn:single_shot_variance} \begin{split}
     \mathrm{Var}[\w_k] = 
    \E{\w_k^2} - \E{\w_k}^2 = \sum_k p_k \w_k^2  - \expval{O}_\rho^2
    = \sum_k \Tr[\rho M_k] \Tr[O D_k]^2 - \expval{O}_\rho^2.
\end{split} \end{equation}
% Throughout this work, the SSV is used as a performance measure for POVM-based estimators. 
Note that the second term $\expval{O}_\rho^2$ depends neither on the POVM nor the dual frame.
However, the first term can be decreased both by adjusting the POVM $\vb{M}$, but also by optimizing the duals $\vb{D}$ (if they are not unique).

The above procedure can be leveraged to estimate several observables from the same underlying measurement data.
Here, we quote a result from Ref.~\cite{stricker2022experimental} to build some intuition. 
As a generic POVM, let's consider a product of POVM of local, single-qubit SIC POVMs (see Sec.~\ref{sec:measurements_theory}).
Let $O_1, \dots, O_L$ be $N$-qubit observables that are all localized to (at most) $k$ qubits and fix $\epsilon, \delta \in (0, 1)$. 
Then, 
\begin{equation}
\label{eq:SIC_POVM_bound}
S \geq \frac{8}{3} 6^k \log(2L/\delta) / \epsilon^2
\end{equation}
measurements are sufficient to $\epsilon$-approximate all expectation values with a success probability of at least $1-\delta$. 
Remarakbly, this is very efficient in $L$, which underscores the power of POVMs to estimate many observables in parallel. 
However, the number of required shots scales exponentially in the weight $k$ of the observables. 
Similar scalings hold for other single-qubit POVMs based on randomized measurements~\cite{huang_predicting_2020}.
The limitation of $k$-locality can be overcome with more powerful, global POVMs, such as randomized Clifford circuit measurements~\cite{huang_predicting_2020}, which, however, require deep circuits. 

The choice of the measured POVM can have a great impact on the variance of the resulting estimator and appropriately chosen POVMs can outperform Pauli grouping strategies for complex observables such as molecular Hamiltonians~\cite{garcia-perez_learning_2021}. 
However, the ideal POVM depends on the state $\rho$, which, in practice, is often not known a priori. 
This motivates optimizing the POVM in an adaptive fashion as shots are collected~\cite{garcia-perez_learning_2021}.
Even without POVM optimization, they can outperform Pauli groupings in settings where sufficiently many observables are required to be measured~\cite{choi2023measurementa}.

\subsection{Relation to classical shadows}
 
Notably, dual operators satisfy $\rho = \sum_k p_k D_k$, so the dual operators themselves serve as unbiased estimators of the state, i.e., $\rho = \mathbb{E}_{k \sim \{p_k\}}[D_k]$.
This observation forms the basis of \emph{shadow tomography} protocols where the dual operators corresponding to measured samples are interpreted as classical representations of the state $\rho$ called \emph{classical shadows}~\cite{huang_predicting_2020}. 
The classical shadows literature usually derives this by inverting the effective channel of randomized measurements without considering the formalism of dual frames. 
This leads to shadows which correspond to the canonical dual frame. 
The POVM formalism with dual frames thus generalizes the principle of classical shadows~\cite{innocenti2023shadow}, which forms the basis of Chap.~\ref{chap:duals_optimization}. 

POVM measurements allow for a separation between the data acquisition phase, which can be carried out without fixing a target observable, and the classical processing and reconstruction stage. 
This has been popularized under the phrase ``\textit{measure first -- ask questions later}''~\cite{elben2023randomized} and led to an abundance of work exploring the use of classical shadows for quantum algorithms beyond the simple task of operator averaging. 
%The power and flexibility of classical shadows led to the development of numerous applications beyond the simple task of operator averaging.
These include the reconstruction of fidelity measures~\cite{struchalin2021} and of genuine quantum properties of states~\cite{vermersch2023many,joshi2022probing,garcia2021scrambling}, the characterization of quantum processes~\cite{levy2021classical}, classical and quantum machine learning~\cite{huang2022provably,jerbi2023shadows,gyurik2023limitations} and error mitigation techniques~\cite{seif2023shadow,filippov2023scalable, jnane2024error}. 
Often, shadows conveniently serve as a bridge between quantum and classical representations such as tensor networks.

\subsection{Heisenberg-limited estimation}

The methods for expectation value estimation discussed so far all lead to a statistical error that scales as $\propto 1/\sqrt{S}$ ( Eqs.~\eqref{eq:epsilon_with_covariances},~\eqref{eqn:monte_carlo_variance},~\eqref{eq:SIC_POVM_bound}) 
This is known as the \emph{standard quantum limit} and applies in all settings where measurements are sampled directly from the state $\ket{\psi}$. 
The optimization of Pauli groupings or POVMs relates to minimizing the prefactor of this scaling.
However, it is possible to improve the scaling up to a fundamental lower bound of $\propto 1/S$ by leveraging more sophisticated quantum algorithms that require additional quantum circuit resources. 
This is known as the Heisenberg limit for estimation. 

For example, the task of expectation value estimation for a Pauli operator can be addressed with \emph{Quantum Amplitude Estimation} (QAE)~\cite{brassard2002quantum} which provides a Grover-like quadratic improvement over direct sampling.
This is achieved by encoding the expectation value into an amplitude and extracting it via a form of quantum phase estimation.
More recent variants, such as iterative QAE~\cite{grinko2021iterative} and Maximum-Likelihood QAE~\cite{suzuki_amplitude_2020}, improve the circuit complexity beyond the naive application of QPE. 
Bayesian techniques~\cite{wiebe_bayesian_phase_estimation} can also saturate the Heisenberg bound in mean squared error given suitable priors and inference schemes.

We note that the quadratic improvement from $\propto 1/\sqrt{S}$ to $\propto 1/S$ relates strictly to the number of queries of the quantum computer and not to the total algorithm runtime or gate count.
Similarly to phase estimation for ground state search (see Sec.~\ref{sec:spectral_calculations}), Heisenberg-limited methods typically far exceed the coherence capabilities of near-term quantum hardware and will thus become more relevant in the fault-tolerant era. 
This again underscores the trade-off between quantum circuit complexity and the number of required circuit executions for quantum applications. 

\section{Error mitigation}
\label{sec:error_mitigation_section}

Present-day quantum computers are subject to noise which leads to imperfect quantum circuit execution.
In the long term, this limitation may be overcome with the advent of fault-tolerant quantum computers that rely on quantum error correction.
However, as discussed in Sec.~\ref{sec:fault-tolerant_QC_perspective} the hardware requirements for fault-tolerance remain out of reach for the time being. 
In the meantime, \emph{error mitigation} aims to reconstruct the quantity of interest from an ensemble of noisy samples (instead of using error correction to only obtain perfect samples). 
Specifically, in this thesis, we are interested in error mitigation of expectation values, see Box~\ref{box:error_mitigation} for a formal problem statement.
%Next, we discuss how to mitigate undesired noise of gates.
In the following, we first present techniques to mitigate gate noise that rely on a model of the noise channels in Secs.~\ref{sec:noise_learning_theory} --~\ref{sec:tensor-network-error-mitigation}. 
Then, in Sec.~\ref{sec:noise-agnostic_EM} we review more application-dependent but noise-agnostic error mitigation methods.
Finally, we discuss the mitigation of readout errors in Sec.~\ref{sec:readout_error_mitigation}.

\begin{problembox}[label={box:error_mitigation}]{Error mitigation of expectation values}
\textbf{Given:} A unitary $U$, observable $O$, shot budget $S$ as in Box~\ref{box:expectation_values} and an imperfect quantum computer subject to noise channels $\{\Lambda_i\}$ resulting in a noisy state $\rho$ with an ``unmitigated'' expectation value $\langle O \rangle^\text{noisy} = \Tr[\rho O]$. \\

\textbf{Task:}
Combine up to $S$ measurements of multiple (potentially different) noisy quantum circuits to construct a statistical estimator $\hat{o}^\text{mit}$ with a reduced bias $\text{Bias}(\hat{o}) \coloneqq | \mathbbm{E}[\hat{o}^\text{mit}] - \langle O \rangle^\text{ideal} | < | \langle O \rangle^\text{noisy} - \langle O \rangle^\text{ideal} |$.
The performance of $\hat{o}^\text{mit}$ can be quantified by its \emph{mean square error} (MSE) which depends on the residual bias and variance as
\begin{equation}
\text{MSE}(\hat{o}^\text{mit}) = \text{Bias}(\hat{o}) + \text{Var}(\hat{o}).
\end{equation}
For explicitly known noise channels $\{\Lambda_i\}$ this is referred to as \emph{noise-model-based error mitigation}, and \emph{noise-agnostic error mitigation} otherwise. 
\end{problembox}

The most generally applicable and powerful error mitigation techniques assume knowledge of the noise channels $\Lambda_i$ in a device. 
We start our review of these techniques by discussing how to obtain realistic models for gate noise from experiments in a scalable way -- a problem referred to as ``noise learning''.  
This forms the basis for our work on improved noise learning protocols presented in Chap.~\ref{chap:gauge_learning}.

\subsection{Noise learning}
\label{sec:noise_learning_theory}
In current quantum hardware, single-qubit gates typically achieve much better fidelities than multi-qubit entangling gates (CNOT or CZ gates) with median infidelities of $\mathcal{O}(10^{-4})$ compared to $\mathcal{O}(10^{-3})$, see Tab.~\ref{tab:IBM_quantum_devices_specs}.
Therefore, we are predominantly concerned with mitigating the noise of entangling gates and will essentially treat single-qubit gates as noiseless in the following. 
This assumption can be relaxed to gate-independent single-qubit noise~\cite{wallman2016noise}, which is justified by the fact that each single-qubit gate is implemented with the same template sequence of $\sqrt{X}$ and $R_z$ gates, see Eq.~\eqref{eq:decomp_RZ_SX}.
We model noise as a channel $\tilde\Lambda$ such that the noisy version of a unitary gate becomes $\tilde{\mathcal{U}} = \mathcal{U} \circ \tilde\Lambda$, where $\mathcal{U}(\rho) = U \rho U^\dagger$ denotes the ideal gate $U$\footnote{Here we pick the convention of noise acting before the unitary part of the gate. Having noise act after the gate is an equally justified convention.}.
We represent $\tilde\Lambda$ in the Pauli transfer matrix (PTM) formalism introduced in Sec.~\ref{sec:quantum_channels_representations}.
In principle, the channel $\tilde\Lambda$ could be inferred from quantum process tomography~\cite{merkel2013self} which reconstructs the full exponentially large PTM.
However, this is limited to few-qubit systems and thus fails to capture the complex crosstalk of parallel gate layers over a full device.
This calls for a restricted model of the noise channel that only depends on a manageable number of parameters but successfully captures the most relevant noise sources.
In the following, we review \emph{sparse Pauli-Lindblad} (SPL) models which have recently become a popular choice for this purpose~\cite{van2023probabilistic}.

 \begin{figure}  
     \centering
     \includegraphics[width=0.8\columnwidth]{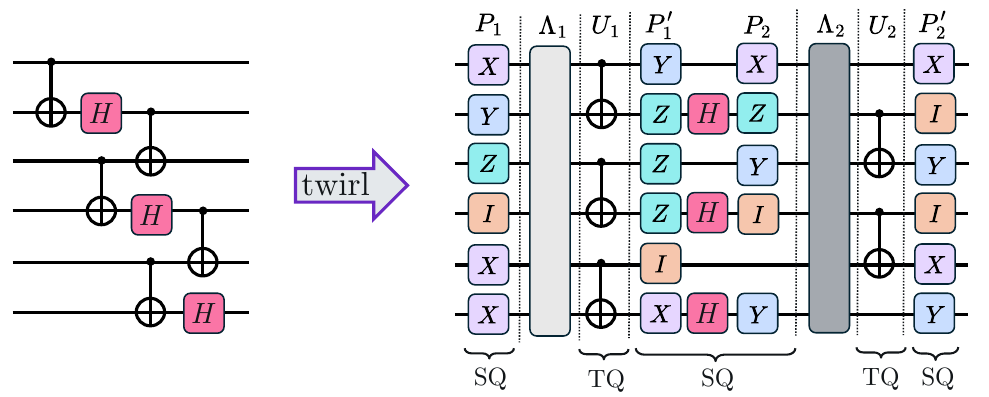}
     \caption[Pauli twirling of two-qubit gates]{ \small
    Schematic of Pauli twirling of two-qubit gates. The target circuit on the left consists of single-qubit (SQ) and two-qubit (TQ) gates. First, the gates are scheduled such that CNOT gates are applied in parallel where possible. In our model, each unique CNOT layer $U_1$ and $U_2$ is accompanied by a distinct sparse Pauli-Lindblad noise channel $\Lambda_i$. Random single-qubit Pauli gates $P$ are inserted before the noise layer, and conjugate Paulis $P^\prime$ after the CNOT layers ensure that the global unitary of the circuit is preserved. Twirling gates can be merged with the original circuit single-qubit gates (here exemplified by $H$ gates), creating a structure of alternating SQ and TQ layers. 
    }
     \label{fig:twirling_schema}
 \end{figure}

\paragraph{Pauli twirling} As a first step, the noise channel is simplified by applying a technique known as \emph{Pauli twirling} or \emph{randomized compiling}~\cite{wallman2016noise}.
Here, single-qubit Pauli gates $P_i$ are sampled uniformly at random and applied before $\tilde{U}$. Afterwards, the action of $P_i$ is undone by applying the \emph{conjugate} Pauli
\begin{equation}
\label{eq:definition_pauli_conjugate}
P_{i}^\prime \coloneqq U P_{i} U^\dagger,
\end{equation}
such that $U = P_i^\prime U P_i$.
When $U$ is a Clifford gate, averaging over many such ``twirling instances'' results in a channel
\begin{align}
\label{eq:twirling_action}
\Lambda: \rho \mapsto \mathbb{E}_i \left[ P_i^\dagger  \tilde{\Lambda}(P_i^{\dagger}\rho P_i) P_i\right]
\end{align}
that is a Pauli channel, i.e., it has a diagonal PTM representation~\cite{geller2013efficient}.
In practice, the expectation value in Eq.~\eqref{eq:twirling_action} is approximated by averaging over a finite number of twirled circuit instances.
The additional twirling gates can be merged with existing twirling gates in the circuit such that this method often comes only with a small gate overhead, if any.
This leads to a layered circuit structure that interleaves parallel, entangled two-qubit gate layers with single-qubit gate layers, as illustrated in Fig.~\ref{fig:twirling_schema}.

\paragraph{Imposing Sparsity}
As a second step, the remaining Pauli channel $\Lambda$ is sparsified.
This is done in a Lindbladian formalism (see Sec.~\ref{sec:time_evolution}) where the channel $\Lambda = \exp(\mathscr{L})$ is generated by a set of Pauli errors as
\begin{equation}
\label{eq:noise_model_def}
\mathscr{L}(\rho) \coloneqq  \sum_{P_i \in \mathcal{P}} \lambda_{i} \left( P_i \rho P_i^\dagger - \rho .\right)
\end{equation}
Here, $\mathcal{P}$ is a set restricted to include only the most physically relevant noise terms.
A common choice is to include all single-qubit Pauli terms and two-qubit Pauli terms of all qubits that are nearest neighbors in the circuit layout on the device.
This captures the most relevant sources of crosstalk which are predominantly nearest-neighbor.
This way, the noise channel $\Lambda$ is fully specified by a number of free parameters $\lambda_i$ that scales linearly with the number of qubits.

 \begin{figure}
     \centering
     \includegraphics[width=0.8\columnwidth]{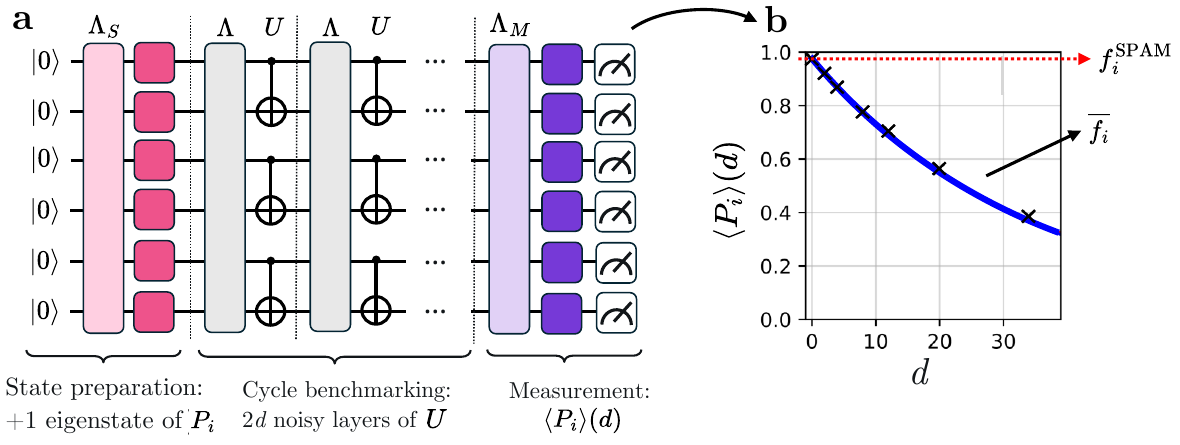}
     \caption[Noise learning with cycle benchmarking]{ \small
    Cycle benchmarking circuits to learn the noise model of a two-qubit layer $U$. 
    \textbf{a)} For a given Pauli $P_i$ from a sparse set $\mathcal{P}$, a $+1$ eigenstate is prepared. This is subject to a state preparation noise channel $\Lambda_S$, which leads to a state preparation fidelity $f_i^{SP}$. An even number of $2d$ noisy layers is applied, where the signal is attenuated by a factor $f_i$ for each odd and $f_i^\prime$ for each even layer. Finally a measurement of $\langle P_i\rangle$(d) is subject to a measurement noise channel $\Lambda_M$ with a fidelity $f_i^M$. \textbf{b)} An exponential fit (blue) to measurements at different depths $d$ yields the pair fidelity $\overline{f}_i$ from the decay constant, and the SPAM fidelity $f^{\textsc{SPAM}}_i = f_i^{SP} f_i^M$ as the y-axis offset. 
    Twirling gates are not shown for simplicity. Measurements are twirled with $X$ gates to symmetrize the measurement error. 
    }
     \label{fig:cycle_benchmarking_schema}
 \end{figure}

\paragraph{Cycle benchmarking}
For sparse Pauli-Lindblad models, the task of noise learning now boils down to characterizing the generator rates $\lambda_i$ of the chosen set of Pauli generators $\mathcal{P}$.
This is commonly done by fitting the SPL model to fidelities (see Eq.~\eqref{eq:Pauli_fidelity_def}) obtained from cycle benchmarking circuits~\cite{erhard2019characterizing, chen2023learnability}, see Fig.~\ref{fig:cycle_benchmarking_schema}.
In these circuits, one first prepares a $+1$ eigenstate of a given Pauli from $P_i \in \mathcal{P}$, then applies the twirled noisy layer $\tilde{U}$ $2d$ times for some set of integers $d$, and finally measures $\langle P_i \rangle$.
For a self-inverse Clifford gate such as CNOT or CZ, every pair of layers in theory applies the identity operator.
The ideal measured expectation values should thus remain at $+1$, while, in practice, the signal decays due to noise as
\begin{equation}
\label{eq:pair_fidelity_decay}
\langle P_i \rangle(d) =  (f_i f_{i^\prime})^d \times f^{\textsc{SPAM}}_i.
\end{equation}
where $f_i$ are the Pauli fidelities of the noise channel and $f^{\textsc{SPAM}}_i$ is the state preparation and measurement fidelity associated with the prepared eigenstate.
% Note that $P_{(i^\prime)^\prime}$ = $P_i$ since the ECR gate is self-inverse.
An exponential fit to experimentally obtained values for $\langle P_i \rangle(d) $ thus retrieves the \emph{pair fidelities} 
\begin{equation}
\label{eq:pair_fidelities_def}
\overline{f_i} \coloneqq \sqrt{f_{i} f_{i^\prime}}.
\end{equation}
We see that this procedure cannot learn the fidelities individually but only as products in pairs (unless the Pauli $P_i$ is self-conjugate for the given unitary).
Previous work in the literature has therefore introduced a \emph{symmetry assumption} setting $f_{i} = f_{i^\prime}$~\cite{van2023probabilistic, kim2023evidence}.
In principle, it is known that fidelities of Pauli operators whose conjugate Pauli has the same Pauli weight can be learned with an ``interleaved'' cycle benchmarking protocol~\cite{chen2023learnability}.
However, the fidelities of Pauli operators that do change weight under conjugation with the layer $U$ remain fundamentally unlearnable in a SPAM-robust way~\cite{chen2023learnability}.
The generator rates $\lambda_i$ can be obtained from a fit of the noise model from Eq.~\eqref{eq:noise_model_def} to the obtained Pauli fidelities.
Let $M$ be a square matrix with entries $M_{ij} = 1, i, j \in \mathcal{P}$ if $\{ P_i, P_j \} = 0$ and $M_{ij} = 0$ otherwise.
Similarly, we define $M^\prime$ with entries  $M^\prime_{ij} = 1$ if $\{ U P_i U^\dagger, P_j \} = 0$ and $M^\prime_{ij} = 0$ otherwise.
We use $\boldsymbol{\lambda}$ to denote the vectors with entries of $\lambda_i$ (and similarly for $f_i$, $f_i^\prime$).
The relationship between generators and fidelities is then given by $ M \boldsymbol{\lambda} = \log(\boldsymbol{f})/2$ and $M^\prime \boldsymbol{\lambda} = \log(\boldsymbol{f^\prime})/2$.
The generators that best describe the obtained fidelities are found by solving the non-negative least-squares problem
\begin{equation}
\label{eq:generator_fidelities_fit}
\boldsymbol{ \lambda}_\text{fit} :=  \operatorname*{arg\,min}_{\lambda_i \geq 0} \; \Bigg\lVert
\left[\begin{array}{c}M \\ M^\prime \end{array}\right] \boldsymbol{\lambda}
+ \frac{1}{2} \log \left[\begin{array}{c} \boldsymbol{f} \\ \boldsymbol{f^\prime} \end{array}\right]  \Bigg\rVert_2^2.
\end{equation}

In practice, the experimental overhead of running cycle benchmarking for all $P_i \in \mathcal{P}$ can be reduced by running the circuits for non-overlapping Paulis in parallel.
For example, for a two-local noise model with nearest-neighbor noise generators, only 9 different combinations of initial state preparations and measurement bases are required~\cite{vandenBerg2024techniqueslearning}.

We have seen that the modeling and learning of sparse Pauli-Lindblad models relies on a series of assumptions that have theoretical justifications but may nevertheless be violated to some extend in reality. 
We summarize these assumptions in Tab.~\ref{tab:noise_model_assumptions}. 

\begin{table}
\begin{center}
\begin{tblr}{l|c|c}
%\hline
\hline \hline
{Assumption} & {Justification}  &  {Potential sources of  violations}  \\
\hline \hline
{Single-qubit noise \\ is gate-independent}  & {Pulse sequence of single- \\ qubit gates only differs\\ in virtual phases} & {Finite resolution of pulse\\ amplitudes and phases} \\
\hline
{Stability in time \\(Markovianity)} & {Experiments have shown \\ stability over several hours~\cite{kim2024error}} & {Two-level system (TLS) defects  \\ can lead to  rapid fluctuations \\ of noise parameters~\cite{kim2024error}}\\
\hline
{Confinement to \\ qubit levels} & {Pulse shaping techniques \\ avoid leakage~\cite{gambetta2011analytic}} & {Leakage can still occur \\ from frequency collisions \\ or induced by measurements} \\
\hline
{Pauli channel} & {Twirled Clifford gates \\lead to Pauli noise channels} & {Insufficient convergence \\ due to finite number of twirls}   \\
\hline
{Nearest-neighbor \\ noise generators} & {Crosstalk mainly affects \\nearest neighbors} & {Higher-weight or non-local \\  correlated noise sources}  \\
\hline
% \SetTblrInner{rowsep=0pt}
{Symmetry of \\ conjugate fidelities} & {Predicted by recent theory  \\ up to leading order~\cite{malekakhlagh2025efficient}} & {Higher-order corrections \\ to the theory}  \\
\hline \hline
\end{tblr}
\caption[Assumptions of sparse Pauli-Lindblad noise model learning]{
\small Assumptions of sparse Pauli-Lindblad noise model learning ordered from general (top) to more specific to sparse Pauli-Lindblad models (bottom).
}
\label{tab:noise_model_assumptions}
\end{center}
\end{table}

\subsection{Probabilistic error cancellation}
\label{sec:PEC}
The most rigorous general-purpose error mitigation method proposed so far is 
\emph{Probabilistic Error Cancellation} (PEC)~\cite{temme2017error}. 
PEC addresses Problem~\ref{box:error_mitigation} by implementing the inverse of the error channels $\{\Lambda_i^{-1}\}$ at the circuit level to completely cancel the effect of noise and obtain in principle unbiased estimators.  
Although the inverse channel $\Lambda_i^{-1}$ is generally not a physical CPTP channel (see Sec.~\ref{sec:quantum_channels_representations}), it can be implemented in expectation by a quasi-probabilistic sampling strategy. 
We illustrate this for the SPL noise models introduced above. 
The inverse of a SPL noise channels takes the form~\cite{van2023probabilistic}
\begin{equation}
\label{eq:inverse_noise_channel}
\Lambda^{-1}(\rho) = \gamma \prod_{i \in \mathcal{P}} \left( w_i \rho - (1 - w_i) P_i \rho P_i^{\dagger} \right)
\end{equation}
with the probabilities $w_i \coloneqq (1 + e^{-2\lambda_i})/2$ and the prefactor $\gamma \coloneqq \exp{(\sum_{i \in \mathcal{P}}) 2 \lambda_i)}$. 
To implement $\Lambda^{-1}(\rho)$, one averages over circuit instances where, for every $P_i \in \mathcal{P}$, the identity gate is sampled with probability $w_i$ and the gate $P_i$ is sampled with probability $1-w_i$, recording a $-1$ factor (all applied gates $P_i$ can be merged to a single Pauli gate). 
The measured sample for a Pauli expectation value is then multiplied by all recorded $-1$ factors and the factor $\gamma$. 
While averaging over such samples fully removes the bias due to noise, the variance of this estimator is increased by a factor of $\gamma^2$ compared to the noiseless case, hence $\gamma^2$ is known as the sampling overhead. 
This applies to every noisy gate layer in the circuit and the respective $\gamma$ overheads compound in a multiplicative way. 
The variance of the mitigated estimator thus increases exponentially with the number of gates in the circuit. 
This is a manifestation of a fundamental limitation of error mitigation techniques as discussed in Sec.~\ref{sec:fault-tolerant_QC_perspective}.
However, the basis of this exponential scaling is itself reduced exponentially with the physical noise rates $\lambda_i$, lending hope to the idea that there is a regime where the exponential cost of PEC is manageable even for comparatively large circuits~\cite{bravyi2022future}.

\subsection{Zero-noise extrapolation}
\label{sec:Zero-noise_extrapolation}

To avoid the exponential overhead of inverting the noisy channels $\{\Lambda_i\}$, it has been proposed to instead artificially \emph{increase} the noise in a controlled way. 
By estimating the noisy expectation value at different noise levels, the limit to the zero-noise case may be taken from a fit to the data. This is known as \emph{zero-noise extrapolation} (ZNE). 
This does not necessarily require a precise model of the noise channels, e.g., the noise can be amplified by inserting noisy identity gates~\cite{he2020zero}. 
However, SPL models allow for a highly controlled way of increasing noise known as \emph{probabilistic error amplification} (PEA). 
Similarly to the sampling technique discussed above for PEC, Pauli gates can be inserted probabilistically to increase (rather than invert) the noise of an SPL model~\cite{kim2023evidence}. 
This represents a physical noise channel and thus comes without the intrinsic sampling overhead of PEC.
ZNE with PEA has enabled quantum simulation experiments at a scale of over $100$ qubits~\cite{kim2023evidence} 

While ZNE-based estimators achieve a smaller variance than PEC, they are heuristic in nature and have a non-zero residual bias. 
This is reflected by a freedom of choice of the analytical function (taken to be either exponential~\cite{endo2018practical} or polynomial~\cite{temme2017error}) used for the extrapolation whose true form is inaccessible. 

\subsection{Tensor-network error mitigation}
\label{sec:tensor-network-error-mitigation}

 \begin{figure}  % let's get rid of 'middle out' and the '1/2' layer annotations in the figure.
     \centering
     \includegraphics[width=\columnwidth]{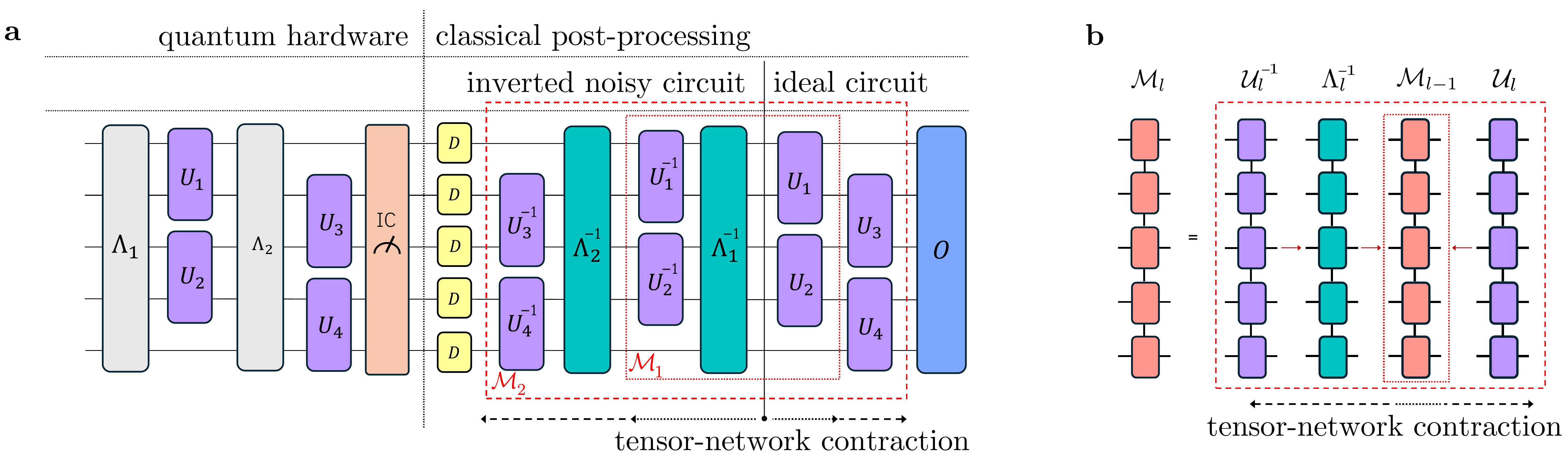}
     \caption[Schematic of tensor-network error mitigation]{ \small
     Schematic of tensor-network error mitigation (TEM).
     \textbf{a}) IC measurements are collected from noisy circuits with Pauli noise channels $\Lambda_l$ acting before the unitary layers $U_l$, assigning a dual operator $D$ to each outcome.
     The TEM mitigation map that undoes the effects of noise is applied in post-processing through a tensor-network representation.
     \textbf{b}) Construction of a single iteration $\mathcal{M}_l$ of the TEM algorithm as a sequence of tnesor-network contractions of the layers inside a portion of the full TEM map. 
%     Steps 1 and 2 define the order in which we contract and compress within a single iteration. 	
    }
     \label{fig:tem_schema}
 \end{figure}
 
 A recent addition to the suite of noise-model-based error mitigation techniques is \emph{tensor-network error mitigation} (TEM) introduced by Filippov et al.~\cite{filippov2023scalable}. 
 TEM addresses Problem~\ref{box:error_mitigation} by applying a map to the noisy quantum circuit which is designed to undo the effects of the noisy channels $\{\Lambda_i\}$ in classical post-processing, thus combining quantum and classical resources. 
 This method is the backbone of quantum simulation experiments presented in Chap.~\ref{chap:dual_unitary_TEM}.
  Our summary here is based on Ref.~\cite{fischer2024dynamical}.

The structure of the quantum circuit, depicted in Fig.~\ref{fig:tem_schema}\textbf{a} consists of a series of ideal unitary layers ${\cal U}_l \coloneqq {U}_l \bullet {U}_l^{\dag}$, each preceded by a noise channel $\Lambda_i$. 
Performing error mitigation on a noisy state $\rho$ in post-processing can be achieved by applying the map
\begin{equation}
\label{eq:TEM_def}
\mathcal{M} \coloneqq  \underbrace{\cdots \circ {\cal U}_2 \circ {\cal U}_1}_{\text{ideal~dynamics}} \circ \underbrace{\Lambda_1^{-1} \circ {\cal U}_1^{-1} \circ \Lambda_2^{-1} \circ {\cal U}_2^{-1} \circ \cdots}_{\text{inverse~of~noisy~dynamics}} 
\end{equation}
to the output density matrix $\rho$ of a noisy computation, such that $\Tr[{\cal M}(\rho) O] = \langle O \rangle^\text{ideal}$ for any observable $O$. 
However, as discussed for PEC above, due to the inclusion of $\Lambda_i^{-1}$ channels, $\mathcal{M}$ is generally not a physical CPTP map and thus cannot be directly implemented on a quantum computer. 
Instead of modifying the noisy state $\rho$, we can instead recover the ideal expectation by modifying the observable as $\Tr[\rho {\cal M}^{\dag}( O)] = \langle O \rangle^\text{ideal}$.
The TEM method exploits this idea by approximating the map $\cal M$ as a tensor network $\tilde{\mathcal{M}}$ and then measuring the ``TEM-modified observable'' $O^\prime \coloneqq \tilde{\mathcal{M}}^{\dag}( O)$ on the noisy circuit.
% Potentially link to VILMA algorithm here and mention that TEM is a special case https://arxiv.org/pdf/2207.01360

Constructing the map $\cal M$ as in Eq.~\eqref{eq:TEM_def} by concatenating the constituent maps layer by layer would lead to a complexity growing exponentially in the number of layers. 
However, as shown in Fig.~\ref{fig:tem_schema}\textbf{b} this can be made efficient via the recurrence relation
\begin{equation}
    {\cal M}_l = {\cal U}_l \circ \Lambda_{l}^{-1} \circ {\cal M}_{l-1} \circ {\cal U}_{l}^{-1},
    \label{Tem_iteration}
\end{equation}
where every unitary layer ${\cal U}_l$ and corresponding inverse noisy layer $(\Lambda_{l}^{-1} \circ {\cal U}_l^{-1})$ approximately cancel each other. 
For example, for a one-dimensional connectivity, both ${\cal U}_l$ and $\Lambda_l^{-1}$ (if modeled in SPL form) allow a tensor network representation in the form of a \emph{matrix product operator} (MPO) of bond dimension $4$~\cite{filippov2023scalable}, so the map ${\cal M}$ is constructed via recurrent conventional tensor network contractions of MPOs, with ${\cal M}_0 = \mathbbm{1}$ being the identity transformation. 
An approximation is made when the bond dimension of ${\cal M}_l$ becomes prohibitively large.
In this case, compression to some maximum bond dimension $\chi$ is done by truncating dimensions corresponding to the smallest singular values, see Refs.~\cite{filippov2023scalable, fischer2024dynamical} for details. 
%Compression of the MPO at each iteration results in the approximate map ${\cal M}'$ capturing the most significant contributions (Supplementary Information III~B).

The TEM-modified observable $O^\prime$ can be estimated from IC POVM measurement outcomes of $\rho$ as outlined in Sec.~\ref{sec:estimators_povm_measurements}.
This has the practical advantage that -- in contrast to PEC -- the noise channels that $O^\prime$ depends on through the inversion map $\tilde{\mathcal{M}}$ do not need to be known before collecting the measurement samples. 
This way, noise learning and application circuits can run on the device in any order or be interleaved. 
Moreover, TEM achieves the universal lower bound for sampling overheads in error mitigation~\cite{filippov2024scalabilityquantumerrormitigation, tsubouchi2023universal}.
The inclusion of classical simulation methods in the form of tensor networks leads to a mitigation overhead that is typically reduced by a factor of $R^2$ compared to PEC and $\approx (1 + 3.59 \log(R))^2$ compared to ZNE with PEA, where $R=\langle O \rangle^\text{ideal} / \langle O \rangle^\text{noisy}$~\cite{filippov2024scalabilityquantumerrormitigation}.
Compared to PEC, this comes at the cost of a slight bias in the TEM estimator that may be introduced by the tensor network truncations. 
 
\subsection{Noise-agnostic error mitigation}
\label{sec:noise-agnostic_EM}

While noise-learning-based methods can provide bias-free error mitigation (with PEC) or a mitigation with a well-controlled approximation error (e.g. with TEM), a myriad of other methods have been proposed to perform error mitigation in a more heuristic way without resorting to noise modeling, see Ref.~\cite{cai2023quantum} for an overview.
These methods typically come with certain limitations such as  being only applicable to certain algorithms or lacking performance guarantees. 
Nevertheless they can be successful in practice. 

\paragraph{Zero-noise extrapolation}
As a first example, zero-noise extrapolation can be performed without knowledge of the noise channels when using other methods than PEA for error amplification. 
This can be achieved by running longer unitarily equivalent circuits. 
For example, the pulses can be stretched out in time~\cite{ZNE_pulse_Stretching} or a single CNOT can be replaced by $3, 5, \dots$ CNOTs~\cite{ZNE_gate_folding}. 

\paragraph{Rescaling methods}
Several methods aim to mitigate expectation values directly in a two-step rescaling procedure. 
First, a noisy circuit $U$ is run to obtain the noisy expectation value for some observable $O$.
Then, $O$ is measured on a second noisy circuit which is supposed to be ``similar'' to $U$ but classically simulable, such that the ideal expectation value is known a priori. The obtained value for $\langle O \rangle$ is then rescaled according to the amount of noise observed in the second circuit. 
For example, the second circuit can be a Clifford relaxation of the original circuit, which is classically tractable (see Sec.~\ref{sec:clifford_circuit_theory}). This is known as \emph{Clifford data regression}~\cite{clifford_data_regression}.
In other variants, the error is assumed to be a fully depolarizing channel such that the error rate can be inferred from measuring a trace-free version of the original observable. This technique is known as \emph{(operator) decoherence rescaling}~\cite{decoherence_rescaling}. 

\paragraph{Post-selection methods}
Another class of techniques is based on post-selecting the individual shots based on some error detection scheme while discarding erroneous shots. 
The exponential overhead of error mitigation manifests in the probability of encountering no errors which is generally exponential in the number of gates. 
Most prominently, this is possible when the problem has symmetries that lead to conserved quantities. 
For example, for molecular Hamiltonians, the number of (spin-up/spin-down) fermions is usually fixed and can be inferred from the number of measured $\ket{1}$-qubits under a Jordan-Wigner mapping. 
%Any shot with the wrong Hamming weight has experienced an error and can thus be discarded. 
This is known under the term \emph{symmetry verification}~\cite{bonetmonroig2018lowcosta}.  
Without symmetries, error detection mechanism can be more sophisticated, such as introducing local Pauli checks with ancilla qubits, similar to syndrome measurements in QEC~\cite{martiel2025low}. 

\paragraph{Purification-based methods}
Noise generally turns a pure state $\ket{\psi}$ into a mixed state $\rho$. 
If the dominant contribution to $\rho$ is the original state (i.e., it's the eigenvector to the largest eigenvalue), then taking powers of the state $\rho^n$ will exponentially boost the contribution of the original state. 
Such purification-based techniques thus aim to measure $\Tr[\rho^n O] / \Tr[\rho^n]$ as an error-mitigated version of $\Tr[\rho O]$. 
An example is \emph{virtual distillation} which entangles several quantum registers that each prepare $\rho$ to access $\Tr[\rho^n O]$\cite{koczor2021exponential}. 

\paragraph{Subspace expansion}
We have encountered subspace expansions as a technique for spectral calculations in Sec.~\ref{sec:spectral_calculations}. 
In the context of a variational minimization, it can also be a versatile tool for error mitigation~\cite{mcclean2017hybrid, colless2018computation}. 
Suppose the base state $\rho_0$ is a noisy implementation of the true ground state, for example as obtained from a VQE optimziation. 
The construction of a subspace around $\rho_0$ can be regarded as an error-mitigation strategy to recover the energy of the true ground state, facilitated, e.g. with Krylov-like expansion operators~\cite{suchsland2021algorithmic, suchsland2022simulating}.
Moreover, using symmetry operators as subspace operators incorporates symmetry verification into this framework~\cite{cai2021quantum}. 
Subspace expansion has also been generalized to include state-dependent excitation operators and can then include the effects of virtual distillation~\cite{yoshioka2022generalized}.

\subsection{Readout error mitigation}
\label{sec:readout_error_mitigation}
The techniques above generally address gate errors. However, state preparation and measurement (SPAM) errors are also significant error sources in superconducting qubits, see Sec.~\ref{sec:common_noise_sources}. 
Here, we review various methods to mitigate SPAM errors.
Traditional approaches have aimed to calibrate a ``confusion matrix'' $M_{ij}$ which is a square matrix that encodes, for each pair of computational basis states $\ket{i}$ and $\ket{j}$ the probability $p(i|j)$ to obtain outcome $\ket{i}$ when $\ket{j}$ was prepared. Once this matrix is constructed, its pseudoinverse is applied to the raw measurement outcomes to obtain a corrected probability distribution~\cite{maciejewski2020mitigation}.
However, this approach suffers from the exponential scaling of the dimension of $M$ with the number of qubits.
This is commonly addressed by limiting the amount of correlations in $p(i|j)$, e.g., down to single-qubit tensor product approximations~\cite{bravyi2021mitigating} or selecting specific subspaces for the construction of $M$~\cite{nation2021scalable}.

\paragraph{Twirled readout error extinction}
For expectation values, readout errors can be mitigated with a lightweight calibration-free technique known as \emph{twirled readout-error extinction} (TREX)~\cite{van2022model, kim2023evidence}. 
TREX applies to a given Pauli observable $P$, while for more complex observables, one relies on a decomposition into Pauli terms and applies TREX for each Pauli individually.
Crucially, each measurement is ``twirled'' by random Pauli $X$ gates to symmetrize the readout error channel. 
That is, before the final computational basis measurements, an $X$ gate is inserted with a probability of $50 \%$, upon which the measured bit is flipped in case the gate was applied. This way the error probabilities $p(0|1)$ and $p(1|0)$ are equal. 
As a result, the readout error on the twirled noisy expectation value $\langle P \rangle_\text{twirled}$ manifests as a multiplicative factor. 
This noise factor can be measured by running a twirled ``depth-zero'' circuit with no entangling gates that measures the same Pauli to obtain $\langle P \rangle_0$. 
The resulting mitigated expectation value is then given by
\begin{equation}
\label{eq:TREX_definition}
\langle P \rangle_\text{TREX} \coloneqq \frac{\langle P \rangle_\text{twirled}}{\langle P \rangle_0}.
\end{equation}
TREX can be thought of as applying the inverse noise map of the readout error channel in post-processing, similar to how TEM applies the inverse gate noise channel, see Sec.~\ref{sec:tensor-network-error-mitigation}.
Due to the particularly simple form of the twirled readout error channel, the post-processing is simply a rescaling.  
In practice, when circuits are Pauli twirled, the distinction of Pauli $Z$/$X$/$Y$ becomes unnecessary and it is only the Pauli weight pattern between $\langle P \rangle_\text{twirled}$ and $\langle P \rangle_0$ that have to match. 
Then, an alternative way to obtain the correction factor $\langle P\rangle_0$ is to take the y-axis offset of a suitable Pauli from cycle benchmarking data, see Fig.~\ref{fig:cycle_benchmarking_schema}. 
Thus, in a setting where a noise model is calibrated for the gate error channels, TREX mitigation requires no further calibration data. 

\subsection{Discussion}
In this chapter we have introduced a selection of error mitigation techniques. 
In practice, it is often convenient to combine several techniques to exploit their respective strengths for mitigating different error sources. 
For example, Clifford data regression has been combined with a form of zero noise extrapolation~\cite{VNCDR} and virtual distillation~\cite{Bultrini2023unifying}, 
and symmetry verification has been combined with probabilistic error cancellation~\cite{papivc2025near}.
However, the fundamental bounds on the scaling of error mitigation predict an exponential overhead for any error mitigation technique~\cite{takagi2022fundamental, tsubouchi2023universal, quek_exponentially_2024}. 
Given this limitation, error mitigation should not be seen as a generally applicable method but as a means to extend the reach of imperfect quantum hardware.
While the exponential scaling is fundamental, reducing the base of this exponential by optimizing error mitigation techniques directly translates to larger available circuit volumes. 

While many popular error mitigation techniques are more heuristic, they can nevertheless work well in practice~\cite{kim2023evidence}. 
In contrast, rigorous error mitigation methods can in principle be bias-free but rely on accurate knowledge of the noise channels. 
In our presentation here, and in much of the existing error mitigation literature, models for gate noise have been discussed individually from readout and state preparation errors. 
This viewpoint necessitates certain assumption on symmetries for learning noise models, which may yield inaccurate error mitigation, as further discussed in Chap.~\ref{chap:dual_unitary_TEM}. 
Moreover, while TREX recovers the noise-free expectation value (in expectation) under the assumption of perfect state preparation~\cite{van2022model}, the denominator in Eq.~\eqref{eq:TREX_definition} may yield systematic errors if state preparation errors are present~\cite{haupt2025statepreparation, chen2025disambiguating}. 
Resolving this issue requires a more sophisticated way of noise learning where state preparation, gate, and readout errors are modelled holistically, which forms the basis for Chap.~\ref{chap:gauge_learning}. 

\chapter{Qudit computation with transmons}
\label{chap:qudit_processing}

\summary{This chapter is based on the articles \\``Ancilla-Free Implementation of Generalized Measurements for Qubits Embedded in a Qudit Space'' by \textbf{Laurin E. Fischer}, Daniel Miller, Francesco Tacchino, Panagiotis Kl. Barkoutsos, Daniel J. Egger, and Ivano Tavernelli., published in Phys. Rev. Research 4.3 033027, 2022~\cite{fischer2022ancillafree} and \\``Universal Qudit Gate Synthesis for Transmons'' by \textbf{Laurin E. Fischer}, Alessandro Chiesa, Francesco Tacchino, Daniel J. Egger, Stefano Carretta, and Ivano Tavernelli, published in PRX Quantum 4.3 030327, 2023~\cite{fischer2023universal}.
\bigskip

\noindent In this chapter, we develop a blueprint for a universal qudit-based quantum computer based on transmons. We show how to implement arbitrary unitary operations with a universal gate set that consists of general single-qudit unitaries and an entangling gate based on the cross-resonance effect. 
As potential applications we show how to leverage the control of qudit levels to implement single-qubit dilation POVMs and how to achieve more efficient circuit compilation compared to the qubit setting.}

In analogy to their classical counterparts, quantum computers usually encode information in binary systems -- known as qubits -- that consist of two physically distinct states.
Many experimental implementations, including trapped ions, superconducting qubits, neutral atoms, and spin qubits, embed the logical qubit subspace in a much larger multi-level Hilbert space~\cite{bruzewicz2019trappedion, clarke2008superconducting, shiquantumlogicentanglement2021, burkard2021semiconductor}. 
This full Hilbert space allows for a richer set of controls that is forgone by confining the computational space to qubits.
By coherently controlling additional states we enrich the set of available operations and make use of $d$-dimensional \emph{qudits} as the local units of information.

Qudits have several conceptual advantages over their qubit counterparts.
The number of qudits needed to reach the same Hilbert space dimension as a system of qubits is reduced by a factor of $\log_2(d)$.
For instance, ququarts, i.e., $d=4$, cut the number of computational units in half.
Moreover, qudits synthesize arbitrary unitaries more efficiently than qubits with regards to the number of required entangling gates~\cite{di2015optimal},
an advantage that already emerges in the qutrit case of $d=3$~\cite{gokhale2019asymptotic}.
This has led to proposals for efficient implementations of quantum algorithms in the qudit space~\cite{nikolaeva2022efficient, wang2020qudits}.
Applications that are formulated in a product space of multivalued units particularly benefit from a qudit encoding. 
These include the quantum simulation of bosonic modes that arise, e.g., in light-matter interaction processes~\cite{tacchino2021proposal,mazzolag2020,miessen2021spin-boson}, lattice gauge theories~\cite{rico2018nuclear,mathis2020,mazzola2021} and chemical vibrations and reactions~\cite{ollitrault_vib2020,macdonell2021analog}, but also classical problems like multivalued integer optimization~\cite{deller2022quantum}.
Moreover, qudit levels can simplify the implementation of qubit gates.
Finally, qudits exhibit more complex entanglement than qubits~\cite{kraft2018characterizing}, which can be leveraged to improve protocols such as superdense coding~\cite{hu2018beating} and quantum error correction (QEC) codes~\cite{prxgirvin}. 
This simplifies the implementation of QEC by strongly reducing the number of controlled multi-qubit operations. 

Qudit-based quantum information processing has recently been explored in trapped ions~\cite{ringbauer2021universal}, photonic systems~\cite{chi2022programmable}, Rydberg atoms~\cite{gonzalezcuadra2022hardware}, ultracold atomic mixtures~\cite{kasper2022universal}, and molecular spins~\cite{chiesa2020}. 
In this chapter, we conceptualize a superconducting qudit quantum processor, where $d$ qudit levels are encoded into the $d$ energetically lowest states of a transmon~\cite{koch2007chargeinsensitive}.
We propose a concrete scheme to transpile arbitrary unitary circuits into a universal set of hardware-native single-qudit and two-qudit gates, which in principle generalizes to any qudit dimension $d$.

Operating transmons as qutrits has already found several applications including multi-qutrit entanglement studies~\cite{blok2021quantum, cerveralierta2022experimental}, realization of multi-qubit gates~\cite{galda2021implementing, egger2019entanglement, nikolaeva2022decomposing},
excited state promotion readout~\cite{elder2020highfidelity, jurcevic2021demonstration}, quantum metrology~\cite{shlyakhov2018quantum}, fast resets~\cite{egger2018pulsed}, and the realization of two-qutrit quantum algorithms~\cite{roy2022realization}.
The ququart case has been considered in the context of single-qudit applications~\cite{kiktenko2015multilevel}, while several experiments have recently reported coherent control of a single ququart~\cite{cao2023emulatinga, seifert2023exploring, liu2023performing}.
However, up to now, it was unclear how to drive general two-qudit unitaries in transmons. 
For qubits, a popular realization of the CNOT gate relies on driving cross-resonance pulses~\cite{rigetti2010fully}, see Sec.~\ref{sec:superconducting_two_qubit_gates}.
Here, we propose a generalization of the echoed cross-resonance (ECR) gate in the qudit space as the fundamental entangling gate between two qudits. 
We study this generalized ECR gate by numerically simulating the time-dynamics of the system, demonstrating that it can reach ququart gate fidelities of $\sim 99\%$ with simple pulse shapes. 
We further show how to transpile an arbitrary $d^n \times d^n$ unitary on a system of $n$ $d$-dimensional transmon qudits into the hardware-native gate set of single-qudit rotations and the ECR gate. 
This constitutes the first practical blueprint for qudit unitary gate synthesis on superconducting qudits.  
As potential applications of this qudit-space toolbox, we show how to use qudit degrees of freedom to implement single-qubit dilation POVMs.

\section{Hardware model}
\label{sec:qudits_hardware_model}
We aim to build a qudit-based architecture that naturally extends existing qubit-based approaches for transmons. 
For now, we impose no explicit limit on $d$ but we discuss practical limits on this in the remainder of this chapter.
In our model, the $d$ levels of a qudit are encoded into the $d$ energetically lowest states of a transmon such that $H_0^{\text{lf}} \coloneqq \sum_{n=0}^{d-1} E_n \ketbra{n}{n}$ denotes the qudit Hamiltonian in its eigenbasis in the laboratory frame (lf).
Each transmon comes with a control line that allows driving the qudit at some frequency $\omega_D$.
We assume that this control frequency -- usually tuned to be at the qubit frequency -- can be modulated to drive transitions between neighboring states in the anharmonic ladder, as indicated in Fig.~\ref{fig:transmon_schema}. 
For the few lowest transitions, this is realistic for standard control stacks\footnote{For example, devices used in this thesis employ qubits with $0 \!\leftrightarrow\! 1$ transition frequencies of $\sim\! 5\,\text{GHz}$ and anharmonicities of $\sim\!-300\,\text{MHz}$. 
Drive pulses are generated by an arbitrary waveform generator with a sampling rate of $4.5\times 10^9\,\text{s}^{-1}$. 
We can thus apply modulations to the carrier frequency of up to approximately $\pm 1\,\text{GHz}$ (still oversampling by a factor of 4.5). 
The carrier frequencies of $\sim 4.7\,\text{GHz}$ and $\sim 4.3\,\text{GHz}$ required to address the $1\!\leftrightarrow\!2$ and $2\!\leftrightarrow\!3$ transitions, respectively, are thus well within the capabilities of the control hardware.}.
However, we do not consider couplings of the states beyond nearest-neighbors or two-photon effects such as sideband transitions. 
% Importantly, we assume no additional control knobs or drive lines compared to what has been established in the field for qubit processing. 
We start out by deriving the action of a single microwave drive on the qudit system building on the transmon model and notation introduced in Chap.~\ref{chap:superconducting_qubits}.
Under the above assumptions, the interaction Hamiltonian of Eq.~\eqref{eq_theo:interaction_Ham_lab_frame} generalizes to 
\begin{align}
\label{eq:qudit_interaction_Ham_lab_frame}
  H_{\text{int}}^{\text{lf}}(t) \coloneqq \epsilon(t)\sum_{n=0}^{d-2} g_n \ket{n}\bra{n+1} + \text{h.c.} \,,
\end{align}
where, as before, $\epsilon(t) \coloneqq \Omega(t)\cos\left(\omega_D t - \delta \right)$ is a drive with envelope $\Omega(t)$, drive frequency $\omega_D$, and phase $\delta$ and $g_n$ denotes the coupling strength to the $n \!\leftrightarrow\! n+1 $ transition for which we assume the relative coupling of a harmonic oscillator, i.e., $g_n \propto \sqrt{n+1}$.
To simplify the expressions, we transform the Hamiltonian $H^{\text{lf}}(t) = H_0^{\text{lf}} + H_{\text{int}}^{\text{lf}}(t)$ into a frame that rotates with the drive (rf) via the transformation $R(t) \coloneqq \sum_{n=0}^{d-1}e^{-i \omega_D n t} \ketbra{n}{n}$, see Appendix~\ref{app:rotating_frame}.
The resulting Hamiltonian in the rotation frame becomes
\begin{align}
    H^\text{rf}(t) &= {R}^\dagger(t) {H^{\text{lf}}}(t) {R}(t) - i\hbar {R}^\dagger(t) \frac{d}{dt} {R}(t) \\
    &= \sum_{n=0}^{d-1} \left(E_{n} - n \omega_D \right) \ketbra{n}{n} + 
    \frac{\Omega(t)}{2}\sum_{n=0}^{d-2} g_n \left( e^{-i\delta} + \underbrace{e^{i(\delta - 2\omega_D t)}}_{\substack{\text{neglected} \\ \text{in RWA}}} \right)  \ketbra{n}{n+1} + \text{h.c.}.
        \label{eq:rotating_frame_derivation}
\end{align}
Because the drive amplitude $\Omega$ is much smaller that the the frequency $\omega_D$, the effective dynamics occur on a much slower timescale than the period of the terms rapidly oscillating at $2\omega_D$. 
We thus neglect these terms which is known as the rotating wave approximation (RWA). 
For $d=2$, Eq.~\eqref{eq:rotating_frame_derivation} simplifies to the qubit case of Eq.~\eqref{eq:qubit_interaction_SQ}.
In the following, $H^\text{rf}$ is the fundamental expression from which we derive a qudit-space gate set 

\section{Single-qudit operations}
\subsection{Givens rotations}
Let us now investigate what gates are implemented when setting the driving frequency on resonance with the $n \!\leftrightarrow\! n+1 $ transition, i.e., $\omega_D= E_{n+1} - E_n$. 
Such a drive is said to be played in a \emph{frame} that consists of a carrier frequency $\omega_D$ and a phase $\delta$. 
When all other transitions are far detuned, evolving $H^\text{rf}(t)$ for a duration $T$ yields the qudit unitary\footnote{We use calligraphic math font for qudit-space unitaries and standard font for two-level unitaries.}
\begin{align}
\label{eq:givens_rotation_hardware}
\mathcal{R}_{n\leftrightarrow n+1}(\theta, \phi) &\coloneqq \mathcal{G}_{n\leftrightarrow n+1}(\theta, \phi)  \times
\text{diag}\left(e^{-i E_0 T}, \ldots, e^{-i (E_{d-1}- (d-1)\omega_D )T} \right).
\end{align}
The first term is a qudit operator $\mathcal{G}_{n\leftrightarrow n+1}(\theta, \phi)$ which applies a \emph{Givens} rotation
\begin{align} 
\label{eq:theory_given_rotation_definition}
 G(\theta, \phi)
 &\coloneqq \begin{pmatrix} 
    \cos(\theta/2) & -i \sin(\theta/2) e^{-i\phi}\\
     -i \sin(\theta/2) e^{i\phi}  & \cos(\theta/2)
    \end{pmatrix}
\end{align}
to the subspace spanned by $\ket{n}$ and $\ket{n+1}$ and acts as the identity everywhere else.
$G(\theta, \phi)$ is a rotation of angle $\theta \sim g_n \int_{0}^T \Omega(t) \text{d}t$ around an axis in the $xy$-plane with a polar angle given by the drive phase $\delta$.
The second term in Eq.~\eqref{eq:theory_given_rotation_definition} is a diagonal matrix that imprints phases on all non-resonant states.
Hence, the drive results in rotations akin to the qubit case between the resonantly-coupled levels but unavoidably comes with phase advances in all other levels. 
To isolate the action of the Givens rotation, we can absorb the undesired phase advances into the frames of subsequent drives.
Let us illustrate this with an example in the ququart space. 
A drive in the $1\!\leftrightarrow\!2$ frame implements the unitary
\begin{align}
\label{eq_app1:given_rotation_example_12}
\mathcal{R}_{1\leftrightarrow 2}(\theta, \phi) &=  e^{i( \omega_{1} - E_1)T} 
\begin{pmatrix} 
    e^{-i (\omega_{1} - \omega_{0})T} &  \begin{matrix}0&\quad0\end{matrix} & 0\\
     \begin{matrix}0\\0\end{matrix} & G(\theta, \phi) & \begin{matrix}0\\0\end{matrix} \\
     0 &  \begin{matrix}0&\quad0\end{matrix}  & e^{-i (\omega_{2} - \omega_{1})T}
\end{pmatrix}.
\end{align}
This results in a relative phase of $\Delta \phi_{0\leftrightarrow 1} = -\alpha_1 T$ between the states $\ket{0}$ and $\ket{1}$ and a relative phase of $\Delta \phi_{2\leftrightarrow 3}= \alpha_2 T$ between the states $\ket{2}$ and $\ket{3}$, where $\alpha_n$ are the anharmonicities (see Chap~\ref{chap:superconducting_qubits}). 
To correct these phases, $\Delta \phi_{0\leftrightarrow 1} $  and $\Delta \phi_{2\leftrightarrow 3}$ have to be subtracted from the phases $\phi$ of all subsequent pulses in the $0\!\leftrightarrow\!1$ and $2\!\leftrightarrow\!3$ frames, respectively. 
Generalizing from this example, under a drive $\mathcal{R}_{n\leftrightarrow n+1}$, the $m$-th level acquires a phase (ignoring global phases) of $\phi_{m\leftrightarrow m+1} \coloneqq \left((m-n)\omega_{n} + E_n - E_m\right)T$ which results in a phase difference of 
\begin{align}
\Delta\phi_{m \leftrightarrow m+1}= \left(\omega_{m} - \omega_{n} \right) T
\end{align}
This defines the necessary phase shift of all following pulses in the $m \leftrightarrow m+1$ frame.

\subsection{Phase gates}
\label{sec:qudit_phase_gates}
In addition to the drive unitaries $\mathcal{R}_{n\leftrightarrow n+1}(\theta, \phi)$, we define generalized $\mathcal{Z}_{n\leftrightarrow n+1}(\varphi)$-rotations, that act as $\text{diag}(e^{-i\frac{\varphi}{2}}, e^{i\frac{\varphi}{2}})$ on the states $\ket{n}$ and $\ket{n+1}$ and as the identity elsewhere.
For qubits, it is common to implement $z$-rotations virtually by adjusting the phases of subsequent drive pulses, see Sec.~\ref{sec:single-qubit_gates}.
We now generalize this concept to the qudit setting for $\mathcal{Z}$-gates.
Just like the phases of the frames can be adjusted to account for phase advances during $\mathcal{R}$-rotations, they can be manipulated to virtually implement $\mathcal{Z}$-gates.
Again, we illustrate this concept by an example of a $\mathcal{Z}$-rotation in a ququart space.
Consider the gate 
\begin{align}
\label{eq_app1:virtual_z_example}
\mathcal{Z}_{1\leftrightarrow 2}(\varphi) = \begin{pmatrix} 
    1 &  0 &  0& 0\\
     0 & e^{-i \frac{\varphi}{2}} & 0 & 0 \\
     0 & 0 & e^{i \frac{\varphi}{2}} & 0 \\
     0 & 0 & 0 & 1
    \end{pmatrix}
\end{align}
which applies a relative phase of $-\varphi$ between states $\ket{1}$ and $\ket{2}$. Therefore, an angle $\varphi$ needs to be  subtracted from the phase of all subsequent pulses played in the $1\!\leftrightarrow\!2$ frame. However, while the above gate leaves the levels $\ket{0}$ and $\ket{3}$ unchanged, it applies a relative phase of $\varphi/2$ between the levels $\ket{0}$ and $\ket{1}$, as well as between $\ket{2}$ and $\ket{3}$. Hence, in addition to affecting all following phases in the $1\!\leftrightarrow\!2$ frame, an angle $\varphi/2$ must be added to all drive phases in the $0\!\leftrightarrow\!1$ and $2\!\leftrightarrow\!3$ frames.

In summary, a sequence of gate instructions consisting of Givens rotations $\mathcal{G}(\theta_{\mathcal{G}}, \phi_{\mathcal{G}}) $ and phase gates  $\mathcal{Z}(\varphi_\mathcal{Z})$ can be implemented in the qudit space through pulses $\mathcal{R}(\theta_{\mathcal{R}}, \phi_{\mathcal{R}})$ where the rotation angles remain unchanged ($\theta_{\mathcal{R}} = \theta_{\mathcal{G}}$) and the phases of the pulses $\phi_{\mathcal{R}} $ depend on the phases $\phi_{\mathcal{G}}$ and $\varphi_\mathcal{Z}$ of all previously implemented gates of the sequence. 
This procedure is summarized as a pseudocode algorithm in Alg.~\ref{alg_app1:summary_of_phase_shifts}.

\begin{algorithm}
\caption[Implementation of a sequence of Givens rotations and phase gates via hardware-native pulses]{Implementation of a sequence of Givens rotations $\mathcal{G}$ and phase gates $\mathcal{Z}$ via hardware-native pulses $\mathcal{R}$ achieved by keeping track of all necessary phase shifts.}
\label{alg_app1:summary_of_phase_shifts}
\begin{algorithmic}
\STATE levels $\gets$ number of levels in qudit space
\STATE {phases} $\gets$ [$0, \dots, 0$]  \COMMENT{list of length {levels}$-1$}
\STATE {gates} $\gets$ sequence of $\mathcal{G}_{n\leftrightarrow n+1}(\theta, \phi)$ and $\mathcal{Z}_{n\leftrightarrow n+1}(\varphi)$ gates
\FOR{{gate} in {gates}}
    \IF{{gate} is of type $\mathcal{G}_{n\leftrightarrow n+1}$}
        \STATE $\theta$ $\gets$ rotation angle of {gate}
        \STATE $\phi$ $\gets$ phase of {gate}
        \STATE $T$ $\gets$ duration of {gate}
        \FOR{$m$ in [0, \dots, $n-1$, $n+2$, \dots, {levels}]}
            \STATE {phases}[$m$] $\gets$ {phases}[$m$] -- $(\omega_{m} - \omega_{n})T$
        \ENDFOR
        \STATE play pulse $\mathcal{R}_{n\leftrightarrow n+1}(\theta,\, {phases}[n]+\phi)$
    \ELSIF{{gate} is of type $\mathcal{Z}_{n\leftrightarrow n+1}$}
        \STATE $\varphi$ $\gets$ rotation angle of {gate}
        \STATE {phases}[$n$] $\gets$ {phases}[$n$] -- $\varphi$
        \STATE {phases}[$n-1$] $\gets$ {phases}[$n-1$] + $\frac{\varphi}{2}$
        \STATE {phases}[$n+1$] $\gets$ {phases}[$n+1$] + $\frac{\varphi}{2}$
    \ENDIF
\ENDFOR
\end{algorithmic}
\end{algorithm}

\subsection{General unitary synthesis}
\label{sec:general_SUD_synthesis}

We now present a scheme to implement any single-qudit unitary gate $U\in SU(d)$ with a sequence of hardware-native $\mathcal{R}$ rotations as defined in Eq.~\eqref{eq:givens_rotation_hardware}.
% gates using Alg.~\ref{alg_app1:summary_of_phase_shifts}. 
This is achieved by adapting an algorithm presented in Ref.~\cite{schirmer2002constructive} that decomposes $U$ (up to remaining phases on the diagonal) into a sequence of Givens rotations $\mathcal{G}_{n\leftrightarrow n+1}(\theta, \phi)$, following a strategy similar to a QR decomposition~\cite{golub1996matrix}. 
We extend this algorithm in two ways. 
First, we add $\mathcal{Z}$-gates to the sequence to fully decompose $U$ (including all relative phases) without increasing the number of pulses.
Second, making use of Alg.~\ref{alg_app1:summary_of_phase_shifts} allows us to replace the inaccessible $\mathcal{G}$-rotations in the decomposition of $U$ with the realistic $\mathcal{R}$-rotations.

For ease of notation, we will present this scheme for the ququart case of $d=4$. 
Let $\mathcal{U}^{(0)}$ be our target unitary. 
We will now iteratively multiply $\mathcal{U}^{(0)}$ with $\mathcal{G}$ or $\mathcal{Z}$ matrices until the resulting matrix is the identity. 
We denote the unitary after the $i$-th gate is applied to $\mathcal{U}^{(0)}$ by $\mathcal{U}^{(i+1)}$. 
The reduction to the identity matrix is accomplished by creating zeros in the off-diagonal entries starting from the top entry in the rightmost column.
We create a zero in the $(0, 3)$ element column of $\mathcal{U}^{(1)}$ with a Givens rotation $\mathcal{G}_{0\leftrightarrow 1}^{(0)}(\theta_1, \phi_1)$ that must satisfy $\mathcal{G}_{0\leftrightarrow 1}^{(0)} \big( \mathcal{U}^{(0)}_{0, 3}, \mathcal{U}^{(0)}_{1, 3}, \mathcal{U}^{(0)}_{2, 3}, \mathcal{U}^{(0)}_{3, 3} \big)^T = \big(0, \mathcal{U}^{(1)}_{1, 3}, \mathcal{U}^{(1)}_{2, 3}, \mathcal{U}^{(1)}_{3, 3}\big)^T $.
If $\mathcal{U}^{(0)}_{0, 3} = r_1 e^{i\delta_1}$ and $\mathcal{U}^{(0)}_{1, 3} = r_2 e^{i\delta_2}$ then the angles of the Givens rotation must be~\cite{schirmer2002constructive}
\begin{align}
    \label{eq_theo_decomp_algo_step_1}
    \theta_1 = 2 \arctan\left(\frac{r_1}{r_2}\right)\text{, and  } 
    \phi_1 = \dfrac{\pi}{2} - \delta_1 + \delta_2.
\end{align}
In the next two iterations, we similarly apply two more Givens rotation such that  $\mathcal{G}^{(1)}_{1\leftrightarrow 2} \big( 0, \mathcal{U}^{(1)}_{1, 2}, \mathcal{U}^{(1)}_{2, 3}, \mathcal{U}^{(1)}_{3, 3} \big)^T = \big(0, 0, \mathcal{U}^{(2)}_{2, 3}, \mathcal{U}^{(2)}_{2, 3}\big)^T$ and $\mathcal{G}^{(2)}_{2\leftrightarrow 3} \big( 0, 0, \mathcal{U}^{(2)}_{2, 3}, \mathcal{U}^{(2)}_{3, 3} \big)^T = \big(0, 0, 0, \mathcal{U}^{(3)}_{2, 3}\big)^T$. 

Due to unitarity, the remaining non-zero entry is a phase factor $\mathcal{U}^{(3)}_{2, 3} = e^{i\beta}$. 
A rotation $\mathcal{Z}^{(3)}_{2\leftrightarrow 3}$ with an angle $\varphi_z = -2\beta$ sets the phase $\beta$ to zero.
This finally results in the matrix
\begin{equation}
\mathcal{U}^{(4)} = \mathcal{Z}^{(3)}_{2\leftrightarrow 3} \mathcal{U}^{(3)} =
\left( 
\begin{array}{c | c} 
  \begin{array}{c c c} 
     \ast & \ast & \ast\\ 
     \ast & \ast & \ast\\ 
     \ast & \ast & \ast
  \end{array} &  
  \begin{array}{c} 
     0 \\ 
     0 \\ 
     0
  \end{array}
  \\ 
  \hline 
  \begin{array}{c c c} 
    0 & 0 & 0
  \end{array}
   & 1 
 \end{array} 
\right)
\end{equation}
that has been reduced to a $3\!\times\!3$ block. 
The above procedure is now repeated for the third column. 
This requires two Givens rotations and one $\mathcal{Z}$-rotation such that
\begin{align}
    \mathcal{Z}^{(6)}_{1\leftrightarrow 2} \mathcal{G}^{(5)}_{1\leftrightarrow 2} \mathcal{G}^{(4)}_{0\leftrightarrow 1} \big( \mathcal{U}^{(4)}_{0,2}, \mathcal{U}^{(4)}_{1, 2}, \mathcal{U}^{(4)}_{2, 2}, 0 \big)^T = (0, 0, 1, 0)^T.
\end{align}
Finally, applying the same strategy once more to the second column results in the identity matrix.
Our initial choice of an $SU(4)$ matrix assures that the final phase of the top left entry vanishes due to $\det(U) = 1$. 
As a result, applying the inverse of all gates in reverse order gives a decomposition of the target unitary $U$ into elementary operations of Givens rotations and (potentially virtual) $\mathcal{Z}$ gates:
\begin{align}
\label{eq_theo:U_final_decomposition_sequence}
U &=\mathcal{G}^{(0)\dagger}_{0\leftrightarrow 1} \,\, 
    \mathcal{G}^{(1)\dagger}_{1\leftrightarrow 2} \,\,
    \mathcal{G}^{(2)\dagger}_{2\leftrightarrow 3}  \,\,
    \mathcal{Z}^{(3)\dagger}_{2\leftrightarrow 3}  \,\,
    \mathcal{G}^{(4)\dagger}_{0\leftrightarrow 1}  \,\,
    \mathcal{G}^{(5)\dagger}_{1\leftrightarrow 2}  \,\,
    \mathcal{Z}^{(6)\dagger}_{1\leftrightarrow 2}  \,\,
    \mathcal{G}^{(7)\dagger}_{0\leftrightarrow 1}  \,\,
    \mathcal{Z}^{(8)\dagger}_{0\leftrightarrow 1} .
\end{align}
The above procedure straightforwardly generalizes to higher qudit dimension $d$, such that a general unitary $U\in SU(d)$ can be implemented with $d (d-1) / 2$ $\mathcal{G}$ gates and $d-1$ $\mathcal{Z}$ gates.
To simplify pulse calibrations, is it convenient to restrict the gate set to virtual $\mathcal{Z}$ gates and the gates $\sqrt{\mathcal{X}}_{n\leftrightarrow n+1} = \mathcal{G}_{n\leftrightarrow n+1}(\theta\! =\! \frac{\pi}{2}, \phi\!=\!0)$ which describe a $\pi/2$-rotation around the $x$-axis between the states $\ket{n}$ and $\ket{n+1}$. 
A general Givens rotation $\mathcal{G}(\theta, \phi)$ can be exactly realized by a sequence of two $\sqrt{\mathcal{X}}$ and three $\mathcal{Z}$ gates
\begin{align}
\label{eq_theo:decomposition_into_SX}
\mathcal{G}(\theta, \phi) = 
 \mathcal{Z}_{}(\phi - \dfrac{\pi}{2})\,
\sqrt{\mathcal{X}}_{}\,
\mathcal{Z}_{}(\pi - \theta)\,
\sqrt{\mathcal{X}}_{}\,
\mathcal{Z}_{}(-\phi - \frac{\pi}{2})
\end{align}
where we have omitted the subscripts $n\!\leftrightarrow\! n+1$~\cite{mckay2017efficient}.
Replacing every $\mathcal{G}$ gate with the decomposition in Eq.~\eqref{eq_theo:decomposition_into_SX} results in a decomposition of $U \in SU(d)$ that contains $d (d-1)$ $\sqrt{\mathcal{X}}$ gates.

\subsection{Experimental benchmarks}
\label{sec:qudits_experimental_benchmarks}

Having laid out the theory of how to drive a single transmon qudit, we now investigate to what extend this can applied in IBM Quantum superconducting hardware which was originally built to be optimized for qubit performance. 
% This gives us a sense of the most dominant error terms (rephrase). 
Since sufficient coherence of all involved states is required to perform information processing with higher-excited states, we present exemplary measurements of both $T_1$-type coherence from decay of higher-excited states and phase coherence related to charge dispersion. 
For all experiments presented in this chapter we use the standard single-qubit $R_x(\pi/2)$-gate that comes with a highly calibrated DRAG-pulse exposed to the user by IBM Quantum systems as the $\sqrt{\mathcal{X}}_{0 \leftrightarrow 1}$-pulse.  
All further pulse-level calibrations are implemented through Qiskit's pulse module~\cite{alexander2020qiskit, mckay2018qiskit}.
For the $1\!\leftrightarrow\!2$ and $2\!\leftrightarrow\!3$ transitions, we first calibrate the transition frequency with spectroscopy after preparing the initial states $\ket{1}$ and $\ket{2}$, respectively.  
For simplicity, we implement the $\sqrt{\mathcal{X}}$-gates on these transitions with Gaussian pulses. 
We choose a duration of $32\,\text{ns}$ for the $\sqrt{\mathcal{X}}_{1 \leftrightarrow2}$- and $14\,\text{ns}$ for the $\sqrt{\mathcal{X}}_{2 \leftrightarrow3}$-pulse. 
These durations are shorter than the $36\,\text{ns}$ standard single-qubit pulse to mitigate charge dispersion in higher-excited states by an increased spectral width. 
Simulations suggest that even shorter pulses are beneficial, see App.~\ref{app:details_simulation}. 
However, we find it more difficult to calibrate them. 
After fixing the pulse duration, we calibrate the angle of the rotations through sinusoidal fits to Rabi oscillations with varying pulse amplitudes.

\subsubsection{Qudit measurement}

We start by investigating how well the standard dispersive measurement (see Sec.~\ref{sec:transmons_measurements}) calibrated for the qubit generalizes to the qudit case. 
To this end, we prepare and measure the states $\ket{0}$, $\ket{1}$, $\ket{2}$, and $\ket{3}$ separately through a sequence of the appropriate $\mathcal{X}$-gates.
We then collect raw measurement data in the IQ plane where each state comes with a distribution of individual shots. 
We obtain a classifier into the four states by performing a quadratic discriminant analysis. 
An example of the resulting measurement data and the classifier decision boundaries is shown in Fig.~\ref{fig:experiment_results_POVM}\textbf{b}.
For each state, we obtain a characteristic signal that clusters in different regions of the IQ-plane.
However, we observe that the separation of higher-excited states in the IQ plane is significantly more difficult than for the qubit states $\ket{0}$ and $\ket{1}$\footnote{The qubit shown in Fig.~\ref{fig:experiment_results_POVM} represents the qubit with the best measurement separation between $\ket{2}$ and $\ket{3}$ we found on the IBM devices available during the time this study was conducted.}.
Specifically, around one quarter of the prepared states in $\ket{3}$ are identified as $\ket{2}$ and vice versa, see Tab.~\ref{tab:readout_assignments}.
To mitigate these misassignment errors, we apply a readout error mitigation based on the inversion of the misassignment matrix for the remaining experiments presented in this chapter.
That is, we recover the ideal outcome probability vector $\vec{p}_\text{ideal}$ by inverting the relationship $\vec{p}_\text{noisy} = M \vec{p}_\text{ideal}$ where $M_{ij} = p(i|j)$ are the probabilities of measuring $\ket{i}$ when $\ket{j}$ was prepared given in Tab.~\ref{tab:readout_assignments}. 
In case of small negative values in $\vec{p}_\text{ideal} = M^{-1} \vec{p}_\text{noisy}$ we use the closest non-negative vector in Euclidean norm as proposed in Ref.~\cite{maciejewski2020mitigation}.

\begin{table}
\begin{minipage}[c]{0.58\textwidth}
    \centering
    \scalebox{0.85}{
    % \begin{ruledtabular}
    \begin{tabular}{@{} c|rrrr @{}}
    \hline \hline
    % \headercell{Measured \\} & \multicolumn{4}{c@{}}{Prepared} \\
    & \multicolumn{4}{c@{}}{Prepared} \\
    Measured & $\ket{0}$ & $\ket{1}$ & $\ket{2}$ & $\ket{3}$ \\ 
    \midrule
    $\ket{0}$ & 98.3\,\% & 4.2\,\% & 0.6\,\% & 0.2\,\% \\
    $\ket{1}$ & 0.5\,\% & 88.8\,\% &  8.8\,\% &  2.1\,\%  \\
    $\ket{2}$ & 0.8\,\% &  6.9\,\% & 59.3\,\% & 22.8\,\% \\
    $\ket{3}$ & 0.4\,\% &  0.1\,\% & 31.3\,\% & 74.9\,\% \\
    \hline \hline
    \end{tabular}
    }
    % \end{ruledtabular}
\end{minipage}\hfill
    \centering
  \begin{minipage}[c]{0.42\textwidth}
    \caption[Readout assignments of higher-excited transmon levels]{\small Measured readout assignment error probabilities when preparing the four qudit states of a transmon. Data taken on qubit 0 of \textit{ibmq\_lima}.   \label{tab:readout_assignments}
    }   
\end{minipage}
\end{table}

\subsubsection{Decay of higher-excited states}
\begin{figure}
  \begin{minipage}[c]{0.55\textwidth}
    \includegraphics[width=\textwidth]{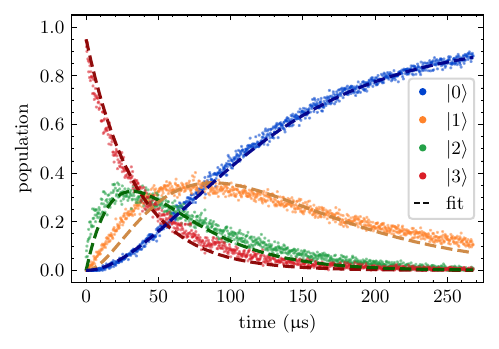}
  \end{minipage}\hfill
  \begin{minipage}[c]{0.43\textwidth}
    \caption[Measurement of population decay in higher-excited transmon levels]{\small Decay of state $\ket{3}$ in a transmon. 1000 measurements are taken at each time step which then undergo readout error mitigation to estimate the state populations.   
Fits are performed with a multi-channel exponential decay model.
Data taken on qubit $24$ of \textit{ibmq\_manhattan}.
    } \label{fig:app_T1_decay}
  \end{minipage}
\end{figure}

In a transmon qudit, higher-excited states have finite lifetimes and decay into lower-excited ones, similar to the $T_1$ decay of qubits (see Sec.~\ref{sec:amplitude_damping}). 
Here, we experimentally probe the decay times of the four lowest levels of a transmon qudit.
We prepare the state $\ket{3}$ by a ladder sequence of $\pi$-pulses $\mathcal{X}_{0\leftrightarrow1}, \mathcal{X}_{1\leftrightarrow2}$, and $\mathcal{X}_{2\leftrightarrow 3}$. 
The system is left to decay for a time $t$ prior to a projective measurement which extracts the populations $\vec{p}(t)$ from 1000 measurements at each time step, see Fig.~\ref{fig:app_T1_decay}. 

To estimate the $T_1$ times, we fit a model based on a multi-channel rate equation  ${\text{d}\vec{p}(t)/\text{d}t = \Gamma^T \vec{p}(t)}$.
$\Gamma$ is a $4\times4$ matrix that contains the decay rates $\Gamma_{ij}$ associated with the decay from $\ket{i}$ to $\ket{j}$ and diagonal entries $\Gamma_{ii} = -\sum_{j=0}^{i-1} \Gamma_{ij}$. We consider only the possible ``downward'' transitions ${\ket{3}\!\shortrightarrow\!\ket{2}}$, ${\ket{3}\!\shortrightarrow\!\ket{1}}$, ${\ket{3}\!\shortrightarrow\! \ket{0}}$, ${\ket{2}\!\shortrightarrow\! \ket{1}}$, ${\ket{2}\!\shortrightarrow\! \ket{0}}$, and ${\ket{1}\!\shortrightarrow\! \ket{0}}$, so $\Gamma_{ij} = 0$ for $i<j$. 
The $T_1$ times arise from all possible decay channels, e.g., ${ T_1^{\ket{3}} = 1/(\Gamma_{32} + \Gamma_{31} + \Gamma_{30}})$. The obtained fit parameters are summarized in Tab.~\ref{tab:measured_decay_times}. 
We find that the non-sequential transitions are strongly suppressed and the qudit mainly decays sequentially, i.e., ${\ket{3}\!\shortrightarrow\!\ket{2}\!\shortrightarrow\!\ket{1}\!\shortrightarrow\!\ket{0}}$. 
Importantly, the $\sim\! 30\,\upmu \text{s}$ $T_1$ times of $\ket{2}$ and $\ket{3}$ indicate that excited states offer lifetimes on the same order of the qubit transition (albeit reduced by a factor $\gtrsim 2$).
The fit in Fig.~\ref{fig:app_T1_decay}, accurately captures the population of the $\ket{0}$ state but deviates slightly for the other states. 
We attribute this to the significant misassignment errors present in the readout stage, which, even after readout error mitigation, remain significant, see above. 

\begin{table}
\begin{minipage}[c]{0.7\textwidth}
    \scalebox{0.8}{
    \centering
    \begin{tabular}{c|ccc|cc|c}
    \hline\hline
 & \multicolumn{3}{c|}{$\ket{3}$} & \multicolumn{2}{c|}{$\ket{2}$} & $\ket{1}$ \\
\hline
 $\Gamma_{ij}$ in & $\ket{3}\!\shortrightarrow\!\ket{2}$ &  $\ket{3}\!\shortrightarrow\!\ket{1}$ &  $\ket{3}\!\shortrightarrow\!\ket{0}$ &  $\ket{2}\!\shortrightarrow\!\ket{1}$ &  $\ket{2}\!\shortrightarrow\!\ket{0}$ &  $\ket{1}\!\shortrightarrow\!\ket{0}$  \\
 $(\upmu\text{s})^{-1}$ & 0.029 & 0.00 & 0.00 & 0.030 & 0.004 &  0.013 \\ 
\hline 
$T_1$ in $\upmu\text{s}$ & \multicolumn{3}{c|}{34.3} &  \multicolumn{2}{c|}{29.7} & \multicolumn{1}{c}{74.5} \\
    \hline\hline
\end{tabular}
    }
\end{minipage}\hfill
  \begin{minipage}[c]{0.3\textwidth}
    \centering
    \caption[Measurement of decay constants of higher-excited transmon levels]{\label{tab:measured_decay_times} \small Experimentally measured decay constants $\Gamma_{ij}$ and $T_1$ times for a transmon qudit extracted from Fig.~\ref{fig:app_T1_decay}.  
    }     
\end{minipage}
\end{table}

\subsubsection{Charge dispersion in state |3⟩}

\begin{figure}
  \begin{minipage}[c]{0.58\textwidth}
    \includegraphics[width=\textwidth]{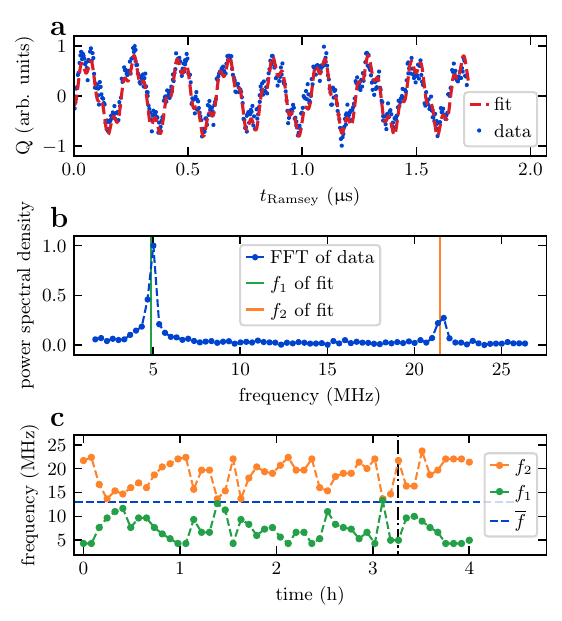}
  \end{minipage}\hfill
  \begin{minipage}[c]{0.4\textwidth}
    \caption[Measurement of charge dispersion in higher-excited transmon levels]{\small Measurement of charge dispersion in the $\ket{3}$ state. \textbf{a)} Ramsey oscillations between states $\ket{2}$ and $\ket{3}$ with a fit of two sinusoidals forming a beating pattern. Each data point is averaged over 1000 shots.  
\textbf{b)} Fourier transform of the signal in \textbf{a} with two contributing frequencies. 
\textbf{c)} Symmetrical fluctuations of the fit frequencies around a mean frequency when repeating the procedure over time. The vertical black line indicates data from panel \textbf{b}. Data taken on qubit 0 of \textit{ibm\_lagos}.
    } \label{fig:ramsey_measurement}
  \end{minipage}
\end{figure}

The theory of transmon outlined in Chap.~\ref{chap:qudit_processing} suggest that the exponential increase of charge dispersion with higher-excited states represents a fundamental limitation for qudit processing. 
Here, we investigate how this manifests in an IBM Quantum device by performing a direct measurement of the charge dispersion of the $\ket{3}$ state via a Ramsey interference experiment on the ${2\!\leftrightarrow\!3}$ transition. 
The experimental sequence consists of a preparation of the $\ket{2}$ state, followed by a $\pi/2$-pulse around the $x$-axis of the ${2\!\leftrightarrow\!3}$ transition, a delay time $t_{\text{Ramsey}}$, and finally a $-\pi/2$-pulse around the $x$-axis of the same transition. 
We measure the signal in the IQ-plane as $t_{\text{Ramsey}}$ is increased. 
This results in oscillations at the difference between the drive frequency and the true ${2\!\leftrightarrow\!3}$ transition frequency, see Fig.~\ref{fig:ramsey_measurement}\textbf{a}.
Transforming the signal into Fourier space reveals that the oscillation is a beating between two contributing frequencies $f_1$ and $f_2$, see Fig.~\ref{fig:ramsey_measurement}\textbf{b}. This can be attributed to quasiparticle tunneling across the qubit
junction~\cite{riste2013millisecond, peterer2015coherence}.
While repeating the Ramsey sequence 50 times, the two frequency components $f_1$ and $f_2$ fluctuate symmetrically around a center frequency $\overline{f}$ of $\sim 13\,\text{MHz}$, see Fig.~\ref{fig:ramsey_measurement}\textbf{c}. This represents the average detuning of the applied drive pulses. 
In total, this data suggests that the true frequency of the $2\!\leftrightarrow\!3$ transition fluctuates by as much as $15$-$20\,\text{MHz}$. 
For this particular qubit with a frequency of $\omega_0/(2\pi) = 5.2\,\text{GHz}$ and an anharmonicity of $\alpha_1/(2\pi)=-340\,\text{MHz}$, a direct diagonalization of the Hamiltonian from Eq.~\eqref{eq:transmon_hamiltonian} predicts a charge dispersion of $\epsilon_3 = 13.9\,\text{MHz}$. 
Our measurements are thus in reasonable agreement with theory.

We conclude that, with current hardware parameters, coherent operation of the full ququart subspace is difficult mainly due to charge noise in the 2-3 transition. 
This motivates us to look for an application of qudit control that requires only little phase coherence time for populations in the higher-excited states compared to the qubit levels. 
This would be the case when higher-excited levels are only populated for a short period of time prior to a projected measurement.
Such an application is given by dilation-POVM measurements, where the higher-excited states play the role of the dilation states instead of ancilla qubits, as will be presented in detail in the following section. 

\section{Application: Ancilla-free POVMs}
\label{sec:ancilla-free-POVMs}

\begin{figure}
  \begin{minipage}[c]{0.5\textwidth}
    \includegraphics[width=\textwidth]{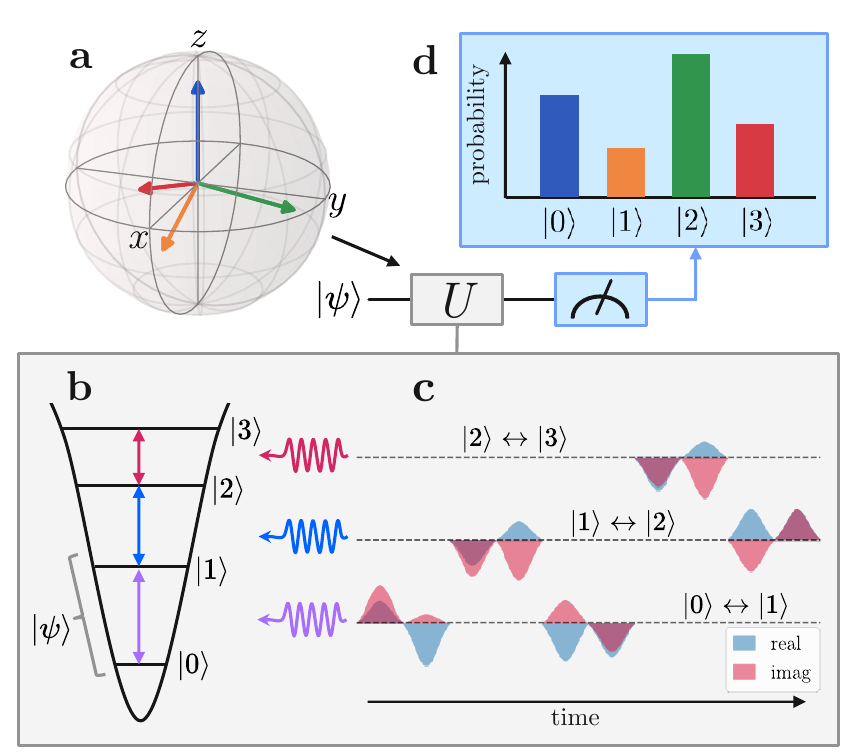}
  \end{minipage}\hfill
  \begin{minipage}[c]{0.48\textwidth}
    \caption[Schematic of a POVM implementation in qudit space]{\small Schematic of a POVM implementation in qudit space. 
        \textbf{a)} The $n=4$ rank-one, single-qubit POVM operators, represented on a Bloch sphere, define a four-dimensional unitary $U$ which encodes the POVM operators. 
        \textbf{b)} We realize this unitary on the qudit space in which the qubit state $\ket{\psi}$ is encoded.
        \textbf{c)} This can be achieved by a sequence of ten $\pi/2$-pulses 
        that couple adjacent levels.
        \textbf{d)} Finally, a projective measurement of the four states yields the outcome probabilities  of the four POVM operators.
    } \label{fig:overview_schematic}
  \end{minipage}
\end{figure}

In Chap.~\ref{sec:estimators_povm_measurements} we have outlined how informationally-complete (IC) POVM measurements are useful for various quantum algorithm subroutines, such as the efficient measurement of expectation values or enabling error mitigation protocols. 
However, the experimental realization of qubit-wise IC-POVMs requires a dilation that couples each qubit representing the trial state to two additional quantum states, see Sec.~\ref{sec:measurements_theory}. 
Traditionally, this is done in a tensor-product extension by coupling each qubit to an ancillary one before readout~\cite{garcia2021learning, brida2012ancillaassisted}.
This approach doubles the number of necessary qubits during the measurement stage, and therefore halves the usable portion of a quantum chip. 
Moreover, the limited connectivity of most quantum architectures leads to a significant SWAP-gate overhead~\cite{weidenfeller2022scaling}. 
To overcome the overhead of ancilla-based POVM implementations we propose to couple qubit state $\rho_\text{S}$ to the next-highest states of the qudit space, denoted $\ket{2}$ and $\ket{3}$ to realize a single-qubit POVM through a direct sum extension.
To realize the measurement of a POVM of operators $ M_0, \dots, M_{n-1}$, a specific unitary $U$ is applied to the ququart space such that the probability distribution of a subsequent 4-outcome projective measurement coincides with the POVM outcome distribution for the original state $\rho_\text{S}$, see Fig.~\ref{fig:overview_schematic}. 
A developed above, the POVM-encoding unitary $U$ on the ququart space can be implemented through a sequence of pulses that couple adjacent levels, as given in Eq.~\eqref{eq_theo:U_final_decomposition_sequence}.
In this context, $U$ can be chosen to have $U_{0, 3} = 0$, such that only five $\mathcal{G}$ gates (or ten $\sqrt{\mathcal{X}}$ gates) are required, see App.~\ref{app:sec_naimark_single_qubit}.

\subsection{Proof-of-principle experiment}
\label{sec:ancilla_free-proofofprinciple}

\begin{figure}
\centering
\includegraphics[width=\textwidth]{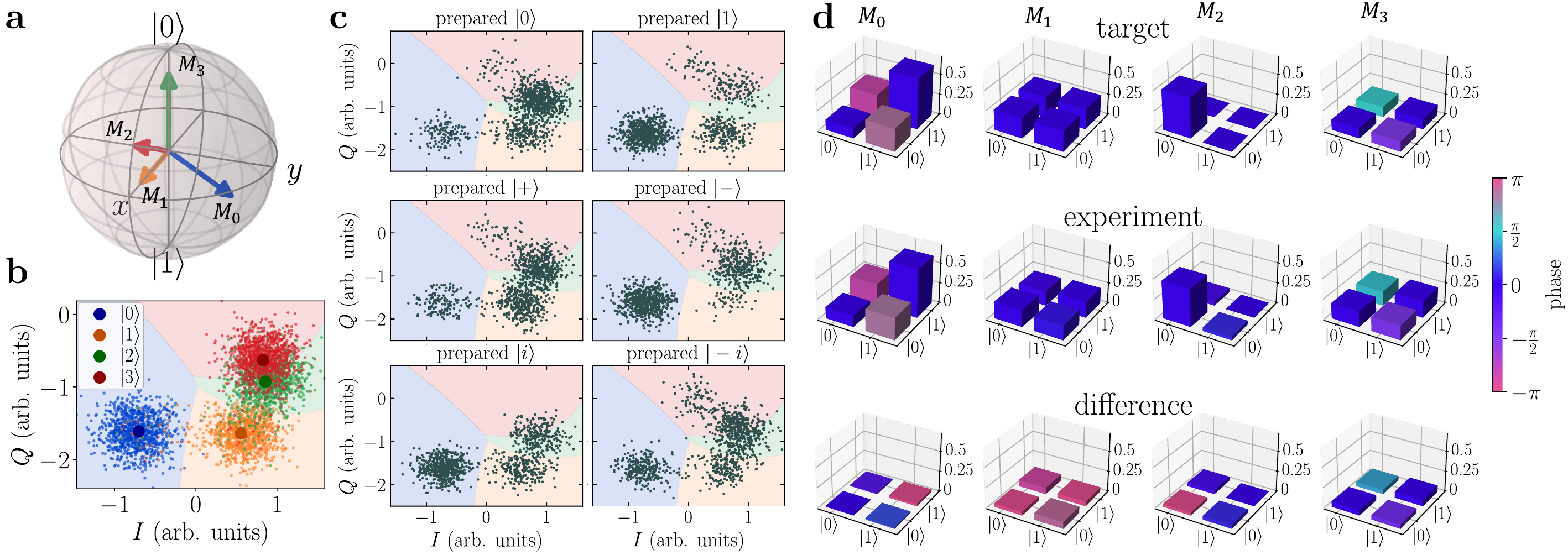}
\caption[Demonstration of a POVM measurement with qudit-space dilation]{\small Experimental realization of a single-qubit informationally complete POVM in the qudit space of a transmon qubit. 
\textbf{a)} Target POVM operators on the Bloch sphere. 
\textbf{b)} Calibration of measurement discrimination in the IQ-plane. Shaded regions show the decision boundaries of the classifier and large circles denote the average over all shots.
\textbf{c)} Raw data of POVM measurement outcomes for the six single-qubit stabilizer states. 
\textbf{d)} Characterization of the experimentally realized POVM operators $M_i$ plotted as matrix histograms. The top row shows
the theoretical target operators, the middle row shows the POVM operators obtained from a maximum-likelihood detector tomography of the experimental data after applying readout error mitigation, and the bottom row shows their difference. Data taken on qubit 0 of \textit{ibmq\_lima} with $E_\text{J}/E_\text{C} \sim 45$.}
\label{fig:experiment_results_POVM}
\end{figure}

We now present a proof-of-principle demonstration of a qudit-space POVM measurement in a superconducting transmon qubit.
We implement a single-qubit IC-POVM which consists of the target POVM operators
\begin{equation}
\label{eq:experimental_POVM_operators}
\begin{aligned}
    M_0 &= \frac{3}{4} \ket{\psi_0}\bra{\psi_0} ,& \quad 
    M_1 &= \frac{1}{2} \ket{+}\bra{+} ,\\ 
    M_2 &= \frac{1}{2} \ket{0}\bra{0} , 
    & \quad M_3 &= \frac{1}{4} \ket{-i}\bra{-i}  \qquad
\end{aligned}
\end{equation}
with $\ket{\psi_0} = \left(\ket{0} + \left(i-2\right)\ket{1}\right)/\sqrt{6}$.
The qudit-space unitary $U$ that encodes our chosen POVM is built up from the  gate sequence
\begin{align}
\label{eq:U_decompositon_exp_POVM}
U = \sqrt{\mathcal{X}}_{1\leftrightarrow2} \, \sqrt{\mathcal{X}}_{2\leftrightarrow3} \, \sqrt{\mathcal{X}}_{0\leftrightarrow1}  \mathcal{Z}_{1\leftrightarrow2}(\pi/2)\, \sqrt{\mathcal{X}}_{1\leftrightarrow2}\,\sqrt{\mathcal{X}}_{0\leftrightarrow1}.
\end{align}
The resulting POVM operators have a simple geometrical interpretation: three of the four operators points along the $x$-, $y$-, and $z$-axis of the Bloch sphere, see Fig.~\ref{fig:experiment_results_POVM}\textbf{a}.
We investigate how well our pulse sequence along with calibrated ququart measurement implements the desired POVM with maximum-likelihood (ML) quantum detector tomography (QDT) as outlined see Sec.~\ref{sec:detector_tomography}.  
Hereby, a set of reference states is prepared and measured by our POVM implementation. 
We choose the set of single-qubit states $\ket{0}$, $\ket{1}$, $\ket{+}$, $\ket{-}$, $\ket{i}$, and $\ket{-i}$ for this purpose.
% From the obtained outcome distributions, shown in Fig.~\ref{fig:experiment_results_POVM}\textbf{c}, the underlying experimental POVM operators can be estimated with a maximum-likelihood (ML) procedure, which guarantees that they form a valid POVM \cite{fiurasek2001maximumlikelihood}, see App. \ref{app:detector_tomography}.
Note that, on the Bloch sphere, the tomography states $\ket{-}$, $\ket{1}$, and $\ket{i}$ lie opposite the POVM operators $M_1$, $M_2$, and $M_3$, respectively.
They should thus have zero measurement probability of the corresponding outcomes, which is attested by a noticeable lack of counts in the respective regions of the IQ-plane in the raw data of Fig.~\ref{fig:experiment_results_POVM}\textbf{c}.
As a result, the operators obtained from the maximum-likelihood detector tomography are in good qualitative agreement with the theoretical target operators, see Fig.~\ref{fig:experiment_results_POVM}\textbf{d}. 
We quantify the fidelity through the \emph{operational distance} $D_\text{OD}$, a measure on the POVM space between the experimentally realized and the target POVM with $0 \leq D_\text{OD} \leq 1$ and $D_\text{OD}=0$ for coinciding POVMs, see Sec.~\ref{sec:distance_measures}.  
The raw measurement data presented in Fig.~\ref{fig:experiment_results_POVM}\textbf{c} yields $D_\text{OD} = 0.22$.
We identify the overlap of the detection regions in the IQ-plane between $\ket{1}$ and $\ket{2}$ and especially $\ket{2}$ and $\ket{3}$ as the main experimental limitation for qudit-based POVM measurements.
Specifically, in our experiments, around one quarter of the prepared states in $\ket{3}$ are identified as $\ket{2}$ and vice versa, see Tab.~\ref{tab:readout_assignments}.
We the readout error mitigation method proposed in Sec.~\ref{sec:qudits_experimental_benchmarks}, we can partially correct the measured raw data 
and achieve an improved $D_\text{OD}$ of $0.15$ between the theoretical and the ML-estimated experimental POVM.

\begin{figure}
  \begin{minipage}[c]{0.44\textwidth}
    \includegraphics[width=\textwidth]{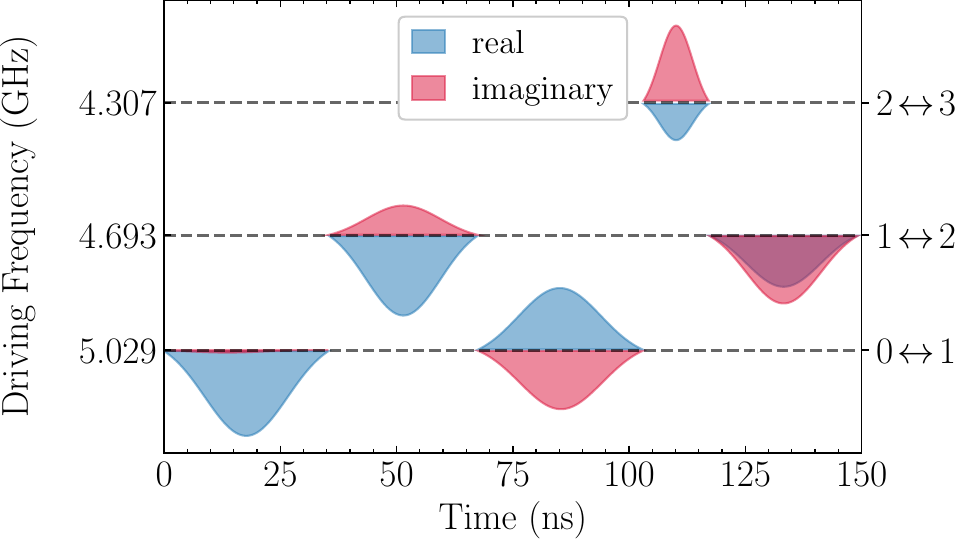}
  \end{minipage}\hfill
  \begin{minipage}[c]{0.54\textwidth}
    \caption[Pulse sequence for qudit-space dilation POVM experiment]{\small Pulse sequence of five $\sqrt{\mathcal{X}}$-gates encoding the experimentally demonstrated POVM operators from Eq.~\eqref{eq:experimental_POVM_operators}.
    The $0\!\leftrightarrow\!1$ drives are \textsc{Drag}-pulses, while the $1\!\leftrightarrow\!2$ and $2\!\leftrightarrow\!3$ transitions are driven with Gaussian pulses. 
    The phases of the drives are represented by real and imaginary parts of the pulse envelopes, depicted in arbitrary units.
    } \label{fig:pulse_sequence_exp_povm}
  \end{minipage}
\end{figure}

The difficulty to reliably distinguish the states $\ket{2}$ and $\ket{3}$ complicates the calibration of the average $2\!\leftrightarrow\!3$ transition frequency. 
At the moment, this renders the implementation of POVMs that require virtual $\mathcal{Z}_{2\leftrightarrow3}$-gates infeasible. 
This motivates the choice of the POVM operators in Eq.~\eqref{eq:experimental_POVM_operators} for our experiments.
The pulse sequence that implements the unitary from Eq.~\eqref{eq:U_decompositon_exp_POVM} is shown in Fig.~\ref{fig:pulse_sequence_exp_povm}. 
The non-trivial phases of the pulses, manifested in non-zero imaginary parts, arise from both the $\mathcal{Z}_{1\leftrightarrow2}$-gate in the sequence as well as from phases acquired during frame changes between different transitions. 
With the lack of $\mathcal{Z}_{2\leftrightarrow3}$-gates in the sequence, this POVM does not represent the most general case from Eq.~\eqref{eq_theo:U_final_decomposition_sequence}. 
Besides this simplification, it exhibits all features of our proposed scheme, thus constituting a reasonable compromise between practical feasibility on hardware that is not tailored for qudit operation and generality of the proof of principle.
The measurement pulses used in our experiment are the default pulses provided by the backend, which are optimized for maximal separation of the $\ket{0}$ and $\ket{1}$ states.
A large-scale implementation of qudit-space POVM measurements would require a more careful calibration of the readout pulses, which optimizes the separation of all four involved basis states. 
This would make the virtual $\mathcal{Z}_{2\leftrightarrow3}$-gates feasible and improve the $\sqrt{\mathcal{X}}_{2\leftrightarrow3}$-gate.

\subsection{Optimal transmon parameter regime}
\label{sec:ancilla-free_optimal_transmon_regime}

In the previous section, we demonstrated a qudit-based POVM measurement on a quantum device designed for optimal qubit operation with an $E_{\text{J}}/E_{\text{C}}$-ratio of $\sim 45$. 
However, the substantial charge dispersion in states $\ket{2}$ and $\ket{3}$ of the transmon suggests that larger $E_{\text{J}}/E_{\text{C}}$-ratios may be advantageous for qudit POVMs. 
This would sacrifice some anharmonicity to decrease charge noise.
We now quantitatively assess this trade-off through numerical pulse-level simulations, which account for both leakage errors due to finite anharmonicity and phase errors due to charge noise.

\begin{figure}
  \begin{minipage}[c]{0.44\textwidth}
    \includegraphics[width=\textwidth]{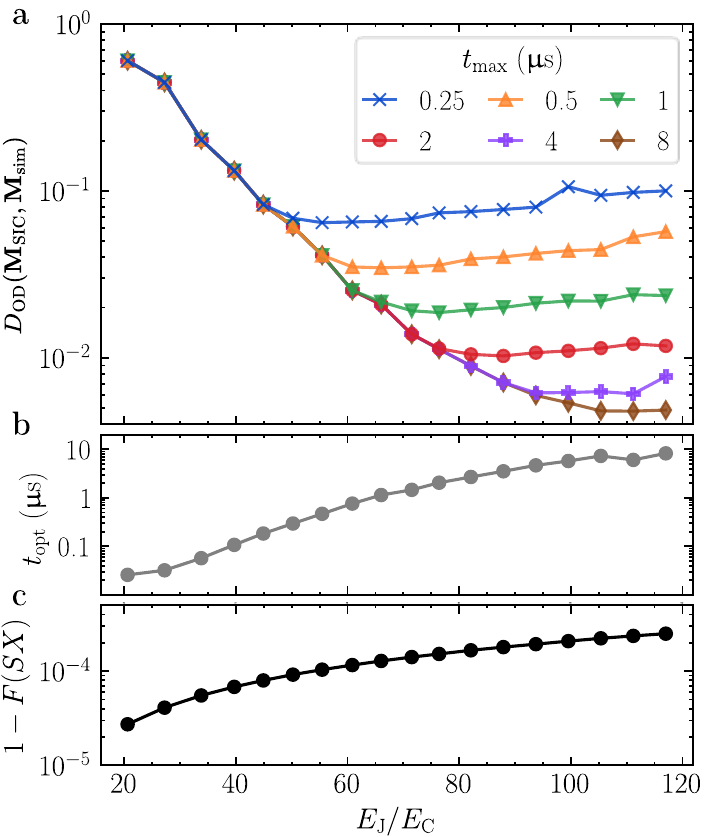}
  \end{minipage}\hfill
  \begin{minipage}[c]{0.54\textwidth}
    \caption[Pulse-level simulations of qudit POVM measurements under charge noise]{\small Simulations of a pulse schedule implementing a SIC-POVM under charge noise for different ratios $E_{\text{J}}/E_{\text{C}}$ of a transmon qubit with a frequency of $5\,\text{GHz}$.
    \textbf{a)} Operational distance $D_\text{OD}$ between theory and simulated POVMs for different maximal durations $t_\text{max}$ of the pulse schedules. 
    \textbf{b)} Optimal total durations $t_{\text{opt}}$ of the POVM pulse schedules that reach the best operational distance. 
    \textbf{c)} Average gate error of a single-qubit $SX$-gate realized through Gaussian pulses with a fixed duration of $36\,\text{ns}$. 
    } \label{fig:transmon_parameter_regime}
  \end{minipage}
\end{figure}

We start by probing how the achievable $D_\text{OD}$ depends on $E_{\text{J}}/E_{\text{C}}$, using a single-qubit symmetric, informationally complete (SIC) POVM $\vb{M}_{\mathrm{SIC}}$ as an example of a generic POVM (see Sec.~\ref{sec:measurements_theory}).
%It consists of four operators $\Pi_{\text{SIC}}^m = \frac{1}{2} \ket{\psi_m}\bra{\psi_m}$ with $\ket{\psi_0} = \ket{0}$ and $\ket{\psi_m} = \left( \ket{0} + \sqrt{2} e^{2\pi i (m-1)/3} \ket{1} \right)/\sqrt{3}$ with $ m\in \{1, 2, 3 \}$ that point towards the corners of a regular tetrahedron, see Fig.~\ref{fig:overview_schematic}\textbf{a}. 
In contrast to the experimentally demonstrated POVM in Eq.~\eqref{eq:experimental_POVM_operators}, $\vb{M}_{\mathrm{SIC}}$ requires implementing the pulse sequence from Eq.~\eqref{eq_theo:U_final_decomposition_sequence} in its full generality.
We simulate this sequence with Gaussian pulse envelopes on a single transmon by numerically integrating the time-dependent Schr{\"o}dinger equation.
For details on how we model charge dispersion and calibrate pulses see App.~\ref{app:details_simulation}.
As the $E_{\text{J}}/E_{\text{C}}$-ratio increases and charge noise becomes less prevalent, 
$D_\text{OD}(\vb{M}_{\text{SIC}}, \vb{M}_{\text{sim}})$
decreases, see Fig.~\ref{fig:transmon_parameter_regime}\textbf{a}. 
While the $D_\text{OD}$ is limited to $0.1$ for $E_{\text{J}}/E_{\text{C}} \sim 40$, it improves to $0.01$ for $E_{\text{J}}/E_{\text{C}} \sim 80$.
The change in anharmonicity with $E_{\text{J}}/E_{\text{C}}$ affects the duration of the pulse sequence that achieves the optimal $D_\text{OD}$, as plotted in Fig.~\ref{fig:transmon_parameter_regime}\textbf{b}.
In the low $E_{\text{J}}/E_{\text{C}}$-regime, short pulses are favored as a broad spectral width is required to cover the large spread of the charge noise, and leakage is minimal due to the large anharmonicity.
Conversely, with increasing $E_{\text{J}}/E_{\text{C}}$, the anharmonicity of the transmon is reduced, which amplifies leakage.
The optimal pulse durations thus increase with the ratio $E_{\text{J}}/E_{\text{C}}$. 

The longer the pulse sequence, the more it is subject to non-unitary processes like decoherence, which are not considered in our simulation. 
Consequently, there is a trade-off between the optimal durations of the pulses under unitary dynamics and noise induced by finite coherence times. 
We therefore limit the total duration of the POVM-encoding pulse sequence to different maximally allowed durations $t_{\text{max}}$, see Fig.~\ref{fig:transmon_parameter_regime}\textbf{a}.
We find that, for fixed $t_{\text{max}}$, the $D_\text{OD}$ improves with increasing $E_{\text{J}}/E_{\text{C}}$ until an optimal ratio is reached after which the $D_\text{OD}$ gradually increases. 
In the parameter regime of current IBM Quantum hardware ($E_{\text{J}}/E_{\text{C}} \sim 35\,\text{--}\,45$), the optimal POVM pulse sequence time is $\sim\! 100\,\text{ns}$. 
On this timescale, we do not expect decoherence to be significant, see Sec.~\ref{sec:qudits_experimental_benchmarks}.
For reference, single-qubit gates typically last $36\,\text{ns}$.
Finally, changing the transmon parameters also affects the conventional gates run in the quantum circuit prior to the POVM measurement. 
This is exemplified by the average gate fidelity $F$ of a single-qubit $36\,\text{ns}$ $SX$-gate, which is shown in Fig.~\ref{fig:transmon_parameter_regime}\textbf{c}. 
As $E_{\text{J}}/E_{\text{C}}$ increases from 20 to 120 the gate fidelity decreases by roughly one order of magnitude due to the reduced anharmonicity. 

The trade-off between anharmonicity and charge noise in a transmon qubit is a complex interplay of many factors, including coherence times, gate fidelities and gate speed~\cite{koch2007chargeinsensitive}.
Our simulations suggest that, when taking qudit POVM fidelities into account, the optimal hardware regime shifts towards higher $E_{\text{J}}/E_{\text{C}}$-ratios.
While this improves the quality of qudit-space POVM measurements, it comes at the expense of either slightly worse gate fidelities or slightly slower gate speeds, whose severity ultimately depend on the available coherence times.
Optimal control methods may alleviate such issues~\cite{werninghaus2021leakage}.

\section{Two-qudit operations}
\label{sec:two_qudit_operations}
In addition to general single-qudit transformations, one entangling two-qudit gate is required to form a complete set of universal qudit operations~\cite{brennen2005criteria}. 
In this section, we investigate whether extensions of the popular two-qubit cross-resonance (CR) gate (see Sec.~\ref{sec:superconducting_two_qubit_gates}) to the qudit space can serve this purpose. 
Previous studies of the CR gate have mostly focused on the qubit subspace~\cite{magesan2020effective, tripathi2019operation, malekakhlagh2020firstprinciples}, while Ref.~\cite{galda2021implementing} uses an echoed cross-resonance (ECR) sequence with the control prepared in $\ket{2}_c$ to enable a pulse-efficient decomposition of the three-qubit Toffoli gate. 
Going beyond this, we now investigate the action of the CR gate in the qudit subspace through numerical simulations of the dynamics of a two-qudit system.  

\begin{figure}
  \begin{minipage}[c]{0.54\textwidth}
    \includegraphics[width=\textwidth]{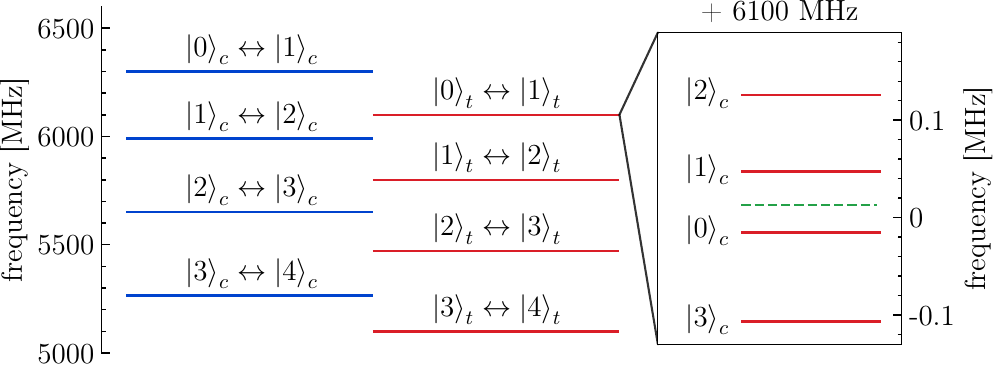}
  \end{minipage}\hfill
  \begin{minipage}[c]{0.44\textwidth}
    \caption[Transition frequencies and spectrum of a two-transmon model]{\small Transition frequencies of a two-transmon model. Blue and red lines denote the control and target, with bare frequencies $\omega_c/(2\pi)=6.3\,\text{GHz}$ and $\omega_t/(2\pi)=6.1\,\text{GHz}$.
The zoomed inset shows the dressing of the bare target frequency due to a capacitive coupling with the average transition frequency of $\overline{\omega}_t/(2\pi)$ in dashed green.
    } \label{fig:two_transmon_level_spectrum}
  \end{minipage}
\end{figure}

As is common for cross-resonance based architectures, we assume a pair of fixed-frequency transmons with an always-on coupling. As we have demonstrated in the previous sections, with the $E_{\text{J}} / E_{\text{C}}$-ratios of current IBM Quantum devices, the charge noise induced fluctuations of around 20 MHz in the $\ket{2}\leftrightarrow\ket{3}$ transition frequency are intolerable for full ququart operation.
Hence, in this chapter, we choose a model with parameters that are more suitable to qudit operation.
Specifically, for the control ($c$) and target ($t$) qudit of the two-qudit system, we choose frequencies of $\omega_c/(2\pi) = 6.3\,\text{GHz}$, $\omega_t/(2\pi) = 6.1\,\text{GHz}$, and anharmonicities of $\alpha_c/(2\pi) = -310\,\text{MHz}$, $\alpha_t/(2\pi) = -300\,\text{MHz}$.
This increases the $E_{\text{J}} / E_{\text{C}}$-ratio to $\sim\!70$, pushing the $\ket{2}\leftrightarrow\ket{3}$ frequency fluctuations down to $\sim 180\,\text{kHz}$. 
Moreover, the chosen parameters avoid crosstalk with a gap of at least $100\,\text{MHz}$ between different transitions, see Fig.~\ref{fig:two_transmon_level_spectrum}, which is smaller than the anharmonicities of each qudit.
% Fortunately, any potential leakage is avoidable by shaping the control pulses~\cite{schutjens2013singlequbit, vesterinen2014mitigating}.
The chosen parameters still result in charge noise of $\sim3\,\text{MHz}$ on the $\ket{3}\leftrightarrow\ket{4}$ transition rendering high-fidelity control of this transition difficult.
We thus focus on the ququart subspace. 

The transmons are coupled by a weak exchange interaction of strength $J/(2\pi)=1.8\,\text{MHz}$ that is routinely achieved in existing systems (see Appendix~\ref{app:details_simulation_two_qudit} for a detailed definition of the model Hamiltonians). 
This always-on coupling $J$ leads to a small shift of the eigenenergies of the joint two-transmon systems.  
From now on, when we denote a basis state as $\ket{n}_c\otimes\ket{m}_t$, we refer to these ``dressed'' basis states, i.e., the eigenstates of the coupled system. 
In this basis, the $\ket{0}_t \leftrightarrow \ket{1}_t$ transition frequency varies by $\sim \pm 100\,\text{kHz}$ depending on the state of the control qudit, see inset of Fig.~\ref{fig:two_transmon_level_spectrum}. 
We therefore set the drive frequency of the CR tones at $\overline{\omega}_t/(2\pi) = \omega_t/(2\pi) + 13\,\text{kHz}$ obtained by averaging over the lowest four states of the control. This keeps the detuning to each transition as small as possible.

\subsubsection{Simulation results}
\label{sec:transmon_qudits_simulation_results}
\begin{figure*}
\includegraphics[width=0.99\textwidth]{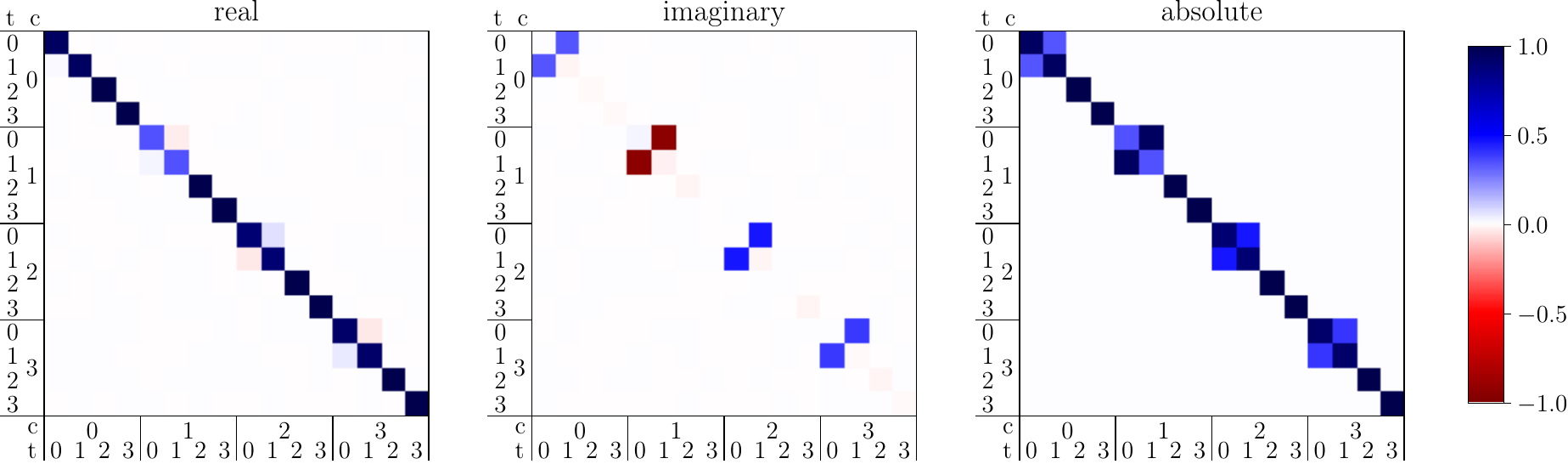}
\caption[Simulations of a single cross-resonance pulse in a two-qudit system]{\small Numerical simulation of the unitary matrix of a single CR pulse in the two-transmon dressed basis. The resulting operation is well described by an $R_x^{01}$ rotation on the target with a direction and angle that depend on the state of the control.}
\label{fig:single_cr_pulse_unitary}
\end{figure*}

We start by analyzing the action of a single CR pulse by numerically integrating the full dynamics of the time-dependent Schr{\"o}dinger equation of the two-transmon system in the $d=4$ subspace. 
We choose a Gaussian-square pulse envelope of amplitude $\Omega/(2\pi)=50\,\text{MHz}$ and duration $\tau$ with a Gaussian rise and fall to suppress leakage out of the $\ket{0}_t - \ket{1}_t$ subspace, as shown in Fig~\ref{fig:ECR_pulse_overview}\textbf{a} and detailed in Appendix~\ref{app:details_simulation_two_qudit}.
The chosen model parameters $\omega_c, \omega_t$, $\alpha_c$, $\alpha_t$, $J$ and $\Omega$ yield effective $Z\otimes X$ and $Z\otimes Z$ interaction strengths in the two-qubit subspace of $\omega_{ZX}/(2\pi) = -1.15\,\text{MHz}$ and $\omega_{ZZ}/(2\pi) = 31\,\text{kHz}$, respectively.
The resulting unitary on the two-ququart basis is shown in Fig.~\ref{fig:single_cr_pulse_unitary}.
We observe that the resulting dynamics create a $R_x^{01}(\varphi)$ rotation on the target, whose rotation angle and direction depend on the state of the control, where we use the notation 
$R_x^{n(n+1)}(\varphi) = \mathbbm{1}_{n} \oplus \exp(-i\tfrac{\varphi}{2} \sigma_x) \oplus \mathbbm{1}_{d-n-2}$
This generalizes the qubit case of the cross-resonance effect discussed in Sec.~\ref{sec:superconducting_two_qubit_gates}.
That is, the unitary is well described by
\begin{align}
\label{eqn:single_Cr_pulse_action}
U_\text{CR}(\vec{\varphi}) \coloneqq &\ketbra{0}{0}_c \otimes R_x^{01}(-\varphi_0) + \ketbra{1}{1}_c \otimes R_x^{01}(\varphi_1) \\
                            + &\ketbra{2}{2}_c \otimes R_x^{01}(-\varphi_2) + \ketbra{3}{3}_c \otimes R_x^{01}(-\varphi_3) \nonumber
\end{align}
where each rotation angle $\varphi_i$ is proportional to the total area under the pulse envelope.
% For the QEC application presented in Sec.~\ref{chap:qec_application}, we require a rotation angle of $\pm \pi$ in the $\ket{0}_c$ case for the echoed sequence. 
For the chosen model parameters, we obtain a maximal relative rotation between $\ket{0}_t$ and $\ket{1}_t$ ($\varphi_0 + \varphi_1 = \pi$) for a pulse duration of $\tau = 289\,\text{ns}$.
Up to local phases on the control and target, the unitary resulting from our pulse simulation reaches an average gate fidelity of $\fidelity = 99.93 \%$ to Eq.~\eqref{eqn:single_Cr_pulse_action} with angles $\vec{\varphi}=\left(\varphi_0, \dots \varphi_3\right) \approx \left(0.22,  0.78, 0.30, 0.26\right)\pi $. 
% This result justifies the intuition behind the schematic illustration in Fig.~\ref{fig:ECR_pulse_overview}\textbf{b}.
Note the difference in the rotation direction between the states $\{\ket{0}_c, \ket{2}_c, \ket{3}_c \}$ and $\ket{1}_c$. 

\begin{figure*}
\centering
\includegraphics[width=0.76\textwidth]{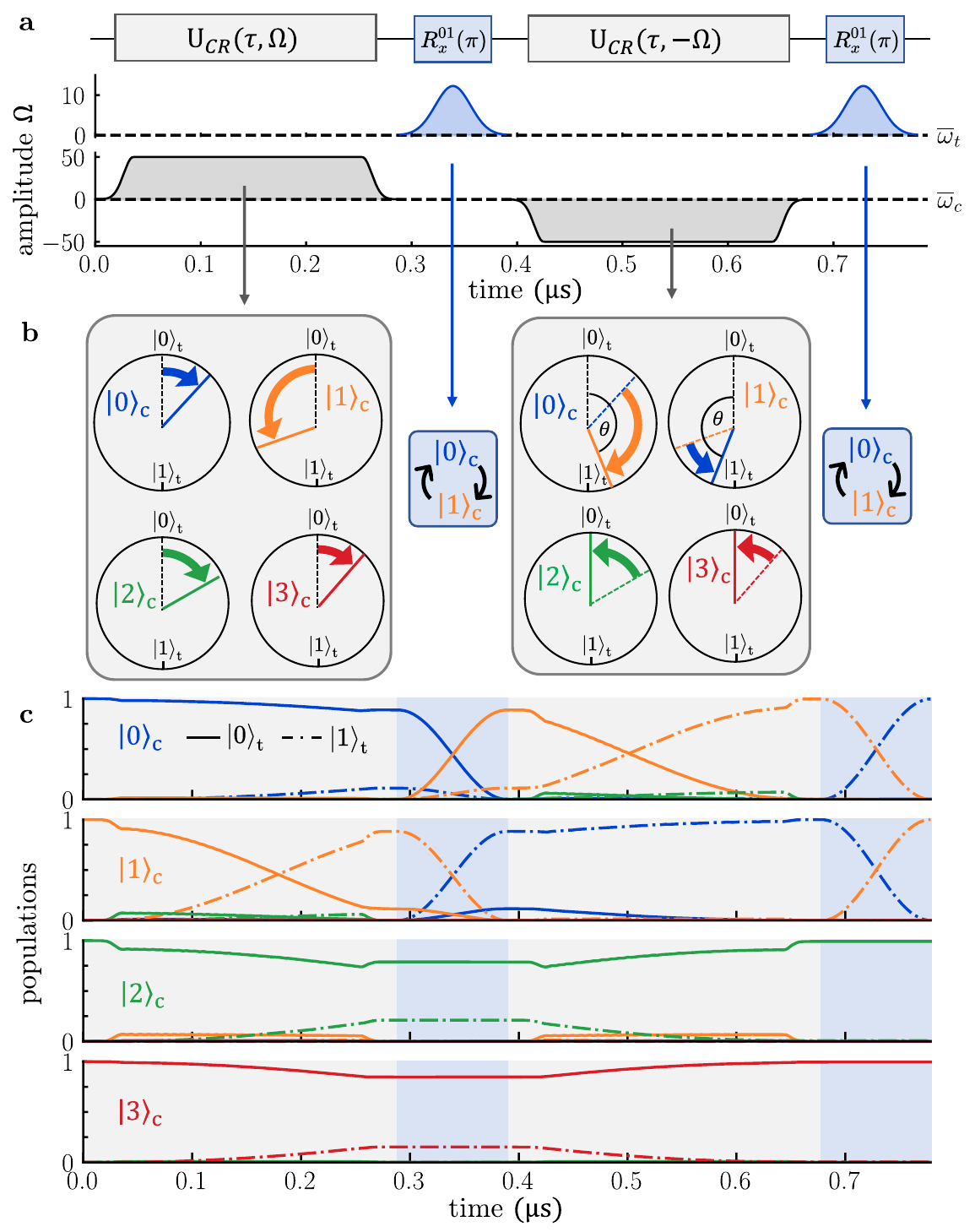}
    \caption[Qudit space action of the echoed cross-resonance gate]{\small Qudit space action of the echoed cross-resonance gate. \textbf{a)} The ECR pulse sequence applied to the control qudit consists of two cross-resonance tones (gray) played at the target qudit frequency $\overline{\omega}_t$  with opposite amplitudes, each followed by a $\pi$-pulse on the control. \textbf{b)} Action of each pulse on the initial target state $\ket{0}_t$ depending on the control state $\ket{\psi}_c$ on the Bloch sphere spanned by $\{\ket{0}_t, \ket{1}_t\}$. By construction of the echo sequence, a $\theta$ ($-\theta$) rotation is applied in the $\ket{1}_c$ ($\ket{0}_c$) case. \textbf{c)} Evolution of the populations starting from $\ket{0}_t$ depending on the control state $\ket{\psi}_c$ for the pulse sequence shown in \textbf{a}. Color denotes the state of the control while line style denotes the state of the target. Pulse durations are calibrated to $\theta = \pi$.
    } \label{fig:ECR_pulse_overview}
\end{figure*}

While we could use the $U_\text{CR}(\vec{\varphi})$ gate directly as a fundamental entangling gate, it depends on four different parameters, which complicates general unitary synthesis with this gate.
We instead want to make use of an echoed CR sequence to cancel unwanted $ZZ$ interactions and obtain a gate that depends only on a single parameter.
We thus introduce an echoing $\pi$-pulse $R_x^{01}(\pi)$ into the pulse sequence which is simulated as a Gaussian at the (average) frequency of the control qudit $\overline{\omega}_c$, see Fig.~\ref{fig:ECR_pulse_overview}.
We fix the pulse duration at $100\,\text{ns}$ and calibrate the amplitude to $12.3\,\text{MHz}$, obtaining a fidelity of $\fidelity = 99.99\%$, for details see Appendix~\ref{app:details_simulation_two_qudit}.
This is slower than current state-of-the-art $X$-gates in transmons to keep leakage minimal. 
As before, for simplicity, we omit a careful calibration of DRAG pulses by which leakage errors and pulse duration could be further reduced~\cite{motzoi2009simple}.
Assuming that the reversed amplitude in the second CR tone of the ECR sequence reverses all rotation angles $\vec{\varphi}$ in Eq.~\eqref{eqn:single_Cr_pulse_action}, we expect that the unitary of the full ECR sequence is
\begin{align}
\label{eqn:ECR_unitary}
U_\text{ECR}(\theta) \coloneqq &\ketbra{0}{0}_c \otimes R_x^{01}(-\theta) + \ketbra{1}{1}_c \otimes R_x^{01}(\theta) \\
+ &\ketbra{2}{2}_c \otimes \mathbbm{1} + \ketbra{3}{3}_c \otimes \mathbbm{1} \nonumber
\end{align}
with $\theta = \varphi_0 + \varphi_1$. 
Indeed, with the CR tones calibrated as described above, our simulation of the entire pulse sequence obtains an average gate fidelity of $99.56 \%$ for the targeted rotation angle $U_\text{ECR}(\theta = \pi)$ (up to local phase gates).
This is the unitary error of the gate. 
To estimate the additional effect of incoherent error channels, we add amplitude damping with a $T_1$-time of $310\,\upmu\text{s}$ and pure dephasing with a $T_2$ time of $170\,\upmu\text{s}$ to the simulation as detailed in Appendix~\ref{app:details_simulation_two_qudit}, which corresponded to median $T_1$ and $T_2$ times of the IBM Quantum device \textit{ibm\_sherbrooke} at the time of writing~\cite{abughanem2025ibm}.
This reduces the fidelity to $98.66\%$. 
With values of $474\,\upmu\text{s}$ for $T_1$ and $666\,\upmu\text{s}$ for $T_2$, corresponding to the best pair of neighboring qubits, the fidelity becomes $99.22\%$.
Here, we are primarily interested in understanding the limits to the unitary error of this gate. 
We thus leave exploring the trade-off between the unitary gate error -- which is minimal for longer gate durations -- and the incoherent gate error that increases with the gate duration for future work. 

The evolution of the populations in the two-transmon system under the echoed CR sequence is shown in Fig.~\ref{fig:ECR_pulse_overview}\textbf{c}, with details given in Appendix~\ref{app:details_simulation_two_qudit}.
The remaining unitary error of the gate originates mainly from the detuning of the CR tones to the target frequency in the cases where the control is in $\ket{2}_c$ and $\ket{3}_c$ (see Fig.~\ref{fig:two_transmon_level_spectrum}).
This leads to a small $Z$-contribution in the effective rotation axis in each respective subspace, which is not fully reversed by the echoed CR sequence. 
These effects could potentially be resolved by adding rotary tones~\cite{sundaresan2020reducing}, including virtual phase gates into the sequence~\cite{mckay2017efficient} or more advanced qudit-based optimal control techniques~\cite{seifert2022timeefficient, simm2023two}.
We find that leakage out of the $\ket{0}_t - \ket{1}_t$ subspace is not a relevant error source with populations of those levels remaining under $10^{-5}$ after the ECR sequence. 

For qudit-based circuit decomposition, the ECR gate from Eq.~\eqref{eqn:ECR_unitary} is particularly convenient, as it only depends on a single parameter $\theta$ which is tunable through the duration of the CR pulses. 
This makes it a suitable candidate to systematically compose arbitrary qudit circuits.
In the following, we present such a decomposition routine that implements general qudit unitaries through the ECR gate and single-qudit gates. 

\section{Universal gate synthesis}
\label{sec:universal_gate_synthesis_qudits}

\subsection{Qudit transpilation}

Any set of arbitrary single-qudit gates combined with a single entangling two-qudit gate is in principle exact-universal~\cite{brylinski2002universal, brennen2005criteria}.
Several constructive decomposition routines exist which rely on different choices of two-qudit gates~\cite{wang2020qudits, brennen2005efficient}. 
For qubits, the quantum Shannon decomposition (QSD) is a powerful tool to synthesize arbitrary unitaries~\cite{shende2006synthesis}. 
We build on the multivalued QSD that generalizes the QSD to the qudit setting~\cite{di2013synthesis}. 
Within this framework, the circuit complexity, quantified by the number of two-qudit gates required to achieve arbitrary $N$-qudit unitaries, is reduced by a factor of $d-1$ compared to the qubit setting of $d=2$, highlighting the comparative efficiency of qudit circuits~\cite{di2015optimal}. 
The multivalued QSD iteratively reduces the desired unitary to block matrices that contain only one-qudit and two-qudit unitaries. The remaining two-qudit blocks consist of singly-controlled gates 
\begin{align}
\label{eq:def_singly_controlled_gate}
C^m[U] \coloneqq \ketbra{m}{m} \otimes U + \sum_{i\neq m} \ketbra{i}{i} \otimes \mathbbm{1}
\end{align}
that apply a unitary $U\in{\rm SU}(d)$ on the target qudit if and only if the control qudit is in the basis state $\ket{m}$~\cite{di2013synthesis}.

\begin{figure}
  \begin{minipage}[c]{0.58\textwidth}
    \includegraphics[width=\textwidth]{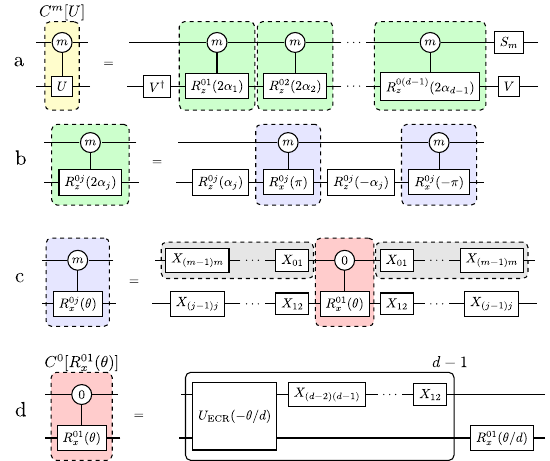}
  \end{minipage}\hfill
  \begin{minipage}[c]{0.4\textwidth}
    \caption[ Gate decomposition to implement a general singly-controlled two-qudit gate]{\small Gate decompositions to implement a general singly-controlled two-qudit gate $C^m[U]$. \textbf{a)} Diagonalization of $C^m[U]$ leads to a sequence of controlled $z$-rotations $R_z^{0j}$ between the $0^{\text{th}}$ and $j^{\text{th}}$ level. \textbf{b)} Each $C^m[R_z^{0j}]$ is implemented through two $C^m[R_x^{0j}]$ rotations and local phase gates. \textbf{c)} Decomposition of $C^m[R_x^{0j}]$ into local permutations and a single $C^0[R_x^{01}]$ gate. Gray shaded gates cancel between consecutive $C^m[R_x^{0j}]$ gates as they arise in \textbf{b}. \textbf{d)} Realization of the $C^0[R_x^{01}(\theta)]$ gate through echoed cross-resonance gates.
    } \label{fig:decomposition_circuits}
  \end{minipage}
\end{figure}

We now present a decomposition routine to realize $C^m[U]$ through single-qudit gates and the $ECR(\theta)$ gate defined in Eq.~\eqref{eqn:ECR_unitary}.
This decomposition holds for any $d$, as long as the $ECR$ gate acts as $\ketbra{n}{n}\otimes \mathbbm{1}$ for all states $n\geq 2$. 
We diagonalize $U = V D V^\dagger$ such that the diagonal matrix $D = e^{i\gamma} \text{diag}(e^{-i(\sum_{j=0}^{d-1} \alpha_j)}, e^{i\alpha_1}, \dots, e^{i\alpha_{d-1}})$ can be decomposed into $R_z^{0j}$ rotations as $ D = e^{i\gamma} \prod_{j=1}^{d-1}  R_z^{0j}(2 \alpha_j)$.
$C^m[U]$ can then be implemented as a product of controlled phase gates between the target's $\ket{0}$ and $\ket{j}$ states
\begin{align}
\label{eqn:D_decomposition_RZ}
C^m[U] = \left(S_m \otimes V \right) \prod_{j=1}^{d-1}  C^m[R_z^{0j}(2 \alpha_j)] \left(\mathbbm{1} \otimes V^\dagger \right)
\end{align}
with a phase gate $S_m = \sum_{j = 0}^{d-1} e^{i\gamma \delta_{jm}}\ket{j}\bra{j}$, see Fig.~\ref{fig:decomposition_circuits}\textbf{a}.
Next, each $C^m[R_z^{0j}(2\alpha_j)]$ gate is expressed through two controlled $x$-rotations, yielding (see Fig.~\ref{fig:decomposition_circuits}\textbf{b})
\begin{align}
\label{eqn:CZ_rot_through_CX_rot}
C^m[R_z^{0j}(2\alpha_j)] = C^m[&R_x^{0j}(-\pi)] \left(\mathbbm{1}\otimes R_z^{0j}(-\alpha_j)\right) \\
\times &C^m[R_x^{0j}(\pi)] \left(\mathbbm{1}\otimes R_z^{0j}(\alpha_j)\right). \nonumber
\end{align}
This shifts all angular dependence into local phase gates we implement virtually. 
The general $m$-controlled $R_x^{0j}$ gates from Eq.~\eqref{eqn:CZ_rot_through_CX_rot} can be realized through $0$-controlled $R_x^{01}$ rotations by applying single-qudit permutation gates on the control and the target, as shown in Fig.~\ref{fig:decomposition_circuits}\textbf{c}. 
Here, $X_{n(n+1)}$ denotes a swap of the levels $i$ and $i+1$ which is equivalent to $R_x^{n(n+1)}(\pi)$ up to virtual local phases. 
Note that the permutation gates on the control qudit to shift the control from $m$ to 0, highlighted in gray in Fig.~\ref{fig:decomposition_circuits}\textbf{c}, need to be applied only once at the beginning and the end of the decomposition sequence.

The final step is now to realize the remaining $C^0[R_x^{01}(\theta)]$ gates with the $ECR$ gate defined in Eq.~\eqref{eqn:ECR_unitary}. 
Since the $ECR$ gate acts non-trivially on the target for both control states $\ket{0}_c$ and $\ket{1}_c$, $C^0[R_x^{01}(\theta)]$ cannot be implemented with a single $U_\text{ECR}(\theta)$ gate.
Instead, we split the rotation into $d-1$ steps of $U_\text{ECR}(-\theta/d)$ and permute the levels of the control between each step, as shown in Fig.~\ref{fig:decomposition_circuits}\textbf{d}. 
This way, the action on the target is $R_x^{01}(\theta(d-1)/d)$ when the control is in $\ket{0}_c$ and $R_x^{01}(-\theta/d)$ when the control is in any other state.  
Applying an $R_x^{01}(\theta / d)$ gate on the target finally recovers the desired $C^0[R_x^{01}(\theta)]$ rotation, since
\begin{align}
\label{eqn:C0RX_to_ECR}
C^0[R_x^{01}(\theta)] = \left(\mathbbm{1} \otimes  R_x^{01}\left(\tfrac{\theta}{d}\right) \right) 
    \times \left(  \left( \prod_{j=2}^{d-1} X_{(j-1)j} \otimes \mathbbm{1} \right) U_\text{ECR}(-\tfrac{\theta}{d})  \right)^{d-1} .
\end{align}

\begin{figure*}
\includegraphics[width=\textwidth]{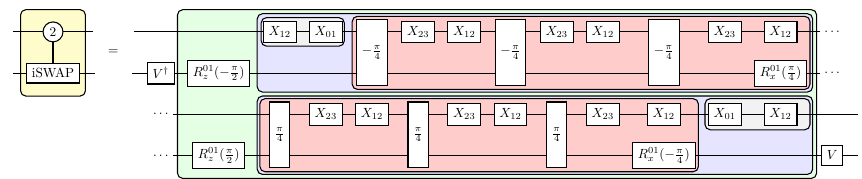}
\caption[Decomposition of an iSWAP gate between two ququarts]{\small
Example application of the general qudit transpiler decomposing a $\ket{2}$-controlled iSWAP gate between two ququarts. 
Colors indicate the different steps of the transpilation shown in Fig.~\ref{fig:decomposition_circuits}, while two-qudit boxes denote $ECR(\theta)$ gates. 
}
\label{fig:Iswap_example}
\end{figure*}

Fig.~\ref{fig:decomposition_circuits} shows the decomposition in the most general case.
In practice, the complexity of a desired two-qudit gate can be much simpler, as illustrated by the decomposition of a $\ket{2}$-controlled iSWAP gate $C^2[R_x^{12}(-\pi)]$, shown in Fig.~\ref{fig:Iswap_example}. 
Following the notation from above, $\alpha_2, \alpha_3$ and $\gamma$ of the diagonal $D$ are zero, and $\alpha_1 = -\pi/2$.
Therefore, only one of the green blocks from Fig.~\ref{fig:decomposition_circuits}\textbf{a} appears in the decomposition.
The single-qudit diagonalization gates $V$ can be realized with two $R_x^{01}$ gates, one $R_x^{12}$ gate and virtual phase gates. 

\subsection{Comparison to qubits}
\label{sec:compiler_comparison_qubits}

In general, the presented decomposition routine requires $O(d^2)$ two-qudit entangling gates and $O(d^3)$ single-qudit gates.
In the $d=4$ case, any $C^m[U]$ unitary can be synthesized with 18 $U_\text{ECR}(\pm \pi/4)$ gates and $56 + 2m$ single-qudit gates (not counting virtually implemented phase gates). 
This construction adds an overhead of single-qudit gates, but makes an efficient use of the entanglement generation rate of the CR effect as the total duration of the ECR pulses is proportional to $\theta (d-1)/d$. 
In comparison, a qubit CNOT gate is locally equivalent to a $U_\text{ECR}(\pm \pi/2)$ pulse sequence, and thus takes roughly twice as long as a $U_\text{ECR}(\pm \pi/4)$ qudit gate.

To benchmark the efficacy of our qudit transpiler, we compare its performance against a state-of-the-art qubit transpiler available in the software package Qiskit~\cite{qiskit2024}.  
Qiskit's transpiler uses a column-by-column decomposition developed in Ref.~\cite{iten2016quantum}.
We focus on the task of unitary synthesis of general $16 \times 16$ matrices $U_\text{target}$. 
For the qudit case, we use two ququarts with a basis gate set made of bidirectional $U_\text{ECR}(\pm \pi/4)$ gates, single-qudit gates and virtual phase gates. 
We make use of an iterative cosine-sine decomposition~\cite{chen2013qcompiler} to synthesize $U_\text{target}$ from a sequence of block-diagonal $C^m[U]$ gates, see Appendix~\ref{app:general_synthesis} for details. 
For the qubit case, we consider four linearly connected qubits and a basis gate set of bidirectional CNOT gates as well as single-qubit $\sqrt{X}$ gates and parametric virtual phase gates.

To compare the gate cost between the qubit and the qudit case, we equate two $U_\text{ECR}(\pi/4)$ gates with one CNOT gate, and decompose all single-qudit gates into $R_x^{n, n+1}(\pi/2)$ gates which we equate to a single-qubit $\sqrt{X}$ gates.
We do not count virtual phase gates since they are implemented purely in software.
We find that the qubit transpiler requires 286 CNOT and 244 $\sqrt{X}$ gates to synthesize random SU$(16)$ gates.
The ququarts transpiler achieves the same unitary with an equivalent of 170 CNOT and 2776 $\sqrt{X}$ gates.
This highlights the central tradeoff between qudit and qubit transpilation: The two ququarts reduce the required number of entangling gates by $\approx 40\%$, at the cost of increasing the number of single-qudit gates by a factor of~$\sim 11$.
Since single-qudit gates are typically an order of magnitude faster with at least an order of magnitude higher fidelity, there might be a regime where the qudit transpilation is favorable despite the larger total gate count.

\begin{figure}
  \begin{minipage}[c]{0.58\textwidth}
    \includegraphics[width=\textwidth]{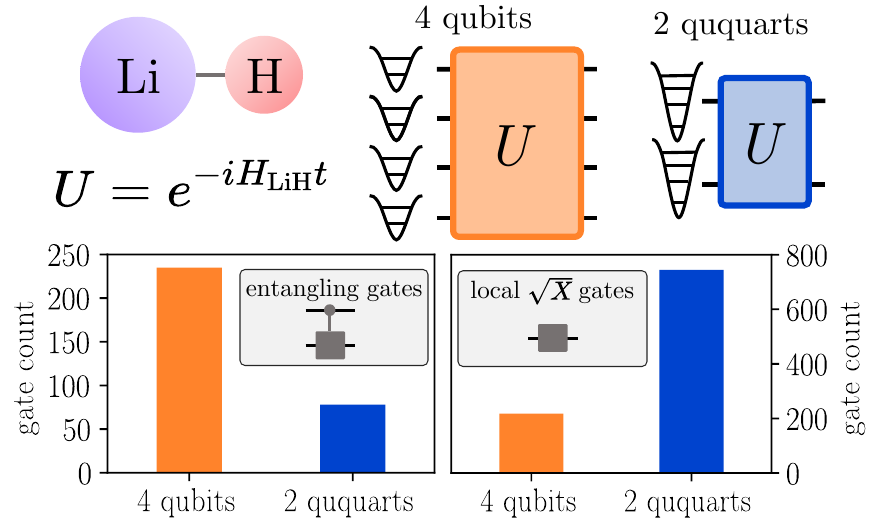}
  \end{minipage}\hfill
  \begin{minipage}[c]{0.4\textwidth}
    \caption[Benchmark study of the qudit transpiler for Hamiltonian simulation]{\small Benchmark study of the qudit transpiler on the Hamiltonian exponentiation of a lithium hydride molecule. 
Two ququarts require only a third of the entangling gates that four linearly connected qubits require, at the expense of 3.4 times as many single-qudit $\sqrt{X}$ gates.
    } \label{fig:gatecount_comparison}
  \end{minipage}
\end{figure}

Besides dense, random unitaries, we consider the physically motivated example of the exponentiation of a Hamiltonian associated to a small molecule. 
We choose the second-quantized Hamiltonian corresponding to the LiH molecule in a minimal Sto-3g basis set, mapped to a four-qubit Hamiltonian $H_\text{LiH}$ with a Jordan-Wigner mapping. 
Next, we synthesize the target unitary $U_\text{target} = e^{-iH_\text{LiH} t} $ for a time $t=10$ and find that qubit transpilation requires 235 CNOTs and 218 $\sqrt{X}$ gates.
The ququart case takes the equivalent of 78 CNOTs and 744 $\sqrt{X}$ gates, as shown in Fig.~\ref{fig:gatecount_comparison}.
This amounts to a reduction by a factor of $3.01$ in entangling gate cost while increasing the single-qudit gate count only by a factor of $3.41$. 
This example illustrates that our qudit transpiler can already rival a heavily optimized qubit transpiler.

\subsection{Qudit gate extensions}

As exemplified above, the qudit-based unitary synthesis creates a substantial overhead in single-qudit gates. 
This is because the entangling gate we use only acts non-trivially when the control is in $\ket{0}_c$ or $\ket{1}_c$ and only affects the target states $\ket{0}_t$ and $\ket{1}_t$, see Eq.~\eqref{eqn:ECR_unitary}. 
This creates the need for single-qudit permutation gates on both the target and the control, such as the single-qudit gates in Fig.~\ref{fig:decomposition_circuits}\textbf{c}. 
We now outline two strategies to overcome this bottleneck. 
(i) The ECR gate acts trivially when the control is in $\ket{2}_c$ and $\ket{3}_c$ due to the single-qudit $R_x^{01}$ gates applied in the echoed sequence, see Fig.~\ref{fig:ECR_pulse_overview}. 
Replacing these gates with $R_x^{12}$ or $R_x^{23}$ rotations creates a gate that instead acts non-trivially on the target when the control is in states $\ket{1}_c / \ket{2}_c$ or  $\ket{2}_c / \ket{3}_c$, respectively.
(ii) We envision that changing the frequency of the ECR drive tones, such that it is resonant with the $\ket{1}_t - \ket{2}_t$ or $\ket{2}_t - \ket{3}_t$ transitions of the target would enable a gate that directly addresses the subspaces spanned by $\{ \ket{1}_t, \ket{2}_t\}$ or $\{ \ket{2}_t, \ket{3}_t\}$, respectively. 
The added flexibility of either approach would strongly increase the effective ``connectivity'' of the different levels of two coupled qudits; the permutation gates would no longer be needed to move states into the $\{\ket{0}, \ket{1}\}$ interaction subspace.

Moreover, the number of gates in our decomposition could potentially be improved by working with a direct (non-echoed) CR gate~\cite{jurcevic2021demonstration}, which, however, increases the complexity as this depends on multiple parameters $\vec{\varphi}$, see Eq.~\eqref{eqn:single_Cr_pulse_action}.
More broadly, our decomposition could be adapted to different interactions as those provided by, e.g., tunable couplers, or frequency-tunable transmons~\cite{goss2022highfidelity, roy2022realization}.
Another approach to improve the single-qudit gate overhead is given by optimal control pulse shaping techniques~\cite{simm2023two}.
Finally, note also that the step in Eq.~\eqref{eqn:CZ_rot_through_CX_rot} introduces controlled $R_x$ gates with a maximal rotation angle of $\pm \pi$, irrespective of the original rotation angle $\alpha$. Thus, the resulting pulse schedules could be shortened by pulse-efficient circuit transpilation techniques as developed in~\cite{earnest2021pulse}, resulting in better gate fidelities.

\section{Feasibility of qudit operation}

Controlling additional qudit levels comes with an increased complexity. 
Let us now summarize the requirements to operate the transmon as a ququart in the presence of charge noise, limited lifetimes, and imperfect ququart readout, from which we can conclude design principles for future qudit-based superconducting hardware.

Firstly, charge noise is exponentially larger in higher-excited states $\ket{n}$~\cite{peterer2015coherence}.
However, charge noise is also exponentially suppressed with increasing $E_{\text{J}} / E_{\text{C}}$ at the cost of a polynomial reduction of the anharmonicity $\alpha$~\cite{koch2007chargeinsensitive}. 
Our simulations suggest that reasonable anharmonicities remain with manageable charge noise for ququart operation when choosing $E_{\text{J}} / E_{\text{C}} \sim 70$.
In fact, recent experiments have even shown operations with up to 12 levels in a transmon with $E_{\text{J}} / E_{\text{C}} > 300$~\cite{Wang_high_EJEC_transmon}. 

Secondly, the lifetime of higher-excited states $\ket{d}$ also generally decreases with $d$.
Fortunately, their decay happens predominantly sequentially, e.g., following $\ket{3} \rightarrow \ket{2} \rightarrow \ket{1} \rightarrow \ket{0}$, which leads to workable coherence times for the lowest-lying qudit levels. 
For example, for devices with qubit lifetimes of $\sim 80\,\upmu\text{s}$, which is below current state-of-the art of $\sim 500\,\upmu\text{s}$, experiments have found lifetimes of $\sim30\,\upmu\text{s}$ for state $\ket{3}$~\cite{fischer2022ancillafree, peterer2015coherence}.
In comparison, our ECR pulse lasts about $1\,\upmu\text{s}$.
Thus, as coherence times of transmons improve further, we also expect sufficient coherence full ququart operation to become viable.

Thirdly, the model parameters of the studied two-transmon system also directly impact the fidelity of the proposed multi-qudit ECR gate.
The transmon frequencies $\omega_c$, $\omega_t$ and $\alpha_c$, $\alpha_t$ define the level structure of the system. 
Through the $E_{\text{J}}/E_{\text{C}}$-ratio, this defines the amount of charge dispersion in each level. 
While the frequency crowding of the system determines the susceptibility to leakage and crosstalk, 
the maximum drive amplitude $\tilde \Omega$ determines the entangling speed of the gate and also has a strong influence on leakage and crosstalk.
With increasing coupling strength $J$ the entangling speed of the gate also increases. 
However, this also increases the dressing of the bare qudit eigenstates, which leads to a dependency of the $\ket{0}_t\leftrightarrow \ket{1}_t$ frequency on the state of the control and limits the unitary of the $ECR$ gate in the idle levels $\ket{2}_c$ and $\ket{3}_c$. 
Thus, coupling multiple ququarts requires a careful design of the control pulses to mitigate leakage due to the weak anharmonicity and the frequency crowding introduced by the additional qudit levels.
Undesired Crosstalk could further be avoided by alternative multi-qudit gates that employ tunable couplers~\cite{multi_qudit_tunable_couplers}.

Finally, transmons are measured with a dispersive readout by coupling to a resonator.  
In our proof-of-principle experiments, this was a limiting factor due to relatively poor distinguishability between the $\ket{2}$ and $\ket{3}$ states. 
A more careful tuning of the readout resonator mode and the employed readout pulses with tailored linewidths, dispersive shifts, and cavity–qudit coupling, can significantly enhance the separation of higher-excited states~\cite{kehrer_2024_qudit_readout, miao2022overcoming}.
For example, a multi-tone readout has been developed that manages to distinguish up to 10 different levels in a high-$E_{\text{J}} / E_{\text{C}}$ transmon~\cite{Wang_high_EJEC_transmon}.

\section{Discussion}

In this Chapter, we have outlined a toolbox to perform universal quantum computation in superconducting transmon qudits.
Our universal basis gate set consists of general single-qudit unitaries and an entangling cross-resonance gate between two qudits. 
With this gate set, we have developed a decomposition routine to realize arbitrary $m$-controlled two-qudit gates $C^m[U]$ as building blocks to synthesize general qudit unitaries. 
This decomposition is the center piece of a general purpose qudit transpiler to map application-level qudit-based quantum circuits to hardware-native instructions.
Compared to a state-of-the-art qubit transpiler, this qudit decomposition reduces the number of entangling gates required to synthesize general unitaries.
We have further proposed several strategies to overcome the single-qudit gate overhead opening future research directions for optimizing qudit unitary synthesis with cross-resonance gates.

The proposed entangling gate is suitable for capacitively coupled transmons. 
It represents a higher-dimensional extension of the echoed cross-resonance gate previously employed for qubits.
Our numerical model that includes charge noise and leakage errors predicts average gate fidelities of up to $99.6 \%$ with simple Gaussian pulse profiles.
The main contribution to the unitary error is the frequency dependence of the dressed target states on the control states.
Open questions are whether a direct cross-resonance driving without echoes benefits qudit operation or whether rotary tones reduce the unitary error, as demonstrated for qubits~\cite{sundaresan2020reducing}.
Finally, we note that cross-Kerr interactions, tunable couplers, or frequency-tunable transmons may offer different qudit gates~\cite{goss2022highfidelity, miao2022overcoming, roy2022realization}.

Qudit operation of transmons is attractive from a theoretical standpoint, as it makes full use of the available quantum resources of the system.
Compared to current standard qubit setups, the proposed gates require no additional microwave drive lines.
Moreover, the fact that higher-excited states suffer increasingly from charge noise can be mitigated by moving the transmon parameters towards higher $E_{\text{J}}/E_{\text{C}}$-ratios than those typically employed for qubit operation.
Our numerical simulations suggest that high-fidelity multi-ququart operations are possible under realistic experimental conditions. 

As an example application that only requires single-qudit control, we have introduced a method to perform general POVM measurements for qubits via a Naimark dilation construction.
This is a particularly hardware-friendly use of qudit levels. It does not require full qudit control, as we only populate the qudit higher-excited levels for a short duration at the measurement stage of the quantum circuit. 
Since the $\ket{3}$ state is only populated once prior to measurement, the phases acquired by $\ket{3}$ do not affect the encoded POVM operators.
Therefore, only modest coherence and pulse fidelities are required.
Compared to traditional dilation POVM implementations, we circumvent the need for ancilla qubits avoiding a considerable SWAP-gate overhead in case of limited device connectivity.
The result is a protocol that is applicable to various qubit architectures including super- and semiconducting qubits, trapped ions, and cold atoms. 

\chapter{Optimized measurement processing}
\label{chap:duals_optimization}

\summary{
This chapter is based on the article ``Dual frame optimization for informationally complete quantum measurements'' by \textbf{Laurin E. Fischer}$^\ast$, Timothée Dao$^\ast$, Ivano Tavernelli, and Francesco Tacchino, published in Phys. Rev. A 109.6 062415, 2024~\cite{fischer2024dual}. \footnotesize{$^\ast$shared first authorship} \normalsize 
\bigskip

\noindent The notion of measurement dual frames generalizes classical shadows to dual operators of POVM effects.
This introduces degrees of freedom in the post-processing stage of randomized measurements that have been neglected by established techniques. 
Here, we leverage dual frames to construct improved observable estimators from informationally complete measurement samples.
We introduce novel classes of parametrized frame superoperators and optimization-free dual frames based on empirical frequencies, which offer advantages over their canonical counterparts while retaining computational efficiency. 
Remarkably, this comes at almost no quantum or classical cost, thus rendering dual frame optimization a valuable addition to the randomized measurement toolbox.
}

Informationally complete dilation-POVMs such as SIC-POVMs are typically considered the gold-standard of quantum measurements, as they contain the full tomographic information in every single shot rather than relying on the randomization of measurement bases, where each shot only carries the information of the selected basis~\cite{stricker2022experimental}. 
However, the implementation of dilation POVMs is challenging. 
They either require an overhead of ancilla qubits or, perhaps more elegantly, additional qudit levels. 
Present-day superconducting hardware does not offer the required connectivity for ancilla-based methods without large SWAP gate overheads and requires a re-design with different device parameters to make qudit computation fully feasible, as discussed in Chap.~\ref{chap:qudit_processing}.
Hence, in practice, randomized measurements (PM-simulable POVMs) are the most readily available type of POVM measurement. 

\begin{figure}
  \begin{minipage}[c]{0.54\textwidth}
    \includegraphics[width=\textwidth]{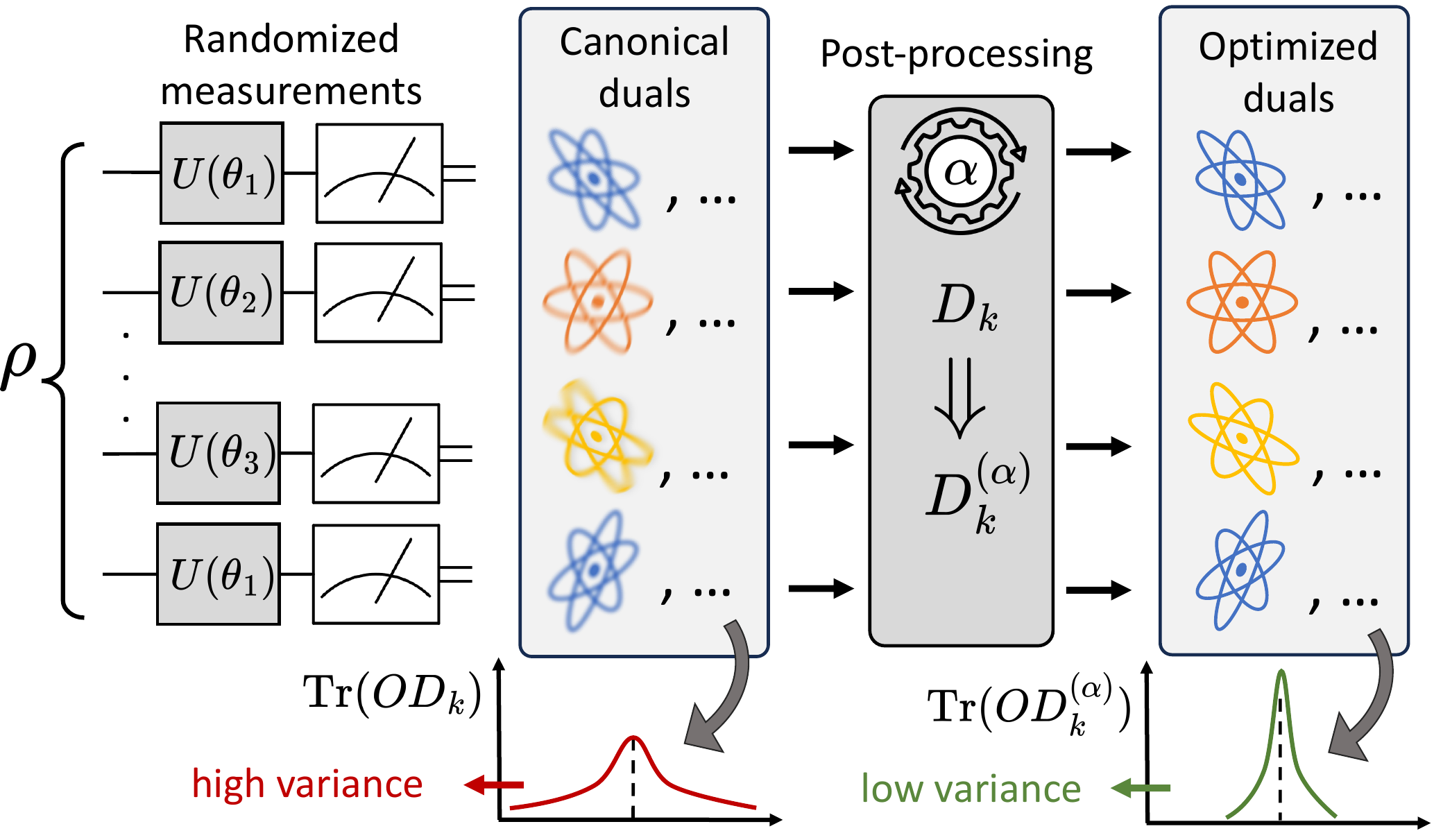}
  \end{minipage}\hfill
  \begin{minipage}[c]{0.44\textwidth}
\caption[Schematic of dual frame optimization]{ \small
Schematic of dual frame optimization.
An overcomplete IC POVM is measured on the system.  
For each outcome $k$, the corresponding canonical dual operator $D_k$ -- also known as the \textit{classical shadow} -- is computed. The expectation value of any observable $O$ can be estimated from samples of dual operators. Leveraging additional degrees of freedom, one can optimize these dual operators in post-processing, reducing the estimation variance. 
} \label{fig:overview_duals_opt}
  \end{minipage}
\end{figure}

In this chapter we develop techniques to optimize the performance of PM-simulable POVMs.
The fundamental expression that we study is the single-shot variance (SSV) of POVM-based estimators derived around Eq.~\eqref{eqn:single_shot_variance}
\begin{equation} 
\label{eqn:single_shot_variance_chap6}
     \mathrm{Var}[\w_k \mid \, \vb{M}, \vb{D}, O, \rho] = \sum_k \Tr[\rho M_k] \Tr[O D_k]^2 - \text{const.},
\end{equation} 
which depends on both the POVM operators $\vb{M}$ and the dual operators $\vb{D}$ (see Sec.~\ref{sec:estimators_povm_measurements} for an introduction of the notation).
Several works have considered the optimization of the POVM operators $\vb{M}$ to improve the statistics of estimators~\cite{hadfield2022measurements, huang_efficient_2021, garcia-perez_learning_2021}. 
However, little emphasis has so far been put on studying how the post-processing stage governs the quality of these estimators. 
Indeed, Eq.~\eqref{eqn:single_shot_variance_chap6} can also be minimized by optimizing the duals $\vb{D}$ (if they are not unique) when the POVM itself remains unchanged, see the schematic in Fig.~\ref{fig:overview_duals_opt}.
Innocenti et al.~\cite{innocenti2023shadow} have recently raised awareness on this point, highlighting the neglected degrees of freedom.
Crucially, the choice of the dual frame can be controlled purely during the post-processing phase and thus comes with no additional cost in the quantum resources.
Moreover, an optimization of the dual operators can also be individually tailored to different observables that one might want to estimate from the same set of IC data. 

In this chapter, we dive deeper into the application of frame theory to IC quantum measurements for digital quantum computing architectures.
We present efficiently computable classes of parametrized dual frames, together with the corresponding optimization routines. 
Indeed, while a number of results are known concerning optimal choices of measurement settings and dual frames for both tomography and observable estimation~\cite{innocenti2023shadow}, these are often impractical or impossible to realize at scale due to inherent technical (e.g., lack of connectivity, device noise) or fundamental (e.g., memory or data processing requirements) limitations. 
By leveraging a product structure, our proposed methods ensure consistent improvements over standard settings while remaining applicable, in principle, up to large sizes of the target qubit registers. 
We support our analytical findings with numerical investigations. 
These suggest that dual frame optimization -- even when subject to certain pragmatic constraints -- can significantly boost the quality of shadow estimators for generic operator averaging tasks. 
In particular, it reduces the performance gap between PM-simulable POVMs and dilation POVMs.

\section{Known optimality results}
\label{sec:duals_previous_results}

For any state $\rho$, the minimal SSV is achieved by performing a projective measurement in the eigenbasis of $O = \sum_k \lambda_k \ketbra{o_k}{o_k}$, in which case $M_k = D_k = \ketbra{o_k}{o_k}$, where $\ket{o_k}$ are the eigenvectors of $O$~\cite{dariano_optimal_2006, hayashi_optimal_2006}.
While this POVM is usually neither known nor easily implementable, it serves as a lower bound for all estimations of $\expval{O}_\rho$ with a $\sqrt{S}$-scaling. 

For a fixed IC-POVM $\vb{M}$ and state $\rho$, the dual frame that minimizes the SSV as a function of $\vb{D}$, irrespective of the observable $O$, is obtained from Eq.~\eqref{eqn:def_canonical_duals} when using a modified frame superoperator given by 
\begin{equation}
\label{eqn:optimal_dual_frame} 
\F_\text{opt} \coloneqq \sum_{k=1}^n \frac{1}{\Tr[\rho M_k]}\kket{M_k}\bbra{M_k}
\end{equation}
and rescaling each dual operator by $1/\Tr[\rho M_k]$~\cite{innocenti2023shadow}.
This, however, requires knowledge of the state $\rho$.
If no prior information on $\rho$ is available, the best choice of duals can be considered the one that minimizes the SSV as a uniform Haar average over all states. 
In that case, the optimal duals are obtained from a modified frame superoperator given by $\F_\text{avg} \coloneqq \sum_{k=1}^n \kket{M_k}\bbra{M_k} / \Tr[M_k]$. 
It is important to point out that this choice is the one generally employed in classical shadows protocols~\cite{elben2023randomized}.
More explicitly, a standard classical shadow $\hat \rho_s$, namely a single-shot estimate of the state obtained by constructing an inverse measurement channel, is equivalent to the ``average-optimal'' dual $D_k$ obtained from inversion of the frame superoperator $\F_\text{avg}$ introduced above.  

A POVM-based estimation of observables is only feasible if the POVM operators themselves as well as the dual operators can be efficiently handled classically. 
Here, we thus restrict the discussion to product POVMs as introduced in Sec.~\ref{sec:measurements_theory}.
This ensures a local structure in the frame superoperator from Eq.~\eqref{eqn:def_frame_superop}, such that it can be constructed and inverted efficiently. 
Moreover, also the duals themselves need to be efficiently processable in order to compute the observable coefficients $\w_k$ via Eq.~\eqref{eqn:observable_POVM_decomp}. 
However, this is not guaranteed by the optimal dual frame $\F_\text{opt}$, even when the POVM operators have a product structure. 
This implies that the optimality results presented above cannot be applied in general.
To fully leverage the ability of optimizing duals for the SSV, we thus require a parametrization of suitable dual frames, such that they remain efficiently processable.  
We provide this through a parametrized frame superoperator in the following. 

\section{Parametrization of duals}
\label{sec:parametrization_of_duals}

We now discuss how to obtain a tractable parametrization of the dual frame in order to optimize the SSV over the duals in Eq.~\eqref{eqn:single_shot_variance_chap6}.
We assume throughout that the POVM $\vb{M}$ has a product structure and thus use a multiindex $\vb{k} = (k_1, k_2, \dots, k_N)$ to label the POVM outcomes. 

\subsubsection{Weighted frame superoperator}
\label{sec:weighted_frame_superop}
If $\vb{M}$ is overcomplete, the set of all valid duals can be explicitly parametrized through a singular value decomposition~\cite{krahmer_sparsity_2013}.
% , see Appendix~\ref{app:svd_param_duals}.
In principle, this parametrization could be used to optimize the dual frame for a minimal SSV. 
However, it is not straightforward to impose a product structure on the dual operators in this way.
For a more practical, albeit non-exhaustive parametrization of the dual frames, we thus define a \emph{weighted frame superoperator} 
\begin{equation} 
\label{eqn:alpha_weighted_framesuperop}
    \mathcal{F}_\alpha \coloneqq \sum_{\vb{k}} \alpha_{\vb{k}} \kket{M_{\vb{k}}} \bbra{M_{\vb{k}}} 
\end{equation}
which resembles the canonical frame superoperator $\F$, but with the contribution of each effect $M_{\vb{k}}$ rescaled by a factor $\alpha_{\vb{k}}\in \mathbb{R}$.
If $\mathcal{F}_\alpha$ is invertible, the effects given by
\begin{equation} 
\label{eqn:alpha_weighted_duals}
    \kket{D_{\vb{k}}} = \alpha_{\vb{k}} \mathcal{F}_\alpha^{-1} \kket{M_{\vb{k}}}
\end{equation}
form a valid dual frame which is invariant under uniform scaling of the coefficients.
Notice that if all the coefficients $\{ \alpha_{\vb{k}} \}$ are positive, then $\mathcal{F}_\alpha$ will be positive definite and invertible.  
We can hence think of the parameters $\{ \alpha_{\vb{k}} \}$ as a probability distribution when restricting them to positive values.

The degree of correlations in the multivariate probability distribution $\alpha_{\vb{k}}$ determines what kind of product structure the resulting duals from Eq.~\eqref{eqn:alpha_weighted_duals} will have.
In the simplest case, when $\alpha_{\vb{k}}$ fully factorizes, i.e., $\alpha_{k_1, k_2, \dots, k_N} = \alpha^{(1)}_{k_1} \alpha^{(2)}_{k_2} \cdots \alpha^{(N)}_{k_N}$, the dual frame will be of product form as in Eq.~\eqref{eqn:povm_product_form}. 
More generally, if $\alpha_{\vb{k}}$ is a product of distributions that each act on at most $m$ qubits, then the duals will be tensor products of terms that act on $m$ qubits. 
In this case, the traces to compute $\omega_k$ in Eq.~\eqref{eqn:coeffs_from_duals} factorize into blocks that involve constructing matrices of at most size $2^m \times 2^m$. 
This way, the complexity of the dual operators in the post-processing can be tuned by imposing restricted correlations in the distribution $\alpha_{\vb{k}}$.

We thus propose the following general procedure to improve statistical estimators based on overcomplete POVMs.
First, a collection of shots $\{ k^{(1)}, \dots, k^{(S)} \}$ is measured from a fixed POVM.  
For a given parametrization of duals through Eq.~\eqref{eqn:alpha_weighted_duals}, the SSV in Eq.~\eqref{eqn:single_shot_variance_chap6} is estimated  with the (corrected) sample variance of the values $\{\omega_{k^{(1)}}, \dots, \omega_{k^{(S)}} \}$.
An optimizer will then minimize the SSV as a function of the parameters entering the weighted frame superoperator, yielding an estimator of $\langle O \rangle_\rho$ with the smallest possible variance. 
This can be repeated independently for each observable of interest starting from the same collection of samples, harnessing the true power of IC measurements.
As this dual optimization does not require changing the quantum circuits to be executed nor increasing the sample size, we consider it to be a ``free lunch'' improvement over standard classical shadows techniques. 

In practice, the optimization landscape of the dual parameters $\alpha_{\vb{k}}$ could be difficult to navigate, due to the complicated dependency of the duals on $\alpha_{\vb{k}}$ in Eq.~\eqref{eqn:alpha_weighted_duals}.
Also, a simultaneous optimization of the duals and the POVM operators themselves can be cumbersome, as it requires a quantum-classical feedback loop. 
As an alternative to optimizing a parametrization of the dual operators, we thus propose the following procedure to obtain suitable dual frames for a fixed and overcomplete IC POVM, which we refer to as \emph{empirical frequencies dual frames}. 

\subsubsection{Empirical frequencies dual frames}
\label{sec:empirical_frequency_dual_frames}
When no knowledge about the state is available, the average-optimal dual frame should be used, as discussed in Sec.~\ref{sec:duals_previous_results}.
As the POVM measurement is repeated and the number of shots $S$ increases, we gain some knowledge about the state, which we can leverage to approximate the optimal dual from Eq.~\eqref{eqn:optimal_dual_frame}.
More precisely, the measured frequencies $f_{\vb{k}} = \#{\vb{k}} / S$ (where $\# {\vb{k}}$ is the number of times the outcome ${\vb{k}}$ was obtained) follow a multinomial distribution and converge to the true measurement probabilities $p_{\vb{k}} = \Tr[\rho M_{\vb{k}}]$ as $\sqrt{p_{\vb{k}}(1-p_{\vb{k}})/S}$.
One could thus replace the outcome probabilities $p_{\vb{k}} = \Tr[\rho M_{\vb{k}}]$ in the optimal dual frame with the empirical frequencies $f_{\vb{k}} = \#{\vb{k}} / S$. That is, we use the \emph{global empirical dual frame}
\begin{equation} \label{eq:empirical_dual_frame} \begin{split}
    \kket{D_{\vb{k}}} &= \frac{1}{f_{\vb{k}}} \mathcal{F}^{-1}\kket{M_{\vb{k}}} \\
    \textrm{where } \mathcal{F} &= \sum_{\vb{k}} \frac{1}{f_{\vb{k}}} \kket{M_{\vb{k}}}\bbra{M_{\vb{k}}}
\end{split} \end{equation}
which is a weighted frame superoperator with $\alpha_k = 1/f_k = S/\# k$.
However, an obvious issue arises if an outcome is not obtained ($f_k =0$).
We address this by adding a regularization to the empirical frequencies with a bias term.
This biases the outcome probabilities with respect to the fully mixed state $\frac{1}{d}\mathbbm{1}$, borrowing an idea from Ref.~\cite{hayashi_optimal_2006}. 
The resulting \emph{biased empirical frequencies} are given by 
\begin{equation}
\label{eqn:biased_empirical_frequencies}
    \Tilde{f}_{\vb{k}}(\{k^{(1)},\dots,k^{(S)}\},S_{\textrm{bias}}) = \frac{\# {\vb{k}} + \Tr[\frac{1}{d}\mathbbm{1} M_{\vb{k}}] S_{\textrm{bias}}}{S + S_{\textrm{bias}}} \, .
\end{equation}
If we assume that all effects are non-null, then $\alpha_{\vb{k}} = 1/\Tilde{f}_{\vb{k}} >0 $, which ensures the frame superoperator is invertible.
Note that, for $S=0$, we recover the average-optimal dual frame, while for $S \rightarrow \infty$ the empirical dual frame converges to the optimal dual frame from Eq.~\eqref{eqn:optimal_dual_frame}.

The global empirical dual frame still suffers from two issues: 
Firstly, for sizable qubit numbers, the number of different POVM outcomes $n$ eventually becomes much larger than the available shot budget $S$. 
In this regime, it is difficult to improve over the average-optimal dual with the above global empirical dual frame.
Secondly, the dual matrices can become exponentially large when the correlations in $\Tilde{f}_{\vb{k}}$ are not restricted, as discussed in Sec.~\ref{sec:weighted_frame_superop}. 
Both of these issues are overcome when relaxing the task from learning the global distribution $p_{\vb{k}}$ to recovering only the most relevant few-qubit correlations of this multivariate distribution. 
In the simplest case, the (potentially biased) outcome probabilities $f_{\vb{k}}$ can be approximated with the product of marginal probabilities 
\begin{equation}
\mathfrak{f}_{\vb{k}} \coloneqq \prod_{i=1}^N \mathfrak{f}^{(i)}_{k_i} \, , \;\;\; \textrm{with} \;\;\; \mathfrak{f}^{(i)}_{k_i} \coloneqq \sum_{\{k_j\}_{j\neq i}} f_{k_1, \dots, k_N} \, .
\end{equation}
While this does not model correlations between POVM outcomes of different qubits, it still presents an advantage over the average-optimal dual frame, while ensuring the dual frame is of product form, see Sec.~\ref{chap:numerics}.

The correlations captured by the empirical frequencies can be systematically tuned up by partitioning $\vb{k}$ into marginals of larger sizes. 
Let $\Sigma = \{ \lambda_1, \dots, \lambda_l \}$ be a partitioning of the qubit indices $\{1,\dots, N\}$ into subsets $\lambda_i$ that each contain up to $m$ terms. 
We can then approximate the global distribution $f_{\vb{k}}$ (or $\tilde f_{\vb{k}}$)  as a product of \emph{$m$-body marginals} $\mathfrak{f}_{\lambda_i}$
\begin{equation}
\label{eqn:m-body-marginals}
\mathfrak{f}_{\vb{k}}^{\Sigma} \coloneqq  \prod_{i=1}^l \mathfrak{f}_{\lambda_i} \, , \;\;\; \textrm{with }  \;\;\; \mathfrak{f}_{\lambda_i} \coloneqq \sum_{\{k_j\}_{j\notin \lambda_i}} f_{k_1, \dots, k_N} \, .
\end{equation}
This leads to dual frame operators that are tensor products of $m$-local terms. 
The question arises how to optimally choose the partitioning $\Sigma$.
Ideally, pairs of qubits whose POVM outcomes are highly correlated should preferably be grouped into the same set. 
We quantify this through the \emph{empirical mutual information} $I(i, j)$ shared by two qubits, given as 
\begin{equation}
\label{eqn:mutual_information}
I\left( i, j \right) \coloneqq \sum_{k_i, k_j} \mathfrak{f}_{\{i, j\}} \log \left(\frac{ \mathfrak{f}_{\{i, j\}}}{   \mathfrak{f}^{(i)}_{k_i} \mathfrak{f}^{(j)}_{k_j}} \right).
\end{equation}
This quantifies the price to pay when approximating the joint distribution $\mathfrak{f}_{\{i, j\}}$ through the product of marginal distributions $ \mathfrak{f}^{(i)}_{k_i}  \mathfrak{f}^{(j)}_{k_j}$, given by their Kullback-Leibler divergence.

In a practical setting, the maximally-allowed degree $m_\text{max}$ should be chosen such that the classical cost in computing the traces of the resulting $2^m \times 2^m$ dual matrices is deemed tolerable, and sufficient statistics are gathered to capture the $m$-body marginals, which becomes exponentially more difficult as $m$ increases.
Once $m_\text{max}$ is chosen, one can define a cost function $\mathcal{C}$ for $\Sigma$
that is given by the sum of the pairwise mutual information over all set, i.e., 
\begin{equation}
\label{eqn:partitioning_score}
\mathcal{C}(\Sigma) \coloneqq \sum_{\lambda_i \in \Sigma} \sum_{\substack{j, k \in \lambda_i \\ j \neq k}} I(j, k).
\end{equation}
While the optimal partitioning can be straightforwardly computed from this cost function for small qubit numbers, this becomes infeasible for larger $N$ due to the super-polynomial scaling of the number of different partitionings. 
In such a setting, one can construct a well-performing partitioning by first computing $I(i, j)$ for all pairs of qubits, and then putting pairs of highest values into the same subset with a greedy allocation strategy.

\section{Numerical benchmarks}
\label{chap:numerics}
\begin{figure}
\centering
\includegraphics[width=1.02\linewidth]{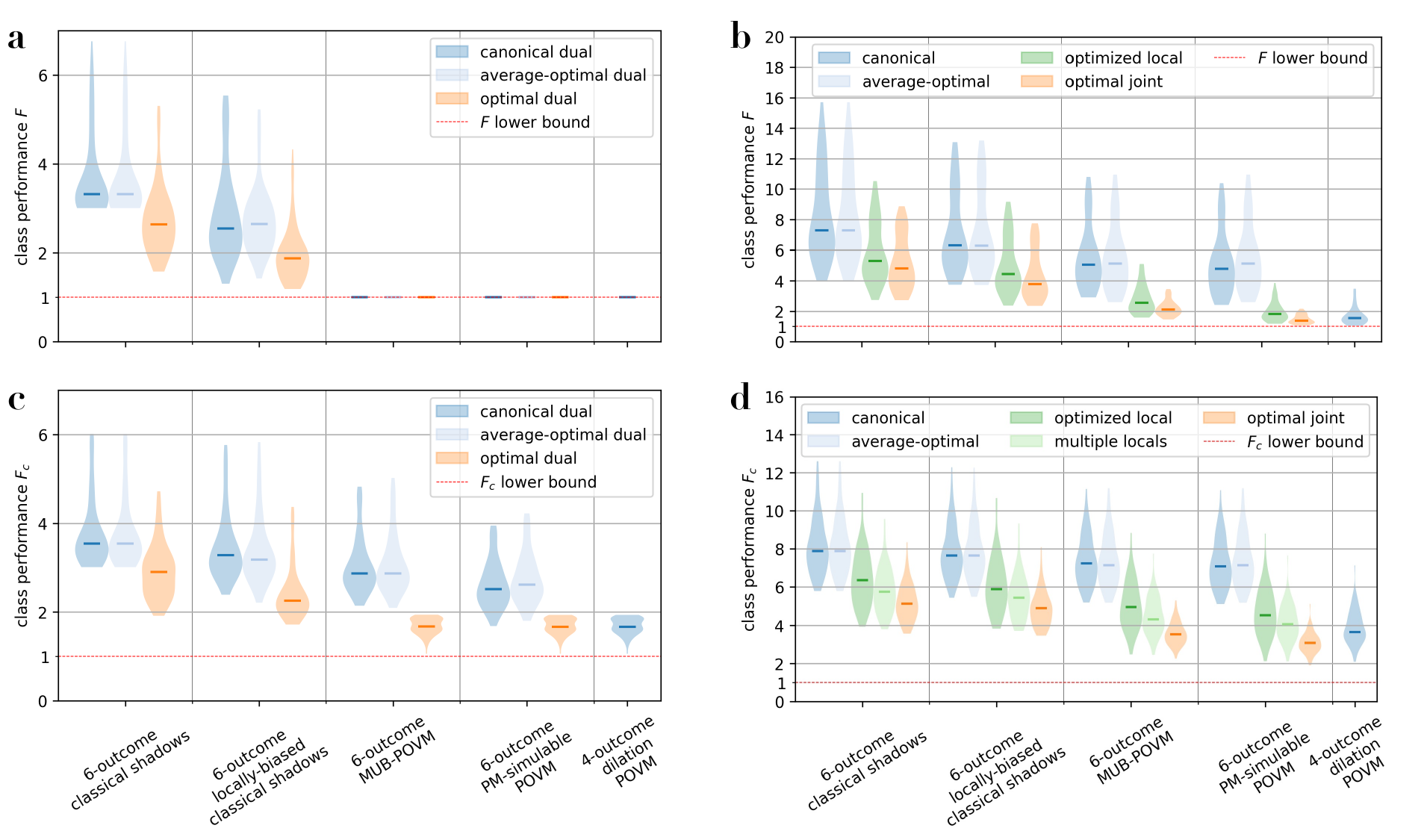} 
\caption[Performance comparison of different POVMs and dual frames]{\small
Performance of different classes of POVMs and dual frames for estimating random observables, shown as violin plots. 
Color indicates the employed class of duals.
Each distribution is built from 200 repetitions of sampling a Haar-random pure state and an observable (or set of observables) and subsequently minimizing the class performance $F$ (or cumulative performance $F_\text{C}$) for each combination of POVM and dual frame.
Horizontal markers show the median of each distribution.
For blue distributions a fixed dual frame is used. 
For orange distributions, the optimal dual frame is used.
The red dashed line represents the optimal lower bound which is saturated for the projective measurement in the eigenbasis of the observable.
\textbf{a)} Single-qubit system with one observable.
\textbf{b)} Two-qubit system with one observable. 
\textbf{c)} Single-qubit system with five observables. 
\textbf{d)} Two-qubit system with five observables.
For two-qubit systems, the class performance is computed for the cumulative variance of all observables and the optimized dual frames are limited to product form for the green distributions. 
For the light-green distributions, the duals are re-optimized for every observable. 
}
\label{fig:povm_class_performances}
\end{figure}

We now showcase our methods for dual frame optimization on numerical examples. 
In the following, in Sec.~\ref{sec:performance_limits_numerics}, we benchmark the performance of different classes of POVM operators and dual frames by optimizing the SSV for generic random states and observables.
Then, in Sec.~\ref{sec:numerical_results_empirical_frequencies}, we demonstrate how an explicit optimization of the operators can be circumvented by using the empirical frequencies of the outcomes to obtain a well-performing dual frame.

\subsection{Performance limit of different POVM classes}
\label{sec:performance_limits_numerics}

Here, we investigate which class of single-qubit POVMs and dual frames yields the best possible estimators. 
In this idealized setting, we assume full knowledge of the underlying state $\rho$. 
Our procedure is the following: 
We first sample a Haar-random pure state $\rho$. 
We construct random observables $O$ by sampling eigenvalues $e_1, \dots, e_d$ uniformly at random from 
$\left[-5, 5\right]$ and applying a Haar-random unitary $U$ that yields $O = U \mathrm{diag}(e_1, \dots, e_d) U^\dagger$.
As discussed in Sec.~\ref{sec:duals_previous_results}, the optimal measurement would be the projective measurement in the eigenbasis with a SSV of $\expval{O^2}_\rho - \expval{O}^2_\rho$.
For a given observable $O$, we therefore evaluate a class of POVMs $\mathcal{M}$ and duals $\mathcal{D}$ by their \emph{class performance}
\begin{equation}
\label{eqn:performance_limit}
    F(\mathcal{M}, \mathcal{D}) \coloneqq \min_{\vb{M} \in \mathcal{M}, \vb{D} \in\mathcal{D} } \left\{\frac{ \mathrm{Var}[\w_k \mid \vb{M}, \vb{D}, O, \rho] }{  \expval{O^2}_\rho - \expval{O}^2_\rho}\right\},
\end{equation}
with $\mathrm{Var}[\w_k]$ given as in Eq.~\eqref{eqn:single_shot_variance}. 
Similarly, we quantify the ability to estimate several observables $\{O_i\}_{i \in \{1,\dots, N_\text{obs}\}}$ from the same IC POVM data through the \emph{cumulative class performance}
\begin{equation}
\label{eqn:cumulative_performance_limit}
    F_\text{C}(\mathcal{M}, \mathcal{D}) \coloneqq \min_{\substack{\vb{M} \in \mathcal{M} \\ \{\vb{D}_i\}_{i \in \mathcal{D} }}} \left\{
    \frac{\sum\limits_{i=1}^{N_\text{obs}}  \mathrm{Var}[\w_k \mid \vb{M}, \vb{D}_i, O_i, \rho] }{ \sum\limits_{i=1}^{N_\text{obs}} \expval{O_i^2}_\rho - \expval{O_i}^2_\rho}
    \right\}.
\end{equation}
Note that the duals $\vb{D}$ in Eqs.~\eqref{eqn:cumulative_performance_limit} and~\eqref{eqn:performance_limit} are implicitly defined through the POVM operators $\vb{M}$ but carry free parameters as in Eq.~\eqref{eqn:alpha_weighted_duals}.
We use a BFGS optimizer to compute $F$ and $F_\text{C}$ for the five classes of single-qubit POVMs detailed in App.~\ref{app:povm_classes}, namely, classical shadows, locally-biased classical shadows, mutually-unbiased bases (MUB) POVMs, and general PM-simulable POVMs (all 6 outcomes each), as well as 4-outcome dilation POVMs.
The distributions of the achieved performance limit for 200 random states $\rho$ state and observables $O$ (or set of $N_\text{obs}=5$ observables $\{O_i\}$ for $F_C$ ) are shown in Fig.~\ref{fig:povm_class_performances} for single-qubit and two-qubit systems. 

In all cases, the class performance improves significantly when moving beyond the canonical dual frame. 
Therefore, the additional degrees of freedom leveraged by our dual optimization improve POVM estimators beyond what can be achieved by optimizing the POVM operators alone. 
Trivially, on a single-qubit observable, the MUB, PM-simulable and dilation POVMs all reach the optimal performance $F=1$, as the eigenbasis projectors are included in this class of POVMs, see Fig.~\ref{fig:povm_class_performances}\textbf{a}. 
However, when estimating several observables from the same POVM data, the cumulative performance is again improved by adapting the dual frame for each observable, see Fig.~\ref{fig:povm_class_performances}\textbf{c}. 
For two-qubit observables, no measurement setting will consistently reach the optimal performance as we restrict ourselves to single-qubit POVM operators, see Fig.~\ref{fig:povm_class_performances}\textbf{b}. 
The optimized local duals perform slightly worse than the optimized global duals but still considerably better than canonical duals. 
When estimating several two-qubit observables, optimizing the dual operators gives a more significant performance improvement than adding more complexity to the measured POVM operators by going from classical shadows to more general PM-simulable POVMs, see Fig.~\ref{fig:povm_class_performances}\textbf{d}.
The optimal local dual frame depends on the observable, hence it should be re-optimized for every observable, offering an additional slight improvement.

A common trend in the results of Fig.~\ref{fig:povm_class_performances} is the following: As more degrees of freedom are optimized in a PM-simulable POVM, the performance gains become increasingly smaller and reach a plateau when using the canonical dual frame (see, e.g., blue violins in Fig.~\ref{fig:povm_class_performances}\textbf{d}). 
However, these gains become increasingly larger when using optimized dual frames. 
In other words, it becomes less and less worth it to add further degrees of freedom to the POVM operators when using the canonical dual frame, which is the opposite to what is observed when using optimized dual frames.
This is especially true when estimating several observables from the same POVM data.
In all cases, PM-simulable POVMs with optimized duals (even local ones) come close to or surpass the performance of optimized dilation POVMs. 
Interestingly, the average-optimal dual frame (see Sec.~\ref{sec:duals_previous_results}) does not offer reliable performance improvements, indicating that this result might not be practical in a realistic setting.

\subsection{Empirical frequencies dual frames}
\label{sec:numerical_results_empirical_frequencies}

\begin{figure}
  \begin{minipage}[c]{0.55\textwidth}
    \includegraphics[width=\textwidth]{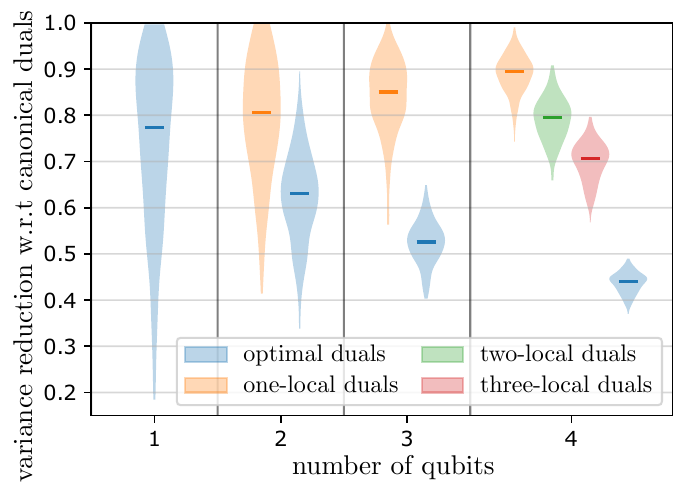}
  \end{minipage}\hfill
  \begin{minipage}[c]{0.43\textwidth}
    \caption[Variance reduction of k-local duals compared to canonical duals]{\small Variance reduction compared to classical shadows with canonical duals for different types of empirical frequencies dual frames. 
Violin plots show the distribution over 200 random pairs of states and observables.
    } \label{fig:marginals_scaling_with_qubit_number}
  \end{minipage}
\end{figure}

\begin{figure}
  \begin{minipage}[c]{0.55\textwidth}
    \includegraphics[width=\textwidth]{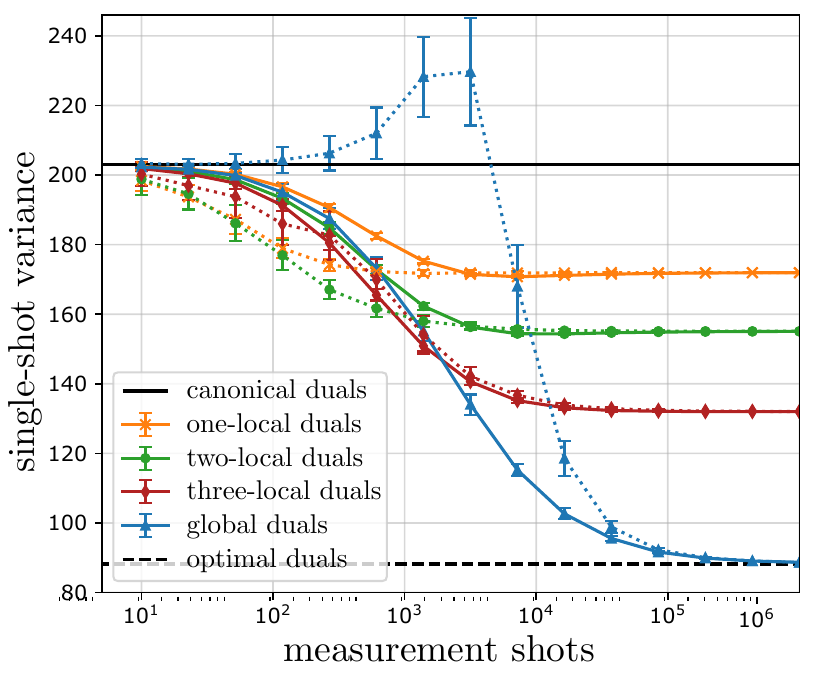}
  \end{minipage}\hfill
  \begin{minipage}[c]{0.43\textwidth}
    \caption[Convergence of the single-shot variance for empirical-frequency dual frames]{\small Convergence of the single-shot variance with increasing shot number for estimators based on empirical frequency dual frames on random four-qubit states and observables. The POVM operators are single-qubit classical shadows. For the solid line data, a bias of $S_\text{bias} = 1296$ is used, while dashed lines are obtained with $S_\text{bias} = 128$. Error bars are the standard deviation over 15 repetitions of sampling the indicated number of shots from the underlying POVM distribution. 
    } \label{fig:marginals_shots_scaling}
  \end{minipage}
\end{figure}

Next, we showcase how to bypass the explicit optimization of the dual frame with the use of empirical frequencies dual frames as introduced in Sec.~\ref{sec:empirical_frequency_dual_frames}.
We first investigate the performance of the $m$-body marginal distributions in the infinite shot limit, i.e., we construct the marginal probabilities in Eq.~\eqref{eqn:m-body-marginals} from the exact outcome distributions $p_{\vb{k}}$. 
In Fig.~\ref{fig:marginals_scaling_with_qubit_number}, we show how the improvement over canonical duals scales with an increase in the system size. 
Here, we plot the ratio of the variance of classical shadows estimators when using optimized duals compared to canonical duals.
Distributions in violin plots are obtained from 200 random samples of states and (single) observables. 
Remarkably, as the system size increases from one to four qubits, the improvement of the optimal global duals becomes more and more pronounced. 
At four qubits, the variance of this optimal estimator is less than half of the variance of the canonical dual estimator. 
This can be understood as a consequence of increasing the classical resources that go into the construction of the dual operators. 
The duals derived from the single-qubit marginal distributions (one-local duals) become less performant as the system size increases. 
This comes as no surprise, as the product of the marginal distributions will capture the true correlated distribution less and less successfully with increasing qubit number.
The performance of the marginal duals can be systematically improved by including higher-order correlations, as shown by the two-local and three-local duals. 
These are constructed by choosing the optimal partitioning of the four qubits into subsets of sizes $(2,2)$ and $(3,1)$ according to Eq.~\eqref{eqn:partitioning_score}. 
Overall, these numerical results show that marginal frequencies dual frames can offer a straightforward improvement over canonical dual frames. 
Their performance can be systematically boosted by introducing higher-order correlations.

Finally, we investigate how well empirical frequencies perform in the practically relevant setting of finite samples.
In Fig.~\ref{fig:marginals_shots_scaling}, we show how the SSV improves with increasing sample size $S$ for different types of marginal distributions when estimating a single four-qubit observable.
Note that we use the biased empirical frequencies introduced in Eq.~\eqref{eqn:biased_empirical_frequencies} to choose the dual frame, 
but plot the true underlying SSV according to Eq.~\eqref{eqn:single_shot_variance} instead of estimating the variance from the finite sample.
To illustrate the role of $S_\text{bias}$, we show the convergence with one lower value of $S_\text{bias}=128$ and one larger value of $S_\text{bias}=1296$ (the number of different POVM outcomes). 
In the regime where $S\leq S_\text{bias}$, the duals remain close to the canonical duals by design. 
As $S$ increases, the empirical frequencies eventually converge to their true underlying values.
The bias controls the rate and stability of this statistical convergence.
The smaller bias is sufficient to give a smooth convergence for the more restricted marginal distributions (dotted lines). 
In fact, the empirical frequencies with the one-local, two-local, and three-local marginals already offer a concrete improvement over the canonical dual variance with only a few hundred measurement samples, which is well below the total number of POVM outcomes. 
However, when approximating the global outcome distribution (blue curves), choosing too small of a bias will render the empirical frequencies unstable (dashed blue line). 
On the other hand, for the one-local and two-local duals, the larger bias comes at the price of a significantly slower convergence, illustrating a tradeoff between stability and speed. 
In practice, $S_\text{bias}$ can nevertheless always be chosen large enough such that the empirical frequencies dual frame gives a performance that is at least as good as the canonical duals.
Our results indicate that a reasonable choice for $S_\text{bias}$ is on the order of the degrees of freedom in the marginalized probability distributions. 

\section{Discussion}
\label{chap:discussion}
In this chapter, we have investigated the connection between frame theory and randomized overcomplete quantum measurements~\cite{innocenti2023shadow}. 
In particular, we developed and tested scalable dual frame optimization strategies that allow us to sharpen the performance of ``measure-first'' schemes while acting entirely at the data post-processing stage. 

Inspired by known analytical results which, albeit optimal, are hardly realizable in practice, we identified a minimal set of constraints that guarantee an efficient and effective computational pipeline. 
More specifically, we proposed a marginalized version of dual frames offering advantages even if restricted to a structure of limited correlations.
This class of duals can, in principle, be parameterized and iteratively refined in combination with, e.g., adaptive POVMs~\cite{garcia-perez_learning_2021}. 
Furthermore, we described an optimization-free dual frame, based on empirical outcome frequencies, that converges (in the statistical sense) to the best $m$-local one. 
This approach does not require any prior knowledge or assumptions on the state being measured, and can be tuned up systematically to capture the most relevant correlations in the measurement outcomes of individual qubits.
%By removing the need to explicitly optimize the dual frame, this solution significantly simplifies the navigation of the measurements space when searching for the most suitable POVM/dual combination. 
Our techniques are especially relevant for use-cases that require estimating several observables for the same state, as the dual operators can be optimized for every observable independently. 
% Beyond reconstructing expectation values, all our proposed methods are applicable to an extensive set of tasks, including reduced state tomography, machine learning. and error mitigation.

To support our analysis, we performed numerical simulations in both the infinite statistics limit and in the finite sample regime. Remarkably, our results suggest that, with a judicious selection of the dual frame, overcomplete PM-simulable POVMs can come close to the best results obtained with dilation ones in the context of operator estimation. 
Due to the inherent simplicity of their implementation, PM-simulable POVMs might then be preferable in practical applications.
Our simulations -- albeit collected at modest scales and for generic, Haar-random states and observables -- also indicate that the advantage unlocked by using global duals increases with the system size, while the relative improvement brought by simple single-qubit marginal dual frames decreases. 
We note that, in most practical settings, observables are seldom dense and are rather linear combinations of Pauli strings with finite weight, e.g., Hamiltonians encountered in condensed matter, lattice problems and quantum chemistry.
For such local Pauli strings, one can construct a dual that acts globally on the qubits that make up the non-trivial support of the observables. Furthermore, physically relevant states most often exhibit a distinctive and restricted correlation structure, which can be exploited in the construction of our proposed mutual-information-based dual frames. 
We leave the in-depth study of these aspects for future work.

As a next step, one could aim at better characterizing the gap between ideal and marginal dual frames in a systematic way, possibly using tools from probability theory~\cite{watanabe_information_1960,fano1961}. 
In parallel, it would also be interesting to focus on the development of alternative classes of efficiently computable dual frames which could help reduce such performance separation. 
Examples might draw inspiration from Clifford~\cite{helsen2022thrifty,hu2023scrambled} and matchgate~\cite{zhao2021fermionic,wan2023matchgate} shadows, or make use of classical techniques such as tensor networks~\cite{filippov2023scalable} and neural network quantum states~\cite{torlai2018neural}. 
Combining estimates obtained with different dual frames, for example through a median of means~\cite{huang_predicting_2020}, could also improve the overall quality of the estimators.

To conclude, our work confirms that dual frames deserve greater attention whenever overcomplete IC techniques are employed to reconstruct properties of quantum states. 
In fact, the freedom they offer only concerns the classical processing of outcomes and can be straightforwardly leveraged to improve any shadow-based protocol without overheads in sampling or quantum circuit complexity. 
We therefore expect that the careful selection of duals will become a standard component of randomized and IC measurement toolboxes, significantly enhancing their performances.
To facilitate this, we have integrated these capabilities into an open-source package built around the Qiskit software development kit, adding to the growing software ecosystem~\cite{povm_toolbox}.
The utility of dual frames is substantiated further through recent work on estimation improvements using dual optimization in the context of Hamiltonian simulation and quantum chemistry~\cite{malmi2024enhanced} and optimized shadow inversion maps~\cite{caprotti2024optimising}.

\bigskip

This concludes our discussion of how to optimally perform expectation value estimation of \textit{individual} observables with IC measurements.
However, the largest benefit of IC measurements lies in the fact that they can be used to sample many observables in parallel. 
In the next chapter, we investigate how to leverage POVM measurements for a quantum algorithm that requires estimating many observables from the same state, namely quantum subspace expansion.

\chapter{Ground state estimation with parallelized subspace expansion}
\label{chap:subspace_expansion}

\summary{
This chapter is based on the article ``Large-scale implementation of quantum subspace expansion with classical shadows'' by \textbf{Laurin E. Fischer}, Daniel Bultrini, Ivano Tavernelli, and Francesco Tacchino, available as a preprint arXiv:2510.25640, 2025~\cite{fischer2025large}. \\
%\bigskip

\noindent Informationally-complete POVMs can be used to speed up measurement-intensive quantum algorithms by simultaneously estimating many expectation values from the same underlying measurements.  
In this chapter, we apply this idea to the family of subspace expansion algorithms presented in Sec.~\ref{sec:spectral_calculations}.
We present a framework called \emph{parallelized subspace expansion} (PSE) where we explicitly manage the statistical covariances that arise from reusing measurement samples for several expectation values. 
We experimentally demonstrate noise-agnostic error mitigation with this technique in ground state preparation circuits of spin models for systems with up to 80 qubits.}

Having discussed how to ideally implement informationally-complete POVMs in Chap~\ref{chap:qudit_processing}, and how to optimally post-process POVM samples for a single expectation value in Chap.~\ref{chap:duals_optimization}, we now explore a different aspect of IC measurements -- the fact that we can use one set of measurement samples from a given state to simultaneously estimate multiple expectation values of that state.
A prominent family of algorithms where this applies is the subspace expansion framework introduced in Sec.~\ref{sec:spectral_calculations}, which is relevant for ground-state and excited-state calculations. 
This addresses a main issue of traditional approaches that have explicitly constructed measurement circuits to estimate the various terms that arise in the construction of the subspaces, which is often a significant roadblock for scaling in practice~\cite{ollitrault2020quantum}. 

The idea to simultaneously estimate all expectation values of a subspace expansion with IC measurements has been explored in the literature before.
Ref.~\cite{choi2023measurementa} highlights that randomized shadow-like measurements can be advantageous for subspace expansion compared to Pauli grouping approaches.
The use of classical shadows to systematically build high-dimensional subspaces spanned by low-weight Paulis has been proposed in Ref.~\cite{boyd2025high}.
Moreover, subspace expansions of logical encodings of a state have been explored with classical shadows~\cite{hu2022logical}.
Here, we approach this problem from a different perspective. 
Instead of solving the underlying generalized eigenvalue problem directly, we formulate subspace expansion as a constrained optimization problem to avoid statistically problematic regions of the subspace.
This allows us to obtain accurate statistical error bars and provides an external tuning knob to control these errors, trading potential bias for variance. 

Subspace expansion is predominantly thought of as a technique to perform spectral calculations. 
However, as discussed in Sec.~\ref{sec:noise-agnostic_EM}, when run on a noisy quantum device, it can also be regarded as a noise-agnostic, but problem-tailored error mitigation technique. 
In Sec.~\ref{sec:PSE_experimental_results}, we investigate the error-mitigation capabilities of PSE in first proof-of-principle experiments. 
We prepare the ground state of two spin models that offer known circuit constructions and extract the error-mitigated ground state energies for system sizes of up to 80 qubits.

\section{Parallelized subspace expansion with IC measurements}
\label{sec:PSE_theory}

Our goal is to perform subspace expansion as presented in Sec.~\ref{sec:spectral_calculations} while using measurements from an IC POVM to parallelize the evaluation of all contributing matrix elements. 
Specifically, we are interested in ground state energy estimation based on potentially noisy ground state $\rho_0$ where the subspace observable $O$ is the corresponding Hamiltonian $H$. 
We build on Eqs.~\eqref{eqn:rho_subspace_expanded}--\eqref{eqn:exp_value_subspace}, where $\{\sigma_i \}$ are the expansion operators.
Given a set of $S$ POVM outcome samples $\{k^{(1)}, \dots , k^{(S)}\}$ measured from the underlying base state $\rho_0$, we aim to simultaneously measure all matrix entries of the projected Hamiltonian $\mathcal{H}_{ij} \coloneqq \Tr[\rho_0 \sigma_j H \sigma_i^\dagger]$ and the overlap matrix $\mathcal{S}_{ij} \coloneqq \Tr[\rho_0 \sigma_j \sigma_i^\dagger]$. 

Let us first construct estimators for $\mathcal{H}$ and $\mathcal{S}$. 
Following Sec.~\ref{sec:estimators_povm_measurements}, a canonical estimator for each matrix element $\mathcal{H}_{ij}$ and $\mathcal{S}_{ij}$ is obtained by simply replacing $\rho_0$ in each expression with the corresponding dual operator $D_k$ (for some fixed dual frame) and then averaging over all shots. 
In principle we could estimate all entries of $\mathcal{H}$ and $\mathcal{S}$ this way and then solve the generalized eigenvalue problem from Eq.~\eqref{eqn:GEVP} by obtaining the lowest eigenvalue of $\mathcal{S}^{-1}\mathcal{H}$. 
However, this direct inversion of the overlap matrix can be numerically unstable in practice due to ill-conditioning of the matrix under shot noise and experimental imperfections, particularly when the expansion operators are non-orthogonal~\cite{epperly2021theory}.
Since we do not estimate the matrix entries independently, there are additional covariances between them which further contribute to statistical errors.
% Need further details in the sentences below
A popular way to address this is through regularization of the overlap matrix, e.g., by discarding dimensions that correspond to small eigenvalues of $\mathcal{S}$~\cite{urbanek2020chemistry}. 

Here, we present a different scheme to avoid ill-conditioning which is tailored to handle the covariances that are unique to PSE.
Let $\hat{H}(\vec{c})$ be an estimator for the energy expectation value of the state at subspace vector $\vec{c}$ and $\hat{\epsilon}(\vec{c})$ be an estimator for the statistical error of $\hat{H}(\vec{c})$. 
We propose to set a maximally tolerated error $\epsilon_\text{max}$ and solve the constrained optimization problem
\begin{equation}
\label{eq:PSE_constrained_optimization}
\underset{\vec{c} \in \mathbbm{C}^L}{\arg\min}\ \hat{H}(\vec{c}) \quad \text{subject to} \quad \hat{\epsilon}(\vec{c}) \leq \epsilon_\text{max}.
\end{equation}
This allows us to explore only those regions of the subspace where the statistical error is well controlled and offers a convenient tuning knob to trade off statistical errors for potentially lower estimates of the energy.  
%\textcolor{green}{could make a figure that illustrates the idea of the constrained minimization given the statistical error. }

We now construct the estimators $\hat{H}(\vec{c})$ and $\hat{\epsilon}(\vec{c})$ based on our set of IC samples. 
Let $x_k \coloneqq \sum_{i,j=1}^L c_i^* c_j \Tr[D_k \sigma_j H \sigma_i^\dagger ]$ and $y_k \coloneqq \sum_{i,j=1}^L c_i^* c_j \Tr[D_k \sigma_j \sigma_i^\dagger ]$ be the random variables of the numerator and denominator of Eq.~\eqref{eqn:exp_value_subspace} as distributions over the POVM samples. 
Then the ratio of $\mu_x \coloneqq \mathbbm{E}[x_k]$ and $\mu_y \coloneqq \mathbbm{E}[y_k]$ is the subspace expectation value
\begin{equation}
\label{eq:subspace_estimator_of_xy}
\langle H \rangle (\vec{c}) = \frac{\mu_x}{\mu_y}.
\end{equation}
An asymptotically-unbiased and consistent estimator of this ratio is given by the ratio of the sample means~\cite{van_kempen_mean_2000}
\begin{equation}
\label{eq:ratio_ostimator_O}
\hat{H}(\vec{c}): (\{k^{(1)}, \dots , k^{(S)}\}) \mapsto \frac{\sum_{n=1}^S x_{k^{(n)}}}{\sum_{n=1}^S y_{k^{(n)}}}.
\end{equation}
While no closed-form expression exists for the variance of this estimator, it can be approximated by a second-order Taylor approximation around $(\mu_x, \mu_y)$ which yields~\cite{van_kempen_mean_2000}
\begin{equation}
\label{eq:ratio_estimator_variance}
\epsilon^2(\vec{c}) = \mathrm{Var}[\hat{H}(\vec{c})]  \approx \frac{1}{S} \left( \frac{\mathrm{Var}[x_k]}{\mu_y^2} +\frac{\mu_{x}^{2}\mathrm{Var}[y_k]}{\mu_{y}^{4}}-\frac{2\mu_{x}\mathrm{Cov}[x_k,y_k]}{\mu_{y}^{3}}\right).
\end{equation}
This approximation becomes very good in the limit of many shots~\cite{huggins2021virtual}.
Note that by choosing $\sigma_0 = \mathbbm{1}$ and $\vec{c} = \left(1, 0, 0, \dots\right)^T$, this reduces to the standard error on the mean from Eq.~\eqref{eqn:monte_carlo_variance}.
Finally we obtain the estimated error $\hat{\epsilon}(\vec{c})$ by inserting the (corrected) sample standard deviations for $x_k$ and $y_k$ and the sample covariance based on the shots $\{k^{(1)}, \dots , k^{(S)}\}$ into Eq.~\eqref{eq:ratio_estimator_variance}. 
Crucially, the covariance term captures the contributions to the statistical error that arise from re-using the same measurement samples to estimate all matrix entries. 

%What to do when repeating shots from a given setting. 
The above statistical error propagation holds when the POVM samples are independent and identically distributed (i.i.d.). 
In practice, for POVMs from randomized measurements such as those discussed in Chap.~\ref{chap:duals_optimization}, it is often experimentally convenient to take multiple measurement shots for each randomized readout basis, which violates the i.i.d. assumption.
In this case, we can re-define the random variables $x_k$ and $y_k$ for the numerator and denominator of the ratio estimator as their averaged value over all repeated shots taken in the same measurement basis. 
This way, the obtained values for $x_k$ and $y_k$ remain i.i.d., and the expressions from Eqs.~\eqref{eq:ratio_ostimator_O}--\eqref{eq:ratio_estimator_variance} are valid to be used as a basis to solve the constrained optimization problem from Eq.~\eqref{eq:PSE_constrained_optimization}. 

Besides the substantial gains in efficiency of evaluating the matrix entries of the subspace expansion in parallel rather than sequentially, our approach based on IC measurements offers two additional advantages over traditional methods. 
Firstly, the expansion operators $\{\sigma_1, \dots, \sigma_L\}$ can be chosen \textit{a posteriori} after measurements have been taken. 
Thus, the same measurements can be reused to iteratively optimize the chosen subspace operators to minimize the energy while the constrained optimization guarantees that the variational principle is respected up the the chosen statistical precision $\epsilon_\text{max}$. 
The estimation of the matrix entries remains feasible if the overall observables remain sufficiently local. 
Secondly, the framework offers the flexibility of estimating additional (local) observables of interest $O$ for the obtained energy-optimized state without additional measurements by constructing the estimator $\hat{O}(\vec{c}_\text{opt})$ equivalently to Eq.~\eqref{eq:ratio_estimator_variance} where $\vec{c}_\text{opt}$ is the solution to Eq.~\eqref{eq:PSE_constrained_optimization}. 
In the following, we showcase these features in a series of proof-of-principle experiments. 

\section{Exact ground state preparation circuits}
\label{sec:PSE_circuits_background}
We benchmark PSE on the task of recovering the ground state energy of Hamiltonians for which there exist known analytical or approximate ground state preparation circuits. 
This allows us to test the method at system sizes that would typically lie beyond the reach of an exact classical reference. 
We borrow two such systems with known ground state preparation circuits from the literature, which are presented in the following. 

\subsection{Bethe ansatz circuits for the XX model}
\begin{figure}[t]
    \centering
    \includegraphics[width=1\textwidth]{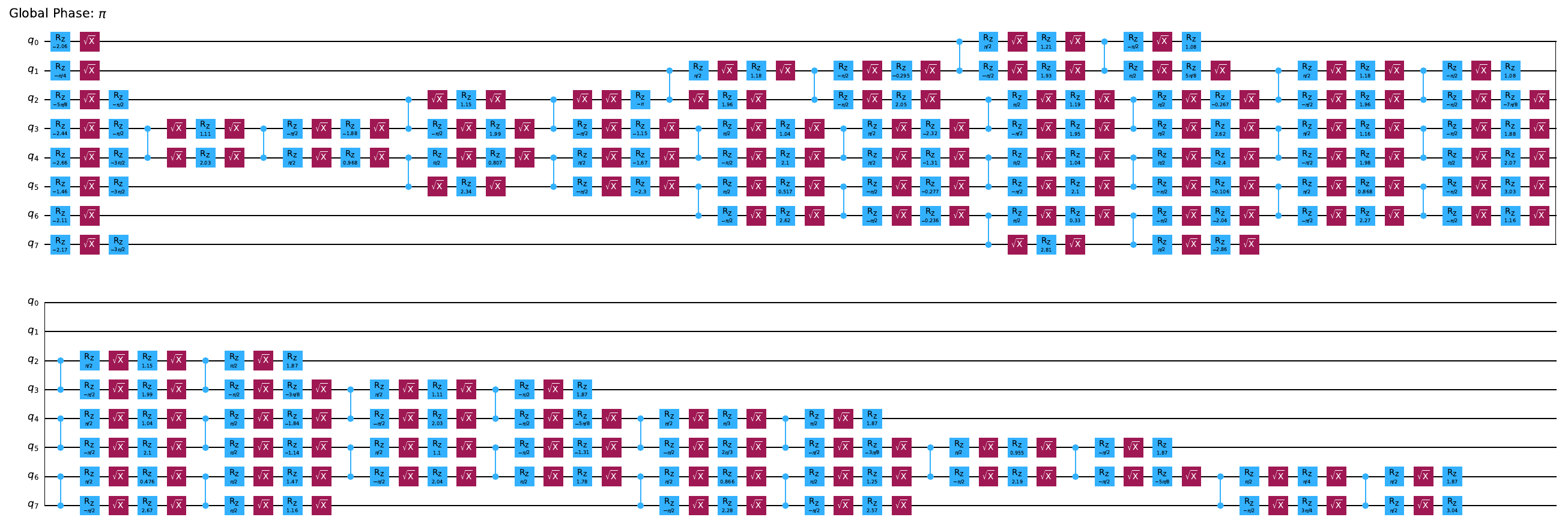}
    \caption[Quantum circuit of the Bethe ansatz state for the XX model]{\small Quantum circuit of the Bethe ansatz state for the XX model of eight qubits following Ref.~\cite{Ruiz2024betheansatzas}. Blue and red boxes indicate single-qubit $R_z$ and $\sqrt{X}$ gates, respectively. 
}
\label{fig:bethe_circ_example}
\end{figure}

Among the most famous analytical constructions of ground states in many-body quantum physics is the Bethe ansatz. 
It was originally introduced to solve the Heisenberg spin-$\frac{1}{2}$ chain~\cite{bethe1931theorie} and is one of the few examples of exact solutions in strongly correlated systems.
Recent works have enabled the state preparation of the algebraic Bethe ansatz states with unitary quantum circuits~\cite{Sopena2022algebraicbethe, Ruiz2024betheansatzas}. 
We employ the circuit construction from Ref.~\cite{Ruiz2024betheansatzas} to implement the exact ground state of the antiferromagnetic $XX$-model defined by the $N$-qubit Hamiltonian
\begin{equation}
\label{eq:bethe_hamiltonian}
H = \sum_{j=1}^N \left( X_j X_{j+1} + Y_j Y_{j+1} \right)
\end{equation}
where we impose periodic boundary conditions (PBC), i.e., $N+1 \equiv 1$. 
This Hamiltonian can be identified with a spinless free-fermion system via a Jordan-Wigner mapping. 
The total magnetization $\sum_{i=1}^N Z_i$ is a conserved quantity, so eigenstates are labeled by their number of spin-down particles, called ``magnons''. 
For even $N$, the ground state corresponds to a Fermi sea of $F = N/2$ magnons with a ground state energy of 
\begin{equation}
\label{eq:bethe_groundstate}
E_0 = \sum_{j = 0}^{N - 1} \cos\left( \frac{2\pi}{N} \left(j -\frac{N}{4}\right) \right).
\end{equation}
The number of two-qubit gates required for the preparation circuit scales as $\mathcal{O}(FN)$~\cite{Ruiz2024betheansatzas}. 
The structure of the circuit is shown for a small-scale example in Fig.~\ref{fig:bethe_circ_example}. 
We see that the circuit relies only on nearest-neighbor CZ gates which makes it hardware-friendly to implement on a one-dimensional chain of qubits. 

\subsection{Symmetry-protected topological phase with MPS ground states}

\begin{figure}[t]
    \centering
    \includegraphics[width=0.88\textwidth]{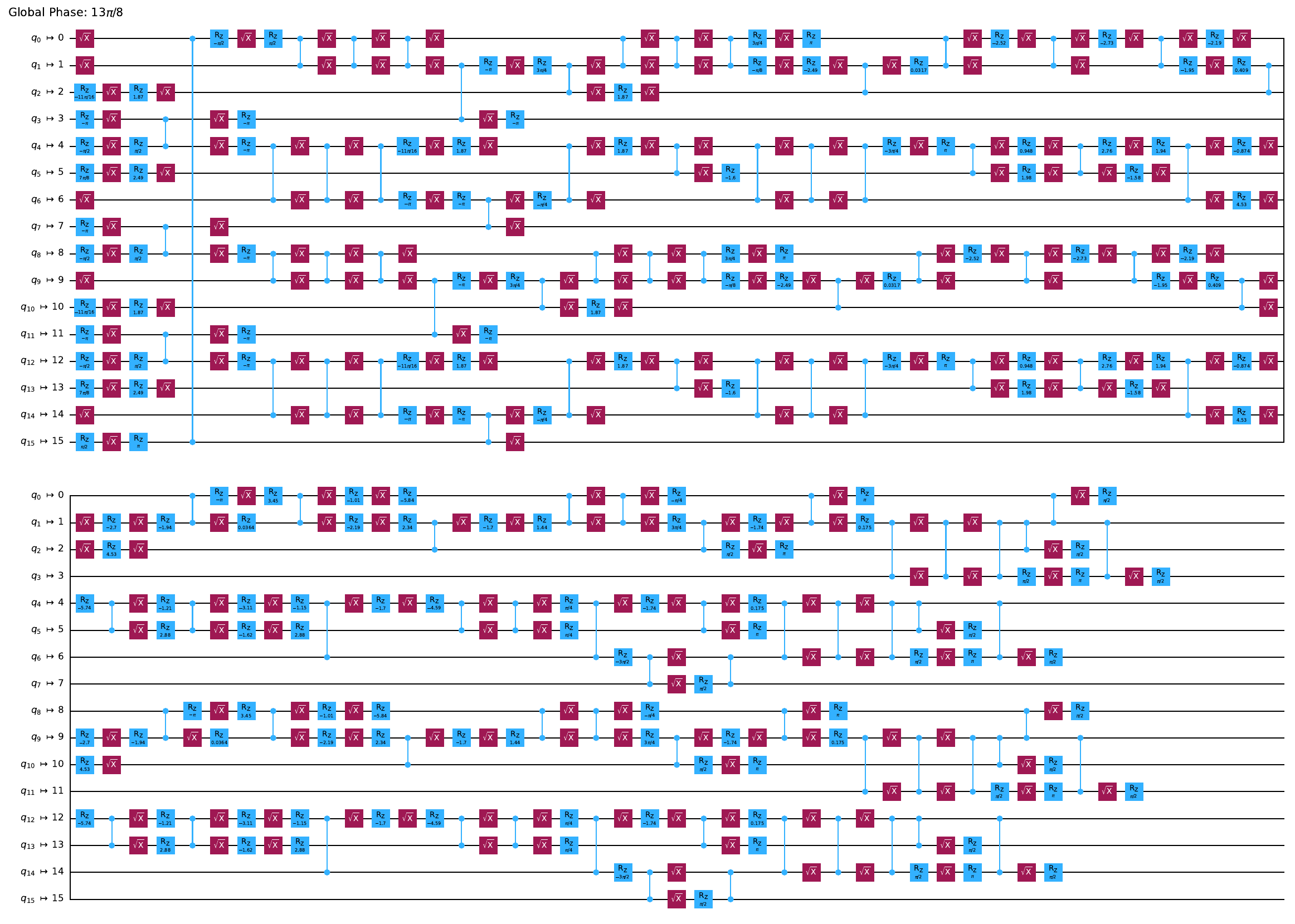}
    \caption[Quantum circuit of the MPS ground state of a ZXZ model]{\small Quantum circuit that prepares an approximate ground state of the Hamiltonian from Eq.~\eqref{eq:Hamiltonian_MPS_toy_model} with $g = -0.7$ through a renormalization-group-based MPS ansatz. 
}
\label{fig:MPS_circuit_example}
\end{figure}

As a second benchmark system we consider another one-dimensional spin chain originally studied in Ref.~\cite{MPS_toy_model_paper} described by the Hamiltonian (with PBC)
\begin{equation}
\label{eq:Hamiltonian_MPS_toy_model}
H = \sum_{i=1}^N \left( - g_{zz} Z_i Z_{i+1} - g_x X_i + g_{zxz} Z_i X_{i+1} Z_{i+2}  \right)
\end{equation}
As shown in Ref.~\cite{MPS_toy_model_paper}, the parameter trajectory
\begin{equation}
\label{eq:phase_trajectpry_MPS}
\langle g_{zz}(g), g_x(g), g_{zxz}(g) \rangle = \langle 2(1- g^2), (1+g)^2, (g-1)^2 \rangle \,,\, g \in \left[ -1, +1 \right]
\end{equation}
describes a cut through the phase space of this model that traverses a quantum phase transition with a critical point at $g=0$ between a symmetry-protected topological (SPT) phase where the $ZXZ$-term dominates, and a trivial phase where the $X$-terms dominate. 
Crucially, along this trajectory, the ground state of the model corresponds to a translation-invariant normal matrix product state (MPS)~\cite{MPS_system_first_demo} with an energy density
\begin{equation}
\label{eq:MPS_exact_gs_energy}
\frac{E_0}{N} = -2\left(g^2+1\right).
\end{equation}
We can thus resort to known techniques for mapping MPS to quantum circuits to prepare the ground state of this system on a quantum computer. 
Here, we employ a recently introduced algorithm based on the renormalization-group transformation that approximately prepares the MPS with an error $\epsilon$ in depth $\mathcal{O}\left(\log(N/\epsilon)\right)$~\cite{MPS_preparation_theory}. 
The approximation is determined by a ``blocking number'' $q$ which navigates a trade-off between circuit depth and the fidelity of the prepared state. 
Here, we opt for a blocking number of $q=4$ which has been shown to be a reasonable choice for current noisy quantum hardware~\cite{moritz_MPS_in_prep}. 
As an example, the resulting quantum circuit for $N=16$ and $g=-0.7$ is shown in Fig.~\ref{fig:MPS_circuit_example}. 
In contrast to the Bethe ansatz circuits, the depth of these MPS circuits does not increase with the system size.
The phase transition at $g=0$ is characterized by order parameters which identify the SPT phase and the trivial phase. 
These serve as interesting additional observables to estimate based on the energy-optimized subspace state.
While SPT phases in general are characterized by non-local order parameters, for the MPS ground states along the trajectory from Eq.~\eqref{eq:phase_trajectpry_MPS}, local string order parameters are sufficient to indicate the phase transition~\cite{MPS_system_first_demo}. 
In that case, the observable 
\begin{equation}
\label{eq:sop_identity}
S^X \coloneqq \frac{1}{N} \sum_{j=1}^N X_j
\end{equation}
signals the trivial phase. It is zero for states with $g < 0$ and goes to $1$ with $g \rightarrow 1$ as $4g / (1+g)^2$. Similarly, the SPT phase is characterized by the observable (with PBC)
\begin{equation}
\label{eq:sop_zy}
S^{ZY} \coloneqq \frac{1}{N} \sum_{j=1}^N Z_j Y_{j+1} X_{j+2} Y_{j+3} Z_{j+4}
\end{equation}
which remains zero for states with $g > 0$ and goes to $1$ with $g \rightarrow -1$ as $-4g / (1-g)^2$.

\section{Experimental results}
\label{sec:PSE_experimental_results}
We implement the MPS ansatz circuits and the Bethe ansatz circuits on the IBM Quantum device \emph{ibm\_fez} of the Heron R2 generation.
This device employs a native gate set of $\{\text{CZ},\sqrt{X},R_z\}$ on a heavy-hexagonal topology.  
While the Bethe ansatz circuits (Fig.~\ref{fig:bethe_circ_example}) already match the native hardware connectivity, the MPS ansatz circuits (Fig.~\ref{fig:MPS_circuit_example}) feature multiple non-nearest-neighbor gates which leads to additional SWAP gates upon transpilation. 
We implement each circuit with system sizes increasing from $N=16$ to $N=80$. 
The resulting numbers of two-qubit gates in the transpiled circuits is summarized in Tab.~\ref{tab:details_subspace_expansion}. 

\begin{table}
\centering
\begin{tblr}{c|ccccc|ccccc}
\hline \hline
    & \SetCell[c=5]{c}MPS ground state circuits &&&&& \SetCell[c=5]{c} Bethe ansatz circuits \\
    \hline\hline
     $N$ & 16 & 32 & 48 & 64 & 80 & 16 & 32 & 48 & 64 & 80 \\
     CZ gates & 108 & 216 & 324 & 432 & 540 & 184 & 753 & 1732 & 3161 & 5041 \\
     IC measurement bases & \SetCell[c=5]{c} 32,768 &&&&& \SetCell[c=5]{c} 16,384 \\
     shots per basis & \SetCell[c=5]{c} 8 &&&&& \SetCell[c=5]{c} 8 \\
     \SetCell[r=2]{c}{\# Pauli traces for \\ PSE Krylov+ ($ \times 10^9$)} & \SetCell[r=2]{c} 27 &\SetCell[r=2]{c} 92 &\SetCell[r=2]{c} 236 &\SetCell[r=2]{c} 474 &\SetCell[r=2]{c} 704 &\SetCell[r=2]{c}  7 &\SetCell[r=2]{c} 25 &\SetCell[r=2]{c} 58 &\SetCell[r=2]{c}  111 &\SetCell[r=2]{c} 187 \\
      &     &     &     &     &     &     &     &     &     &     \\
\hline \hline
\end{tblr}
\caption[Experimental details on circuits used to demonstrate PSE]{\small Experimental details on circuits used to demonstrate PSE.}
\label{tab:details_subspace_expansion}
\end{table}

\begin{figure}[t]
    \centering
    \includegraphics[width=1\textwidth]{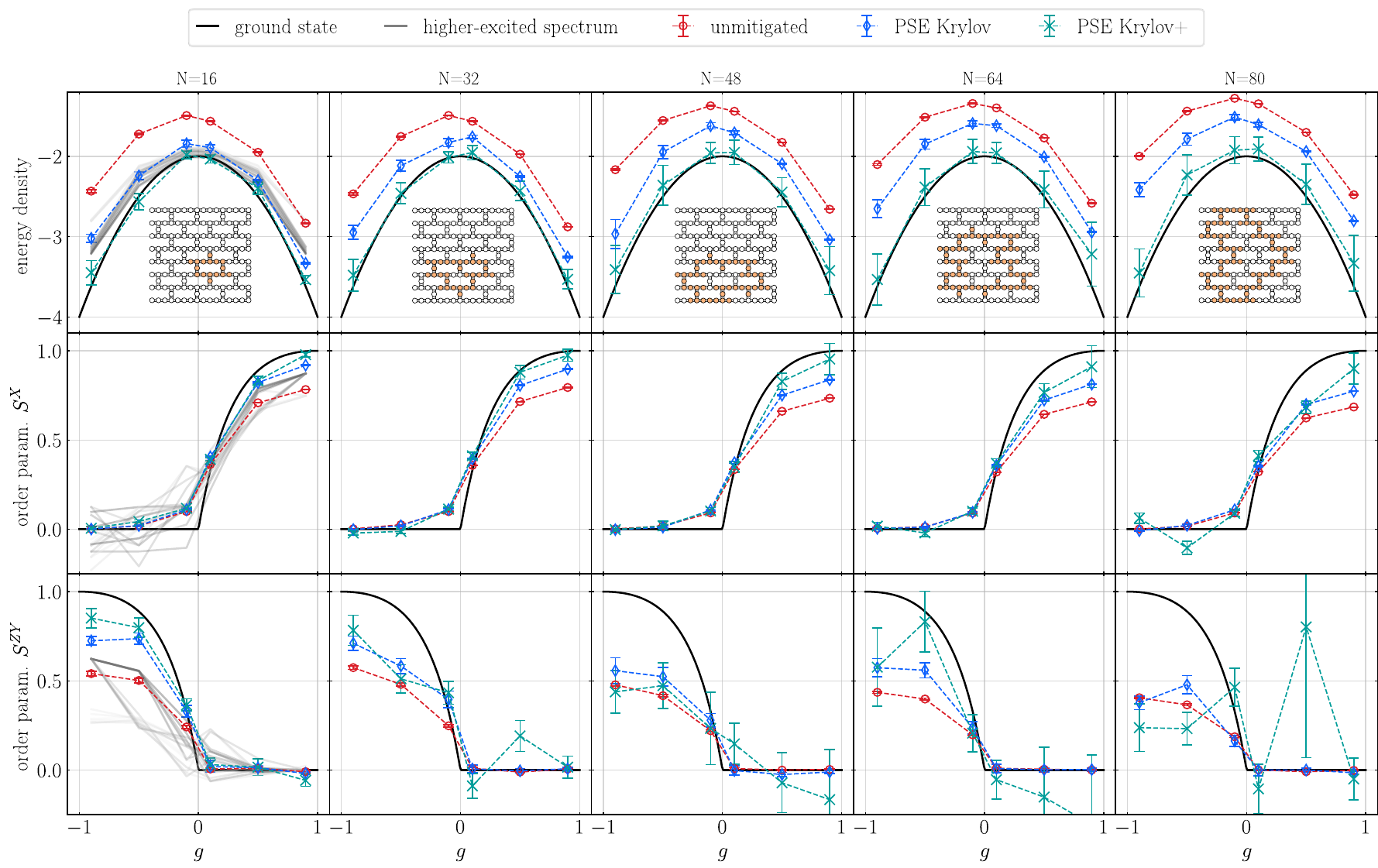}
    \caption[Parallelized subspace expansion results across a phase transition]{\small Error mitigation with parallelized subspace expansion for a phase transition across an SPT phase with MPS ground states. 
    The qubit number increases from left to right from $N=16$ to $N=80$ while the three rows show the energy density and the order parameters of the traversed phase transition. 
    Inlays show the chosen physical qubits of the 156-qubit device \textit{ibm\_fez}. 
    For $N=16$, grey shaded curves show the theoretical values for the lowest 25 excited states obtained with exact diagonalization, with a color gradient from black for the ground state to lighter shades for progressively higher-excited states.
}
\label{fig:PSE_MPS_results}
\end{figure}

For maximum flexibility in the post-processing, we measure an IC POVM that consists of uniformly randomized single-qubit $X$, $Y$, and $Z$ measurements (standard classical shadows with canonical duals).
The number of randomized readout bases and repeated shots per basis are also summarized in Tab.~\ref{tab:details_subspace_expansion}.
In principle, we could now choose any set of sufficiently local observables as expansion operators $\sigma_i$ of the subspace. 
We first select $\sigma_1 = \mathbbm{1}$ to ensure that the original state $\rho_0$ is part of the subspace such that the subspace-optimized energy can never be worse than the unmitigated one. 
Next, we choose $\sigma_2 = H$ to build a Krylov subspace which boosts the ground state population as discussed in Sec.~\ref{sec:spectral_calculations}.
We would ideally like to add more powers of the Hamiltonian as expansion operators. 
However, the inclusion of $\sigma_2 = H$ already requires estimating a term $\mathcal{H}_{2, 2} = \Tr[\rho_0 H^3]$ which is an observable with $2.14 \times 10^6$ Pauli terms for the SPT model with $N=80$. 
The trace of each of these Pauli terms with each measured dual operator (here $S=262,144$) needs to be evaluated. 
The inclusion of $\sigma_3 = H^2$ would result in a term $\mathcal{H}_{3, 3} = \Tr[\rho_0 H^5]$ with $35\times 10^9$ Pauli terms and strings with weight up to 15. 
This both exceeds our available classical processing resources and is beyond the scope of estimation with single-qubit classical shadows. 
We hence limit ourselves to one Krylov dimension here and define the expansion set $\text{PSE Krylov}\,\, \coloneqq \{\mathbbm{1}, H\}$.

Besides the Hamiltonian, we systematically add low-weight expansion operators that help reduce the energy with the following procedure, which is similar to a strategy proposed in Ref.~\cite{boyd2025high}.
For Pauli weights $w \in \{1, 2, 3, 4, 5\}$, we sample 300 Pauli strings of weight $w$ uniformly at random to form a pool $\mathcal{P}_w$. 
To assess which of these Pauli strings adds a useful dimension to the subspace, we perform a small subspace expansion with operators $\{\mathbbm{1}, P_i \}$ for each $P_i \in \mathcal{P}_w$. 
We sort the Pauli strings in each pool by how much they have reduced the energy compared to the unmitigated estimate when optimizing for the lowest upper error $ \arg\min_{\vec{c}}\, \hat{H}(\vec{c}) + \hat{\epsilon}(\vec{c})$. 
Finally, we include the five best-performing Paulis of each weight into the set of expansion operators. 
Together with $\sigma_1 = \mathbbm{1}$ and $\sigma_2 = H$, this results in a 27-dimensional subspace which we refer to as ``PSE Krylov+''.
The total numbers of Pauli traces between dual operators of measurement outcomes and the expansion operators of the subspace are given in Tab.~\ref{tab:details_subspace_expansion}.
For both the ``PSE Krylov'' and the ``PSE Krylov+'' subspaces, we solve the constrained optimization problem from Eq.~\eqref{eq:PSE_constrained_optimization} with the \emph{constrained optimization by linear approximation} (COBYLA) algorithm~\cite{Powell1994}.
The allowed maximum errors $\epsilon_\text{max}$ are shown as error bars on the PSE curves in Figs.~\ref{fig:PSE_MPS_results} and~\ref{fig:PSE_MPS_results} and were chosen a posteriori to optimally navigate the inherent bias/variance trade-off.

The results for the error-mitigated energy densities of the MPS ansatz circuits are shown in the top row of Fig.~\ref{fig:PSE_MPS_results}.
Across all studied system sizes, the ``PSE Krylov'' approach already significantly improves on the unmitigated energies, while the ``PSE Krylov+'' approach manages to accurately recover the true ground state expectation values from Eq.~\eqref{eq:MPS_exact_gs_energy} while respecting the variational principle within statistical uncertainties. 
As the system size increases, the statistical errors of the unmitigated results become progressively worse which leads to larger error bars of the mitigated values.
For the smallest system size of $N=16$ we can perform an exact diagonalization of the Hamiltonian to get the full low-energy (density) spectrum. 
This reveals that several states lie between the unmitigated and ``PSE Krylov'' values and the true ground state. 
However, the ``PSE Krylov+'' values lie well below the first excited state, indicating that the optimized subspace state indeed manages to produce significant overlap with the true ground state. 

We reuse the IC measurement samples to also estimate the string order parameters (Eqs.~\eqref{eq:sop_identity}--\eqref{eq:sop_zy}) that characterize the phase transition. 
The single-qubit order parameter $S^X$ is shown in the second row of Fig.~\ref{fig:PSE_MPS_results}. 
Here, the energy-optimized subspace states consistently improve over the unmitigated values for high $g$ values. 
The weight-five observable $S^{ZY}$ shown in the final row of Fig.~\ref{fig:PSE_MPS_results} is more difficult to mitigate. 
While the results recover the theoretical behavior well for $N=16$, the statistical uncertainties in the ``PSE Krylov+'' values dominate at larger system sizes. 
Form this, we conclude that, while the energy remains well-mitigated throughout all system sizes, the recovered subspace state does not provide sufficient overlap with the ground state to accurately reflect complex observables like $S^{ZY}$. 

\begin{figure}
  \begin{minipage}[c]{0.55\textwidth}
    \includegraphics[width=\textwidth]{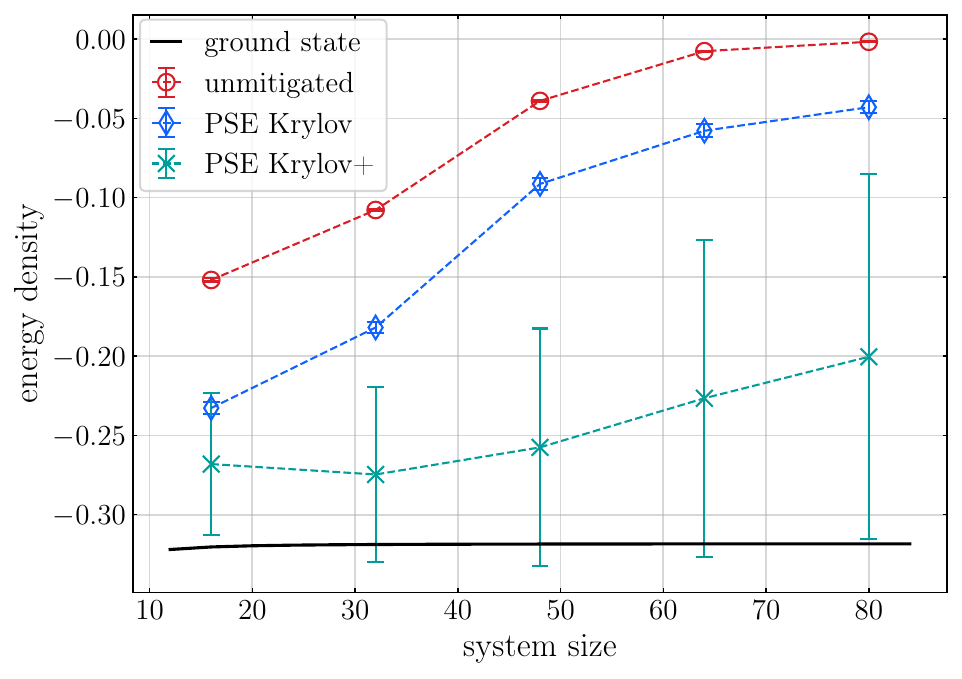}
  \end{minipage}\hfill
  \begin{minipage}[c]{0.43\textwidth}
    \caption[Parallelized subspace expansion results for the Bethe ansatz circuit]{\small 
    Error mitigation with parallelized subspace expansion for exact Bethe ansatz ground state circuits of the XX model. 
    } \label{fig:PSE_Bethe_results}
  \end{minipage}
\end{figure}

The results for the energy density of the  Bethe ansatz circuits are shown in Fig.~\eqref{fig:PSE_Bethe_results}. 
While ``PSE Krylov'' and ``PSE Krylov+'' do provide lower energies than the unmitigated values, they remain well above the theoretical ground state. 
Only when increasing the allowed $\epsilon_\text{max}$ of the ``PSE Krylov+'' approach can we match the ground state within error bars, where our statistical error treatment accurately respects the variational principle. 
We note that the depths and gate counts of these circuits are significantly larger than the MPS ansatz circuits and thus accumulate more errors. 
The large mitigation error bars indicate that these circuits currently lie beyond the capabilities of PSE.

\section{Discussion}
\label{sec:subspace_discussion}

We have developed a formalism to parallelize the processing of quantum subspace expansion with IC measurements which we refer to as \emph{parallelized subspace expansion} (PSE). 
Besides avoiding the measurement overhead of other subspace techniques, this offers the great flexibility of exploring various expansion operators a posteriori in the spirit of the ``measure first -- ask questions later'' paradigm~\cite{elben2023randomized}. 
Moreover, besides the subspace-optimized observable (usually the energy), expectation values of additional observables can be evaluated on the subspace state without the need for additional measurements.
Our main technical contribution is a rigorous treatment of the statistical error in the presence of covariances that arise from reusing the same measurement samples for all contributing expectation values.

As a proof-of-principle demonstration, we use PSE as an error-mitigation method to obtain ground state energies for two spin models with known ground state preparation circuits. 
While we were able to obtain accurate energies for circuits with up to 80 qubits and over 500 entangling gates, deeper circuits with $>1000$ two-qubit gates proved too noisy for accurate mitigation. 
When mitigating order parameters of a phase transition as additional observables, we have found the method successful for a low-weight observable but more challenging for a higher-weight observable. 

Our experiments serve as an example of how quantum and classical resources can be effectively combined.
For reference, the classical post-processing of the obtained datasets involved computing over one trillion individual traces between dual operators from measurement shots and Pauli operators from observables.
With measurements in 32.768 randomized bases for the largest dataset, this constitutes one of the most advanced classical shadow measurements to date. 

There are several avenues to further generalize and improve the PSE method. 
Firstly, in the processing of our experiments, we have used the canonical dual frame for estimating all the matrix entries. 
It would be desirable to integrate the duals optimization procedure from Chap.~\ref{chap:duals_optimization} into PSE. 
In principle, one would like to optimize the dual frame for the estimation of each matrix entry $\mathcal{H}_{ij}$ and $\mathcal{S}_{ij}$. 
However, this would violate the i.i.d. condition of the individual samples for the random variables $x_k$ and $y_k$ which is not compatible with our statistical error propagation in its current form. 
We thus leave the exploration of this aspect for future work. 
Secondly, the locality of observables we can process is limited for our above choice of local classical shadow measurements. 
This could be improved with more powerful classes of POVMs like dilation POVMs, shallow~\cite{bertoni2024shallowshadows} or even global Clifford randomized unitaries~\cite{huang2020predicting}.

Moreover, while we have focused on expansion operators inspired by Krylov methods~\cite{suchsland2021algorithmic} and selected Pauli excitations~\cite{boyd2025high}, PSE is in principle ready to incorporate other subspace methods such as symmetry verification~\cite{bonetmonroig2018lowcosta} or even methods based on powers of the prepared state~\cite{yoshioka2022generalized}. 
Finally, our processing is based on using the sample means to estimate ratio of Eq.~\eqref{eq:subspace_estimator_of_xy}. 
While this allows our straightforward treatment of the statistical errors, more sophisticated estimators based on medians-of-means could be more powerful in suppressing outliers~\cite{huang2020predicting}. 

We emphasize that the error-mitigation capabilities of PSE come at no additional quantum resource costs given the ability to sample IC measurement data. 
However, compared to other error-mitigation methods it is only applicable to variational minimization rather than arbitrary quantum circuits. 
In the next chapter, we explore error-mitigation in a more general context with a method that is also enabled by IC measurements but relies on a calibrated noise model.

\chapter{Dynamical simulations of many-body quantum chaos}
\label{chap:dual_unitary_TEM}

\summary{
This chapter is based on the article ``Dynamical simulations of many-body quantum chaos on a quantum computer'' by \textbf{Laurin E. Fischer}$^\ast$, Matea Leahy$^\ast$, Andrew Eddins, Nathan Keenan, Davide Ferracin, Matteo A. C. Rossi, Youngseok Kim, Andre He, Francesca Pietracaprina, Boris Sokolov, Shane Dooley, Zoltán Zimborás, Francesco Tacchino, Sabrina Maniscalco, John Goold, Guillermo García-Pérez, Ivano Tavernelli, Abhinav Kandala, and Sergey N. Filippov, published in Nature Physics (2026)~\cite{fischer2024dynamical}. \footnotesize{$^\ast$shared first authorship} \normalsize 
\bigskip

\noindent Quantum circuits with local unitaries have emerged as a rich playground for the exploration of many-body quantum dynamics of discrete-time systems. 
Their intrinsic locality makes them particularly suited to run on current quantum processors.
In this chapter, we study a special class of maximally chaotic circuits known as dual-unitary circuits --- exhibiting unitarity in both space and time --- that are known to have exact analytical solutions for certain correlation functions. 
With advances in noise learning and the implementation of tensor-network error mitigation, we accurately simulate these correlators with up to 91 qubits. 
We then probe dynamics beyond exact verification, by perturbing the circuits away from the dual unitary point, and compare our results to classical approximations with tensor networks. 
These results cement error-mitigated digital quantum simulation on pre-fault-tolerant quantum processors as a trustworthy platform for the exploration of quantum many-body phases.
}    

In this Chapter, we explore large-scale digital quantum simulations with error mitigation. 
Our main goal is to scale such simulations up to system sizes that are beyond brute-force classical methods. 
The choice of the quantum model that we simulate is guided by the following desiderata:
Firstly, the circuit volumes need to be within the reach of state-of-the-art error mitigation techniques which extends towards $\mathcal{O}(1000)$ two-qubit gates~\cite{kim2023evidence}.
Secondly, we aim to measure a quantity that carries a clear physical meaning, going beyond the more arbitrary observables measured in previous large-scale computations~\cite{kim2023evidence}. 
Finally, we would like to have a way to verify the accuracy of the quantum simulation. 
Here, we simulate the Floquet evolution of a chaotic kicked Ising system with a special dual unitary property which satisfies all the above criteria.

Traditionally, the study of many-body quantum dynamics has been that of continuous time processes. 
In this context, digital quantum simulation algorithms decompose a continuous evolution into elementary, discrete steps, e.g., through product formulas (see Sec.~\ref{sec:quantum_dynamics})~\cite{miessen2023quantum}.
However, as hardware platforms matured and became capable of executing large-scale quantum circuits, a different paradigm emerged. 
In this new scenario, the building blocks of quantum circuits themselves---local unitary gates and measurements---directly give rise to non-equilibrium, discrete-time phenomena. 
Crucially, these protocols can be implemented exactly at any circuit depth, as they are by definition not subject to the Trotterization errors of product formulas (see Sec.~\ref{sec:quantum_dynamics}).
A remarkable example is represented by the simulation of stroboscopic Floquet dynamics, which offers a wealth of new possibilities to probe universal and emergent phases. 
This includes the investigation of random quantum circuits~\cite{fisher2023random}, computational sampling problems~\cite{hangleiter2023computational}, measurement induced criticality~\cite{fisher18,skinner19}, the emergence of time-crystalline order~\cite{mi2022time}, the existence of many-body localized phases~\cite{mbl1,mbl2,mbl3}, and integrable circuits~\cite{integrable}. 

Among the many advancements that this ``digitalization'' of quantum dynamics brought in the theory of many-body quantum physics and quantum chaos~\cite{chan18, ber-18a, ber-21a}, a key development has been the identification of a class of models known as \emph{dual unitary} (DU) circuits~\cite{ber-19a}.
These are composed of gates that exhibit unitarity in both the temporal and spatial dimensions.
This unique characteristic allows for the exact computation of certain system properties that would typically be exceedingly challenging to evaluate~\cite{gopalakrishnan-2019, pir-20a, ber-20b, ipp-21a, suz-22}.
DU circuits act as rapid scramblers of quantum information, with two-time correlation functions and out-of-time correlators propagating at their maximum possible velocities~\cite{ber-19a, cla-20a, ber-20a}, a signature that has already been simulated, e.g., in Ref.~\cite{chertkov2022holographic}.
For this reason, these circuits are often described as ``maximally chaotic''~\cite{ber-18a, ber-21a}.
Similarly, for certain solvable initial states, it has been shown that entanglement growth occurs at the maximum rate~\cite{ber-19a, ber-19b, zho-22a}.

The ability to simulate Floquet dynamics---including DU circuits---with local gates and short-depth quantum circuits makes them particularly suitable to explore with current, pre-fault-tolerant quantum computers.
Although advances in scale and quality have already enabled the exploration of increasingly complex quantum simulation~\cite{mi2021information,keenan2023evidence,kim2023evidence, shinjo2024unveiling, physrevd.109.114510, shtanko2023uncovering} on these processors, their accuracy is still impacted by noise, which is why they rely on error mitigation (see Sec.~\ref{sec:error_mitigation_section}). 
However, a natural question then emerges: in general, how does one build trust in error-mitigated quantum computations at these scales?
While Clifford circuits are a powerful benchmarking tool~\cite{kim2023evidence}, they may not be representative of performance at parameter regimes of interest~\cite{govia2024bounding}.
We argue that dual-unitary Floquet models like the one studied here can serve as relevant benchmarks in this context, producing non-Clifford circuits of non-trivial scales with analytical solutions.

\section{The kicked Ising model and dual-unitary circuits}
\label{sec:TEM_dual_unitary_setup}

\begin{figure}
    \centering
    \includegraphics[width=\textwidth]{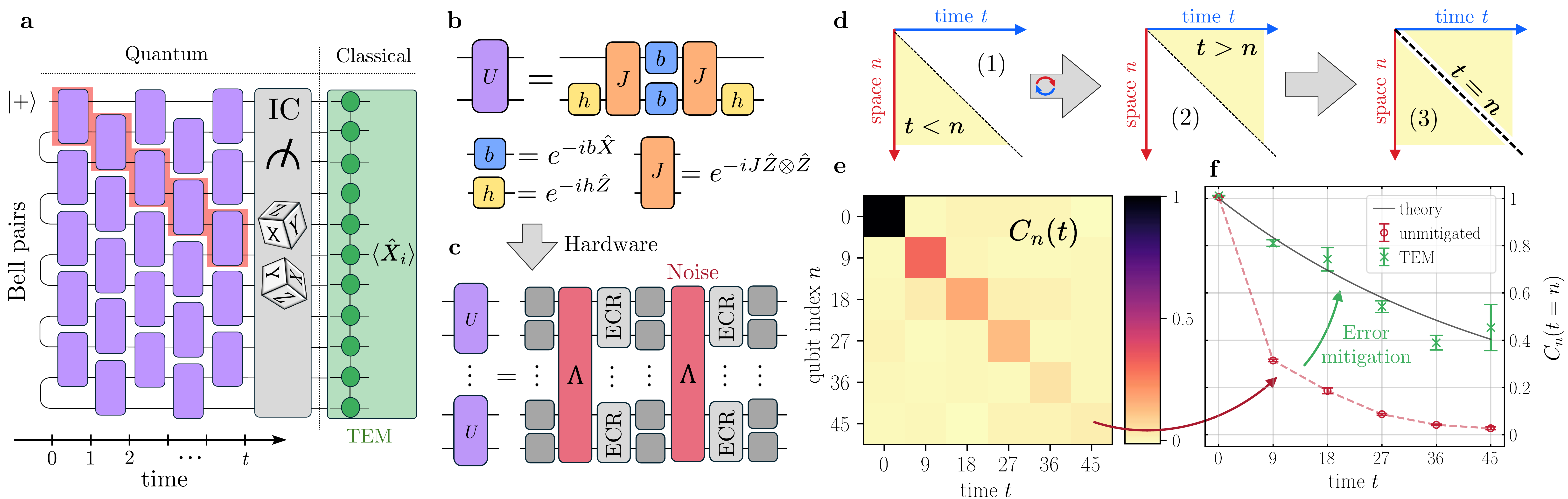}
    \caption[Simulating dual-unitary circuits with tensor-network error mitigation]{ \small
    Simulating dual-unitary circuits with tensor-network error mitigation.
    \textbf{a})~Brickwork circuits of dual unitary blocks implement the Floquet evolution of a kicked Ising model. 
    The single-qubit $ X_i$ observable on the light cone boundary (red shaded region) yields the infinite-temperature autocorrelation function $C_n(t)$. 
    We take randomized, informationally complete single-qubit measurements. 
    Measured samples are classically post-processed by the TEM algorithm, which inverts the effects of noise in the circuits.
    \textbf{b})~Building blocks of the two-qubit gate ${U}$, which is dual unitary for $\abs{J} = \abs{b} = \pi/4$. 
    \textbf{c})~Transpiled to the hardware, one time step consists of two entangling two-qubit ECR layers alternating with single-qubit gates. 
  	We model noise as Pauli channels $\Lambda$ associated with every unique layer of ECR gates.
    \textbf{d})~In brickwork circuits, information spreads in a light cone with zero correlations between $(n_0, t_0) = (0, 0)$ and $(n, t)$ for $t < n$ (1). 
    Similarly, dual unitarity limits information spread in the spatial direction such that correlations vanish for $n < t$ (2). As a result, non-zero correlations are only found on the $t=n$ light cone boundary (3). 
    \textbf{e})~Unmitigated measurements of $C_n(t)$ for a dual unitary circuit for $N=91$ qubits and $h=0.1$.
    \textbf{f})~Error mitigation recovers the correct decay of the autocorrelator.
    }
    \label{fig:TEM_Fig1}
\end{figure}

Given a two-qubit unitary ${U} = \sum_{i,j,k,l=0}^1 U_{ij}^{kl} \ket{k}\bra{i}\otimes\ket{l}\bra{j}$, one defines a dual operator\footnote{Note that this notion of duality is different from the ``dual frames'' used in Chap.~\ref{chap:duals_optimization}.} ${U}_D = \sum_{ijkl} U_{ij}^{kl} \ket{j}\bra{i}\otimes\ket{l}\bra{k}$ through a shuffling of some input/output subsystems of ${U}$ (exchange of the bra/ket indices $j \leftrightarrow k$). 
If the dual operator is itself unitary, i.e., if ${U}_D^{\dagger} {U}_D = {U}_D {U}_D^{\dagger} = {\mathbbm{1}}$, then the gate ${U}$ is called \emph{dual-unitary}. 
This can be thought of as a unitarity in space as well as in time. 
Dual-unitary circuits consist of $N$ qubits (labeled $n = 0,1,\hdots,N-1$) evolving by a ``brickwork'' pattern of dual-unitary gates ${U}_{n,n+1}$.
The brickwork is an even layer of dual-unitary gates ${\mathbb{U}}_e = \bigotimes_{j=0}^{(N-1)/2-1} {U}_{2j,2j+1}$ followed by an odd layer ${\mathbb{U}}_o = \bigotimes_{j=1}^{(N-1)/2} {U}_{2j-1,2j}$, repeated periodically, see Fig.~\ref{fig:TEM_Fig1}\textbf{a} for a quantum circuit sketch. 

We simulate the time evolution of a kicked Ising model, which is described by the time-dependent Hamiltonian
\begin{equation} 
\label{eq:kickising} 
{H}_{KI}(t) \coloneqq {H}_{I} + \sum_{m \in \mathbb{Z}} \delta(t-m){H}_{K} , 
\end{equation} 
where ${H}_{I} \coloneqq J \sum_{n=0}^{{N-2}}{Z}_{n} {Z}_{n+1} + h \sum_{n=0}^{N-1} {Z}_{n}$ is the Ising Hamiltonian, and the system is periodically kicked by the transverse field Hamiltonian ${H}_{K} \coloneqq b \sum_{n=0}^{N-1}{X}_{n}$.
The Floquet unitary for the stroboscopic evolution (see Sec.~\ref{sec:quantum_dynamics}) generated by ${H}_{KI}(t)$ is 
\begin{equation} 
\label{eq:Ising_Floquet_operator} 
{\mathbb{U}}_{KI} \coloneqq \mathcal{T}\exp \Big[ -i\int^{1}_{0}{H}_{KI}(t) dt \Big] = e^{-i{H}_{K}}e^{-i{H}_{I}}.
\end{equation} 
Conveniently, this Floquet unitary can be related to a brickwork circuit with the two-qubit gate
\begin{equation} 
\label{eq:two_qubit_gate} 
{U}_{n,n+1} \coloneqq e^{-ih{Z}_{n}} e^{-iJ{Z}_{n} {Z}_{n+1}} e^{-ib( {X}_{n} + {X}_{n+1}) } e^{-iJ{Z}_{n} {Z}_{n+1}} e^{-ih{Z}_{n}}
\end{equation}  
as shown in Fig.~\ref{fig:TEM_Fig1}\textbf{b}. 
Specifically, the kicked Ising Floquet unitary ${\mathbb{U}}_{KI}$ and the unitary of two brickwork layers ${\mathbb{U}} \coloneqq {\mathbb{U}}_o {\mathbb{U}}_e$ are related as ${\mathbb{U}} = {\Sigma} {\mathbb{U}}_{KI}^2 {\Sigma}^\dagger$ where ${\Sigma} \coloneqq \prod_{n {\rm ~odd}}e^{-iJ{\sigma}_n^z {\sigma}_{n+1}^z - ih{\sigma}_n^z}$. 
Since ${\mathbb{U}}^t = {\Sigma} {\mathbb{U}}_{KI}^{2t} {\Sigma}^\dagger$, this shows that, up to a layer ${\Sigma}^\dagger$ of unitary gates at the beginning of the circuit and a layer ${\Sigma}$ at the end of the circuit, each brickwork Floquet period corresponds to two Floquet periods of the kicked Ising model. 

Importantly, for the parameter choice $|J| = |b| = \pi/4$ in Eq.~\eqref{eq:Ising_Floquet_operator}, the corresponding brickwork circuit is dual-unitary. 
We refer to this as the ``dual-unitary point''. 
If, in addition, $h=0$, the model is integrable as it can be mapped to free fermions, and the corresponding circuit is composed of Clifford gates. 
However, for a general choice of $h$ the model is non-integrable.

We measure the one-point function $\bra{\Psi (0)} {X}_n (t) \ket{\Psi (0)}$ after a kicked Ising quantum quench from the initial state $\ket{\Psi(0)} \coloneqq \ket{+}_0 \otimes \ket{\psi_{\rm Bell}}^{\otimes \lfloor(N-1)/2 \rfloor}$ that prepares the $0$-th qubit in $\ket{+}_0=(\ket{0}_0+\ket{1}_0)/\sqrt{2}$ and all other qubits in a product of Bell pairs $\ket{\psi_{\rm Bell}} = (\ket{00}+\ket{11})/\sqrt{2}$ (for all qubit pairs $i, i+1$, with $i=1,3, \dotsc, N-2$), as illustrated in Fig.~\ref{fig:TEM_Fig1}\textbf{a}.
This is motivated by the observation that, at the dual unitary point, this observable offers an intuitive physical interpretation -- it corresponds to the infinite-temperature autocorrelation function of the form 
\begin{equation}
C_n(t) \coloneqq \Tr[{\rho}_{\infty} {X}_{0}(0){X}_{n}(t)],
%C(t)=\langle{X}_{0}{X}_{n}(t)\rangle
\label{eq:autocorr}
\end{equation}
where $\rho_{\infty}$ is the infinite-temperature (maximally-mixed) initial state ${\rho}_{\infty}=\mathbbm{1}/2^N$, and $t$ denotes the number of time steps. 
Exploiting the dual-unitary property, the autocorrelation function can be calculated exactly~\cite{ber-19a} for our model as 
\begin{equation} 
C_n (t) = \begin{cases} \left[\cos(2h)\right]^t \, & \text{if } n=t \\ 
            0 \,&\text{otherwise}
		 \end{cases}
\label{eq:C_exact} 
\end{equation} 
for $t \leq (N-1)/2$ (where $N$ is odd), see Appendix~\ref{app:dual_unitary_circuits}. 
As dual unitarity limits causality to within not only a temporal but also a spatial light cone, the autocorrelation function vanishes outside of the light cone boundary of $n=t$, see Fig.~\ref{fig:TEM_Fig1}\textbf{d}. 
On the light cone, the autocorrelator is constant at the integrable Clifford point $h=0$, but otherwise decays exponentially in time. 
A measurement of $C_n(t)$ for various qubits $n$ and time steps $t$ is shown in Fig.~\ref{fig:TEM_Fig1}\textbf{e}. 
As predicted analytically, we observe a negligible signal for $t\neq n$ and finite signal is only measured along the light-cone boundary for $t=n$. 

However, as a consequence of noise, the measured values on the light cone boundary decay quicker with $t$ than the exact evolution from Eq.~\eqref{eq:C_exact}. 
We mitigate gate noise with tensor-network error mitigation (TEM)~\cite{filippov2023scalable} (see Sec.~\ref{sec:tensor-network-error-mitigation}) and readout errors with TREX (see Sec.~\ref{sec:readout_error_mitigation}).
TEM employs a MPO representation of the noise inversion map $\tilde{\mathcal{M}}^\dagger$ which we find to be well converged with a maximum bond dimension of $\chi = 70$.
We estimate the TEM-modified observable $\tilde{\mathcal{M}}^\dagger(X_n)$ with IC samples from randomized single-qubit measurements.
We sample the measurement bases uniformly at random for each qubit, except for the ``signal qubit'' $i=t$, where the observable is biased towards $ X$ ($80\%$ probability for $ X$, $10\%$ for $ Y$, and $10\%$ for $ Z$), as the estimated observable is dominated by the $ X$-contribution for that qubit.
Finally, the estimation of the mitigated signal $\Tr[\rho\tilde{\mathcal{M}}^\dagger(X_n)]$ is obtained via tensor network machinery following Ref.~\cite{filippov2023scalable}. 

\section{Experimental details}
The experiments presented in this chapter are performed on the IBM Quantum Eagle processor \textit{ibm\_strasbourg}. 
This device consists of 127 fixed-frequency transmons arranged in a heavy-hexagonal lattice (see qubit layout in Fig.~\ref{fig:Fig2_DU_results}), with a median $T_1$ time of $315\,\upmu s$ and median $T_2$ time of $187\,\upmu s$.
All the quantum circuits that we execute are decomposed into layers of parallel single-qubit gates and layers of parallel two-qubit entangling echoed cross-resonance (ECR) gates. 

We implement the kicked Ising evolution circuits with different system parameters $J, b, h$ and time steps $t$, see Tab.~\ref{tab:circuit_parameter_settings} for an overview of all considered parameter settings.
Moreover, we run all experiments at the scale of 51, 71 and 91 qubits. 
Before choosing the physical qubit layout, we perform calibrations of the single-qubit state preparation and measurement (SPAM) fidelities, $T_1$ times, and two-qubit Bell-pair preparation fidelities.
We then select 1d-chains of qubits that avoid outliers in these metrics.

\begin{table}
\begin{center}
\scalebox{0.8}{
\begin{tabular}{|c||c|c|c|c|c|c|c|} 
 \hline
$N$ & \makecell{dual unitary \\ $J = b =\pi/4$ \\$ h = \{0, 0.05, 0.1, 0.15\}$} & \makecell{non dual unitary \\ $J=\pi/4,\, t= N$ \\$ h = \{0, 0.05, 0.1, 0.15\}$} &  \makecell{ circuit \\ twirls} & \makecell{shots \\ per twirl} & \makecell{$R_z(\theta)$ \\ gates } & \makecell{execution \\time} & \makecell{sampling \\ rate [kHz]} \\ %[0.5ex] 
 \hline\hline
 51 & $ t = \{0, 5, 10,  15, 20, 25\}$ & \multirow{3}{10em}{$b -\frac{\pi}{4} = \{ -0.15, -0.1, \dots, 0.15\}$} & \multirow{3}{1.5em}{256} & \multirow{3}{2em}{1024} & 7956 & 2h 18min & 2.10 \\ 
 % \hline
 71 & $ t = \{0, 7, 14, 21, 28, 35\}$ & & & & 15336  & 2h 55min & 1.75 \\
 % \hline
 91 & $ t = \{0, 9, 18, 27, 36, 45\}$ & & & & 25116 & 3h 24min & 1.57\\
 \hline
\end{tabular}
}
\caption[Summary of parameter settings for kicked Ising circuits]{\small Summary of parameter settings for quantum hardware execution.
The number of twirls and measurements (``shots'') per twirl are implemented for each combination of the model parameters $\{J, b, h, t\}$. 
The circuit randomizations include both gate twirling and the sampling of (twirled) randomized readout bases. 
The number of $R_z$ gates are for the circuits at maximum depth.
The reported execution time for each dataset includes noise learning circuits (see Sec.~\ref{sec:TEM_noise_learning}), and additional benchmark circuits (see Sec.~\ref{sec:additional_circuits}). 
The sampling rate is the total number of shots taken in each dataset divided by the wall clock run time.}
\label{tab:circuit_parameter_settings}
\end{center}
\end{table}
%}

We perform uniform Pauli twirling (see Sec.~\ref{sec:noise_learning_theory}) of the ECR gate layers to suppress coherent errors and obtain a Pauli noise channel.
That is, for every parameter set, we run several instances of circuits that implement the same global unitary but differ in their single-qubit gate layers.
Similarly, we twirl measurements by inserting a Pauli $ X$ or $ I$ gate (sampled uniformly at random) prior to the readout, which we correct for in post-processing. 
This symmetrizes the noise channel of the readout~\cite{van2022model}.

If done naively, gate twirling and randomized measurements create a circuit compilation overhead which can become prohibitively large for high circuit volumes and number of twirls. 
We alleviate this overhead by leveraging a recently introduced parametric circuit compilation and parameter binding pipeline facilitated by the \texttt{Sampler} primitive within the IBM Qiskit runtime service~\cite{qiskit2024}.
Each sequence of consecutive single-qubit gates on a given qubit (originating from the circuit itself, twirling, or readout basis rotation) is merged and implemented on the device with a sequence $R_z(\theta_3) \times \text{SX} \times R_z(\theta_2) \times \text{SX} \times R_z({\theta_1})$, parametrized by three angles $\theta_i$ (see Sec.~\ref{sec:single-qubit_gates}). 
Hence, the twirled and randomized circuits are instances of the same parametrized circuit template and only differ in their $\theta$ angles.
With parametric compilation, we only need to create this underlying template circuit once, alongside the array of angles $\theta$ that represent the different twirled instances of the circuit. 
Our computational pipeline is thus significantly more efficient (both in terms of memory and execution time) than building the full circuit anew for every set of angles.
Nonetheless, the cost of resampling twirling and measurement configurations remains non-negligible, which is why we opt to collect multiple shots per setting. 
We take 1024 shots each for 256 randomized circuits per model parameter settings for a total of 262,144 shots per data point, see Tab.~\ref{tab:circuit_parameter_settings}. 
In this way, we achieve a sampling rate ranging from $2.1\,\text{kHZ}$ (51-qubit dataset) to $1.57\,\text{kHZ}$ (91-qubit dataset). 

\section{Noise learning beyond symmetry assumptions}
\label{sec:TEM_noise_learning}

\begin{figure}
    \centering
    \includegraphics[width=\textwidth]{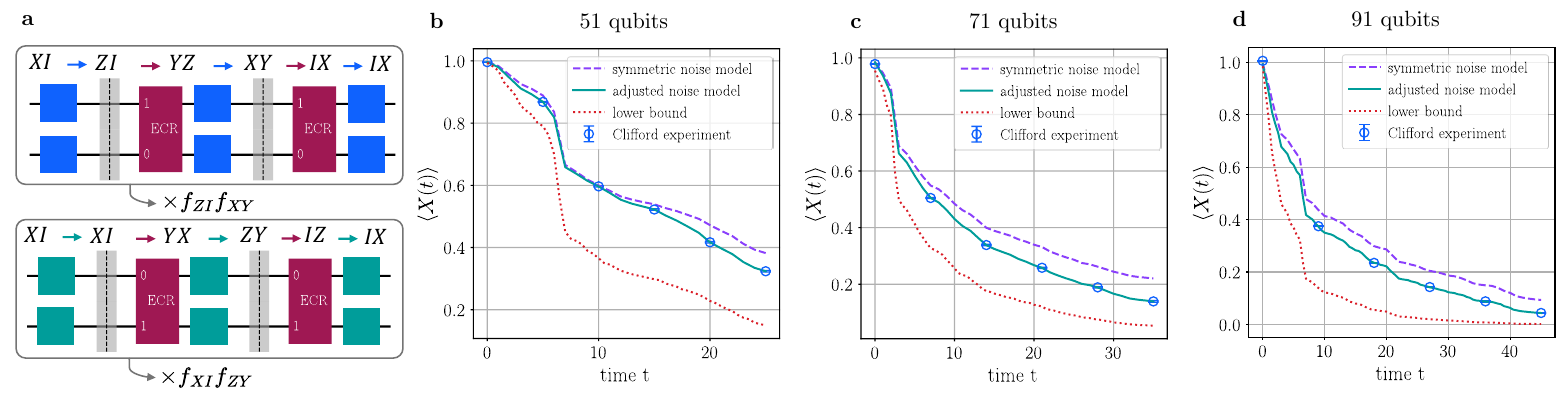}
    \caption[Fine-tuning of the noise model with Clifford kicked Ising circuits]{\small
    Fine-tuning of the noise model with kicked Ising circuits at the Clifford point.
    \textbf{a)} The single-qubit $\langle X \rangle$ observable is affected by specific Pauli fidelities as the signal propagates through the noisy Clifford circuit. Here we show the contributing fidelities that depend on the direction of the ECR gate within the relevant two-qubit blocks (negative signs are omitted). 
    Note our convention of the noise occurring before the ECR gates, as represented by dashed lines. 
    The contributing single-qubit (two-qubit) Pauli fidelities are turned into two-qubit (single-qubit) fidelities and are thus not learnable in isolation by standard protocols. \\
    \textbf{b)} -- \textbf{d)} The noisy signal of the circuit is sensitive to the symmetry assumption between a given Pauli fidelity and its conjugate. 
    Traditionally, a symmetric split between these is assumed which predicts values (dashed lines) that do not match our experiments (round markers). 
    We thus adjust the underlying degrees of freedom to obtain a noise model that matches the experiment (solid lines). 
    This is well within the region of physically allowed values indicated by the lower bound (dotted lines).
    }
    \label{fig:tweaked_fidelities_with_ki}
\end{figure}

Our quantum circuits consist of two unique entangling ECR layers (labelled ``odd'' and ``even'') for which we accurately calibrate the noise in order to mitigate it.
For both ECR layers, we learn the associated sparse Pauli-Lindblad noise channel $\Lambda$ as defined in Eq.~\eqref{eq:noise_model_def} building on the techniques developed in Ref.~\cite{van2023probabilistic} and summarized in Sec.~\ref{sec:noise_learning_theory}.
We implement cycle benchmarking with various depths $d \in \{0, 2, 6, 12, 20, 34\}$, each with 64 twirling instances and 32 shots per instance to retrieve pair fidelities (Eq.\eqref{eq:pair_fidelities_def}). 
By measuring non-overlapping Paulis in parallel, a total of 9 different initial state settings are sufficient to cover the sparse basis $\mathcal{P}$ of nearest-neighbor terms~\cite{van2023probabilistic}.

As discussed in Sec.~\ref{sec:noise-agnostic_EM}, previous literature has relied on the symmetry assumption that the fidelities of a given Pauli and its conjugate Pauli are equal. 
Let us now discuss how this limitation affects our kicked Ising Floquet circuits.  
In Fig.~\ref{fig:tweaked_fidelities_with_ki}\textbf{a} we show the propagation of Paulis for the desired $X$-observable of the two-qubit dual unitary circuit blocks at the Clifford point ($h=0$). 
Depending on the control-target direction of the \textsc{ECR} gate, the observable is affected by a factor of $f_{ZI} f_{XY}$ or $f_{XI} f_{ZY}$. 
However, the conjugate fidelities of these are $f_{ZI^\prime} = f_{YZ}$, $f_{XY^\prime} = f_{IX}$, $f_{XI^\prime} = f_{YX}$, and $f_{ZY^\prime} = f_{IZ}$. 
Hence the Clifford signal is a product of fidelities that can not be individually learned by standard cycle benchmarking circuits. 
As a result, the kicked Ising circuits are highly sensitive to the underlying symmetry assumptions of the noise model.
Fig.~\ref{fig:tweaked_fidelities_with_ki}\textbf{b} -- \textbf{d} shows the comparison of the Clifford point experiment (after readout error mitigation) with the measured pair fidelities. 
Indeed, the product of characterized pair fidelities deviates from the measured values for $\langle X(t) \rangle$ indicating that noise models based on the symmetry assumption do not reflect the noise of the device with sufficient accuracy. 
This motivates us to introduce asymmetric weights $\alpha_i$ for every fidelity $f_i$ that contributes to the $\langle X(t) \rangle$ signal at the Clifford point. 
We define re-weighted fidelities as $f_i(\alpha_i) \coloneqq \alpha_i \overline{f_i}$ and $f_{i^\prime}(\alpha_i) \coloneqq \overline{f_i} / \alpha_i$.
This way, the pair fidelities $\sqrt{f_i(\alpha_i) f_{i^\prime}(\alpha_i)}$ remain independent of $\alpha_i$.
Our goal is to find parameters $\alpha_i$ such that the product of the relevant fidelities $\prod_i f_i(\alpha_i)$ matches the measured value in the kicked Ising experiment at the Clifford point. 
For a physical (positive and trace-preserving) channel $\Lambda$, Pauli fidelities are bounded as $f_i \leq 1$. 
To ensure this, $ \alpha_i^\text{min} \leq \alpha_i \leq \alpha_i^\text{max}$ with $\alpha_i^\text{min} = \overline{f^C_i}$ and $\alpha_i^\text{max} = 1/\overline{f^C_i}$ must hold. 
We can thus more effectively parametrize the weighted fidelities by $\delta_i \in \left[ 0, 1 \right]$ as $\alpha_i (\delta_i) = \delta_i \alpha_i^\text{min} + (1-\delta_i) \alpha_i^\text{max}$. 

Since there are more relevant weights $\alpha_i$ than available data points of the Clifford observable, their choice when fitting the noise model to the experiment values is not unique. 
Our procedure for obtaining $\alpha_i$ is then the following: 
starting from the Clifford data point at lowest depth $>0$, we choose a uniform $\delta$ for all $\delta_i$ that affect the data point, such that the noise model prediction matches that value. 
We then iteratively move to the next data point, choosing a new uniform $\delta$ for the fidelities that enter the signal between that and the previous data point, until all data points are in agreement with the chosen fidelity splits.
This procedure ensures that the resulting values for $\alpha_i$ avoid edge cases where one of the fidelities becomes $\approx 1$. 
For those fidelities that are not probed by the Clifford experiments, we continue to assume a symmetric split ($\alpha_i = 1$).

The question arises of how exhaustively we must exploit the range of possible values for $\alpha_i$ to match the Clifford experiment. 
We show the lower bound on the measured observable obtained from $\prod f^C_i(\alpha_i^\text{min})$ as a dotted red line in Fig.~\ref{fig:tweaked_fidelities_with_ki}\textbf{b} -- \textbf{d}. 
This confirms that the chosen fidelity splits are well within their allowed physical regions. 
However, we observe a consistent trend that the splits need to be chosen such that the contributing fidelities become lower (and the conjugate ones higher). 
This indicates that there either is a systematic physical mechanism that causes the fidelity splits to fall on this side or that certain noise sources are present that are not fully captured by the sparse Pauli-Lindblad model. 
This point is further investigated in Sec.~\ref{sec:additional_circuits}. 

Finally, we find the generators of the sparse Pauli-Lindblad models that best describe the obtained fidelities by solving the least-square problem from Eq.~\eqref{eq:generator_fidelities_fit} with the adjusted fidelities $\boldsymbol{f}(\boldsymbol{\alpha})$ and conjugate fidelities $\boldsymbol{f^\prime}(\boldsymbol{\alpha})$. 

\section{Results}
\label{sec:TEM_results}

\subsection{Dual-unitary circuits with exact verification}

\begin{figure}
    \centering
    \includegraphics[width=1.01\textwidth]{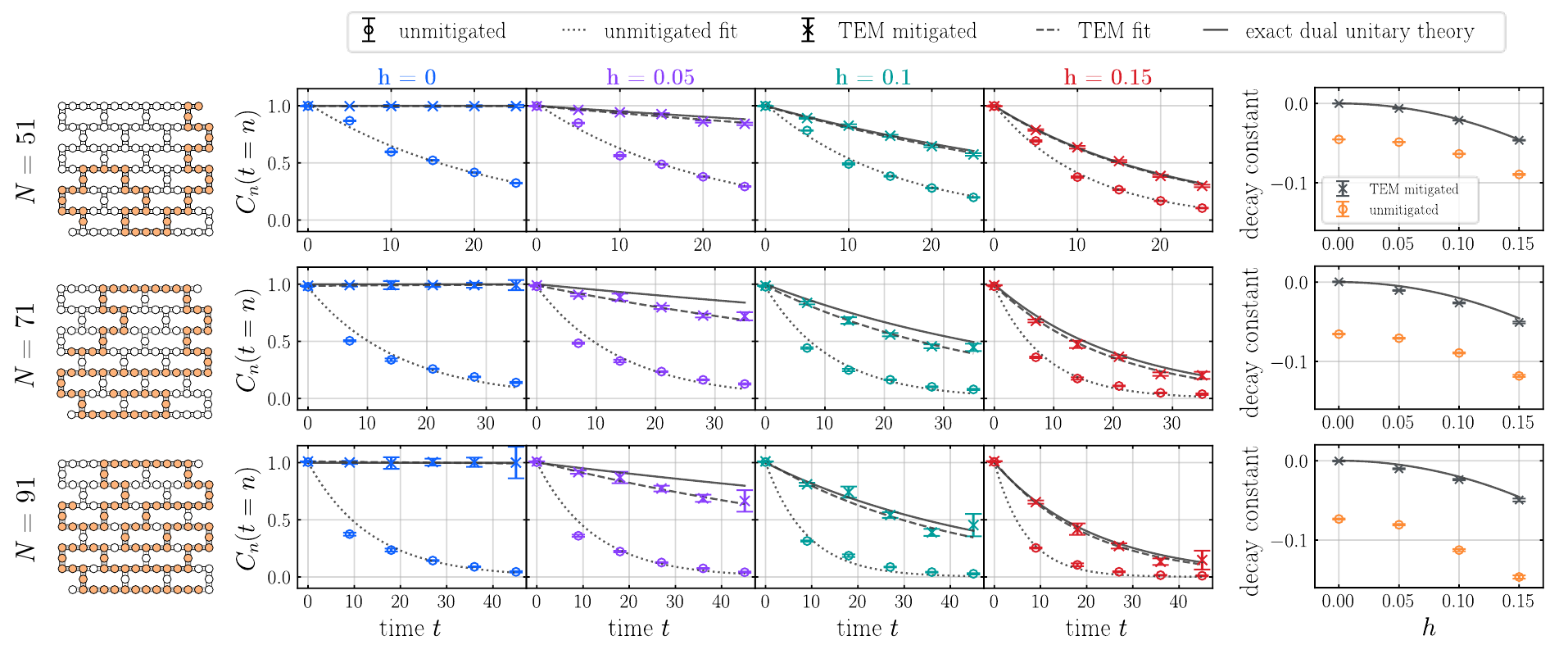}
    \caption[Error mitigation results of dual-unitary circuits]{ \small
    Measured autocorrelation functions at the dual unitary point on the light cone $C_n(t=n)$ for increasing values of $h$ and different system sizes $N$ (qubit layout shown on the left). 
    In each plot, we show the unmitigated (circles) and error-mitigated signals (x-marks), with error bars indicating one standard error, alongside the theoretical curve (black solid lines). 
    For the Clifford point $h=0$, the mitigated signal matches the theoretical curves nearly exactly---as expected, given that the noisy Clifford signal is used to calibrate the noise model (see Sec.~\ref{sec:TEM_noise_learning}).
    For $h > 0$, the mitigated points show good agreement with the theoretical curve, albeit with some deviations, particularly for $h=0.05$. 
    We further assess the quality of the results by experimentally inferring the decay rate in each case. 
    We fit exponential curves to the unmitigated and mitigated data (dotted and dashed lines, respectively), and compare the resulting decay rates of the autocorrelation function against the theory in the rightmost column. 
    }
    \label{fig:Fig2_DU_results}
\end{figure}

Using the approach outlined above for noise learning and error mitigation, we first simulate the infinite-temperature autocorrelation function at various dual unitary points, see Fig.~\ref{fig:Fig2_DU_results}. 
Even when integrability is broken for $h > 0$, we are able to closely recover the expected behavior for all considered values of $h$. 
At larger system sizes, we report small deviations in the mitigated results.
As the system size increases, note that the circuit depth also increases, and at these larger circuit volumes, errors in the noise model can accumulate, leading to a residual bias in the mitigated results~\cite{govia2024bounding}. 
These are likely a consequence of, for instance, imperfections in noise learning, model violations due to incorrect model assumptions, or even the increased instability in the noise from the longer runtimes associated with larger circuit volumes~\cite{kim2024error}. 
Benchmarking the accuracy of the measured noise model in predicting the experimental noisy data for other families of Clifford circuits reveals small systematic errors, that are particularly prominent at longer depths (see Sec.~\ref{sec:additional_circuits}).

The decay rate of $C_{n}(t)$ as a function of $h$ given in Eq.~\eqref{eq:C_exact} is a universal quantity independent of the system size. 
We demonstrate that with error mitigation we can accurately recover the exact prediction of this decay constant for all system sizes studied. 
This not only showcases the effectiveness of our approach in studying high-temperature autocorrelation functions of large-scale quantum chaotic circuits but also provides a valuable benchmark of system performance for non-Clifford circuits.

\subsection{Non-dual-unitary circuits beyond exact verification}

\begin{figure}
    \centering
    \includegraphics[width=1.01\textwidth]{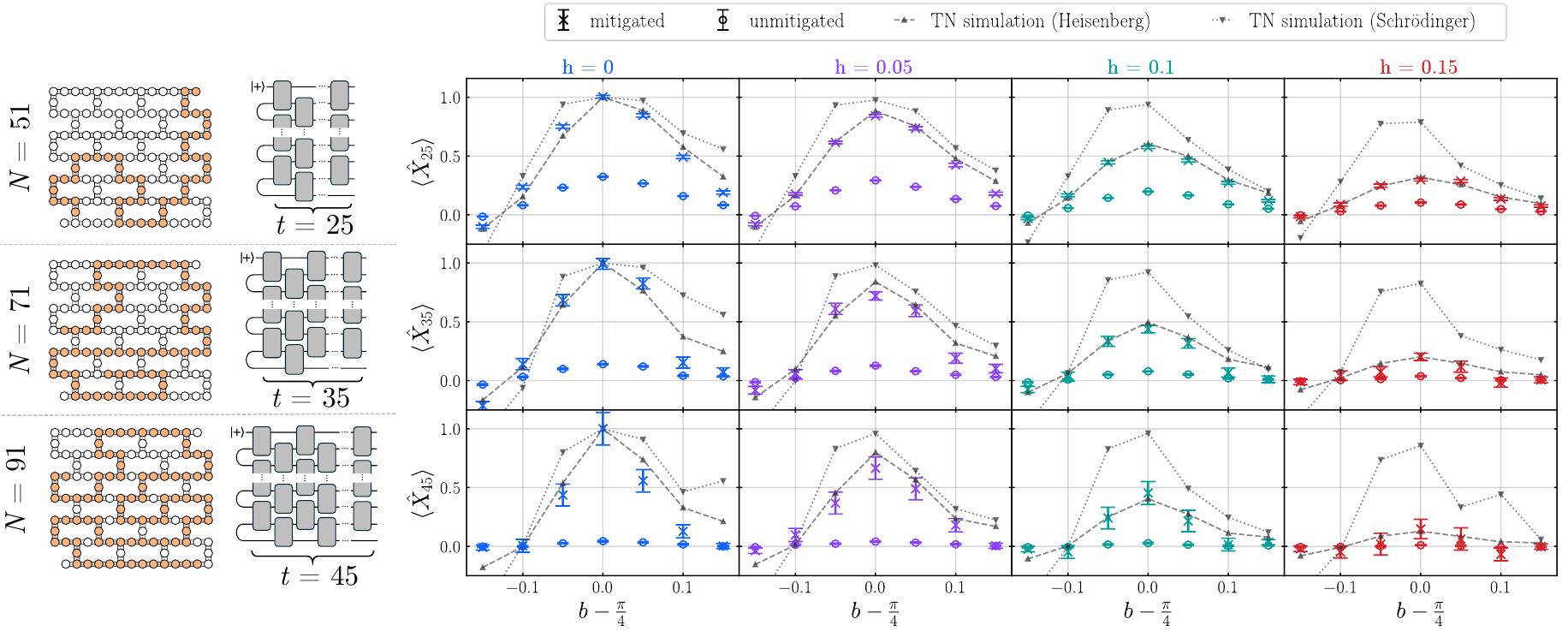}
    \caption[Error mitigation results of non-dual-unitary circuits]{ \small
    {Non-dual-unitary circuits beyond exact classical verification.} Each plot shows the evolution of $\langle {X}_t (t) \rangle$, with $t = (N - 1) / 2$, as the transverse field $b$ is swept away from dual-unitarity, for a different value of $h$ and system size $N$.
    The dual unitary points $b = \pi / 4$ correspond to the right-most points in Fig.~\ref{fig:Fig2_DU_results}. 
    No analytical solution exists for $b \neq \pi / 4$ and a brute-force statevector simulation is not available either given the scale of the quantum circuits. 
    Instead, we compare our results against classical tensor-network simulations in the Schr\"odinger (dotted line, $\chi=$1500) and the Heisenberg (dashed line, $\chi$=500) pictures. 
    }
    \label{fig:Fig3_beyond_DU}
\end{figure}

After assessing the accuracy of the mitigated results for analytically solvable DU circuits, we perturb away from the DU parameters. 
We note that, while working with the same initial state and observable as before, the local expectations values $\expect{{X}_n(t)}$ lose their interpretation as autocorrelation functions away from the DU point. 
In Fig.~\ref{fig:Fig3_beyond_DU}, at each of the previously considered values of $h$,we report the change of $\expect{{X}_n(t)}$ for the final simulation time $n = t =(N-1)/2$ as we perturb the transverse field $b$ away from dual unitarity. 
We reiterate that in the absence of exact analytical solutions and at a scale beyond brute-force classical simulation, these computations can only be compared to approximate classical methods. 

We benchmark the noise-mitigated results against purely classical approximate simulations of the noiseless circuits in both the Schr\"{o}dinger and Heisenberg picture by using tensor network techniques (see Sec.~\ref{sec:tensor_networks_intro}). 
All tensor-network post-processing and simulation methods were implemented in Python and Julia using a combination of the Quimb~\cite{gray2018quimb} and ITensor~\cite{itensor} libraries. 
Simulations were run on the Karolina, Leonardo HPC clusters and Microsoft Azure cloud virtual machines.
For the Schr\"{o}dinger-picture simulations, the compression of the evolved MPS and the contraction with the final observable are carried our according to Ref.~\cite{schollwock201196}.
For the Heisenberg-picture simulations, the compression of the MPO-MPS contractions use the procedure outlined in Ref.~\cite{dmrg_2007}.

Across the different parameters, the experimental data show strong agreement with the Heisenberg simulations with some deviations arising at larger circuit volumes, but large disagreements with the Schrödinger-picture simulations. 
Simulations in the Schrödinger picture are particularly demanding and inefficient, because the initially prepared local correlations in the form of Bell pairs quickly become highly non-local due to the entangling nature of the circuit. 
These simulations do not seem to be reliable even with the high bond dimensions employed ($\chi=1500$). 
In contrast, the Heisenberg-picture simulations in Fig.~\ref{fig:Fig3_beyond_DU} employ bond dimension $\chi=500$. 
They display a faster convergence rate with more moderate resources: the absolute difference between results with bond dimension \(\chi\) and \(\chi+100\) is below \(10^{-2}\) for \(100\leq\chi\leq500\) and below \(10^{-3}\) for \(500\leq\chi\leq 900\). 
Note that the bond dimensions required to reach these convergence levels are still considerably higher than the bond dimension of the noise inversion tensor network used for TEM ($\chi= 70$). 
Overall, while dynamics in the Heisenberg picture are converging on classical computers, simulations in the Schr\"odinger picture become unaffordable at the scale of our experiments. 
Our results thus emphasize the progress of error-mitigated quantum computing in becoming increasingly competitive with widely used classical algorithms in regimes where brute-force exact solutions are unavailable.

\section{Additional benchmark circuits}
\label{sec:additional_circuits}

In this Section, we further investigate how well the noise model generalizes to different benchmark observables for circuits built from the same even and odd ECR layers for which the noise was characterized. 
{We also investigate the effect of statistical noise on the learned noise models from finite shot numbers in the underlying learning circuits.}

\subsection{Repeated odd/even kicked Ising layers}
\label{sec:repeated_DU_layers}

\begin{figure}
    \centering
    \includegraphics[width=\textwidth]{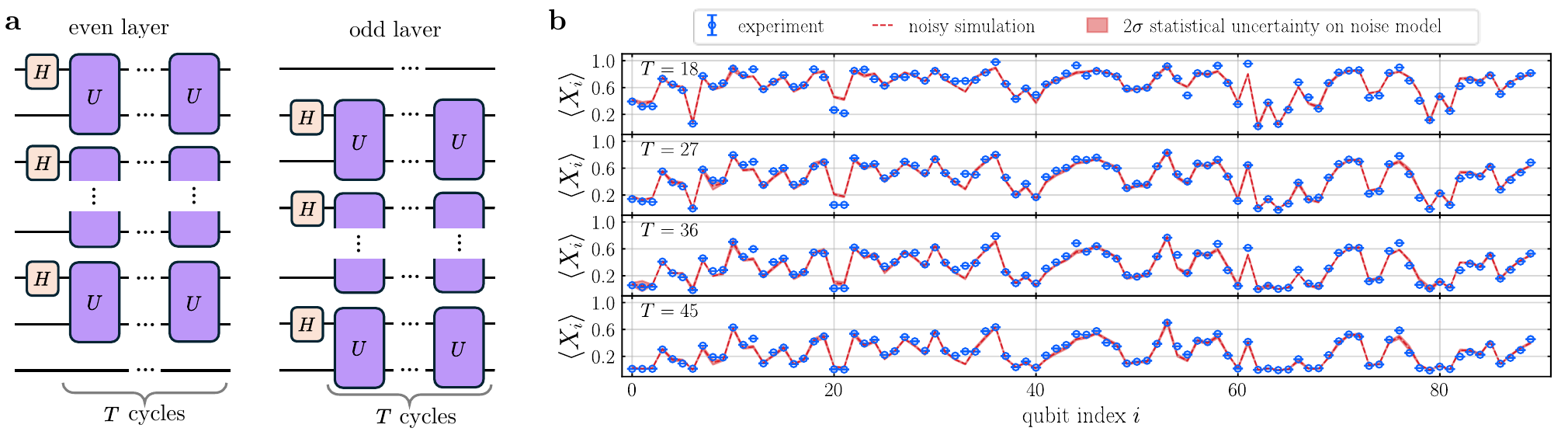}
    \caption[Repeated odd/even kicked Ising layer benchmark experiments]{\small Repeated odd/even kicked Ising layer benchmark experiments. \textbf{a)} Benchmark circuits consist of repeated layers of parallel two-qubit kicked Ising blocks at the Clifford point, applied on all even and odd qubit pairs for different cycle depths $T$, respectively. $H$ denotes the Hadamard gate.  
    \textbf{b)} Measured $\langle X_i\rangle$ expectation values compared to a stabilizer simulation given the learned noise model. {The red shaded region shows two standard deviations of the statistical error on the noise model prediction obtained from bootstrapping.}
    Note that the ECR layer depth is $2T$.}
    \label{fig:repeated_DU_blocks}
\end{figure}

First, we examine the two-qubit building blocks of the kicked Ising model as defined in Eq.~\eqref{eq:two_qubit_gate}.
Specifically, we run two sets of quantum circuits where we implement repetitions of the even dual unitary layer ${\mathbb{U}}_e$ and the odd dual unitary layer ${\mathbb{U}}_o$ at the Clifford point ($J=b=\pi/4, h=0$, respectively).
For the repeated even (odd) layer circuits, we initialize every even (odd) qubit in the $\ket{+}$ state, see Fig.~\ref{fig:repeated_DU_blocks}\textbf{a}. 
After $T$ cycles of the repeated even (odd) layers, we measure the single-qubit $\langle X_i \rangle$ observable for every even (odd) qubit index $i$ if $T$ is even or for every odd (even) qubit index $i$ if $T$ is odd. 
Combining the results from both circuits, we obtain $N-1$ expectation values $\langle X_i \rangle$, where $N$ is the number of qubits. 
As for the other kicked Ising experiments, we take 256 twirling randomizations of the circuit with 1024 shots per circuit. 

Fig.~\ref{fig:repeated_DU_blocks}\textbf{b} shows the measured $\langle X_i \rangle$ values of different depths $T$ for the 91-qubit data set, where readout error mitigation has been applied as described in Sec.~\ref{sec:readout_error_mitigation}. 
We compare these values to a noisy simulation of the circuits given the noise model learned for the ECR layers. 
Since the circuit consists of Clifford gates and the noise channels are Pauli error channels, we can simulate this efficiently in the stabilizer formalism by propagating the Heisenberg-evolved observable backwards through the circuit.
The prediction of the learned noise model matches the experimental values well. 
The main difference of these repeated layer circuits to the noise learning circuits is the additional single-qubit gates in between the ECR layers. 
This experiment can thus be seen as an interleaved cycle benchmarking run whose Pauli cycle consists of the particular pair fidelities that enter the two-qubit kicked Ising blocks when the $X$ operator propagates through it.
Our noise model predicts these decays well up to a depth of 22.5 cycles ($T=45$, ECR depth 90). 

{We investigate the statistical errors on the learned noise model by a simple bootstrapping strategy.
First, we resample all expectation values that go into the noise learning by re-drawing an equal number of raw samples from Bernoulli distributions with $p_0$ and $p_1$ according to the originally measured noisy values.
Then, we carry out the fitting procedure of the model parameters $\boldsymbol{ \lambda}$ (Eq.~\eqref{eq:generator_fidelities_fit}). 
We repeat this process 50 times and compute the model prediction for our benchmark observables of this section for each bootstrapped noise model.
Two standard deviations of these model predictions are shown as a shaded region around the mean in Fig~\ref{fig:repeated_DU_blocks}. 
We find that the statistical errors are negligibly low. 
}

\subsection{Mirror circuits of Floquet evolution}
\label{sec:mirror_circuits}
\begin{figure}
    \centering
    \includegraphics[width=\textwidth]{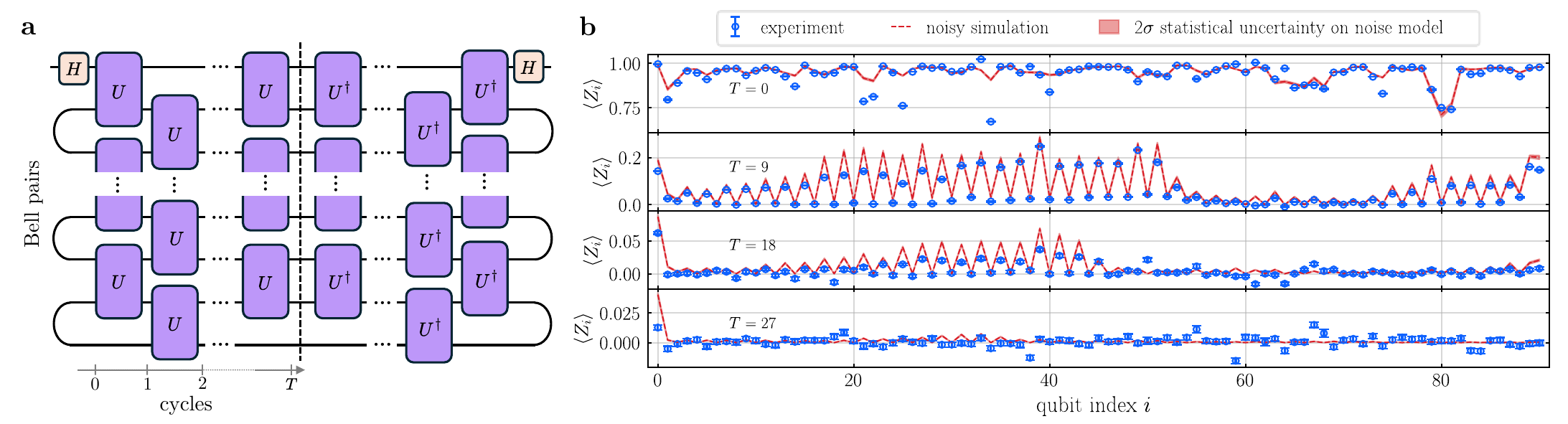}
    \caption[Mirrored kicked Ising benchmark experiments]{\small Mirrored kicked Ising benchmark experiments. \textbf{a)} The circuits consist of the forward-time kicked Ising evolution for $T$ Floquet cycles with Bell pair initialization at the Clifford point, followed by the inverse gates in reverse order to create a mirrored identity circuit. 
    \textbf{b)} Measured $\langle Z_i\rangle$ expectation values compared to a stabilizer simulation given the learned noise model. {The red shaded region shows two standard deviations of the statistical error on the noise model prediction obtained from bootstrapping.}
    Note the different vertical scales across panels, and that the ECR layer depth is $4T + 2$.}
    \label{fig:mirror_circuits}
\end{figure}

Another class of circuits we run is the Clifford point of the kicked Ising experiment followed by a mirrored ``uncomputation'' of the entire circuit, i.e., applying the inverse unitary of each gate in reverse order, see Fig.~\ref{fig:mirror_circuits}. 
These circuits thus first implement the forward-time evolution of the Floquet dynamics for some depth $T$, then run the reverse time evolution and finally undo the initial state preparation, ideally recovering the reference $\ket{0}^{\otimes N}$ state. 
Since the entangling layers are self-inverse, this circuit still only consists of the two unique ECR layers and single-qubit gates. 
We measure single-qubit $\langle Z_i \rangle$ expectation values for every qubit. 
Note that these observables are intrinsically insensitive to the pair fidelity weights $\boldsymbol{\alpha}$ used to fit the noise model in Eq.~\eqref{eq:generator_fidelities_fit}.
This is because the uncomputation gates always pick up the conjugates of the fidelities that enter the noisy signal during the forward evolution. 
These circuits thus form a benchmark of the noise model that is independent of the assumptions on symmetry in Pauli noise learning (and SPAM mitigation). 

Fig.~\ref{fig:mirror_circuits}\textbf{b} shows the obtained $\langle Z_i \rangle$ expectation values (readout error mitigated) alongside a noisy Clifford simulation. 
For even qubit indices (except $i=0$) the values decay quickly with increasing $T$ as the Pauli weight of the contributing fidelities grows linearly in the forward evolution for these observables.
In contrast, for odd $i$, the Pauli weight of the contributing fidelities never grows beyond four, which explains the zig-zag shapes of the measured values. 
This pattern is qualitatively well reflected in the measured data. 
However, quantitatively, some of the measured expectation values fall below the prediction of the noise model. 
{As before, we investigate whether statistical errors can account for this effect via bootstrapping of the noise models. 
We observe that the systematic mismatches between noisy experiments and noise model predictions are larger than the minor statistical errors on the noise model. This showcases that the learned model is well converged statistically. }

Interestingly, the repeated odd/even layer circuits from Sec.~\ref{sec:repeated_DU_layers} are not affected by an underestimation of noise. 
Indeed, in contrast to the $\langle Z_i \rangle$ observables of the mirror circuit, the Pauli fidelities that contribute to the noise of $\langle X_i \rangle$ in those circuits are confined to two-qubit strips (which also explains why they do not decay as quickly).
This is closer to the fidelity cycles measured in the noise learning circuits which extend at most to four-qubit strips. 
Our data thus suggests that -- when the Pauli weight pattern of the observable traverses larger regions of the qubit lattice -- there are additional noise sources that our noise model does not account for.
Candidates for this include higher-order or non-nearest-neighbour noise generators, residual coherent errors, as well as leakage, i.e., transmon states with population outside of the qubit subspace. 
We leave a more thorough investigation of these effects and their implications on error mitigation for future work. 
{We also conclude that the model accuracy seems more limited by these small systematic errors than by the statistical noise level, and we would not expect the performance of the model to increase further if we increased the shot numbers in the learning circuits.}

\subsection{Timing of experiments and stability of the noise model}
For noise learning based error mitigation techniques, temporal drifts of the device noise model may cause imperfections in the mitigated results.
The question arises if this effect may explain the slight mismatch between the noise model prediction and the observed expectation values for the mirrored circuits shown in Sec.~\ref{sec:mirror_circuits}. 
The order in which the experiments are run on the device is the following: We first run the circuits of repeated odd/even kicked Ising layers from Sec.~\ref{sec:repeated_DU_layers}, followed by the mirror circuit experiments. 
Next, we perform the main experiments of all kicked Ising circuits summarized in Tab.~\ref{tab:circuit_parameter_settings} and finally run the noise learning circuits. 
Hence, the repeated odd/even kicked Ising layers are the circuits most separated from the noise learning circuits in time. 
Yet they match the prediction of the obtained noise model accurately, indicating that the noise model is stable in time over the duration of the experiments. 
We thus believe that the minor mismatch observed for the mirror circuits is a more systematic effect rather than caused by temporal drifts.
This is further corroborated by the fact that the observed mismatch systematically tends towards lower values (underestimation of the noise) rather than spreading into both directions, which we would expect from stochastic drifts in time.

\section{Discussion}
\label{sec:TEM_discussion}

The framework, methodology and results displayed in this chapter highlight the utility of pre-fault-tolerant quantum processors for studying models at the forefront of quantum many-body physics. 
First and foremost, we demonstrate the capability to accurately simulate the decay of autocorrelators at the dual unitary point of the kicked Ising model. 
We believe that our work will inspire further experiments of condensed matter physics where the same autocorrelation functions can be used to extract transport properties~\cite{prosen_bal} and predict the existence of localized phases~\cite{mbl_phenom}.
Our method could give access to a range of observables $O = P_k P_l$, $P = I,X,Y,Z$, which includes two-body correlators, imbalance, and the two-qubit mutual information.
Moreover, by leveraging the analytic tractability of dual unitary circuits, we demonstrate how these systems can serve as performance benchmarks for non-Clifford circuits. 

Finally, and perhaps most importantly, we advance the boundaries of quantum simulation on multiple technical fronts. 
Central to our approach is the integration of quantum and classical resources, achieved through the implementation of TEM~\cite{filippov2023scalable}. 
To this end, we run an accurate characterization of the device noise channels.
By leveraging a novel circuit compilation pipeline, we reach sampling rates well above $1\,\text{kHz}$ for circuit volumes of up to 4095 entangling two-qubit gates, which is a significant improvement over the $\mathcal{O}(10\,\text{Hz})$ rates reported in previous experiments~\cite{kim2023evidence}. 
Our experiment adds to the growing body of work that harnesses classical computation to extend the reach of near-term quantum processors~\cite{filippov2023scalable,robledomoreno2025,eddins2024,robertson2024}. 
As quantum hardware advances towards lower error rates~\cite{stehlik2021tunable}, more stable noise~\cite{kim2024error} and faster speeds~\cite{wack2021,rajagopala2024}, our approach could open up the path to the first class of quantum simulations of many-body dynamics on universal quantum processors that surpass classical simulators already before the advent of fault-tolerance.

Our experiments have pointed out flaws in previously established noise characterization techniques~\cite{van2023probabilistic, kim2023evidence} regarding unjustified assumptions on unconstrained parameters of the noise models. 
We have successfully addressed this with a problem-tailored approach that treats the Clifford point of the studied model as additional learning circuits. 
In Chapter~\ref{chap:gauge_learning}, we build on these insights to develop a generally applicable noise learning and error mitigation framework that overcomes these limitations for arbitrary circuits.

\chapter{Gauge-consistent error mitigation}
\label{chap:gauge_learning}

\summary{This chapter is based on the article ``Disambiguating Pauli noise in quantum computers'' by Edward H. Chen$^\ast$, Senrui Chen$^\ast$, \textbf{Laurin E. Fischer}$^\ast$, Andrew Eddins, Luke C.G. Govia, Brad Mitchell, Andre He, Youngseok Kim, Liang Jiang, and Alireza Seif available as a preprint arXiv:2505.22629, 2025~\cite{chen2025disambiguating}. \footnotesize{$^\ast$shared first authorship} \normalsize 
\bigskip

\noindent Fundamental limitations prevent the unique identification of a device noise model. 
This raises the question of whether these limitations impact the ability to predict noisy dynamics and mitigate errors. 
Here, we show, both theoretically and experimentally, that when learnable parameters are self-consistently characterized, the \textit{unlearnable} (gauge) degrees of freedom do not impact error mitigation. 
We use the recently introduced framework of gate set Pauli noise learning to efficiently and self-consistently characterize and mitigate noise of a complete gate set, including state preparation, measurements, single-qubit gates and multi-qubit entangling Clifford gates. 
We validate our approach through experiments with up to 92 qubits and show that while the gauge choice does not affect error-mitigated expectation values, optimizing it reduces the associated sampling overhead. 
}    

\begin{figure}[t]
    \centering
	\includegraphics[width=1.0\textwidth]{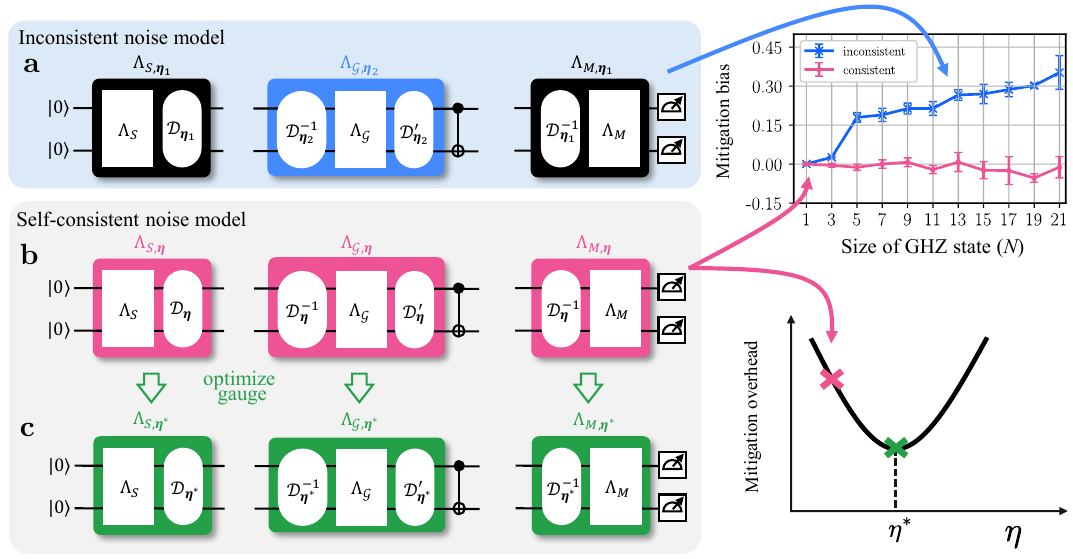}
	\caption[Schematic of error mitigation with gauge-consistent noise models]{\small
		Noise-model-based error mitigation methods rely on accurate knowledge of hardware error channels. 
		However, hardware noise fundamentally cannot be fully determined by experiments due to unlearnable ``gauge'' degrees of freedom. 
        \textbf{a)} Without accounting for this indeterminacy, previous experiments implicitly used an \textit{inconsistent} set of the gauge parameters $\{\mathcal{D}_{\boldsymbol{\eta}_1}, \mathcal{D}_{\boldsymbol{\eta}_2} \}$ across the quantum gate set, e.g., different gauge choices for state preparation and measurement and two-qubit gates (top, blue).
        \textbf{b)} We show that a self-consistent set of gauge parameters (middle, pink) is necessary for unbiased quantum error mitigation, as exemplified here in the mitigation bias for the global $\langle X^{\otimes N}\rangle$ observable of a 21-qubit GHZ state preparation experiment, see Sec.~\ref{sec:gauge_GHZ_experiments} for details.
       \textbf{c)} Furthermore, the choice of a consistent gauge can be optimized to reduce the sampling-cost overhead of error mitigation. 
	}
	\label{fig:gauge_fig1}
\end{figure}

In this chapter, we dive deeper into the task of noise learning to enable rigorous error mitigation. 
Previous literature on noise-model-based error mitigation has largely assumed that SPAM errors and gate errors can be characterized and mitigated independently~\cite{van2023probabilistic, vandenBerg2024techniqueslearning, kim2023evidence, van2022model, gupta2024probabilistic, ferracin2024efficiently}.
A widespread approach to learning noise models of gates in a SPAM-independent way is \emph{cycle benchmarking}~\cite{erhard2019characterizing, calzona2024multi} where twirling ensures Pauli noise channels, as detailed in Sec.~\ref{sec:noise_learning_theory}. 
However, this leads to unlearnable degrees of freedom in the noise model~\cite{chen2023learnability}. 
These ambiguities of noise model parameters are in practice circumvented by imposing a ``symmetry assumption''~\cite{van2023probabilistic,kim2023evidence,ferracin2024efficiently}. 
For example, consider a Clifford gate $\mc G$ that satisfies $\mc G^2 = I$, e.g., a CNOT gate. 
For any $P\in \mathbbm{P}_N$, $P^\prime = \mc G(P)$, whenever $P\neq P^\prime$ up to a sign, the symmetry assumption imposes equal Pauli fidelities $f^{\mc G}_P=f^{\mc G}_{P^\prime}$, e.g., $f^{\text{CNOT}}_{XI}=f^{\text{CNOT}}_{XX}$, see our discussion around Eq.~\eqref{eq:pair_fidelities_def}. 
Similarly, the readout error mitigation method TREX (see Sec.~\ref{sec:readout_error_mitigation}) relies on the readout fidelity $f^M_P$ of a measured Pauli $P$ which cannot be separated from state preparation errors in the typically employed depth-zero ``prepare-and-measure'' circuits. 
In practice, the rescaling factor used in TREX thus becomes $\langle P \rangle_0 = f^M_P f^S_P$ where $f^S_P$ is the state preparation fidelity. 
Applying readout error mitigation in post-processing with TREX thus implicitly assumes perfect state preparation~\cite{van2022model}. 

It has recently been pointed out that the limitations to the learnability of Pauli noise channels can be rigorously understood through gauge degrees of freedom in a unified framework that encompasses SPAM and gate noise~\cite{chen2024efficient}.
This is conceptually similar to \emph{gate set tomography} (GST)~\cite{nielsen2021gate,blume2013robust}, but specialized to Pauli noise for scalability and efficiency.
While the gauge parameters determine the unlearnable degrees of freedom of the noise model, their choice has no influence on any conceivable circuit built from the gate set. 
From this perspective, the assumption of a perfect state preparation corresponds to a specific choice of the gauge parameters, while the assumption of symmetric conjugate fidelities in gate noise corresponds to a \emph{different} choice of the gauge. 
When combined, this leads to inconsistencies in the predictions of the error model, see Fig.~\ref{fig:gauge_fig1}\textbf{a}. 
This insight also explains the shortcomings of previous noise learning schemes in predicting the noisy dynamics of our quantum simulation experiments of chapter~\ref{chap:dual_unitary_TEM}, see Sec.~\ref{sec:TEM_noise_learning}. 
Instead, in Sec.~\ref{sec:gauge_model_theory}, we show that when gauge parameters are handled self-consistently, the fundamental inability to identify them does not impact the success of error mitigation, see Fig.~\ref{fig:gauge_fig1}\textbf{b}. 

We remark that the idea of combining GST with error mitigation to address gauge ambiguity has been discussed in the literature~\cite{endo2018practical}. However, due to the extreme complexity and resource cost of GST, it is unclear how to apply such protocols beyond a few qubits. 
Instead, our method builds on the recently proposed gate set Pauli noise learning framework~\cite{chen2023learnability,chen2024efficient}, which enables explicit and efficient parameterization of all learnable and gauge parameters under a practical quasi-local noise assumption. 
In Sec.~\ref{sec:gauge_experiments}, we present the first experimental demonstration of self-consistent error mitigation, with comparable scalability as state-of-the-art QEM protocols~\cite{van2023probabilistic,kim2023evidence,ferracin2024efficiently}.
We show that the self-consistent model removes the bias of symmetric models for GHZ state preparations of up to 21 qubits. 
Moreover, we observe reduced biases in specifically tailored benchmark circuits on up to 92 qubits. 
Lastly, we show, surprisingly, that despite not affecting expectation values, changing the gauge parameters can have a significant impact on the overhead of required shots for error mitigation, see Fig.~\ref{fig:gauge_fig1}\textbf{c}.
Building on this insight, we propose and demonstrate a scalable method for identifying the gauge parameters needed to minimize this sampling overhead in Sec.~\ref{sec:gauge_opt}. 

\section{Theory of self-consistent noise channels}
\label{sec:gauge_model_theory}

\subsection{Noisy gate sets}
Our work considers a ``gate set''~\cite{nielsen2021gate} comprised of state preparation, measurement, single-qubit gates, and entangling gates (See Fig.~\ref{fig:gauge_fig1}).
Concretely, let the ideal initial state be $\rho_0=\ketbra{0}{0}^{\otimes N}$, the measurement be the projection onto the computational basis $\mc M_Z$, the entangling gates be a finite collection of Clifford gates $\{\mc G\}$, and the single-qubit gates be arbitrary $\{\mc U=\bigotimes_{i=1}^N\mc U_i\}$.
In practice, the gate set is noisy. 
We use a Pauli noise model to describe the noisy gate set, where state preparation, measurement, and entangling gates are subject to Pauli noise channels,
\begin{equation}
    \tilde{\rho}_0 \coloneqq \Lambda^S(\rho_0),\quad \tilde{\mc M}_Z \coloneqq \mc M_Z\circ\Lambda^M,\quad \tilde{\mc G} \coloneqq \mc G\circ\Lambda^{\mc G}.
\end{equation}
We further assume that the single-qubit gates have negligible noise (which can be relaxed to gate-independent noise~\cite{wallman2016noise}),
and that the SPAM noise channels $\Lambda^S$ and $\Lambda^M$ are generalized depolarizing channels, which are Pauli channels whose Pauli error rates only depend on the support of the corresponding Pauli operators.
%, and thus contain only $2^N-1$ degrees of freedom.
As before (see Sec.~\ref{sec:noise_learning_theory}), these assumptions rely on Pauli twirling of gates and measurements. 

%\paragraph{Gauge degrees of freedom}
The notion of a gauge freedom is a well-known phenomenon in gate set tomography~\cite{nielsen2021gate}. 
Even in an idealized setting, the initial state, unitary layer, and measurement operators can only be inferred up to specific gauge degrees of freedom from any conceivable experiment. 
It has recently been pointed out that gate set noise models introduced above are subject to similar gauge freedom~\cite{chen2024efficient}.
That is, the Pauli noise channels $\Lambda^S, \Lambda^G, \Lambda^M$ can be transformed in a way such that no circuits that consist of a state preparation step, some layers of gates, and a final measurement layer can distinguish whether the original or the transformed channels were applied. 
Ref.~\cite{chen2024efficient} shows that these gauge transformations of the Pauli noise model can be expressed as
\begin{equation}
\label{eq:gauge_trans}
\begin{aligned}
    &\Lambda^S\mapsto\Lambda_{\bm\eta}^S \coloneqq \mc D_{\bm\eta}\circ\Lambda^S,\\ &\Lambda^M\mapsto\Lambda^M_{\bm\eta} \coloneqq  \Lambda^M\circ\mc D_{\bm\eta}^{-1},\\ & \Lambda^{\mc G}\mapsto\Lambda^{\mc G}_{\bm\eta} \coloneqq  \mc D_{\bm\eta}'\circ\Lambda^{\mc G}\circ\mc D_{\bm\eta}^{-1},
\end{aligned}
\end{equation}
where $\mc D_{\bm\eta}$ is any generalized depolarizing map, written as
\begin{equation}\label{eq:generalized_dep}
    \mc D_{\bm\eta}(\rho) \coloneqq \sum_{a\in \mathbbm{P}_N}e^{-\eta_{\mr{pt}(a)}} P_a\tr[P_a\rho]/2^N
\end{equation}
with a real $2^N$-dimensional vector $\bm\eta$ we refer to as the gauge parameters. 
Here, $\mr{pt}(a)$ is the ``pattern'' of a Pauli $a$ which is a bitstring of length $N$ that is 0 for $I$ and 1 for $X/Y/Z$, and $\eta_{0\dots0}=0$ by the trace-preserving condition. 
$\mc D_{\bm\eta}'$ is the gauge channel conjugated by the entangling Clifford gate defined by $\mc D_{\bm\eta}' \coloneqq  \mc G^{-1}\circ\mc D_{\bm\eta}\circ\mc G$, such that $\mathcal{D}_{\bm\eta} \circ \mathcal{G} = \mathcal{G} \circ \mathcal{D}_{\bm\eta}^{'}$. 
Note further note that $\mc D_{\bm\eta}$ commutes with any single-qubit gates. 
The transformations in Eq.~\eqref{eq:gauge_trans} preserve any experimental outcome, and also the Pauli form of the noise channels.
The remaining consideration is the positivity of the transformed channels -- excluding Pauli channels on the boundary of the set of positive maps, any sufficiently small $\bm\eta$ yield physical channels~\cite{chen2023learnability}.
Furthermore, in the context of error mitigation, it is perfectly acceptable to work with $\Lambda_{\bm\eta}$ that are not positive. 
Thus, a noisy gate set can be learned up to the $2^n-1$ gauge parameters parameterized by $\bm\eta$~\cite{chen2024efficient}.

\subsection{Gate set noise learning}
To learn the gate set noise models self-consistently, we first define the logarithm of the Pauli fidelities (see Eq.~\eqref{eq:Pauli_fidelity_def}) $x_a = -\log f_a$ for all the Pauli channels $\{\Lambda^S,\Lambda^M,\{\Lambda^{\mc G}\}\}$. 
We will design a set of experiments to learn $\bm x$, where each experiment consists of a sequence of up to $T$ Clifford gates and a Pauli observable $P^\text{meas}$ measured at the end.
Due to the Cliffordness, the expectation value then satisfies 
\begin{equation}
\label{eq:pauli_path_def}
\langle P^\text{meas}\rangle = f_{a_0}^S f_{a_1}^{\mc G_1}\cdots f_{a_T}^{\mc G_T} f_{a_{T+1}}^M,
\end{equation}
i.e., it is a product of Pauli fidelities known as a Pauli path~\cite{aharonov2023polynomial}.
Taking the negative logarithm on both sides yields a linear equation of $\bm x$
\begin{equation}
    -\log\langle  P^\text{meas} \rangle = x_{a_0}^S+x_{a_1}^{\mc G_1}+\cdots+x_{a_T}^{\mc G_T}+x_{a_{T+1}}^M.
\end{equation}
Combining the linear equations from all experiments, we arrive at
\begin{equation}
\label{eq:design_matrix}
    \bm b = F\bm x,
\end{equation}
where $b_j=-\log{\langle P^\text{meas}_j \rangle}$ is the (log) expectation value for the $j$-th measured Pauli observable on the $j$-th circuit, and $F$ is called the \emph{design matrix}. 
Note that this matrix connects fidelities of state preparation, gates, and readout in a holistic way.

Our first goal is to collect enough experiments such that $F$ has the maximal possible rank. 
That is, the dimension of the null space of $F$ equals the number of free gauge parameters, $2^N-1$. 
This can be achieved by including some depth-$0$ experiments that contain no entangling gates (as discussed in the context of TREX, Sec.~\ref{sec:readout_error_mitigation}) with $b=x_{a_0}^S+x_{a_1}^M$ or depth-$1$ experiments that contain one layer of entangling gates  such that $b=x_{a_0}^S+x^{\mc G}_{a_1}+x_{a_2}^M$.
To improve the estimated precision of model parameters, we can also include experiments that concatenate multiple layers of entangling gates (e.g., depth-$k$ experiments, $b=x_{10}^S+k\,x^{\mr{CNOT}}_{ZI}+x_{10}^M$), similar to cycle benchmarking circuits, see Sec.~\ref{sec:noise_learning_theory} and Fig.~\ref{fig:cycle_benchmarking_schema}.
Finally, the gate set noise model is obtained by solving $\|F\bm{x}_0-\bm b\|\le\epsilon$, where $\epsilon$ is chosen based on the tolerable amount of residual errors, resulting in a solution $\bm x_0$.
The final estimate yields $\bm x_{\bm\eta}=\bm x_0 + \bm y_{\bm\eta}$, where $\bm y_{\bm\eta}$ is a gauge vector depending on the gauge parameters $\bm\eta$, associated with the kernel of $F$.

The formalism of gauged noise models can also be applied when sparsifying the Pauli error channels, for example with sparse Pauli-Lindblad models as introduced in Sec.~\ref{sec:noise_learning_theory}. 
Specifically, as proven in Ref.~\cite{chen2024efficient}, for the sparse Pauli noise models considered throughout this thesis (up to two-local nearest-neighbor generators), the gauge parameters can be completely described by $N$ single-qubit depolarizing channels, leading to a reduction in the number of gauge parameters from $2^N-1$ to $N$.
To learn such a quasi-local Pauli noise model up to gauge parameters, we can similarly construct a linear system of equations $\bm b = F\bm x$ such that the design matrix $F$ reaches the maximal rank determined by the number of gauge parameters.

Finally, we remark that the self-consistency achieved by our framework comes with a more subtle experimental drawback.
Original cycle benchmarking protocols were motivated by the fact that they isolate and \emph{amplify} specific error terms, such that they can be learned accurately, i.e., to relative precision~\cite{calzona2024multi}. 
In contrast, our method relies on estimating individual fidelities that fundamentally cannot be amplified from depth-one circuits, hence they can only be learned to absolute precision. 
As error rates become lower, this can only be combated with an increase of shots for the estimation of isolated fidelities.

\newpage
\subsection{Self-consistent error mitigation}

\begin{figure} 
	\centering
	\includegraphics{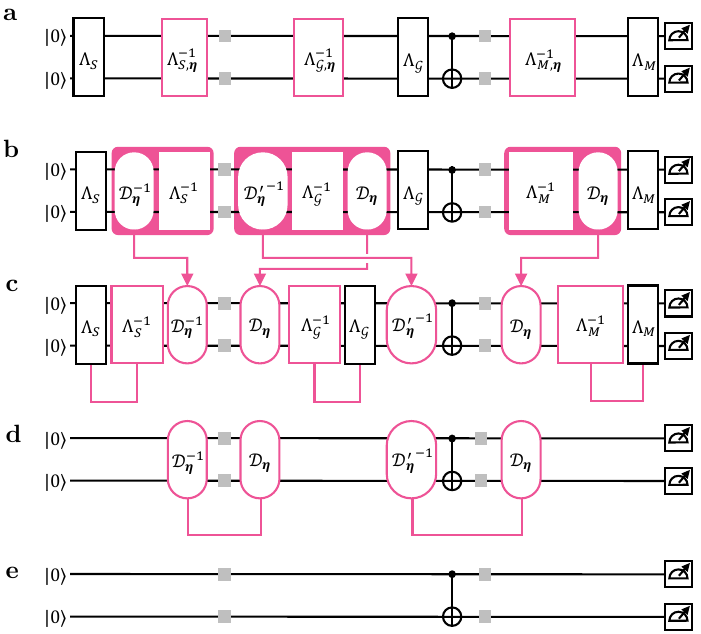}
	\caption[Graphical proof that self-consistent error mitigation is bias-free]{\small Graphical proof that a complete gate set $\bm\Lambda_{\bm\eta}=\{\Lambda^S_{\bm\eta}, \Lambda^M_{\bm\eta},\Lambda_{\bm\eta}^{\mc G}\}$ learned in a self-consistent manner leads to bias-free error mitigation.
	\textbf{a)} Error mitigation methods like PEC or TEM effectively apply the inverse channels $\Lambda_{S/\mathcal{G}/M,\bm \eta}^{-1}$. For simplicity, we show a circuit with a single CNOT gate as the two-qubit gate. This same argument applies to any other choice of the two-qubit Clifford gate and to arbitrary numbers of qubits.
	\textbf{b)} The learned noise model $\bm\Lambda_{\bm\eta}$ is related to the true noise channels $\bm\Lambda =\{\Lambda^S, \Lambda^M,\Lambda^{\mc G}\}$ by a gauge transformation with some choice of $\bm \eta$.
	Substituting the inverse of the noise channels according to Eq.~\eqref{eq:gauge_trans} leads to the operations in the red boxes.
	\textbf{c)} Reordering the gauge channels leads to cancellations (pink brackets). Note that Pauli channels $\Lambda$ and the generalized depolarizing channels $\mc D$ commute.
	\textbf{d)} The gauge channels and their inverses also compose to identity channels (pink brackets). Note that the generalized depolarizing channels commute with any single-qubit gates (gray squares). Also recall $\mathcal{D}_{\bm\eta}^{'}$ is defined to be the gauge channel conjugated by the entangling Clifford gate ($\mathcal{G} \circ \mathcal{D}_{\bm\eta}^{'} = \mathcal{D}_{\bm\eta} \circ \mathcal{G}$).
	\textbf{e)} Finally, the resulting, mitigated circuit shows a noise-free operation of a CNOT gate on two qubits. 
%	Figure courtesy of Edward Chen and Senrui Chen. 
	}
	\label{fig:noise_learning_visual_proof} 
\end{figure}

Due to the gauge degrees of freedom, a learned gate set noise model does not reflect the underlying ``ground truth'' noise channels of the hardware. 
This raises the question whether such a noise model is truly suitable to perform noise-model-based error mitigation, e.g. with PEC (Sec.~\ref{sec:PEC}) or TEM (Sec.~\ref{sec:tensor-network-error-mitigation}). 
Indeed, a self-consistently learned noise model leads to bias-free mitigation, see Box.~\ref{box:gauge_learning_theorem} for formal statement. We provide an intuitive visual proof in Fig.~\ref{fig:noise_learning_visual_proof} and refer to Ref.~\cite{chen2025disambiguating} for a formal mathematical proof. 

\begin{problembox}[label={box:gauge_learning_theorem}]{Error mitigation with self-consistent noise models}

Consider the following set of operations and their noisy implementation
\begin{enumerate}
    \item Initialization: $\ketbra{0}{0}\mapsto \tilde\rho_0 = \Lambda_S(\ketbra{0}{0})$.
    \item Measurement: $\{\ketbra{b}{b}\}_{b\in\{0,1\}^N}\mapsto\{\tilde E_b = \Lambda_M(\ketbra{b}{b})\}_{b\in\{0,1\}^N}$
    \item Layer of single-qubit unitaries: $\mc U=\otimes_{k=1}^N\,\mc U_k$, implemented without noise
    \item Layer of multi-qubit Cliffords: $\mc G\mapsto\tilde{\mc G}={\mc G}\circ\Lambda_{\mc G}$, for all $\mc G$ from a finite set $\mf G$
\end{enumerate}
where $\Lambda_{\mc G}$ are Pauli channels, and $\Lambda_S,\Lambda_M$ are generalized depolarizing channels.\\

Let $\bm\Lambda_{\bm\eta}=\{\Lambda^S_{\bm\eta}, \Lambda^M_{\bm\eta},\{\Lambda_{\bm\eta}^{\mc G}\}_{G \in \mf G}\}$ be a collection of Pauli noise channels learned in a self-consistent way. 
They are then equivalent to the true noise channels $\bm\Lambda =\{\Lambda^S, \Lambda^M,\{\Lambda^{\mc G}\}_{G \in \mf G}\}$ up to a gauge transformation $\bm\eta$. \\
Using $\bm\Lambda_{\bm\eta}$ instead of $\bm\Lambda$ to invert hardware noise with bias-free error mitigation (e.g. PEC) yields an unbiased estimator of the noiseless observable for any circuit composed of initial state preparation, gate layers, and final measurement.
\end{problembox}

\section{Experimental demonstration}
\label{sec:gauge_experiments}

We now report two experiments that demonstrate the importance of self-consistent noise learning for error mitigation with increasing complexity of the noise models. 
In both experiments, we learned two noise models. 
The first model represents the previous state-of-the-art~\cite{van2023probabilistic, kim2023evidence, van2022model} which imposes the symmetry assumption between conjugate Pauli eigenvalues and assumes ideal state preparation for TREX readout error mitigation. 
We refer to this as the ``inconsistent'' noise model. 
The second model learns all noise channels in a self-consistent way and is referred to as the ``consistent'' noise model. 
Then, we compare the performance of both models in predicting noisy expectation values in the corresponding circuits, which is numerically tractable due to the Clifford nature of the chosen circuits. 
We divide the measured noisy observables of the target circuits by the values predicted by each model. 
This results in ``pseudo-mitigated'' values which correspond to the expectation value one would obtain with PEC in the infinite-shot limit.

\subsection{GHZ preparation with restricted Pauli basis}
\label{sec:gauge_GHZ_experiments}

First, we examine the impact of self-consistent noise learning for a circuit that prepares a highly-entangled GHZ state. 
The GHZ state on $N$ qubits is a stabilizer state that is specified by being the simultaneous +1 eigenstate of a set of generators, which includes a full-weight term $\langle X^{\otimes N}\rangle$ that forms our target observable. 
However, rather than relying on well-known preparation circuits which require learning $O(N)$ unique layers of entangling gates, we prepare the state using only two alternating dense layers of parallel CNOT gates, restricting noise learning to a gate set with only two unique gate layers. 
Specifically, we first chose a fixed set of 21 qubits on the device \textit{ibm\_strasbourg}. 
We then employ a SAT-solver to prepare GHZ states on $N=1,3,\dots,21$-qubit subsets of those 21 qubits with two unique layers of entangling gates covering all 21 qubits~\cite{gavrielov2024linear,yoshioka2024diagonalization}. 

We examine the impact of the self-consistent learning approach on the $\langle X^{\otimes N}\rangle$ observable of the target GHZ state by only learning the Pauli eigenvalues which contribute to the final observable for each of the two template layers. 
That is, we restrict the set $\mathcal{P}$ over which the noise model is defined to the $N-1$ Paulis in the Pauli path of the observable as defined in Eq.~\eqref{eq:pauli_path_def}.
Of course this is only tractable because our target circuit is a Clifford circuit.
In this sense, this is a \textit{restricted} noise model because we do not learn the full Pauli noise channel for both layers, but allow for the possibility of \textit{nonlocal} noise by not imposing any locality constraints.
%In Sec.~\ref{supp:constructing_design_matrix}, we include an example for learning the noise of template `a' used in preparing the $n=21$ GHZ state.

The weight of the Paulis in the path of the $\langle X^{\otimes N}\rangle$ observable increases linearly from 1 for the first Pauli towards a global Pauli by the end of the circuit. 
Therefore, the observable is essentially only affected by state preparation errors from one qubit but is sensitive to measurement errors from all $N$ qubits. 
This asymmetry creates a challenge for the inconsistent model. 
The TREX rescaling factor mistakenly contains any potential state preparation errors, which may lead to an overcorrection in the rescaling factor of Eq.~\eqref{eq:TREX_definition}.
Indeed for GHZ states up to 21 qubits, we observe an increasing bias using the inconsistent noise model reaching 35.2\%$\pm$6.5\%, while we observe statistically insignificant -1.2\%$\pm$4.1\% biases using the self-consistent noise model for the largest depths, see the plot in Fig.~\ref{fig:gauge_fig1}.
The error bars of the pseudo-mitigated values are the result of averaging over seven separate experimental runs, interleaving the learning experiments and the target experiments over the course of 18 hours.
Note that we cannot strictly conclude from the consistent model whether indeed state preparation errors are the root cause for the failure of the inconsistent model (rather than asymmetries in the gate noise), because we do not know which gauge corresponds to the true hardware noise. 
However, we believe that the linear increase in the bias of the inconsistent model is a strong indicator that indeed state preparation errors are a main cause for the observed deviations. 

\subsection{Ring circuits with two-local basis}
\label{sec:ring_experiments}

While the noise model in the previous experiment was limited to the set of Paulis relevant to the observable, we now perform self-consistent noise learning in a scalable way that is applicable to general circuits. 
As discussed around Eq.~\eqref{eq:noise_model_def}, we achieve this by restricting the set of Paulis for each layer to single-qubit and nearest-neighbor two-qubit terms, follwoing Ref.~\cite{chen2024efficient}. 
For $N$ qubits on a ring, we consider a gate set consisting of two gate layers $G_{\rm{even}}= {\rm{CNOT}}_{1,2}\otimes{\rm{CNOT}}_{3,4}\otimes\dots\otimes{\rm{CNOT}}_{N-1,N}$ and $G_{\rm{odd}}= {\rm{CNOT}}_{2,3}\otimes{\rm{CNOT}}_{4,5}\otimes\dots\otimes{\rm{CNOT}}_{N,1}$. 
As shown in Ref.~\cite{chen2024efficient}, this model has $28N$ parameters with a fully local gauge. 
That is, there are $27N$ learnable parameters and $N$ gauge parameters corresponding to $N$ single-qubit depolarization maps. 
Due to the locality of the noise model, only local expectation values are needed to learn the model parameters. 
This allows many parameters to be estimated in parallel. 
As a result, the number of measurement settings needed to learn the complete noise model remains constant and does not increase with system size. 

\begin{figure} 
	\centering
	\includegraphics[width=0.74\textwidth]{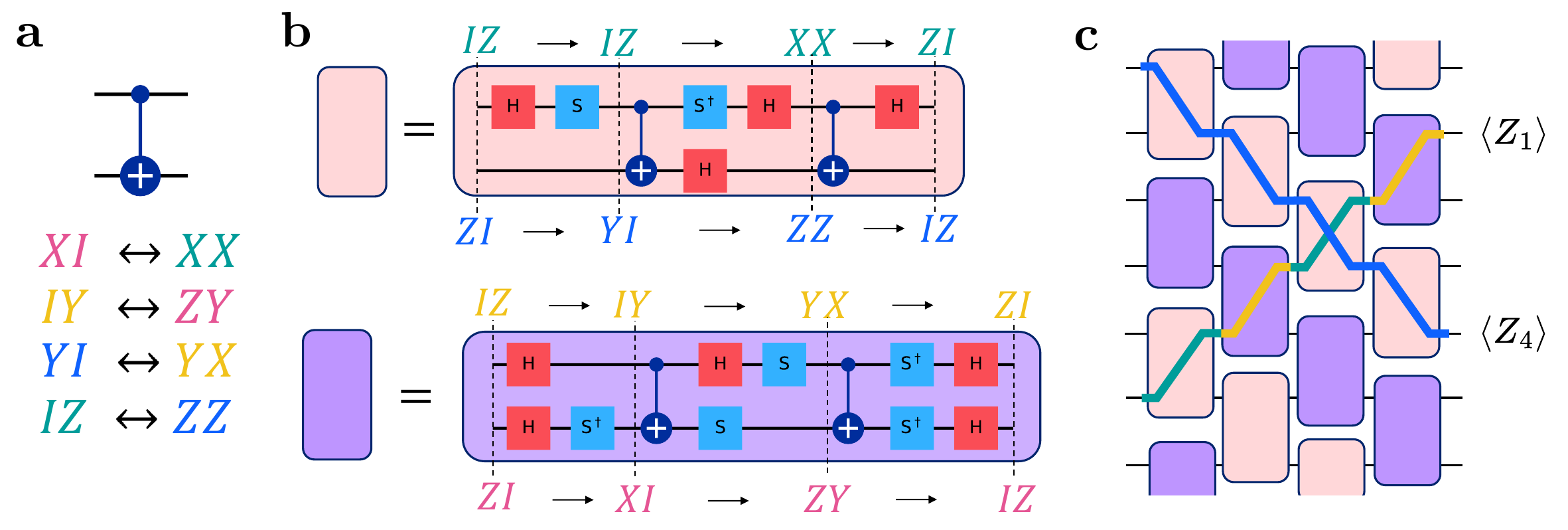} 
	\caption[Circuit for ring experiments with weight-1 observables]{\small Circuit for ring experiments with local $Z$ observables.
	\textbf{a)} A single CNOT gate has four conjugate Pauli fidelity pairs that change pattern under conjugation. Conventional noise learning techniques learn the product of these pairs but not the individual fidelities. 
	\textbf{b)} Two-qubit circuit blocks that transform $ZI \leftrightarrow IZ$ with their Pauli paths shown above/below the gates. The circuits are chosen such that the noisy Pauli fidelities that affect each observable (before the CNOT gates) originate from different conjugate pairs, as indicated by color. 
    \textbf{c)} Arranging the two-qubit blocks from \textbf{b} in an alternating pattern, each single-qubit $Z$ observable propagates in a staircase shape, such that every degenerate fidelity from each CNOT gate shown in \textbf{a} is probed by one observable. 
	}
	\label{fig:methods_ring_circuit} 
\end{figure}

\paragraph{Experiment design}
We applied the learned noise model to a target circuit where we measure local $Z$ observables for every qubit on a closed ring of a circuit built from the odd and even CNOT layers. 
The circuit and observables are designed to probe all fidelity pairs of the CNOT gates which change Pauli weight pattern under conjugation.
For a single CNOT gate, there are four such conjugate Pauli pairs, and in (interleaved) cycle benchmarking experiments their fidelity splits remain unlearnable, see Sec.~\ref{sec:noise_learning_theory} and Ref.~\cite{chen2023learnability}.
These are $IZ \leftrightarrow ZZ$, $XI \leftrightarrow XX$, $ZY \leftrightarrow IY$, and $YX \leftrightarrow YI$, as illustrated in Fig.~\ref{fig:methods_ring_circuit}\textbf{a}. 
We start by designing two different two-qubit blocks with two noisy CNOT gates each, such that all single-qubit $Z$ observables are sensitive to two Pauli fidelities that originate from different degenerate cycles. 
These two-qubit blocks are shown in Fig.~\ref{fig:methods_ring_circuit}\textbf{b}. 
Note that they shift the support of the $IZ$ observable to the other qubit.  
Hence, when arranging the blocks in the pattern shown in Fig~\ref{fig:methods_ring_circuit}\textbf{c}, each $Z$-observable propagates in a ``staircase''-like trajectory.
After four layers of the alternating pattern (eight total layers of CNOTs), every fidelity split of each participating CNOT connection is probed by one observable.
There are only two unique layers of CNOT gates, as the two-qubit blocks only differ in their single-qubit gate structure. 
With this construction, we ensure that the $Z$ observables in this experiment are maximally sensitive to asymmetries in the gate noise fidelities. 
Moreover, every observable is affected by the state preparation noise from one qubit and the measurement noise of a different qubit. 
Hence, these observables also expose flaws in the inconsistent model when state preparation errors are not uniform across the ring of qubits. 

\begin{figure} 
	\centering
	\includegraphics[width=1\textwidth]{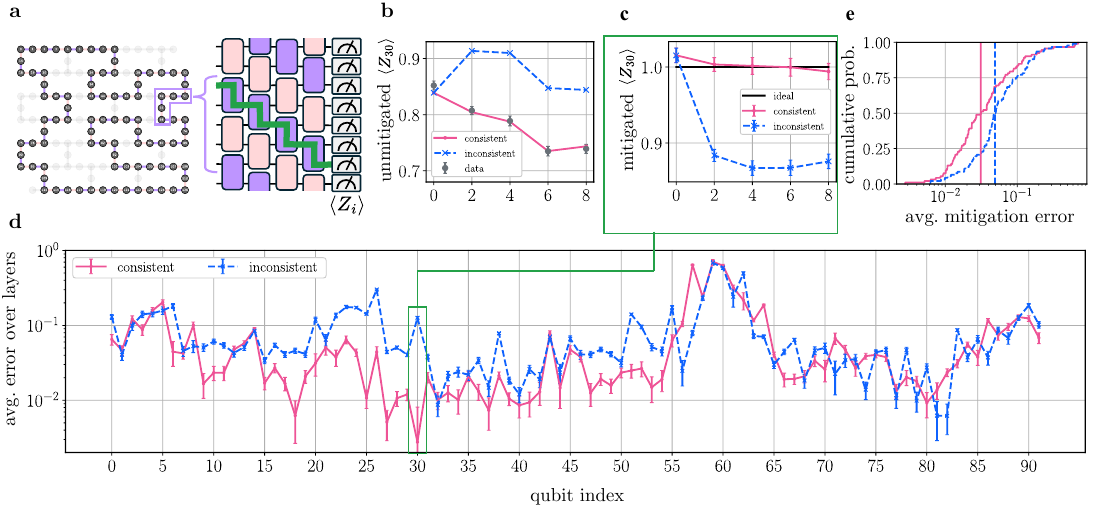} 
	\caption[Self-consistent noise learning and error mitigation on a ring of 92 qubits]{\small Scalable (quasi-local) self-consistent noise learning and mitigation of local $Z$ observables on a ring of 92 qubits.
	\textbf{a)} Closed loop of 92 qubits on a 127-qubit device, \textit{ibm\_strasbourg}. Boxed section of ring shows a section of the 92 qubits with a Clifford circuit designed specifically to propagate all 92 $Z_i$ observables in a ``staircase'' (green) fashion such that the initial and final qubit support of the observable falls on a different qubit.
	\textbf{b)} Experimentally measured (filled gray circle) expectation values versus number of circuit layers compared against the self-consistent (solid red) and symmetric noise predictions (dashed blue).
	\textbf{c)} Predicted values divided by the measured values yield a mitigated value for the data set in \textbf{b}, and boxed in \textbf{a}.
	\textbf{d)} Average mitigation error up to four circuit layers on all 92 qubits calculated in the same manner as for a single qubit as seen in \textbf{c}.
    Error bars depict one standard deviation of the shot noise on the unmitigated data. 
	\textbf{e)} Cumulative distribution of mitigation errors \textbf{d} between the experimentally measured and predicted outcomes. The median mitigation error, denoted by vertical lines, shows a reduction from 4.9\% (dashed blue) to 3.1\% (solid red) bias.
	}
	\label{fig:7} 
\end{figure}

\paragraph{Results}
We use a ring of 92 of the 127 qubits available on \textit{ibm\_strasbourg} shown in Fig.~\ref{fig:7}\textbf{a}. 
We highlight one of the 92 available observables, and show how it evolves for different circuit depths in Fig.~\ref{fig:7}\textbf{b}. 
Then, we compare the experimental outcomes against the predicted outcomes based on the consistent and inconsistent noise models by computing the mitigated values as before.
In Fig.~\ref{fig:7}\textbf{c}, we show one specific example where the bias using the inconsistent model reaches 12\%$\pm$0.5\% whereas the consistent model shows no statistically significant bias of 0.3$\pm$0.5\%. 
Applying the same analysis as described above across all 92 qubits, we find that the consistent noise model generally yields mitigation errors at or below that predicted with the inconsistent noise model, see Fig.~\ref{fig:7}\textbf{d}. 
In fact, the median mitigation error is reduced from 4.9\% to 3.1\% (Fig.~\ref{fig:7}\textbf{e}). 

For most observables, a residual bias remains also under the self-consistent noise model. 
This is an indication that there are error sources present in the device which even the consistent model does not accurately account for. 
Candidates for such errors could be leakage out of the qubit subspace, temporal drifts of the noise model between the learning circuits and the target circuits, remaining coherent errors, or non-nearest-neighbor correlated noise sources. 
Finally, few individual observables in Fig.~\ref{fig:7}\textbf{d} (e.g. qubit index 57) show a larger bias under the self-consistent model than the symmetric model. 
We note that this occurs predominantly when there are severe outliers in the individual Pauli or SPAM fidelities affecting the respective observables. 
This could be caused, e.g., by the presence of two-level systems (TLS). 
These lead to strong fluctuations in the noise parameters on short time scales, to which the self-consistent learning protocol is particularly vulnerable due its dependence on depth-one circuits. 
We thus expect that our learning protocol will further benefit from recent techniques to stabilize the noise~\cite{kim2024error}. 

\begin{table}
\begin{center}
\begin{tblr}{|l|c|c|}
\hline 
Experiment & {GHZ preparation \\ (see Sec.~\ref{sec:gauge_GHZ_experiments})} & {ring circuit \\ (see Sec.~\ref{sec:ring_experiments})} \\
\hline \hline
Number of qubits & 21 & 92 \\ 
\hline
Mitigated observables & $X^{\otimes n}$ & $Z_i, i \in \{0, \dots, 91\}$ \\
\hline
Number of twirls & 100 & 100 \\
\hline
Shots per twirl & 256 & 150 \\
\hline
{Even-depth learning layers \\ for symmetric model} & $d = \{2, 4, 8 \}$ & $d = \{4, 12, 24 \}$ \\
\hline
{Model Pauli basis} & {restricted to Paulis  \\ relevant for observable} & {all one- and \\ two-local Paulis} \\
\hline
{Design matrix dimensions \\ of self-consistent model} & (56, 46) & (9108, 2576) \\
\hline
Model locality assumption & None & two-local \\
\hline
\end{tblr}
\caption[Experimental details on self-consistent noise learning demonstrations]{\small
Details on the noise learning experiments for GHZ and ring circuits. 
}
\label{tab:gauge_experiment_settings}
\end{center}
\end{table}

\subsection{Further experimental details}

For all executed circuits, we employ uniform Pauli twirling of the respective two-qubit gate layers and measurements to suppress coherent errors and justify the assumption of a Pauli noise channel. 
For learning the inconsistent model, we follow the formalism originally established in Ref.~\cite{van2023probabilistic}: For each noisy layer, we implement a given number of even-depth learning circuits for a basis of Paulis as specified in Table~\ref{tab:gauge_experiment_settings}.
The TREX mitigation factors are inferred from a prepare-$\ket{0}$ circuit (under measurement twirling). 
The noisy estimate of the reference observables $\bra{0}O\ket{0}$ was performed with the same number of twirls and shots per twirl as stated in Table~\ref{tab:gauge_experiment_settings}.
Finally, the set of learning circuits for the self-consistent noise models comprises the same even-depth learning circuits used for the inconsistent model as well as additional depth-one learning circuits for the respective Pauli basis of the model chosen to be a minimal set of experiments to ensure a full-rank design matrix $F$. 
This leads to the design matrix dimensions stated in Tab.~\ref{tab:gauge_experiment_settings}.
We emphasize that the final column for the 92-qubit experiment is the \textit{scalable} approach, where the total number of learning circuits remains independent of the system size. 
This is because the noise is assumed to be two-local and thus long-range noise terms are not being considered. 
The design matrices can be reconstructed using code provided in Ref.~\cite{chen2024efficient}.

\section{Gauge optimization}
\label{sec:gauge_opt}
Finally, recall that error mitigation generally comes with an exponential sampling overhead.  
In PEC the sampling overhead takes the form of
\begin{equation}
\label{eq:gauge_sampling_overhead}
\Gamma \coloneqq \prod_j\gamma_j^2, \quad \gamma_j = \exp(\sum_a \text{max}(2\lambda^j_a, 0))
\end{equation}
where $\gamma_j$ is associated with the $j$-th circuit layer, and $\lambda_a^j$ are the generator rates of the sparse Lindblad model\footnote{Here, generator rates may be negative due to the gauge transformations of Eq.~\eqref{eq:gauge_trans}.}, see Sec.~\ref{sec:PEC}. 
Box~\ref{box:gauge_learning_theorem} suggests that, by assuming the true noise model to be any of the gauge-equivalent models $\bm\Lambda_{\bm\eta}$, parametrized by the gauge parameters $\bm\eta$, PEC yields unbiased estimators.
Interestingly, while different $\bm\Lambda_{\bm\eta}$ all yield the same observable outcomes, different gauge parameters do not result in the same sampling overhead.
This motivates us to conduct gauge optimization -- searching for ${\bm\eta^*}$ that minimizes the PEC overhead. 

Suppose we have a design matrix $F$ and estimation of $\hat{\bm b}$ from experiments.
A naive approach to obtain the optimal gauge would be to minimize the sampling overhead by first performing a pseudo-inverse of the design matrix $\hat{\bm{x}}_0={F}^+ \hat{\bm{b}}$ (which fixes the residual error $\epsilon = \left\lVert {F} \hat{\bm{x}}_0 - \hat{\bm{b}} \right\rVert$), followed directly by a second optimization step over gauge parameters $\bm\eta$ on the overhead in Eq.~\eqref{eq:gauge_sampling_overhead}. 
However, such an approach unnecessarily restricts the gauge optimization procedure without taking into consideration that the residual errors can vary depending on the statistical fluctuations of the observed outcomes.  

Rather, we introduce a strategy where the possible $\bm{x}$ parameters are searched in a self-consistent manner with a constrained residual error $\epsilon$ chosen \textit{a priori}, which leads to the following optimization problem:
\begin{equation}
\min_{{\bm{x}}}  
	\left\{
		\sum_{a\in \mathcal{P},~\text{layer}}
		\text{max} 
		\left(2\lambda_a^{\text{layer}}, 
			0
		\right) 
	\right\} ~~\text{such that}~  \left\lVert {F} {\bm{x}} - \hat{\bm{b}} \right\rVert \le \epsilon,
	\label{eq:gauge_opt_1step} % See Def 3.3 in 2410.03906
\end{equation}
where the generators $\lambda_i$ depend on the fidelities ${\bm x}$ as discussed in Sec.~\ref{sec:noise_learning_theory}\footnote{For consistency with other chapters of this thesis covering sparse Pauli-Lindblad models, here we have parametrized the noise models through the generator rates $\lambda_i$ and log-fidelities $x_i$. 
For sparsified models in the gauge formalism a lower-dimensional parametrization through a vector $\bm{r}$ given by the Möbius transform of $\bm{x}$~\cite{wagner2023learning} (referred to as the ``reduced parameter'' in Ref.~\cite{chen2024efficient}) is more practical.}.
Note that we only consider optimizing the overhead of mitigating the gate errors as these errors can accumulate through the computations, but the SPAM errors need to be mitigated only once. 
This can be efficiently minimized over large system sizes $N$ with standard convex optimization solvers~\cite{cvxpy}. 

We apply gauge optimization to the noise models learned for the 92-qubit ring circuit experiments shown in Fig.~\ref{fig:7}.
We observe a large difference between $\Gamma_0$ (overhead without gauge optimization) and $\Gamma({\bm\eta^*})$, that corresponds to $233\times$ smaller sampling overhead per CNOT layer.
However, we note that the overhead without gauge optimization corresponds to a rather arbitrary point in the space of gauge parameters. 
Hence, the impact of this reduction in overhead is better understood by comparing against experiments with previous noise learning procedures~\cite{van2023probabilistic}. 
We compare $\Gamma({\bm\eta^*})$ against $\Gamma^{\text{inc}}$, where $\Gamma^{\text{inc}}$ is calculated using the inconsistent sparse Pauli-Lindblad model with symmetric gate fidelities, obtaining a $40 \%$ reduction in sampling overhead per CNOT layer.
Note that this accumulates exponentially as more layers are added to the circuit.  
To summarize, we have observed that self-consistent learning not only improves mitigation bias, but also reduces mitigation overheads compared to previous approaches to learning noise.

\section{Discussion}

Traditional noise characterization methods have struggled with inherent ambiguities in noise learning and thus relied on assumptions about noise, such as perfect state preparation and certain symmetries in gate noise.
We have showed that it is possible to self-consistently learn all relevant features of Pauli noise across an entire gate set consisting of state preparation, gate, and measurement errors up to unlearnable ``gauge'' degrees of freedom. 
We have experimentally validated this protocol on specifically tailored benchmark circuits of up to 92 superconducting qubits.
While we found that our method improves the resulting mitigation bias compared to previous state-of-the-art methods, non-negligible biases can remain in practice.
They originate from shortcomings of the employed Pauli models to accurately describe the hardware noise. 
We anticipate that the self-consistent formalism will become a valuable tool to systematically detect out-of-model errors such as higher-weight noise generators or leakage. 

Besides enabling more accurate error mitigation, we point out that the associated experimental overhead itself depends on the gauge parameters. 
With a clever choice of the gauge parameters, the overhead of error mitigation can thus be significantly reduced.
This represents a paradigm shift from treating gauge freedom as a nuisance to treating it as a resource, which opens new directions for optimizing device performance.
Our discussion has focused on PEC for error mitigation.
We leave the exploration of how the gauge parameters affect the overhead of other noise-model-based error mitigation techniques, such as TEM (see Chap.~\ref{chap:dual_unitary_TEM}), for future work.  

Our formalism as presented here only applies to circuits with an initial state preparation and a final measurement at the end.
It would be intriguing to extend this self-consistent noise formalism to \emph{dynamic} circuits where subsequent classical operations can depend on the outcomes of mid-circuit measurements~\cite{zhang2025generalized, hines2025pauli}. 
Such dynamic circuits are promising for preparing and simulating interesting states with significantly smaller circuit depths~\cite{tantivasadakarn2023hierarchy, buhrman2024state} and may reduce the issue of barren plateaus in variational circuits~\cite{deshpande2024dynamic}. 
More accurate noise models of such non-unitary operations are also essential for optimizing the performance of decoders needed to {actively} correct errors in large-scale, fault-tolerant quantum computers~\cite{PhysRevLett.128.110504,bausch2024learning}.
A promising avenue would be to combine the self-consistent learning framework with recently introduced techniques to learn the true hardware state-preparation errors, circumventing the gauge indistinguishability by coupling to a higher-dimensional qudit space~\cite{haupt2025statepreparation}. 

To summarize, our work provides a practical solution to a longstanding issue in quantum noise characterization. 
By addressing the ambiguities in learning and mitigating quantum noise, our method improves both the bias -- by removing inconsistencies -- and the variance -- by optimizing the gauge -- of previous error mitigation schemes.

\chapter{Conclusion}
\label{chap:conclusion}

\section{Main contributions} 

In this thesis, we have explored strategies  to maximize the computational utility of superconducting quantum computers in the presence of hardware constraints and noise. 
Our contributions spanned hardware-level control, algorithmic improvements, and refined error mitigation strategies, pushing the boundaries of digital quantum simulation in the pre-fault-tolerant era.

To make the most of the available quantum resources in superconducting circuits, we have developed a framework that leverages higher-excited levels beyond the qubit subspace to perform universal qudit-based computations with transmons.
We have proposed a qudit-space extension of the entangling cross-resonance gate and showed how general multi-qudit unitaries can be synthesized, resulting in fewer entangling gates than comparable qubit alternatives.
Our numerical simulations guide the design of future superconducting qudit hardware: 
While they suggest that multi-qudit operation with current qubit-optimized hardware remains impractical due to increased charge noise in qudit levels, they predict that high-fidelity qudit operations become feasible up to the ququart case when tuning deeper into the transmon regime.

Another key question we have addressed is how to extract information from quantum processors in the most efficient way. 
A flexible and powerful framework for this is given by informationally-complete (IC) POVM measurements. 
We have developed a novel method to realize such IC measurements based on a dilation using qudit levels, avoiding the ancilla overhead of previous approaches. 
In parallel, we have implemented POVMs based on randomized measurements, also known as classical shadows, at unprecedented scale and efficiency. 
Crucially, classical shadows offer previously neglected degrees of freedom as they overparameterize the space of observables. 
We have shown that optimizing the decomposition of an observable, formalized through the dual frame of overcomplete POVMs, enables more efficient expectation value estimation without requiring additional quantum resources. 

IC measurements enable the interfacing of quantum and classical computational resources through the classical post-processing of quantum measurement data. 
We have explored how this can be used to extend the capabilities of quantum processors through error mitigation. 
In particular, we have implemented two error-mitigation methods based on IC data in large-scale experiments ($\geq 80$ qubits), demonstrating that accurate quantum computations can be performed at scales beyond brute-force, state-vector classical simulations. 
In the first example, we have leveraged IC samples to parallelize the estimation of operators that span a subspace tailored to recover ground state energies. 
In the second example, we have made use of tensor networks -- a leading classical circuit simulation technique -- to recover noiseless expectation values, a technique coined as \emph{tensor-network error mitigation} (TEM).
We have used TEM to simulate the dynamics of an autocorrelation function in a chaotic quantum system which features special non-integrable dynamics that are nevertheless analytically solvable due to the dual-unitarity of the generating gates. 
This feature allowed us to verify the accuracy of our error-mitigated quantum computation in otherwise complex circuits. 

Error mitigation methods such as TEM rely on accurate models of the device noise. 
Fundamental gauge degrees of freedom have forced previous noise learning strategies to introduce biases that originate from an artificial separation of models for gate, state preparation and measurement noise. 
As a final contribution of this thesis, we showed that a self-consistently learned holistic noise model can overcome these limitations. 
We verified in experiments with up to 92 qubits that our method can reduce the bias of error mitigation compared to previous learning approaches. 
Moreover, the overhead associated with error mitigation can be minimized as a function of the gauge parameters. 
Hence, our method can improve both the bias and variance of error mitigation.

In summary, this thesis has contributed a series of theoretical proposals and experimental hardware demonstrations that highlight the potential of pre-fault-tolerant quantum computers as valuable tools for exploring digital quantum simulation of physically relevant systems. 

\section{Outlook}

The field of digital quantum simulation with quantum computers currently stands at a crossroads where experiments are reaching scales that are starting to rival classical simulations~\cite{anand2023classical,  abanin2025constructive, haghshenas2025digital}.
We conclude this thesis by examining key considerations for advancing the field toward achieving a practical quantum advantage.

A crucial aspect is the identification of problems that are simultaneously hard to simulate classically, suitable to current quantum hardware constraints such as limited connectivity and circuit depth, and of practical interest. 
While early interest in quantum simulation centered around solving the electronic structure problem in chemistry~\cite{mcardle2020quantum}, the high complexity of the corresponding qubit Hamiltonians has prompted a shift toward alternative systems, such as lattice models from condensed matter and high-energy physics, which offer simpler time-evolution circuits and more tractable energy estimation~\cite{dalzell2023quantum}.
The systems studied in this thesis were mainly one-dimensional spin systems with short-range interactions which fit naturally on quantum hardware but can be solved well by approximate classical methods. 
It is expected that genuine two-dimensional systems will present significantly harder challenges for classical methods~\cite{haghshenas2025digital, rudolph2025simulating}. 

Moreover, as highlighted by the techniques developed in this thesis, quantum advantage is not merely about surpassing classical computing with quantum processors, but creating synergies between both in a framework coined as ``quantum-centric supercomputing''~\cite{lanes2025framework}. 
Recent years have shown that as quantum simulations mature, new classical simulation paradigms emerge, often designed to exploit the ``weaknesses'' of specific quantum circuits~\cite{rudolph2025pauli}. 
We thus expect a back and forth between quantum advantage claims and classical refutations, as were already witnessed in previous declarations of quantum supremacy~\cite{arute_quantum_2019, larose2024brief}.

It remains an open question if error mitigation can extend the reach of noisy quantum hardware into a genuine quantum advantage regime~\cite{zimboras2025myths}. 
The largest roadblock towards this goal is the fundamental exponential overhead of measurement samples required for error mitigation. 
As quantum hardware has continuously improved throughout the progression of this thesis -- both in terms of size and quality, but also speed~\cite{abughanem2025ibm} -- we expect future improvements of these metrics to significantly reduce the error mitigation overhead. 
Our framework of self-consistent noise learning also contributes to this effort by optimizing the gauge parameters for the purpose of error mitigation and has solved, at least in theory, the long-standing issue of how to obtain accurate noise models for bias-free error mitigation. 
However, this framework still relies on a layered circuit structure with few unique two-qubit gate layers.  
It would thus be enticing to design noise models that are less tailored to a specific circuit structure, e.g., by breaking down a full device noise model into the pieces relevant to a given circuit. 
However, this may be challenging for our gauge formalism as it does not reveal the ground truth of the hardware noise model.
Recent works have suggested that additional resources in the form of qudit states can help alleviate this issue~\cite{chen2025enhancing, haupt2025statepreparation}.

Finally, one might wonder if research focused on error mitigation will become obsolete once the first fault-tolerant quantum computers become available. 
We believe that this is not the case. 
For example, error correction relies on device noise manipulations such as twirling to achieve Pauli channels and characterizations thereof to assess the performance of a given quantum error correction code in practice. 
Moreover, error mitigation may be used in combination with error correction to lower the error threshold or to mitigate remaining logical errors~\cite{aharonov2025importance}. 
Thus, we anticipate the methods developed in thesis to remain relevant throughout the transition from error-mitigated to fault-tolerant hardware.

% use to end the last part if the thesis is composed of parts
\addtocontents{toc}{\vspace{\normalbaselineskip}}
\cleardoublepage
\bookmarksetup{startatroot}

%%%%%%%%%%%%%%%%%%%%%%%%%%%%%%%%%%%%%%%%%%%%%%
%%%%% TAIL: Bibliography, Appendix, CV
%%%%%%%%%%%%%%%%%%%%%%%%%%%%%%%%%%%%%%%%%%%%%%
\appendix

\chapter{Frame theory for generalized measurements}
\label{app:chap_frame_theory}
Here, we summarize frame theory, which forms the mathematical basis for informationally-complete POVMs.
This section is based on Appendices B and C of Ref.~\cite{fischer2024dual}.

\section{Vectorized double-ket notation}
\label{app:sec_double_ket_notation}
Before discussing frame theory, we outline a widespread notation that allows for more compact expressions for the space $\text{Lin}(\HS)$ of linear operators on $\mathcal{H}$, where $\HS$ is a $d$-dimensional Hilbert space of a quantum system~\cite{gilchrist2009vectorization}.
With the Hilbert-Schmidt inner product $O_1 \cdot O_2 \coloneq \Tr( O_1^\dagger O_2)$, $\text{Lin}(\HS)$ forms a Hilbert space of its own. 

Given some basis $\ket{i}_{i \in\{1,\dots, d\}}$ in which an operator $O \in \text{Lin}(\HS)$ takes the form $O =  \sum_{i, j =1}^d o_{ij} \ketbra{i}{j}$, we can associate associate $O$ with a vector in the $d^2$-dimensional product space written as a \emph{double ket} 
\begin{equation}
    \label{eq:double_ket_def}
    \kket{O} = \sum_{i, j=1}^d o_{ij} \ket{i} \otimes \ket{j}, \quad 
\end{equation}
This can be thought of as stacking the rows of the matrix form of $O$ on top of each other to instead create a vector of length $d^2$.
Similarly, the \emph{double bra} becomes 
\begin{equation}
    \label{eq:double_bra_def}
    \bbra{O} = \sum_{i,j = 1}^d o_{i,j}^* \bra{i} \otimes \bra{j}
\end{equation}
such that the inner product can be conveniently expressed as $\Tr( O_1^\dagger O_2) = \bbrakket{O_1}{O_2}$.

\section{Frames and dual frames}
For finite-dimensional vector spaces $V$, a \emph{frame} $F = \{{f}_k\}_{k\in \mathcal{K}} \subset V$ is a generating system of $V$, i.e., $F$ spans the space $V$ (where $\K$ is taken to be finite).
If all $f_k$ are linearly independent the frame is called \emph{minimally-complete} (and forms a basis), else the frame is \emph{overcomplete}.
For any frame $F$, there exists at least one \emph{dual frame} $D = \{{v}_k\}_{k\in \mathcal{K}} \subset V$~\cite{li_general_1995} which is defined through the relation
\begin{equation}
\label{eq:dual_frame_def}
    {v} = \sum_{k \in \K} \inprod{{v}}{{d}_k} {f}_k = \sum_{k \in \K} \inprod{{v}}{{f}_k} {d}_k \ \textrm{ for all } {v} \in V.
\end{equation}
The dual frame is unique when $F$ is minimally-complete, but infinitely many dual frames exist for overcomplete frames.
With dual frames, we can find the decomposition of any $v \in V$ into elements of the frame as 
\begin{equation}
\label{eq:frame_decomposition}
    {v} = \sum_{k \in \K} c_k {f}_k \, ,  \textrm{ with } c_k = \inprod{{v}}{{d}_k} \textrm{ for all } k \in \K.
\end{equation}
% \section{Frame operators}
Let us now construct a dual frame explicitly. 
Consider the linear operator  
\begin{equation} 
\label{eq:frame_superoperator_def}
    \F : V \to V, \, {v}  \mapsto \sum_{k \in \K} \langle v, f_k \rangle {f}_k
\end{equation}
which is known as the \emph{canonical frame operator}. 
$\F$ is invertible and can be used to transform between a frame and a dual frame. 
That is, the set $\{ {d}_k \in V \mid {d}_k = \F^{-1}({f}_k), \, k \in \K \}$ forms a valid dual frame known as the \emph{canonical dual frame}.

\section{Application to POVMs}

Any IC-POVM on a Hilbert space $\HS$ can be regarded as a frame on $\herm$. 
The corresponding dual frames can be utilized to find the decomposition of any $O \in \herm$ into the POVM operators, which is the crucial subroutine for building estimators for $\langle O \rangle$, see Sec.~\ref{sec:measuring_observables}.
Using the vectorized notation introduced in Sec.~\ref{app:sec_double_ket_notation} to express the POVM as $M = \{ \kket{M_1}, \dots,  \kket{M_n}\}$, the condition for a dual frame $D = \{ \kket{D_1}, \dots, \kket{D_n} \}$ from Eq.~\eqref{eq:dual_frame_def} can be expressed as
\begin{equation}
    \label{eq:dual_frame_condition_vectorized}
    \sum_{k=1}^n  \kket{M_k} \bbra{D_k} =\sum_{k=1}^n  \kket{D_k} \bbra{M_k} = \mathbbm{1}.
\end{equation}
Similarly, the canonical frame superoperator from Eq.~\eqref{eq:frame_superoperator_def} becomes
\begin{equation}
    \label{eq:frame_superop_vectorized}
    \F = \sum_{k=1}^n \kket{M_k}\bbra{M_k}
\end{equation}
such that the elements of the canonical dual frame are obtained as  
\begin{equation}
    \label{eq:canonical_dual_frame_vectorized}
    \kket{D_k} = \mathcal{F}^{-1} \kket{M_k}.
\end{equation}
For a given dual frame, the decomposition of $O \in \herm$ is given by
\begin{equation}
    \label{eq:decomp_O_vectorized}
    \kket{O} =   \bbrakket{O}{D_k} \bbra{M_k}   .  
\end{equation}

\chapter{Single-qubit POVMs}

\section{Classes of single-qubit POVMs}
\label{app:povm_classes}

Here, we  summarize different classes of increasingly complex single-qubit IC POVMs that can be used to realize product POVMs as introduced in Eq.~\eqref{eqn:povm_product_form}. 

\subsubsection{Classical shadows}
The simplest way to construct an informationally complete single-qubit POVM is to measure a qubit randomly in the $Z$, $X$ or $Y$ basis with equal probabilities. 
Formally, this POVM is described by the operators 
\begin{equation}
    \frac{1}{3} \vb{P}_X \uplus \frac{1}{3} \vb{P}_Y \uplus \frac{1}{3} \vb{P}_Z
\end{equation}
where $\vb{P}_Z \coloneqq \{\ketbra{0}{0}, \ketbra{1}{1} \}$, $\vb{P}_X \coloneqq \{\ketbra{+}{+}, \ketbra{-}{-} \}$, and $\vb{P}_Y \coloneqq \{\ketbra{+i}{+i}, \ketbra{-i}{-i} \}$. 
In the literature, this six-outcome POVM is known as the Pauli-6 POVM but has in recent years also gained popularity under the term \emph{classical shadows} measurements~\cite{huang_predicting_2020, elben2023randomized}.

\subsubsection{Locally-biased classical shadows}
Classical shadows can be generalized by treating the probabilities of measuring a qubit with $\vb{P}_X$, $\vb{P}_Y$, or $\vb{P}_Z$ as additional degrees of freedom~\cite{hadfield2022measurements}.
We refer to the ensuing single-qubit POVM
\begin{equation}
    q_X \vb{P}_X \uplus q_Y \vb{P}_Y \uplus q_Z \vb{P}_Z
\end{equation}
as \emph{locally-biased classical shadows} (LBCS).
This POVM carries two free parameters $q_X$ and $q_Y$, with $q_Z = 1 - q_X - q_Y$. 

\subsubsection{Mutually-unbiased-bases POVM}
Consider a set of different orthonormal bases for a $d$-dimensional Hilbert space.
These bases are said to be \emph{mutually unbiased bases} (MUB) if for any pair of bases $\{\ket{a_k}\}_k, \{\ket{b_k}\}_k$ in the set, we have $\abs{\langle a_k | b_{k'}\rangle}^2 = \frac{1}{d} \ \forall k,k'$. 
For a set of MUB, each basis $\{\ket{\psi^i_{k}}\}_k$ induces a projective measurement $\vb{P}_i = \{\ketbra{\psi^i_{k}}\}_k$.
We refer to a POVM which is a convex combination $\vb{M} = \biguplus_i q_i \vb{P}_i$ of PMs induced by mutually unbiased bases as a \emph{MUB-POVM}.
It can be shown that any single-qubit MUB-POVM can be obtained by applying a fixed unitary operator $U$ to the effects of the LBCS POVMs~\cite{zhu_quantum_2012}, resulting in the POVM\footnote{The notation $U^\dagger \vb{P}_i  U$ is understood element-wise, e.g., $U^\dagger \vb{P}_X  U =  \{U^\dagger\ketbra{+}{+}U, U^\dagger\ketbra{-}{-}U\}$.}
\begin{equation} \begin{split}
q_X U^\dagger \vb{P}_X  U \uplus q_Y  U^\dagger \vb{P}_{Y} U \uplus q_Z U^\dagger \vb{P}_{Z} U  
\end{split} \end{equation}
In addition to the two degrees of freedom of the LBCS POVMs, MUB POVMs thus carry three parameters that define the single-qubit unitary $U$ (see Eq.~\ref{eq:decomp_RZ_SX} for an explicit parametrization of $U$). 

\subsubsection{General 6-outcome PM-simulable POVM}
Finally, the most general type of single-qubit PM-simulable POVM we consider in this thesis is obtained similarly to the MUB POVMs, with a different unitary rotating each projective measurement, resulting in the POVM 
\begin{equation} \begin{split}
q_X U_X^\dagger \vb{P}_X  U_X \uplus q_Y  U_Y^\dagger \vb{P}_{Y} U_Y \uplus q_Z U_Z^\dagger \vb{P}_{Z} U_Z.
\end{split} \end{equation}

\subsubsection{4-outcome dilation POVM}
For a single qubit, any four-outcome POVM of linearly independent operators represents a minimal IC measurement. 
In practice, it can be realized by applying a two-qubit dilation unitary $U$ to the targeted qubit in state $\rho$ and a second ancillary qubit that is prepared in a fixed state, typically $\ket{0}$. 
Then, both qubits are measured in the computational basis, and the four possible outcomes will occur with the probability $\text{Tr}[\rho M_k]$ of the four POVM effects $M_k$. 
This reduction of a POVM measurement to a projective measurement on a higher-dimensional space is known as a \emph{Naimark dilation} and similar in spirit to the purification of mixed states and the Stinespring dilation of quantum channels discussed in Chap.~\ref{chap:theory_of_qc}. 
We provide more details in the following section. 

\section{Naimark construction for single-qubit POVMs}
\label{app:sec_naimark_single_qubit}

Here, we detail the connection between a unitary $U$ applied to a four-dimensional extended Hilbert space $\mathcal{H}_\text{ext}$ and the POVM operators realized on the single-qubit space $\mathcal{H}_\text{S}$ through a Naimark dilation construction. 
In a tensor product extension (TPE), the four basis states of $\mathcal{H}_\text{ext}$ are formed with an ancilla qubit as
${\ket{0}_\text{ext} = \ket{0}_\text{S} \otimes \ket{0}_\text{A}}$, 
 $\ket{1}_\text{ext} = \ket{1}_\text{S} \otimes \ket{0}_\text{A}$, 
$\ket{2}_\text{ext} = \ket{0}_\text{S} \otimes \ket{1}_\text{A}$, and
$\ket{3}_\text{ext} = \ket{1}_\text{S} \otimes \ket{1}_\text{A}$.
In contrast, in a direct sum extension (DSE), the four states $\ket{i}_\text{ext}$ form a qudit space, where the qubit is encoded in the states $\ket{0}_\text{ext} \equiv \ket{0}_\text{S}$ and $\ket{1}_\text{ext} \equiv \ket{1}_\text{S}$.

For simplicity, we assume a pure state of the system qubit $\ket{\psi}_\text{S} = \alpha \ket{0}_\text{S} + \beta \ket{1}_\text{S}$.
A Naimark construction for both a TPE and a DSE applies a unitary $U$ to the initial state $\ket{\psi}_\text{init} = \alpha \ket{0}_\text{ext} + \beta \ket{1}_\text{ext}$, to create the final state
\begin{align}
U  \ket{\psi}_\text{init} 
= \sum_{k=0}^3 \left(\alpha U_{k, 0} + \beta U_{k, 1} \right) \ket{k}_\text{ext} .
\end{align}
Measuring $U  \ket{\psi}_\text{init} $ in $\mathcal{H}_\text{ext}$ produces an outcome $m \in \{0, 1, 2, 3\}$ with a probability $p_k = \left| U_{k, 0}\alpha + U_{k, 1} \beta \right|^2$. 
This is equal to the probabilities $p_k = \text{Tr}[M_k \ket{\psi}_\text{S} \bra{\psi}_\text{S} ]$ associated with a POVM of four rank-1 operators
\begin{align}
\label{eq_theo:POVM_ops_from_unitary}
M_k = \Gamma_k \ket{\pi_k}\bra{\pi_k}
\end{align}
acting on $\mathcal{H}_\text{S}$, which are proportional to projectors along the states 
\begin{align}
\label{eq_theo:POVM_ops_state}
\ket{\pi_k} &=  \frac{1}{\sqrt{\Gamma_k}} \left( U_{k, 0}^* \ket{0}_\text{S} + U_{k, 1}^* \ket{1}_\text{S} \right)  
\end{align}
with normalization factors $\Gamma_k = | U_{k, 0}|^2 + |U_{k, 1}|^2$.
Through Eqs.~\eqref{eq_theo:POVM_ops_from_unitary} and~\eqref{eq_theo:POVM_ops_state}, the unitary $U$ applied to $\mathcal{H}_{\text{ext}}$ can emulate the measurement of any POVM with four rank-one operators on $\mathcal{H}_{\text{S}}$. 

Since the desired POVM only defines the first two columns of $U$, i.e., $U_{k,0}$ and $U_{k,1}$, we find the remaining columns with a Gram-Schmidt procedure to ensure $U$ is unitary.
Without loss of generality, for the decomposition algorithm presented in Sec.~\ref{sec:general_SUD_synthesis}, it is convenient to choose the top right element of $U$ to vanish, i.e., $U_{0, 3} = 0$.

\chapter{Rotating frame transformation}
\label{app:rotating_frame}

The dynamics of periodically-driven systems, such as superconducting qubits under microwave drives, are more conveniently expressed in an interaction picture known as the \emph{rotating frame}.
Here, since the frame of reference rotates with the drive, the interaction picture transformation is explicitly time-dependent. 
Since this time-dependence is often not explicitly covered in textbooks, we review the interaction-picture expressions for states and operators in the rotating frame in this Appendix. 
The time-dependent Schrödinger equation in the lab frame is
$i\hbar \frac{d}{dt} |\psi(t)\rangle_{\text{lf}} = H_{\text{lf}}(t) |\psi(t)\rangle_{\text{lf}}$.
Let $U(t)$ be an explicitly time-dependent transformation of the reference frame, such as the rotating frame.
The interaction-picture state is then given by
\begin{align}
|\psi(t)\rangle_{\text{rf}} = {U}^\dagger(t) |\psi(t)\rangle_{\text{lf}}.
\end{align}
We can derive the form of the interaction-picture Hamiltonian from the effective Schrödinger equation by taking the time derivative of the interaction-picture state:
\begin{align}
i \hbar \frac{d}{dt} |{\psi}(t)\rangle_{\text{rf}} &= i \hbar \frac{d {U}^\dagger(t)}{dt} |\psi(t)\rangle_{\text{lf}} + i \hbar {U}^\dagger(t) \underbrace{\frac{d}{dt} |\psi(t)\rangle_{\text{lf}}}_{= - \frac{i}{\hbar} {H_{\text{lf}}(t) |\psi(t)\rangle}} \\
&=  i \hbar \underbrace{\frac{d {U}^\dagger(t)}{dt} U(t)}_{= -  U^\dagger(t)  \frac{d {U}(t)}{dt} } |{\psi}(t)\rangle_{\text{rf}} + {U}^\dagger(t) H_{\text{lf}}(t) U(t) |{\psi}(t)\rangle_{\text{rf}} \\ 
&= \underbrace{\left[ {U}^\dagger(t) {H_{\text{lf}}}(t) {U}(t) - i\hbar {U}^\dagger(t) \frac{d}{dt} {U}(t) \right]}_{ = H_{\text{rf}}(t)}  |{\psi}(t)\rangle_{\text{rf}}
\end{align}
Here, we have used the product rule, the Schrödinger equation in the lab frame, and the unitarity of $U(t)$ in the three steps, respectively.
We find that the transformed Hamiltonian in the rotating frame $H_{\text{rf}}(t)$ carries a second term that describes a fictitious force that arises purely as a consequence of the time-dependence of the frame transformation. 
This is a quantum analogue of frame-induced effects like the centrifugal force in classical mechanics. 

\chapter{Details on pulse-level numerical simulations}
\label{app:details_simulation}

\section{Single-qudit simulations}
\label{app:details_simulation_single_qudit}

\begin{figure*}[t]
\centering
\includegraphics[width=\textwidth]{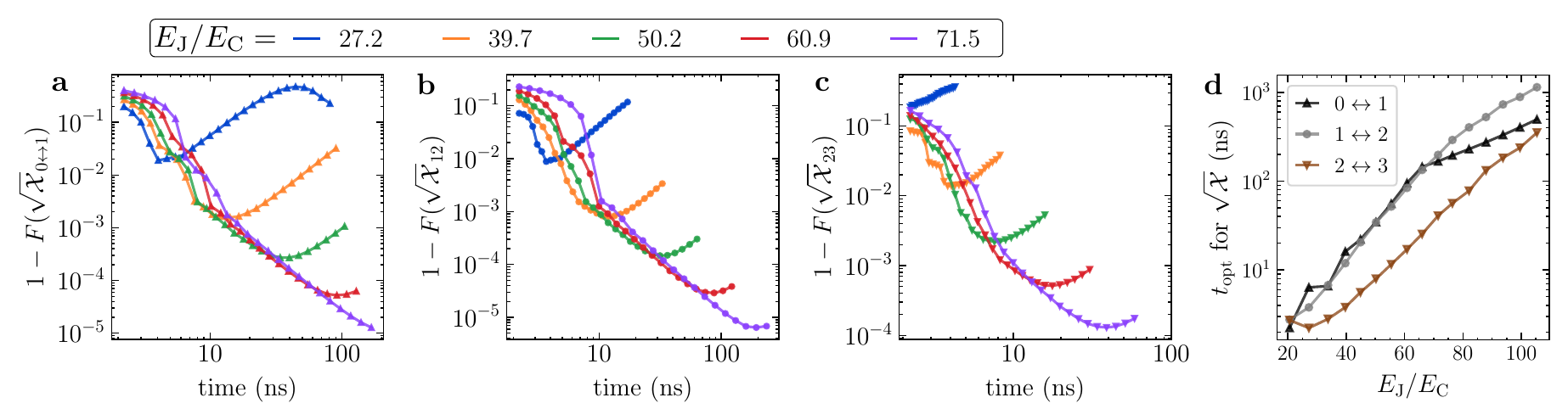}
\caption[]{Calibration of $\sqrt{\mathcal{X}}$-pulses through numerical simulations of the pulse-level dynamics. \textbf{a) -- c)} Average gate infidelities $1-F(\sqrt{\mathcal{X}})$ in the qudit space relevant for the POVM encoding as a function of the gate duration for selected values of $E_\text{J} / E_\text{C}$. \textbf{d)} Optimal times $t_{\text{opt}}$ that maximize $F(\sqrt{\mathcal{X}})$ for different $E_\text{J} / E_\text{C}$.}    \label{fig:app_simulation_SX_calibrations}
\end{figure*}

In this appendix, we summarize the technical details of the numerical simulations from Sec.~\ref{sec:ancilla-free_optimal_transmon_regime}. 
Transmons are modeled by the Hamiltonian in Eq.~\eqref{eq:transmon_hamiltonian}. 
This Hamiltonian is diagonalized in the charge representation after truncating to 20 Fourier modes in the superconducting phase $\hat{\phi}$ to obtain the low-energy spectrum~\cite{gambetta2013quantum}. 
The parameters $E_{\mathrm{J}}$ and $E_{\mathrm{C}}$ are then adjusted to fix the base frequency of the qubit (at $n_g=0$) at $5\,\text{GHz}$. 
The dynamics of the system under a drive are modeled by the interaction Hamiltonian from Eq.~\eqref{eq:rotating_frame_derivation}, where we assume the relative coupling of a harmonic oscillator, i.e., $g_n \propto \sqrt{n+1}$, and truncate the system at $d=5$ levels.

To implement a desired POVM, we decompose the corresponding target unitary into a sequence of $\sqrt{\mathcal{X}}$-pulses with virtual $\mathcal{Z}$-gates.
For simplicity, we employ Gaussian pulse envelopes with a standard deviation of one quarter of the pulse duration. 
The pulses feature a piece-wise constant envelope with a sample duration of $222\,\text{ps}$, matching IBM control hardware. 
The transition frequencies depend on $n_g$, whose exact value fluctuates from one experimental run to another.
We model $n_g$ to be uniformly distributed as $p(n_g) \sim \text{Uni}([0, 1])$, which is sufficient due to the periodicity of the eigenenergies with $n_g$, see Fig.~\ref{fig:transmon_properties}a.
Each pulse is played at the average transition frequency $\overline{\omega_n} = \int_0^1 p(n_g) \omega_n(n_g) \text{d}n_g$. 
The pulse amplitudes are chosen such that the resulting rotation angles for the average transition frequency are $\pi/2$. 

We model the quantum dynamics of a state $\rho$ under a pulse sequence by an effective channel 
\begin{align}
\label{eq_app:channel_charge_noise}
\mathcal{E}: \rho \longmapsto \int_0^1 p(n_g)  U(n_g) \rho U(n_g)^\dagger \text{d}n_g  
\end{align}
where $U(n_g)$ are the unitary dynamics for a fixed offset charge $n_g$. 
We obtain  $U(n_g)$ under a sequence of pulses with an integrator of the time-dependent Schr{\"o}dinger equation provided by QuTip~\cite{johansson2012qutip}.
The channel $\mathcal{E}$ is numerically approximated by computing $U(n_g)$ for 20 values of $n_g$ equally spaced between 0 and 1.

To calibrate $\sqrt{\mathcal{X}}$-pulses in the simulation, we keep the amplitude fixed while varying the duration of the pulses.
The target unitary of such a pulse is given by the implemented rotation $U^{\text{tar}} = \mathcal{R}(\varphi=\pi/2, \gamma=0)$, defined in Eq.~\eqref{eq:givens_rotation_hardware}, which includes phases that are accumulated in the idle levels. 
As a figure of merit, we compute the average gate fidelity $F(\mathcal{E}, U^{\text{tar}})$ between the target unitary and the channel of the simulated unitary under charge noise~\cite{nielsen2002simple}.
We hereby restrict the computation of $F(\mathcal{E}, U^{\text{tar}})$ to the subspace that is relevant for the POVM pulse sequence. 
Recall that the POVM-encoding unitary is always realized with pulses that couple adjacent levels in the order $0\!\leftrightarrow \!1$, $1\!\leftrightarrow\!2$, $0\!\leftrightarrow\!1$, $2\!\leftrightarrow\!3$, and finally $1\!\leftrightarrow\!2$.
Since the $\ket{3}$ state is only populated once prior to measurement, the phases acquired by $\ket{3}$ during a $0\!\leftrightarrow\!1$ and $1\!\leftrightarrow\!2$ gate do not affect the encoded POVM operators. 
The fidelities of $\sqrt{\mathcal{X}}_{0\leftrightarrow1}$ and $\sqrt{\mathcal{X}}_{1\leftrightarrow2}$ are thus only computed over the subspaces spanned by $\ket{0}, \ket{1}$, and $\ket{2}$. 
Similarly, only the $\ket{1}, \ket{2}, \ket{3}$ subspace is considered for the fidelities of $\sqrt{\mathcal{X}}_{2\leftrightarrow3}$ since no $0\!\leftrightarrow\!1$ pulses are applied after the $2\!\leftrightarrow\!3$ pulse.

The average gate fidelities for different hardware parameters as a function of the pulse duration are shown in Fig~\ref{fig:app_simulation_SX_calibrations}.
For short durations, the broad spectral range of the pulse leads to leakage errors. 
In contrast, for long pulse durations, the phases accumulated over time by the idle levels become difficult to track due to charge noise.
The infidelities $1-F(\sqrt{\mathcal{X}})$ thus typically show a distinct minimum where these two effects are traded off optimally.   
As $E_\text{J}/E_\text{C}$ increases, this optimum shifts towards longer gate durations, see Fig.~\ref{fig:app_simulation_SX_calibrations}\textbf{d}.
For reference, the default single-qubit $SX$-gate of the hardware used in experiments is carried out with DRAG pulses of a duration of $36\,\text{ns}$. 
We find that it is important to employ much shorter pulses when including the phase uncertainty of a neighboring state, despite our use of simple Gaussian pulse envelopes, which are not specifically designed to correct for leakage errors (especially for the $\sqrt{\mathcal{X}}_{2\leftrightarrow3}$-gate). 
This suggests that phase uncertainties in higher excited states have an overall bigger impact on the qudit gate fidelities than leakage. 
The remaining leakage errors could be further reduced by a careful calibration of \textsc{Drag} pulses.
For current hardware ($E_{\mathrm{J}}/E_{\mathrm{C}} \sim 35\,\text{--}\,45$), our simulations suggest achievable gate fidelities in the relevant qudit spaces that reach up to $99.9\%$ for the $\sqrt{\mathcal{X}}_{0\leftrightarrow1}$- and $\sqrt{\mathcal{X}}_{1\leftrightarrow2}$-gates, and $98\%$ for $\sqrt{\mathcal{X}}_{2\leftrightarrow3}$. 
This can be improved by over an order of magnitude by tuning deeper into the transmon regime (e.g., $E_{\mathrm{J}}/E_{\mathrm{C}} \sim60$), at the expense of increased gate durations.  

For our simulation of the full POVM pulse sequences in Sec.~\ref{sec:ancilla-free_optimal_transmon_regime} we employ the durations of the $\sqrt{\mathcal{X}}_{n\leftrightarrow n+1}$-pulses that maximize their respective fidelities. 
When limiting the total duration of the sequence as in Fig.~\ref{fig:transmon_parameter_regime}\textbf{a}, we incrementally shorten those pulses whose fidelity is affected the least. This is repeated until a pulse sequence is obtained which is at most as long as the desired total length.
From the implemented channel $\mathcal{E}$ of the pulse sequence, we finally obtain an effective POVM $\Pi_{\text{sim}}$ as the average over the POVM operators encoded by the unitaries $U(n_g)$. 

\section{Two-qudit simulations}
\label{app:details_simulation_two_qudit}

Here, we summarize the technical details of the numerical simulations presented in Sec.~\ref{sec:two_qudit_operations}.
Our analysis of the two-transmon system is largely based on Ref.~\cite{malekakhlagh2020firstprinciples}.
The transmon Hamiltonian from Eq.~\eqref{eq:transmon_hamiltonian} is commonly simplified in the Kerr approximation by expanding the cosine term to fourth order. 
Taking also the next-highest order into account, the Hamiltonian in its eigenbasis truncated to the $d=4$ subspace becomes~\cite{malekakhlagh2020firstprinciples}
\begin{align}
\label{eq:single_qudit_approx_hamiltonian}
H^\prime \coloneqq \omega \ketbra{1}{1} + \left(2\omega + \alpha\right) \ketbra{2}{2} 
+ 3 \left( \omega + \alpha - \frac{E_{\text{C}}}{8E_{\text{J}}} \sqrt{2E_{\text{C}}E_{\text{J}}} \right) \ketbra{3}{3}
\nonumber
\end{align}
with the qubit frequency $\omega$ and the anharmonicity $\alpha$. 
The capacitive coupling Hamiltonian between control (c) and target (t) is given by $H_J = J y_c \otimes y_t$. 
It is expressed up to $O(\epsilon^3)$ in the unitless anharmonicity measure $\epsilon = \sqrt{2 E_{\text{C}} / E_{\text{J}}} \sim 0.168$ through the unitless charge operator $y = -i(b - b^\dagger)$ with 
\begin{equation}
\label{eqn:lowering_operator}
b \coloneqq \begin{pmatrix}
0 & b_{01} & 0      & b_{03} \\
0 & 0      & b_{12} & 0      \\
0 & 0      & 0      & b_{23} \\
0 & 0      & 0      & 0
\end{pmatrix}
\end{equation}
and
\begin{subequations}
\begin{align}
b_{01} &\coloneqq 1 - \frac{\epsilon}{8} - \frac{11 \epsilon^2}{256}, \quad
&b_{12} \coloneqq \left(1 - \frac{\epsilon}{4} - \frac{73 \epsilon^2}{512}\right)\sqrt{2}, \\
b_{23} &\coloneqq \left(1 - \frac{3\epsilon}{8} - \frac{79 \epsilon^2}{256}\right)\sqrt{3}, \quad
&b_{03} \coloneqq - \frac{\sqrt{6}\epsilon}{16} - \frac{5\sqrt{6} \epsilon^2}{128}.
\end{align}
\end{subequations}
This always-on coupling leads to a dressing of the basis states of the two-transmon system. 
The states $\ket{n}_c \otimes \ket{m}_t$ referred to in the context of the coupled system, denote the dressed eigenstates of the time-independent Hamiltonian
\begin{equation}
\label{eqn:H0}
H_0 \coloneqq H^\prime_c \otimes \mathbbm{1} + \mathbbm{1} \otimes H^\prime_t + H_J.
\end{equation}
In this notation, similar to the derivation from Sec.~\ref{sec:qudits_hardware_model}, an external microwave drive with a carrier frequency of $\omega_{\text{d}}$ and pulse envelope $\Omega(t)$ leads to an interaction Hamiltonian (in the rotating wave approximation) of
$H_{\text{int}}(t) \coloneqq \frac{\Omega(t)}{2} \left( b e^{i \omega_{\text{d}} t} + b^\dagger e^{-i\omega_{\text{d}} t} \right)$.
We simulate the action of a pulse applied to the control by numerically solving the time-dependent Schr{\"o}dinger equation for the full Hamiltonian $H_{\text{tot}} \coloneqq H_0 + H_{\text{int}}(t)\otimes \mathbbm{1}$ with the master equation solver provided by QuTip~\cite{johansson2012qutip}. 

\begin{figure*}
\includegraphics[width=\textwidth]{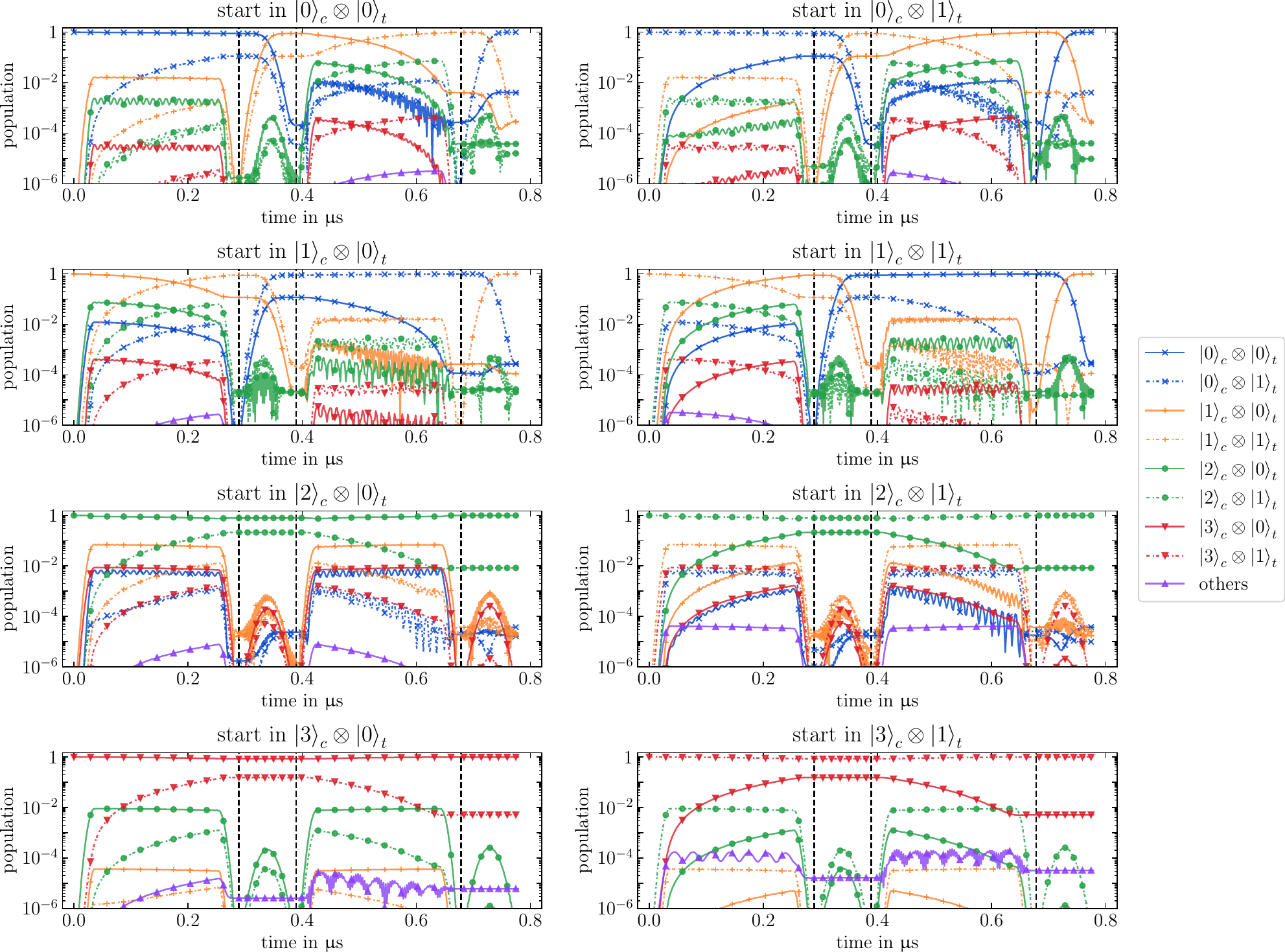}
\caption[]{Evolution of the populations in the two-transmon dressed basis during the echoed CR sequence for different initial states. 
Color indicates the state of the control, while the linestyle indicates the state of the target.
Vertical dashed black lines form four regions that correspond to the four total pulses of the ECR sequence. }
\label{fig:ecr_dynamics_log}
\end{figure*}

All CR pulses we apply to the control are played at a frequency of $\omega_{\text{d}} = \overline{\omega}_{\text{t}}$, i.e., the target qudit frequency in the dressed basis averaged over the four lowest levels of the control qudit. 
We choose a Gaussian-square pulse shape which consists of a plateau of duration $\tau_s$ between a Gaussian rise and fall with duration $\tau_g$ and standard deviation $\sigma$. 
The pulse envelope is thus
\begin{equation}
\label{eqn:gaussian_square_pulse}
\Omega(t) = \tilde\Omega \cdot
\begin{cases}
     \dfrac{e^{-\frac{1}{2}\frac{(t - \tau_g)^2}{\sigma^2}} - \chi}{1  - \chi} &,\,0 < t \leq \tau_g \\
    1 &,\,\tau_g  < t < \tau_g + \tau_s \\
     \dfrac{e^{-\frac{1}{2}\frac{(t - \tau_g - \tau_s)^2}{\sigma^2}} - \chi}{1  - \chi} &,\,\tau_g + \tau_s \leq t < \tau \\    
\end{cases}
\end{equation}
where $\tilde\Omega$ is the maximal amplitude, $\chi \coloneqq e^{-\frac{1}{2}\frac{(1+\tau_g)^2}{\sigma^2}}$ is a rescaling constant, and $\tau \coloneqq \tau_s + 2\tau_g$ is the total duration. 
We choose parameter values of $\tilde\Omega/(2\pi) = 50\,\text{MHz}$, $\tau_g=36\,\text{ns}$, and $\sigma = \tau_g/4$ for the simulated CR pulses, as we empirically find that this leads to a tolerable amount of leakage, see Figs.~\ref{fig:single_cr_pulse_unitary},\ref{fig:ecr_dynamics_log}. 
To calibrate the rotation angles, we fix $ \tilde\Omega $, $\tau_g$, and $\sigma$, adjusting only the width $\tau_s$, and denote the obtained unitary as $R_{\text{CR}}(\tau)$.

Finally, the qudit eigenenergies $E_n$ are subject to fluctuations with $n_g$, see Sec.\ref{sec:transmon_qubits}.
To account for charge noise, we estimate the deviation $\Delta \omega^{n,n+1}$ of the transition frequencies $\omega^{n,n+1} = E_{n+1}(n_g) - E_{n}(n_g)$ from their mean as a uniform average over $n_g$. 
For the chosen transmon model parameters with $E_{\text{J}}/E_{\text{C}}\sim70$ we obtain 
$\Delta \omega^{01}/(2\pi) = 0.2\,\text{kHz}$, 
$\Delta \omega^{12}/(2\pi) = 7.7\,\text{kHz}$, and
$\Delta \omega^{23}/(2\pi) = 185\,\text{kHz}$.
To account for this uncertainty caused by charge dispersion, we apply $\Delta \omega^{n,n+1}$ as a detuning to each pulse played on the $n\leftrightarrow n+1$ transition. 
This simplifies the explicit treatment of charge noise as given in Eq.~\eqref{eq_app:channel_charge_noise}.

We benchmark how well the action of a CR pulse is described by the conditional $R_x^{01}$ rotations from Eq.~\eqref{eqn:single_Cr_pulse_action} (denoted $U_\text{CR}(\vec{\varphi})$) through the average gate fidelity $\fidelity$ as defined in Eq.~\eqref{eq:avg_gate_fidelity}. 
Under the time evolution of the CR pulse, each basis state $\ket{n}_c \otimes \ket{m}_t$ acquires a phase $e^{-i\alpha_{nm}}$, which we assume to be uncorrelated, i.e., $\alpha_{nm} = \alpha_n \alpha_m \,\forall n, m \in \{0, 1, 2, 3\} $.
We apply local phases after the action of the pulse to obtain the phase-corrected unitaries
\begin{align}
\label{eqn:local_phase_correction}
\tilde{R}_{\text{CR}}(\tau) \coloneqq [&\text{diag}(e^{i \alpha_{c_0}}, e^{i \alpha_{c_1}}, e^{i \alpha_{c_2}}, e^{i \alpha_{c_3}})\,\, \otimes \\
&\text{diag}(e^{i \alpha_{t_0}}, e^{i \alpha_{t_1}}, e^{i \alpha_{t_2}}, e^{i \alpha_{t_3}}) ] R_{\text{CR}}(\tau). \nonumber
\end{align}
In practice, these can be applied virtually on each qudit, as discussed in Sec.~\ref{sec:qudit_phase_gates}.
For a duration of $\tau_\pi = 289\,\text{ns}$, we numerically optimize the angles $\vec{\varphi}$ and phases $\vec{\alpha_c}$, $\vec{\alpha_t}$, obtaining an optimal fidelity $\fidelity(\tilde{R}_{\text{CR}}(\tau), U_\text{CR}(\vec{\varphi}))  = 99.93\%$.
Such a high fidelity is possible since the optimized unitary $\tilde{R}_{\text{CR}}(\tau)$ has a uniform phase structure with vanishing imaginary parts on the diagonal and little leakage, see Fig.~\ref{fig:single_cr_pulse_unitary}, . 
The duration $\tau_\pi$ is calibrated such that $\varphi_0 + \varphi_1 = \pi$, which leads to an echoed CR sequence with a rotation angle of $\theta = \pi$.

The full ECR sequence from Fig.~\ref{fig:ECR_pulse_overview} requires additional $R_x^{01}(\pm \pi)$ gates. 
We simulate this with a Gaussian envelope and fix the duration of the pulse to $ 100\,\text{ns}$.
The sign and amplitude $\tilde{\Omega}$ of the pulses is then tuned such that the area under the envelope matches the desired rotation angle. 
Within this model, we achieve average gate fidelities of $>99.99 \%$ (unitary error) in the ququart subspace for all single-qudit pulses. 

Let us now take a closer look at simulations of the full ECR sequence.
To investigate the dominant contributions to the unitary error, we plot the evolution of all populations throughout the pulse sequence in Fig.~\ref{fig:ecr_dynamics_log}. This is the same data as shown in Fig.~\ref{fig:ECR_pulse_overview}(c), plotted on a logarithmic ordinate to highlight the small contributions in the undesired levels. 
We identify small non-zero off-diagonal entries in the $\ket{2}_c$ and $\ket{3}_c$ subspaces, which manifest in remaining populations of $0.005$ -- $0.01$ for $\ket{1}_t$ and $\ket{0}_t$ in the bottom two panels of Fig.~\ref{fig:ecr_dynamics_log}.
This means that the second CR pulse does not fully reverse the rotations that were applied to these states by the first CR pulse. 
We attribute this to the fact that the driving frequency $\overline{\omega}_t$ is detuned by $120\,\text{kHz}$ to the $\ket{2}_c \ket{0}_t \leftrightarrow \ket{2}_c \ket{1}_t$ transition and by $-110\,\text{kHz}$
to the $\ket{3}_c \ket{0}_t \leftrightarrow \ket{3}_c \ket{1}_t$ transition~[see Fig.~\ref{fig:two_transmon_level_spectrum}].
Therefore, both the $x$-rotation of the first CR pulse and the $-x$-rotation of the second CR pulse carry a small $Z$-component, which results in a misalignment between the rotation axis of the two CR pulses. 
The populations of the higher-excited states of the target, shown in purple in Fig.~\ref{fig:ecr_dynamics_log}, remain below $10^{-4}$ from which we conclude that the chosen pulse shape successfully limits leakage.

We model non-unitary dynamics under a Markovian noise approximation with a Lindblad master equation as defined in Eq.~\eqref{eq:lindblad_general}.
We consider two types of errors for ququarts, pure dephasing and amplitude damping. 
Firstly, we model pure dephasing with a jump operator $L_0 = \sqrt{2/T_2} \sum_{m=0}^3 \ketbra{m}{m}$.
The choice of $T_2$ is motivated by the fact that Eq.~\eqref{eq:lindblad_general} with only pure dephasing leads to a quantum channel $\mathcal{E}(t)$ whose fidelity $\fidelity(\mathcal{E}(t), \mathbbm{1})$ in the qubit subspace decays exponentially in time with a time constant of $T_2$. 
Secondly, our model of amplitude damping contains three individual jump operators of the form $L_n = \sqrt{n/T_1}\ketbra{n-1}{n}$ for $n\in \{1, 2, 3\}$.
These lead to an exponential decay of the population in the $n$-th level, with a time constant given by $T_1/n$.
We include the jump operators on each qudit individually, i.e., adding operators $L_j \otimes \mathbbm{1}$ and $\mathbbm{1}\otimes L_j$ to Eq.~\eqref{eq:lindblad_general}.

\chapter{Qudit gate decompositions}

\section{Details on decompositions of singly-controlled unitaries}
\label{app:transpiler}
Here, we point out technical details relating to  the decomposition of a general unitary into the ECR and single-qudit gates and distinguish our work from previous state-of-the-art.  
Our transpilation routine presented in Sec.~\ref{sec:universal_gate_synthesis_qudits} builds on the multivalued quantum Shannon decomposition as developed in Ref.~\cite{di2013synthesis} where it is shown how to synthesize arbitrary unitaries with a generalized $m$-controlled $X$ gate
\begin{equation}
\label{eq:def_gcx}
U_\text{GCX} \coloneqq C^m[\ketbra{0}{1} + \ketbra{1}{0} + \sum_{k=2}^d \ketbra{k}{k}]
\end{equation}
as the fundamental entangling gate.   
One might assume that existing implementations of the CNOT gate naturally generalize to the GCX gate. 
However, this is not the case for both direct and echoed CR gates, as discussed below.
This necessitates the modification we propose in Sec.~\ref{sec:universal_gate_synthesis_qudits} to the transpilation provided in Ref.~\cite{di2013synthesis}.

The cross-resonance effect applies an $R_x^{01}(\varphi_j)$ rotation to the target qudit, where $\varphi_j$ depends on the control state $\ket{j}_c$ and the gate duration, see Eq.~\eqref{eqn:single_Cr_pulse_action}.
To implement a \emph{direct} CNOT, the angles are tuned such that $\varphi_0 - \varphi_1 = \pi$. 
In the qubit subspace, the resulting unitary is equivalent to the CNOT up to local operations. 
However, in contrast to Eq.~\eqref{eq:def_gcx}, the gate acts non-trivially on higher-excited states of the control; an effect that can not be reversed by single-qudit gates. 
For qutrits, this can be compensated for by tuning the drive strength such that $\varphi_0 = \varphi_2$, as demonstrated in Ref.~\cite{blok2021quantum}.
This approach does not scale to higher qudit dimension. 

The action of higher-excited control states can be canceled by an echoed CR sequence, leading to the $U_\text{ECR}(\theta)$ gate defined in Eq.~\eqref{eqn:ECR_unitary}. 
While $U_\text{ECR}(\pi/2)$ is local-Clifford equivalent to the CNOT for qubits, it is again not possible to transform $U_\text{ECR}(\pi/2)$ to the GCX gate with local operations. 
As shown in Sec.~\ref{sec:universal_gate_synthesis_qudits}, $d-1$ $U_\text{ECR}(-\pi/d)$ gates can be combined to synthesize a $C^m[R_x^{01}(\pi)]$ gate.
However, there is still a subtle difference to the GCX gate. 
A $C^m[R_x^{01}(\pi)]$ gate introduces a relative phase factor of $i$ between the qubit subspace and higher-excited states, as
\begin{equation}
\label{eq:CRX_GCX_comparison}
R_x^{01}(\pm \pi) =\mp i \ketbra{0}{1} \mp i \ketbra{1}{0} + \sum_{k=2}^d \ketbra{k}{k}.
\end{equation}
Cancelling this relative phase on the target requires a controlled phase gate.
The unitary synthesis proposed in Ref.~\cite{di2013synthesis} is thus not suitable for architectures that implement $m$-controlled $x$ rotations, such as the ECR gate. 
We circumvent this issue by using pairs of $C^m[R_x^{01}(\pi)]$ and $C^m[R_x^{01}(-\pi)]$ gates instead of the GCX gate. 

\section{Ququart unitary synthesis with singly-controlled unitaries}
\label{app:general_synthesis}

In Sec.~\ref{sec:universal_gate_synthesis_qudits} we present a decomposition of $C^m[U]$ gates into ECR pulses.
However, making a quantum computer based on ququarts requires realizing general ${\rm SU}(16)$ gates.
We now show how to do this with bidirectional $C^m[U]$ gates.
We decompose a general unitary $U$ into a product of $C^m[U]$ gates with an iterative cosine-sine decomposition (CSD)~\cite{chen2013qcompiler}.
In general, a cosine-sine decomposition of a $2^n\times 2^n$ matrix $U$ is
\begin{align}
    U=
    \begin{pmatrix}
    u  & \\
    & v
    \end{pmatrix}
    \begin{pmatrix}
    C  & -S \\
    S & C
    \end{pmatrix}
    \begin{pmatrix}
    x  & \\
    & y
    \end{pmatrix},
\end{align}
with $C={\rm diag}(\cos\theta_1, ..., \cos\theta_{2^{n-1}})$, $S={\rm diag}(\sin\theta_1,$ $ ..., \sin\theta_{2^n-1})$.
Here, $u$, $v$, $x$, and $y$ are $2^{n-1}\times 2^{n-1}$ unitary matrices that can be further decomposed.
We label two quqaurts according to $\textit{control}\,\otimes\,\textit{target}$. 
Applying the CSD five times to a dense two-ququart unitary $U$ yields
\begin{align}
\label{eq:cosinesine_decomp}
U =  
&\left( \arraycolsep=0pt \begin{array}{cccc} 
    \setlength{\fboxsep}{1.5pt} \fbox{$U_0$}  &  &  & \\ 
       & \setlength{\fboxsep}{1.5pt} \fbox{$U_1$} &  & \\
       &  &  \setlength{\fboxsep}{1.5pt} \fbox{$U_2$} & \\ 
       &  &  & \setlength{\fboxsep}{1.5pt} \fbox{$U_3$}  
    \end{array} \right) 
\left( \arraycolsep=0pt \begin{array}{cc} 
    \dboxed{
    \arraycolsep=2pt \begin{array}{|c|c|} 
    \hline
    C_0 & -S_0  \\ \hline 
    S_0 & C_0  \\ \hline
     \end{array}
    }{2} & \\
     & \dboxed{
    \arraycolsep=2pt \begin{array}{|c|c|} 
    \hline
    C_1 & -S_1  \\ \hline 
    S_1 & C_1  \\ \hline
     \end{array}
    }{2} \end{array} \right) 
\left( \arraycolsep=0pt \begin{array}{cccc} 
    \setlength{\fboxsep}{1.5pt} \fbox{$U_4$}  &  &  & \\ 
       & \setlength{\fboxsep}{1.5pt} \fbox{$U_5$} &  & \\
       &  &  \setlength{\fboxsep}{1.5pt} \fbox{$U_6$} & \\ 
       &  &  & \setlength{\fboxsep}{1.5pt} \fbox{$U_7$}  
    \end{array} \right) 
\left(  \vspace{-15pt}\arraycolsep=0pt \begin{array}{cc} 
    \dbox{\begin{minipage}[t][20pt][t]{22pt}\vspace{5pt}\large{$\,\,C_2$}\end{minipage}}  & \dbox{\begin{minipage}[t][20pt][t]{22pt}\vspace{5pt}\large{$-S_2$}\end{minipage}}     
    \vspace{-5pt} \\ 
    \dbox{\begin{minipage}[t][20pt][t]{22pt}\vspace{5pt}\large{$\,\,S_2$}\end{minipage}}  & \dbox{\begin{minipage}[t][20pt][t]{22pt}\vspace{5pt}\large{$\,\,C_2$}\end{minipage}}  
    \end{array} \right) \\ \nonumber
    \times
&\left( \arraycolsep=0pt \begin{array}{cccc} 
    \setlength{\fboxsep}{1.5pt} \fbox{$U_8$}  &  &  & \\ 
    & \setlength{\fboxsep}{1.5pt} \fbox{$U_{9}$} &  & \\
    &  &  \setlength{\fboxsep}{1.5pt} \fbox{$U_{10}$} & \\ 
    &  &  & \setlength{\fboxsep}{1.5pt} \fbox{$U_{11}$}  
    \end{array} \right) 
\left( \arraycolsep=0pt \begin{array}{cc} 
    \dboxed{
    \arraycolsep=2pt \begin{array}{|c|c|} 
    \hline
    C_3 & -S_3  \\ \hline 
    S_3 & C_3  \\ \hline
     \end{array}
    }{2} & \\
     & \dboxed{
    \arraycolsep=2pt \begin{array}{|c|c|} 
    \hline
    C_4 & -S_4  \\ \hline 
    S_4 & C_4  \\ \hline
     \end{array}
    }{2} \end{array} \right) 
\left( \arraycolsep=0pt \begin{array}{cccc} 
    \setlength{\fboxsep}{1.5pt} \fbox{$U_{12}$}  &  &  & \\ 
    & \setlength{\fboxsep}{1.5pt} \fbox{$U_{13}$} &  & \\
    &  &  \setlength{\fboxsep}{1.5pt} \fbox{$U_{14}$} & \\ 
    &  &  & \setlength{\fboxsep}{1.5pt} \fbox{$U_{15}$}  
\end{array} \right). 
\end{align} 
Here, solid and dashed boxes indicate $4\times4$ and $8\times8$ subspaces, respectively.  
$C_i$ and $S_i$ are diagonal cosine and sine matrices defined by  $C_i = \text{diag}(\cos \theta_{i_0}, \dots, \cos \theta_{i_{N-1}})$ and $S_i = \text{diag}(\sin\theta_{i_0}, \dots, \sin\theta_{i_{N-1}})$ for some angles $\theta_j$ with $N=4$ for $i\in\{0, 1, 3, 4\}$ and $N=8$ for $i=2$.
Crucially, the $U_i$ are arbitrary four-dimensional unitaries and the block-diagonal matrices of which they are blocks are directly realized by a sequence of $C^m[U]$ gates as derived in the main text.
For example, the first block-diagonal matrix in Eq.~\eqref{eq:cosinesine_decomp} is implemented by $\prod_{i=0}^3 C^i[U_i]$.

Next, we implement the cosine and sine matrices with $R_y^{ij}$ rotations applied to the control qubit depending on the state of the target.
This requires a bidirectional ECR gate such that the role of control and target can be reversed. 
For example, the first block of the second matrix in Eq.~\eqref{eq:cosinesine_decomp} is 
\begin{equation}
\label{eq:example_cosinesine}
    \dboxed{
        \arraycolsep=2pt \begin{array}{|c|c|} 
        \hline
        C_0 & -S_0  \\ \hline 
        S_0 & C_0  \\ \hline
         \end{array}
        }{2}
    = \sum_{n=0}^3 R_y^{01}(\theta_{0_{n}}) \otimes \ketbra{n}{n}_t
\end{equation}
while the center matrix is
\begin{align}
\vspace{-15pt}\arraycolsep=0pt \begin{array}{cc} 
    \dbox{\begin{minipage}[t][15pt][t]{17pt}\vspace{5pt}{$\,\,C_2$}\end{minipage}}  & \dbox{\begin{minipage}[t][15pt][t]{17pt}\vspace{5pt}{$-S_2$}\end{minipage}}     
    \vspace{-5pt} \\ 
    \dbox{\begin{minipage}[t][15pt][t]{17pt}\vspace{5pt}{$\,\,S_2$}\end{minipage}}  & \dbox{\begin{minipage}[t][15pt][t]{17pt}\vspace{5pt}{$\,\,C_2$}\end{minipage}}  
    \end{array} 
    = \sum_{n=0}^3 R_y^{02}(\theta_{2_{n}}) R_y^{13}(\theta_{2_{4+n}}) \otimes \ketbra{n}{n}_t .
\end{align}
Up to a phase on the ECR drive pulses, these controlled $R_y$ rotations are equivalent to the $C^m[R_x^{0j}]$ rotations derived in Sec.~\ref{sec:compiler_comparison_qubits}, and can be accomplished through the decomposition shown in Fig.~\ref{fig:decomposition_circuits}(c).
Since our ECR only operates in the $\ket{0}/\ket{1}$-subspaces of the control and the target, this necessitates single-qudit permutation gates on both the control and target. 
These additional permutation gates needed to shift the control gates are accounted for in the gate counts presented in Sec.~\ref{sec:compiler_comparison_qubits}.

The CSD of an arbitrary ${\rm SU}(16)$ gate shown here serves as the building block for an arbitrary quantum circuit that is executed on a ququart-based hardware.
It requires a total of 340 ECR pulses and, as discussed in Sec.~\ref{sec:compiler_comparison_qubits}, is equivalent to 170 CNOT gates in a qubit-based architecture.

\chapter{Dual-unitary kicked Ising circuits}
\label{app:dual_unitary_circuits}

%\section{Details on decompositions of singly-controlled unitaries}
%\label{app:transpiler}

\begin{figure}
    \centering
    \includegraphics[width=0.7\columnwidth]{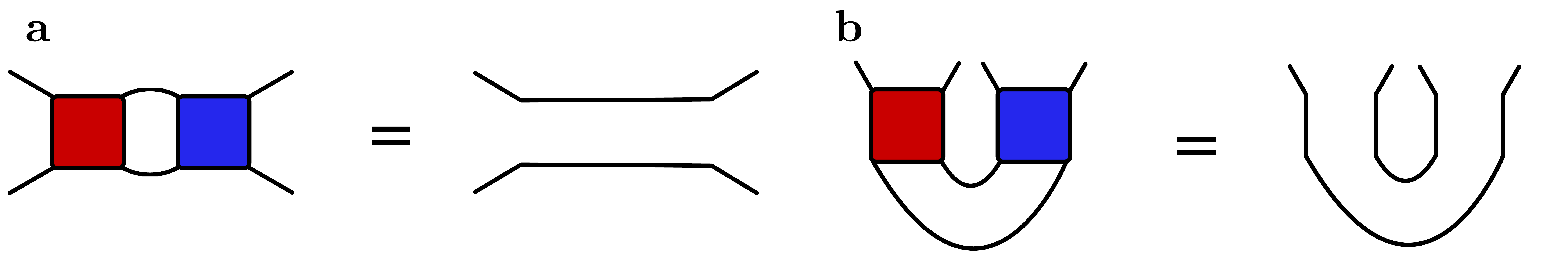}    
    \caption[]{\small Contraction rules for dual unitary gates. Let the red tensor be a two-qubit dual unitary operator $\hat{U} = \sum_{i,j,k,l=0}^1 U_{ij}^{kl} \ket{k}\bra{i}\otimes\ket{l}\bra{j}$ and the blue tensor its Hermitian conjugate $\hat{U}^\dagger$. \textbf{a)} Diagrammatic representation of the unitary relation $\hat{U} \hat{U}^{\dagger} = \hat{\mathbbm{1}}$. \textbf{b)} Diagrammatic representation of the dual unitary relation $\hat{U}_D \hat{U}_D^{\dagger} = \hat{\mathbbm{1}}$ where $\hat{U}_D = \sum_{ijkl} U_{ij}^{kl} \ket{j}\bra{i}\otimes\ket{l}\bra{k}$.
    See Ref.~\cite{ber-19a} for an introduction to the diagrammatic representation. 
    }
    \label{fig:unitary}
\end{figure}

In this Appendix, we show that the expectation value $\langle \Psi(0)|\hat{X}_n(t)|\Psi(0)\rangle$ where $\ket{\Psi(0)} = \ket{+_0} \otimes \ket{\psi_{\rm Bell}}^{\otimes (N-1)/2 }$ (with $N$ being odd) corresponds to the infinite-temperature correlator $C_n(t) = \Tr\bigl(\hat{X}_0\hat{X}_n(t) \bigr)/2^N$ at the dual unitary point through a diagrammatic representation. 
We start with the case where $n=t$, from the expression 
\begin{equation}
\centering
    2^{\frac{N+1}{2}}\langle \Psi(0)|\hat{X}_n(t)|\Psi(0)\rangle = \adjincludegraphics[valign=c, width=0.2\columnwidth]{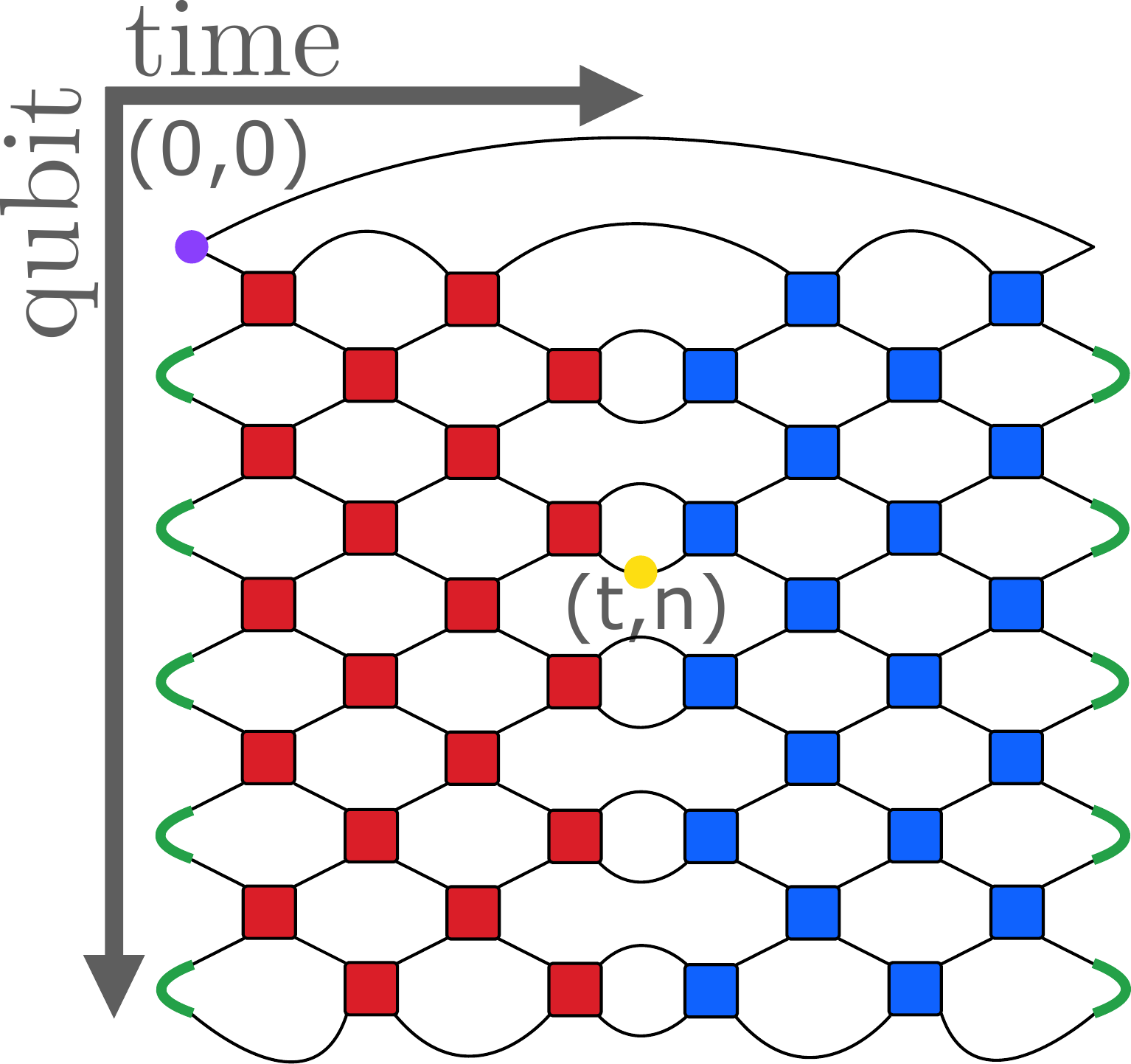} = \quad \adjincludegraphics[valign=c, width=0.2\columnwidth]{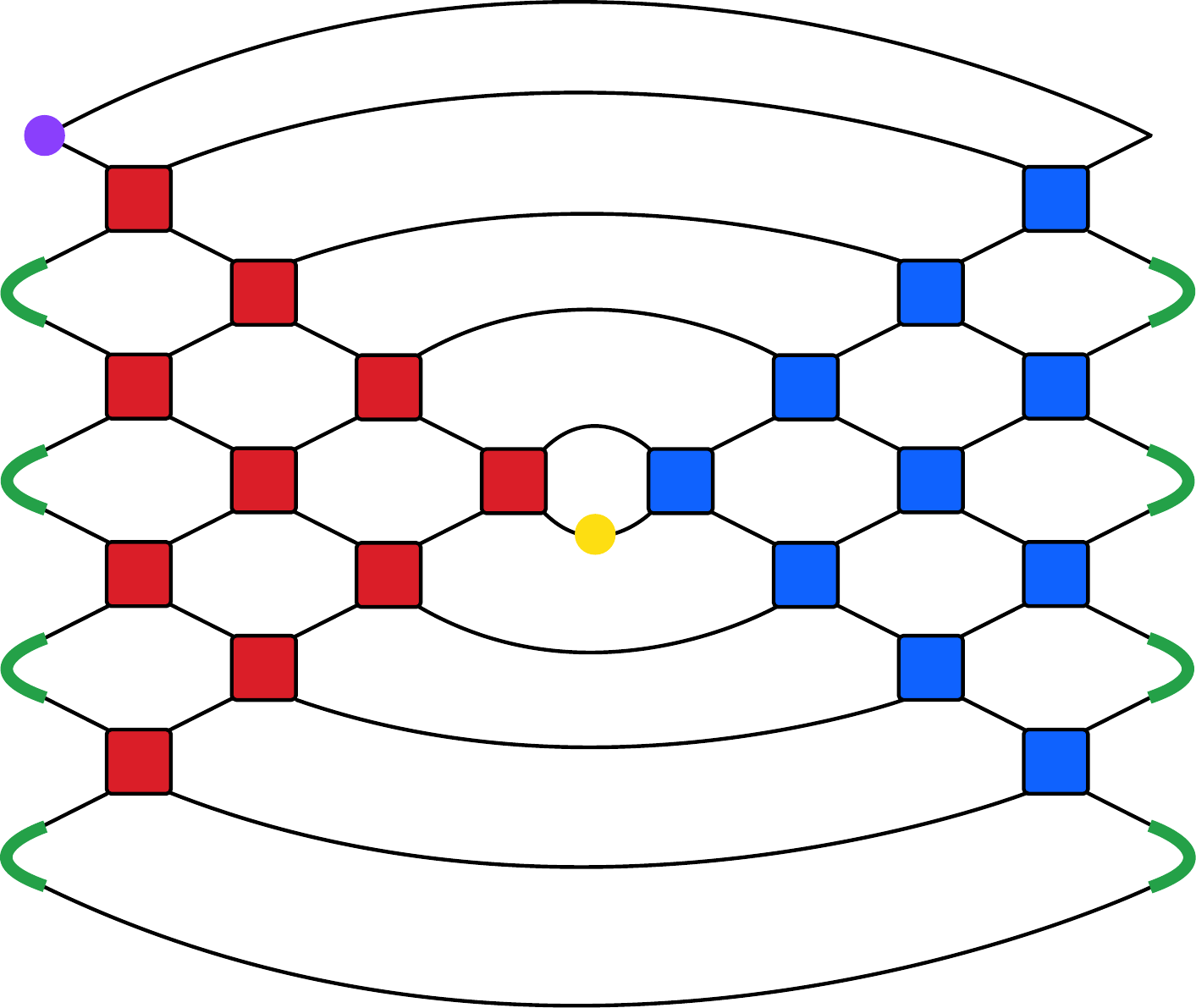}
\label{eq:Bell_expect}
\end{equation}
where the yellow tensor at coordinate $(t, n)$ denotes the $\hat{X}$ observable, and the purple tensor at coordinate $(0, 0)$ denotes the state $\ket{+}\bra{+}$. 
The Bell pairs in $\ket{\Psi(0)}$ are shown capping off the left and right sides of the diagram in green. 
We employ the unitary and dual unitary relations shown in Fig.~\ref{fig:unitary} to simplify this expression.
The step shown in Eq.~\eqref{eq:Bell_expect} immediately follows from exhausting all of the unitary contractions. 
This simplification can be seen as a statement of causality, such that gates outside of the observable light cone do not affect the expectation value.

\begin{figure}
    \centering
    \includegraphics[width=\columnwidth]{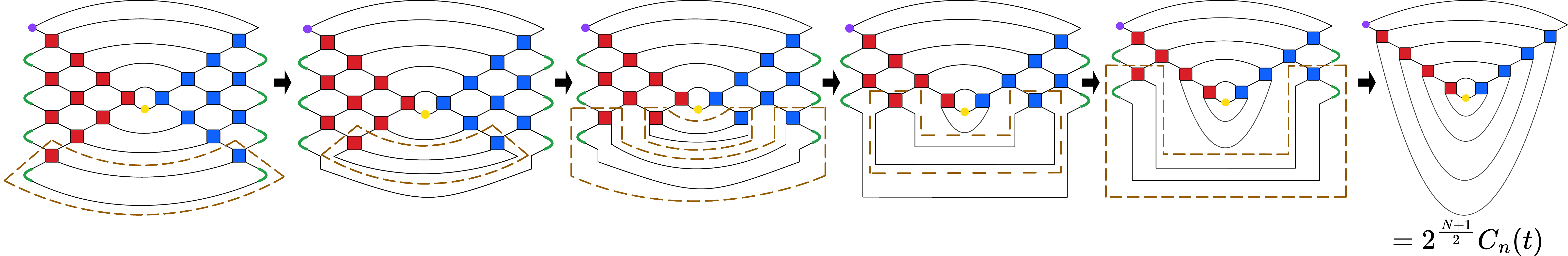}
    \caption[]{\small Simplifying the expectation value through dual unitary contractions. We start from the final expression of Eq.~\eqref{eq:Bell_expect}. 
    Dual-unitary contractions as indicated by dashed lines simplify the light cone structure on the left to the expression on the right which only includes gates on the light cone boundary.}
    \label{fig:Bell_proof_2}
\end{figure}

To further simplify the expression in Eq.~\eqref{eq:Bell_expect}, we employ dual unitary contractions as shown in Fig.~\ref{fig:Bell_proof_2}.
We note that the final expression we arrive at is identical to the one given in Ref.~\cite{ber-19a}, where identity (i.e., infinite temperature) initial states are used in place of our Bell pairs{, provided $N \geq 2t + 1$ (otherwise the equivalence does not hold)}. 
Hence our expectation value is equivalent to the infinite-temperature autocorrelator $C_n(t)$.
Using similar contractions, it is easy to show that the results also match between the two different initial states in the case when $\hat{X_n}$ is not placed on the boundary of the lightcone of the $0$-th qubit -- both evaluate to zero.

With the end result of Fig.~\ref{fig:Bell_proof_2}, we can evaluate the tensor diagram by defining the following map on the local operator space: 
\begin{equation} 
\mathcal{M}_U[\hat{a}] \coloneqq \quad \frac{1}{2}\adjincludegraphics[valign=c, width=0.2\linewidth]{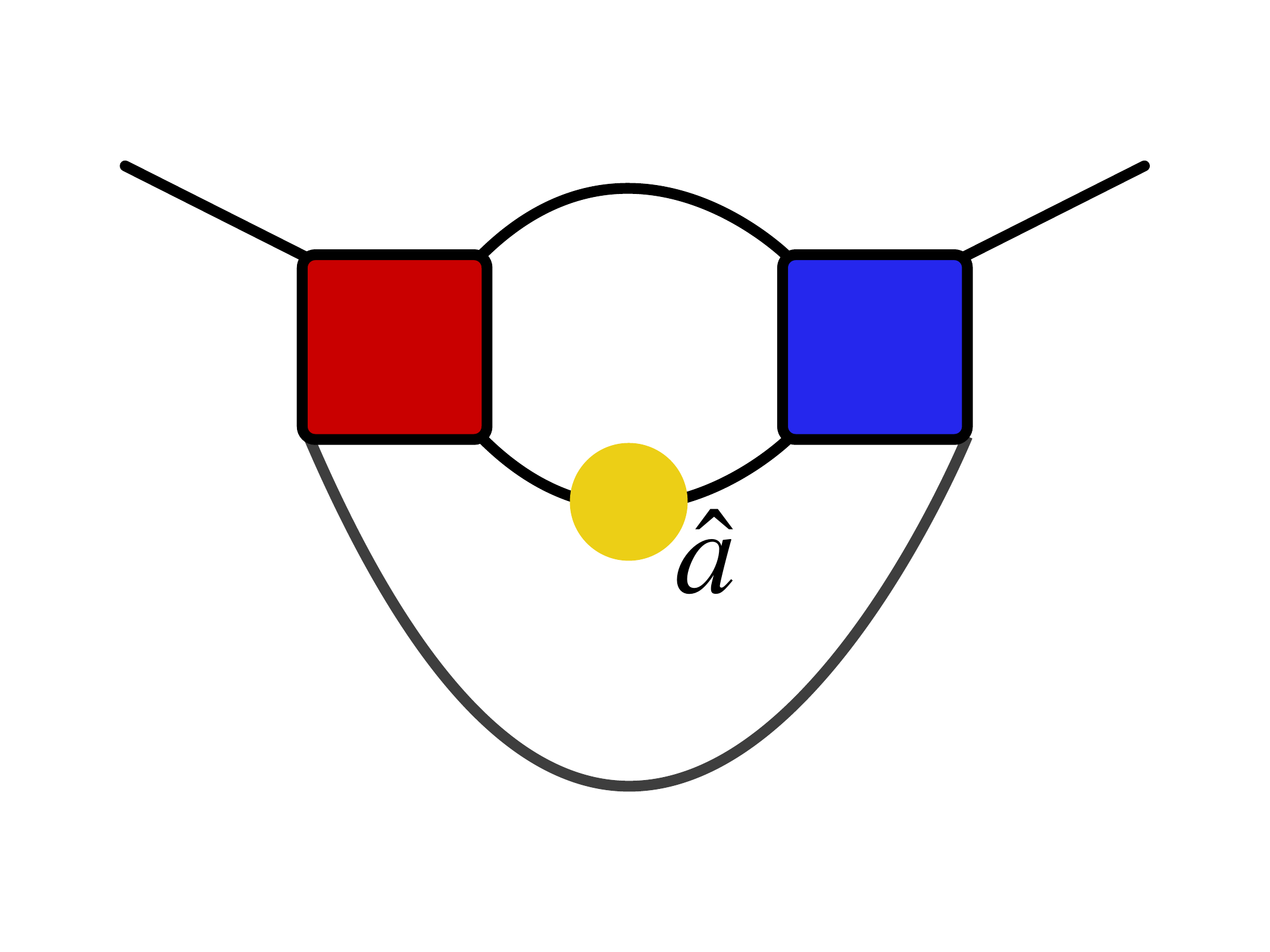} \quad = \frac{1}{2}\Tr_2\left(\hat{U}^\dagger(\hat{\mathbbm{1}}\otimes \hat{a}) \hat{U}\right). 
\end{equation}
Keeping in mind that $|+\rangle\langle +| = (\hat{X} + \hat{\mathbbm{1}})/2$ and $\Tr(M^k_U[\hat{X}]) = 0$ for any $k=1, 2, ...$, we may interchange $|+\rangle\langle +|$ and $\hat{X}$ in the role of $\hat{a}$ above provided we keep track of the factor of $2$.
Hence the correlator can be written as 
\begin{equation} C_n(t) = \Tr\left(|+\rangle\langle +|\mathcal{M}_U^n[\hat{X}]\right).
\end{equation}
Given the two-qubit unitary $U = e^{-ih\hat{Z}\otimes\hat{\mathbbm{1}}} e^{-iJ\hat{Z}\otimes\hat{Z}} e^{-ib(\hat{X}\otimes\hat{\mathbbm{1}}+ \hat{\mathbbm{1}}\otimes\hat{X}) } e^{-iJ\hat{Z}\otimes\hat{Z}} e^{-ih\hat{Z}\otimes\hat{\mathbbm{1}}}$ (see Eq.~\eqref{eq:two_qubit_gate}), we construct the matrix form of $\mathcal{M}_U$ in the Pauli basis $\{\hat{\mathbbm{1}}, \hat{X}, \hat{Y}, \hat{Z}\}$: \begin{equation}
\mathcal{M}_U = \begin{bmatrix}
        1 & 0 & 0 & 0\\
        0 & \cos(2h) & 0 & 0\\
        0 & \sin(2h) & 0 & 0\\
        0 & 0 & 0 & 0
    \end{bmatrix}.\end{equation}
We finally arrive at the following result along the boundary of the lightcone (when $n=t$): \begin{equation} C_n(t) = \cos^t(2h).
\end{equation}

\backmatter
\cleardoublepage
\phantomsection
% \bibliographystyle{apalike}
% \bibliographystyle{unsrtnat}
% \bibliographystyle{abbrvnat}
% \bibliography{tail/bibliography}
% \bibliographystyle{unsrt}
\setcounter{biburlnumpenalty}{9000}
% \emergencystretch=1em

\printbibliography

@article{fischer2024dynamical,
  title={Dynamical simulations of many-body quantum chaos on a quantum computer},
  author={Fischer, Laurin E and Leahy, Matea and Eddins, Andrew and Keenan, Nathan and Ferracin, Davide and Rossi, Matteo AC and Kim, Youngseok and He, Andre and Pietracaprina, Francesca and Sokolov, Boris and Dooley, Shane and Zimborás, Zoltán and Tacchino, Francesco and Maniscalco, Sabrina and Goold, John and García-Pérez, Guillermo and Tavernelli, Ivano and Kandala, Abhinav and Filippov, Sergey N.},
  journal={Nature Physics},
  pages={},
  year={2026},
  publisher={Nature Publishing Group UK London},
  doi={https://doi.org/10.1038/s41567-025-03144-9}
}

@article{fischer2025large,
  title={Large-scale implementation of quantum subspace expansion with classical shadows},
  author={Fischer, Laurin E and Bultrini, Daniel and Tavernelli, Ivano and Tacchino, Francesco},
  journal={arXiv preprint arXiv:2510.25640},
  year={2025},
  doi={https://arxiv.org/abs/2510.25640}
}

@article{MPS_system_first_demo,
  title = {Crossing a topological phase transition with a quantum computer},
  author = {Smith, Adam and Jobst, Bernhard and Green, Andrew G. and Pollmann, Frank},
  journal = {Phys. Rev. Res.},
  volume = {4},
  issue = {2},
  pages = {L022020},
  numpages = {8},
  year = {2022},
  month = {Apr},
  publisher = {American Physical Society},
  doi = {10.1103/PhysRevResearch.4.L022020},
  url = {https://link.aps.org/doi/10.1103/PhysRevResearch.4.L022020}
}

@article{ferracin2024efficiently,
  title={Efficiently improving the performance of noisy quantum computers},
  author={Ferracin, Samuele and Hashim, Akel and Ville, Jean-Loup and Naik, Ravi and Carignan-Dugas, Arnaud and Qassim, Hammam and Morvan, Alexis and Santiago, David I and Siddiqi, Irfan and Wallman, Joel J},
  journal={Quantum},
  volume={8},
  pages={1410},
  year={2024},
  publisher={Verein zur F{\"o}rderung des Open Access Publizierens in den Quantenwissenschaften},
  doi={https://doi.org/10.22331/q-2024-07-15-1410}
}

@article{blume2013robust,
  title={Robust, self-consistent, closed-form tomography of quantum logic gates on a trapped ion qubit},
  author={Blume-Kohout, Robin and Gamble, John King and Nielsen, Erik and Mizrahi, Jonathan and Sterk, Jonathan D and Maunz, Peter},
  journal={arXiv preprint arXiv:1310.4492},
  year={2013}, 
  url={https://arxiv.org/abs/1310.4492}
}

@article{wagner2023learning,
  title={Learning logical pauli noise in quantum error correction},
  author={Wagner, Thomas and Kampermann, Hermann and Bru{\ss}, Dagmar and Kliesch, Martin},
  journal={Physical review letters},
  volume={130},
  number={20},
  pages={200601},
  year={2023},
  publisher={APS},
  doi={https://doi.org/10.1103/PhysRevLett.130.200601}
}

@article{cvxpy,
  author       = {Steven Diamond and Stephen Boyd},
  title        = {{CVXPY}: A {P}ython-Embedded Modeling Language for Convex Optimization},
  journal      = {Journal of Machine Learning Research},
  note         = {To appear},
  url          = {https://stanford.edu/~boyd/papers/pdf/cvxpy_paper.pdf},
  year         = {2016},
}

@article{gavrielov2024linear,
  title={Linear Circuit Synthesis using Weighted Steiner Trees},
  author={Gavrielov, Nir and Garion, Shelly and Ivrii, Alexander},
  journal={Quantum Information and Computation},
  year={2024},
  url={https://arxiv.org/abs/2408.04060}
}

@article{yoshioka2024diagonalization,
  title={Diagonalization of large many-body Hamiltonians on a quantum processor},
  author={Yoshioka, Nobuyuki and Amico, Mirko and Kirby, William and Jurcevic, Petar and Dutt, Arkopal and Fuller, Bryce and Garion, Shelly and Haas, Holger and Hamamura, Ikko and Ivrii, Alexander and others},
  journal={arXiv preprint arXiv:2407.14431},
  year={2024},
  url={https://arxiv.org/abs/2407.14431}
}

@article{MPS_preparation_theory,
  title = {Preparation of Matrix Product States with Log-Depth Quantum Circuits},
  author = {Malz, Daniel and Styliaris, Georgios and Wei, Zhi-Yuan and Cirac, J. Ignacio},
  journal = {Phys. Rev. Lett.},
  volume = {132},
  issue = {4},
  pages = {040404},
  numpages = {9},
  year = {2024},
  month = {Jan},
  publisher = {American Physical Society},
  doi = {10.1103/PhysRevLett.132.040404},
  url = {https://link.aps.org/doi/10.1103/PhysRevLett.132.040404}
}

@article{decoherence_rescaling,
  title = {Mitigating Depolarizing Noise on Quantum Computers with Noise-Estimation Circuits},
  author = {Urbanek, Miroslav and Nachman, Benjamin and Pascuzzi, Vincent R. and He, Andre and Bauer, Christian W. and de Jong, Wibe A.},
  journal = {Phys. Rev. Lett.},
  volume = {127},
  issue = {27},
  pages = {270502},
  numpages = {6},
  year = {2021},
  month = {Dec},
  publisher = {American Physical Society},
  doi = {10.1103/PhysRevLett.127.270502},
  url = {https://link.aps.org/doi/10.1103/PhysRevLett.127.270502}
}

@article{clifford_data_regression,
  title={Error mitigation with Clifford quantum-circuit data},
  author={Czarnik, Piotr and Arrasmith, Andrew and Coles, Patrick J and Cincio, Lukasz},
  journal={Quantum},
  volume={5},
  pages={592},
  year={2021},
  publisher={Verein zur F{\"o}rderung des Open Access Publizierens in den Quantenwissenschaften},
  doi={https://doi.org/10.22331/q-2021-11-26-592}
}

@INPROCEEDINGS{ZNE_gate_folding,
  author={Majumdar, Ritajit and Rivero, Pedro and Metz, Friedrike and Hasan, Areeq and Wang, Derek S.},
  booktitle={2023 IEEE International Conference on Quantum Computing and Engineering (QCE)},
  title={Best Practices for Quantum Error Mitigation with Digital Zero-Noise Extrapolation},
  year={2023},
  volume={01},
  number={},
  pages={881-887},
  doi={10.1109/QCE57702.2023.00102}}

@article{ZNE_pulse_Stretching,
  title = {Benchmarking noise extrapolation with the OpenPulse control framework},
  author = {Garmon, J. W. O. and Pooser, R. C. and Dumitrescu, E. F.},
  journal = {Phys. Rev. A},
  volume = {101},
  issue = {4},
  pages = {042308},
  numpages = {5},
  year = {2020},
  month = {Apr},
  publisher = {American Physical Society},
  doi = {10.1103/PhysRevA.101.042308},
  url = {https://link.aps.org/doi/10.1103/PhysRevA.101.042308}
}

@article{tensor_networks_review,
  title={Tensor network algorithms: A route map},
  author={Ba{\~n}uls, Mari Carmen},
  journal={Annual Review of Condensed Matter Physics},
  volume={14},
  number={1},
  pages={173--191},
  year={2023},
  publisher={Annual Reviews},
  doi = {https://doi.org/10.1146/annurev-conmatphys-040721-022705}
}

@article{shor1995scheme,
  title={Scheme for reducing decoherence in quantum computer memory},
  author={Shor, Peter W},
  journal={Physical review A},
  volume={52},
  number={4},
  pages={R2493},
  year={1995},
  publisher={APS},
  url={https://doi.org/10.1103/PhysRevA.52.R2493}
}

@article{beguvsic2025simulating,
  title={Simulating quantum circuit expectation values by Clifford perturbation theory},
  author={Begu{\v{s}}i{\'c}, Tomislav and Hejazi, Kasra and Chan, Garnet Kin},
  journal={The Journal of Chemical Physics},
  volume={162},
  number={15},
  year={2025},
  publisher={AIP Publishing},
  doi={https://doi.org/10.1063/5.0269149}
}

@article{magic2022,
  title = {Many-Body Quantum Magic},
  author = {Liu, Zi-Wen and Winter, Andreas},
  journal = {PRX Quantum},
  volume = {3},
  issue = {2},
  pages = {020333},
  numpages = {18},
  year = {2022},
  month = {May},
  publisher = {American Physical Society},
  doi = {10.1103/PRXQuantum.3.020333},
  url = {https://link.aps.org/doi/10.1103/PRXQuantum.3.020333}
}

@article{rudolph2025pauli,
  title={Pauli propagation: A computational framework for simulating quantum systems},
  author={Rudolph, Manuel S and Jones, Tyson and Teng, Yanting and Angrisani, Armando and Holmes, Zo{\"e}},
  journal={arXiv preprint arXiv:2505.21606},
  year={2025},
  url={https://arxiv.org/abs/2505.21606}
}

@article{cicero2024simulation,
  title={Simulation of Quantum Computers: Review and Acceleration Opportunities},
  author={Cicero, Alessio and Maleki, Mohammad Ali and Azhar, Muhammad Waqar and Kockum, Anton Frisk and Trancoso, Pedro},
  journal={arXiv preprint arXiv:2410.12660},
  year={2024},
    url={https://arxiv.org/abs/2410.12660}
}

@article{bravyi2022future,
	title = {The future of quantum computing with superconducting qubits},
	volume = {132},
	issn = {0021-8979},
	url = {https://doi.org/10.1063/5.0082975},
	doi = {10.1063/5.0082975},
	number = {16},
	journal = {Journal of Applied Physics},
	author = {Bravyi, Sergey and Dial, Oliver and Gambetta, Jay M. and Gil, Darío and Nazario, Zaira},
	year = {2022},
	pages = {160902},
}

@article{he2020zero,
  title={Zero-noise extrapolation for quantum-gate error mitigation with identity insertions},
  author={He, Andre and Nachman, Benjamin and de Jong, Wibe A and Bauer, Christian W},
  journal={Physical Review A},
  volume={102},
  number={1},
  pages={012426},
  year={2020},
  publisher={APS},
  url={https://doi.org/10.1103/PhysRevA.102.012426}
}

@article{bravyi2021mitigating,
  title={Mitigating measurement errors in multiqubit experiments},
  author={Bravyi, Sergey and Sheldon, Sarah and Kandala, Abhinav and Mckay, David C and Gambetta, Jay M},
  journal={Physical Review A},
  volume={103},
  number={4},
  pages={042605},
  year={2021},
  publisher={APS},
  url={https://doi.org/10.1103/PhysRevA.103.042605}
}

@article{nation2021scalable,
  title={Scalable mitigation of measurement errors on quantum computers},
  author={Nation, Paul D and Kang, Hwajung and Sundaresan, Neereja and Gambetta, Jay M},
  journal={PRX Quantum},
  volume={2},
  number={4},
  pages={040326},
  year={2021},
  publisher={APS},
  url={https://doi.org/10.1103/PRXQuantum.2.040326}
}

@article{merkel2013self,
  title = {Self-consistent quantum process tomography},
  author = {Merkel, Seth T. and Gambetta, Jay M. and Smolin, John A. and Poletto, Stefano and C\'orcoles, Antonio D. and Johnson, Blake R. and Ryan, Colm A. and Steffen, Matthias},
  journal = {Phys. Rev. A},
  volume = {87},
  issue = {6},
  pages = {062119},
  numpages = {9},
  year = {2013},
  publisher = {American Physical Society},
  doi = {10.1103/PhysRevA.87.062119},
  url = {https://link.aps.org/doi/10.1103/PhysRevA.87.062119}
}

@article{malekakhlagh2025efficient,
  title={Efficient Lindblad synthesis for noise model construction},
  author={Malekakhlagh, Moein and Seif, Alireza and Puzzuoli, Daniel and Govia, Luke CG and Berg, Ewout van den},
  journal={arXiv preprint arXiv:2502.03462},
  year={2025},
  url={https://arxiv.org/abs/2502.03462}
}

@article{geller2013efficient,
  title = {Efficient error models for fault-tolerant architectures and the Pauli twirling approximation},
  author = {Geller, Michael R. and Zhou, Zhongyuan},
  journal = {Phys. Rev. A},
  volume = {88},
  issue = {1},
  pages = {012314},
  numpages = {7},
  year = {2013},
  publisher = {American Physical Society},
  doi = {10.1103/PhysRevA.88.012314},
  url = {https://link.aps.org/doi/10.1103/PhysRevA.88.012314}
}

@article{zhang2025generalized,
  title={Generalized Cycle Benchmarking Algorithm for Characterizing Midcircuit Measurements},
  author={Zhang, Zhihan and Chen, Senrui and Liu, Yunchao and Jiang, Liang},
  journal={PRX Quantum},
  volume={6},
  number={1},
  pages={010310},
  year={2025},
  publisher={APS},
  doi={https://doi.org/10.1103/PRXQuantum.6.010310}
}

@article{hines2025pauli,
  title={Pauli noise learning for mid-circuit measurements},
  author={Hines, Jordan and Proctor, Timothy},
  journal={Physical Review Letters},
  volume={134},
  number={2},
  pages={020602},
  year={2025},
  publisher={APS},
  doi={https://doi.org/10.1103/PhysRevLett.134.020602}
}

@article{tantivasadakarn2023hierarchy,
  title={Hierarchy of topological order from finite-depth unitaries, measurement, and feedforward},
  author={Tantivasadakarn, Nathanan and Vishwanath, Ashvin and Verresen, Ruben},
  journal={PRX Quantum},
  volume={4},
  number={2},
  pages={020339},
  year={2023},
  publisher={APS},
 doi={ https://doi.org/10.1103/PRXQuantum.4.020339}
}

@article{buhrman2024state,
  title={State preparation by shallow circuits using feed forward},
  author={Buhrman, Harry and Folkertsma, Marten and Loff, Bruno and Neumann, Niels MP},
  journal={Quantum},
  volume={8},
  pages={1552},
  year={2024},
  publisher={Verein zur F{\"o}rderung des Open Access Publizierens in den Quantenwissenschaften},
  doi={https://doi.org/10.22331/q-2024-12-09-1552}
}

@article{deshpande2024dynamic,
  title={Dynamic parameterized quantum circuits: expressive and barren-plateau free},
  author={Deshpande, Abhinav and Hinsche, Marcel and Najafi, Sona and Sharma, Kunal and Sweke, Ryan and Zoufal, Christa},
  journal={arXiv preprint arXiv:2411.05760},
  year={2024},
  url={https://arxiv.org/abs/2411.05760}
}

@article{PhysRevLett.128.110504,
  title = {Calibrated Decoders for Experimental Quantum Error Correction},
  author = {Chen, Edward H. and Yoder, Theodore J. and Kim, Youngseok and Sundaresan, Neereja and Srinivasan, Srikanth and Li, Muyuan and C\'orcoles, Antonio D. and Cross, Andrew W. and Takita, Maika},
  journal = {Phys. Rev. Lett.},
  volume = {128},
  issue = {11},
  pages = {110504},
  numpages = {7},
  year = {2022},
  month = {Mar},
  publisher = {American Physical Society},
  doi = {10.1103/PhysRevLett.128.110504},
  url = {https://link.aps.org/doi/10.1103/PhysRevLett.128.110504}
}

@article{bausch2024learning,
  title = {Learning High-Accuracy Error Decoding for Quantum Processors},
  author = {Bausch, Johannes and Senior, Andrew W. and Heras, Francisco J. H. and Edlich, Thomas and Davies, Alex and Newman, Michael and Jones, Cody and Satzinger, Kevin and Niu, Murphy Yuezhen and Blackwell, Sam and Holland, George and Kafri, Dvir and Atalaya, Juan and Gidney, Craig and Hassabis, Demis and Boixo, Sergio and Neven, Hartmut and Kohli, Pushmeet},
  year = {2024},
  month = nov,
  journal = {Nature},
  volume = {635},
  number = {8040},
  pages = {834--840},
  issn = {1476-4687},
  doi = {10.1038/s41586-024-08148-8},
  url = {https://doi.org/10.1038/s41586-024-08148-8}
}

@article{chen2025disambiguating,
	title = {Disambiguating {Pauli} noise in quantum computers},
	url = {http://arxiv.org/abs/2505.22629},
	author = {Chen, Edward H. and Chen, Senrui and Fischer, Laurin E. and Eddins, Andrew and Govia, Luke C. G. and Mitchell, Brad and He, Andre and Kim, Youngseok and Jiang, Liang and Seif, Alireza},
  journal	= {arXiv preprint arXiv:2505.22629},
  year		= {2025},
  doi={https://doi.org/10.48550/arXiv.2505.22629}
}

@article{haupt2025statepreparation,
	title = {State-preparation and measurement error mitigation with non-computational states},
	url = {http://arxiv.org/abs/2506.09145},
	author = {Haupt, Conrad J. and Vazquez, Almudena Carrera and Fischer, Laurin E. and Woerner, Stefan and Egger, Daniel J.},
  journal	= {arXiv preprint arXiv:2506.09145},
  year		= {2025},
  doi={https://doi.org/10.48550/arXiv.2506.09145}
}

@Article{	  wang2020qudits,
  title		= {Qudits and {high}-{dimensional} {quantum} {computing}},
  volume	= {8},
  issn		= {2296-424X},
  url		= {https://www.frontiersin.org/articles/10.3389/fphy.2020.589504/full},
  doi		= {10.3389/fphy.2020.589504},
  urldate	= {2022-07-20},
  journal	= {Frontiers in Physics},
  author	= {Wang, Yuchen and Hu, Zixuan and Sanders, Barry C. and
		  Kais, Sabre},
  month		= nov,
  year		= {2020},
  pages		= {589504}
}

@InCollection{	  brylinski2002universal,
  title		= {Universal quantum gates},
  author	= {Brylinski, Jean-Luc and Brylinski, Ranee},
  booktitle	= {Mathematics of quantum computation},
  pages		= {117--134},
  year		= {2002},
  publisher	= {Chapman and Hall/CRC},
  url		= {https://doi.org/10.1201/9781420035377 }
}

@Article{	  brennen2005criteria,
  title		= {Criteria for exact qudit universality},
  volume	= {71},
  issn		= {1050-2947, 1094-1622},
  url		= {https://link.aps.org/doi/10.1103/PhysRevA.71.052318},
  doi		= {10.1103/PhysRevA.71.052318},
  number	= {5},
  urldate	= {2022-07-21},
  journal	= {Physical Review A},
  author	= {Brennen, Gavin and O’Leary, Dianne and Bullock,
		  Stephen},
  month		= may,
  year		= {2005},
  pages		= {052318}
}

@article{pascuzzi2024quantum,
  title={Quantum-centric Supercomputing for Physics Research},
  author={Pascuzzi, Vincent R and C{\'o}rcoles, Antonio},
  journal={arXiv preprint arXiv:2408.11741},
  year={2024},
  url={https://arxiv.org/abs/2503.20870}
}

@article{scholten2024assessing,
  title={Assessing the benefits and risks of quantum computers},
  author={Scholten, Travis L and Williams, Carl J and Moody, Dustin and Mosca, Michele and Hurley, William and Zeng, William J and Troyer, Matthias and Gambetta, Jay M},
  journal={arXiv preprint arXiv:2401.16317},
  year={2024},
  url={https://arxiv.org/abs/2401.16317}
}

@Article{jie2024quntum-centric,
author ="Liu, Jie and Ma, Huan and Shang, Honghui and Li, Zhenyu and Yang, Jinlong",
title  ="Quantum-centric high performance computing for quantum chemistry",
journal  ="Phys. Chem. Chem. Phys.",
year  ="2024",
volume  ="26",
issue  ="22",
pages  ="15831-15843",
publisher  ="The Royal Society of Chemistry",
doi  ="10.1039/D4CP00436A",
url  ="http://dx.doi.org/10.1039/D4CP00436A",
}

@article{beverland2022assessing,
  title={Assessing requirements to scale to practical quantum advantage},
  author={Beverland, Michael E and Murali, Prakash and Troyer, Matthias and Svore, Krysta M and Hoefler, Torsten and Kliuchnikov, Vadym and Low, Guang Hao and Soeken, Mathias and Sundaram, Aarthi and Vaschillo, Alexander},
  journal={arXiv preprint arXiv:2211.07629},
  year={2022},
  url={https://arxiv.org/abs/2211.07629}
}

@article{dalzell2023quantum,
  title={Quantum algorithms: A survey of applications and end-to-end complexities},
  author={Dalzell, Alexander M and McArdle, Sam and Berta, Mario and Bienias, Przemyslaw and Chen, Chi-Fang and Gily{\'e}n, Andr{\'a}s and Hann, Connor T and Kastoryano, Michael J and Khabiboulline, Emil T and Kubica, Aleksander and others},
  journal={arXiv preprint arXiv:2310.03011},
  year={2023},
  url={https://arxiv.org/abs/2310.03011}
}

@Article{	  tacchino2021proposal,
  title		= {A proposal for using molecular spin Qudits as Quantum
		  Simulators of Light\textendash Matter Interactions},
  author	= {Tacchino, Fracesco and Chiesa, Alessandro and Sessoli,
		  Roberta and Tavernelli, Ivano and Carretta, Stefano},
  year		= {2021},
  journal	= {Journal of Materials Chemistry C},
  volume	= {9},
  number	= {32},
  pages		= {10266--10275},
  issn		= {2050-7526, 2050-7534},
  doi		= {10.1039/D1TC00851J},
  url		= {http://xlink.rsc.org/?DOI=D1TC00851J},
  urldate	= {2022-11-08},
  langid	= {english}
}

@Article{	  ringbauer2021universal,
  title		= {A Universal Qudit Quantum Processor with Trapped Ions},
  author	= {Ringbauer, Martin and Meth, Michael and Postler, Lukas and
		  Stricker, Roman and Blatt, Rainer and Schindler, Philipp
		  and Monz, Thomas},
  year		= {2022},
  month		= sep,
  journal	= {Nature Physics},
  volume	= {18},
  number	= {9},
  pages		= {1053--1057},
  issn		= {1745-2473, 1745-2481},
  doi		= {10.1038/s41567-022-01658-0},
  url		= {https://www.nature.com/articles/s41567-022-01658-0},
  urldate	= {2022-11-09},
  langid	= {english}
}

@Article{	  nikolaeva2022efficient,
  title		= {Efficient realization of quantum algorithms with qudits},
  url		= {http://arxiv.org/abs/2111.04384},
  urldate	= {2022-07-21},
  publisher	= {arXiv},
  author	= {Nikolaeva, Anna S. and Kiktenko, Evgeniy O. and Fedorov,
		  Aleksey K.},
  month		= jun,
  year		= {2022},
  journal	= {arXiv:2111.04384}
}

@Article{	  cerveralierta2022experimental,
  title		= {Experimental {high}-{dimensional}
		  {Greenberger}-{Horne}-{Zeilinger} {entanglement} with
		  {superconducting} {transmon} {qutrits}},
  volume	= {17},
  issn		= {2331-7019},
  url		= {https://link.aps.org/doi/10.1103/PhysRevApplied.17.024062},
  doi		= {10.1103/PhysRevApplied.17.024062},
  number	= {2},
  urldate	= {2022-08-04},
  journal	= {Physical Review Applied},
  author	= {Cervera-Lierta, Alba and Krenn, Mario and Aspuru-Guzik,
		  Alán and Galda, Alexey},
  month		= feb,
  year		= {2022},
  pages		= {024062}
}

@Article{	  galda2021implementing,
  title		= {Implementing a {ternary} {decomposition} of the {Toffoli}
		  {gate} on {fixed}-{frequency transmon} {qutrits}},
  url		= {http://arxiv.org/abs/2109.00558},
  urldate	= {2022-01-28},
  journal	= {arXiv:2109.00558},
  author	= {Galda, Alexey and Cubeddu, Michael and Kanazawa, Naoki and
		  Narang, Prineha and Earnest-Noble, Nathan},
  month		= sep,
  year		= {2021}
}

@Article{	  malekakhlagh2020firstprinciples,
  title		= {First-principles analysis of cross-resonance gate
		  operation},
  volume	= {102},
  issn		= {2469-9926, 2469-9934},
  url		= {https://link.aps.org/doi/10.1103/PhysRevA.102.042605},
  doi		= {10.1103/PhysRevA.102.042605},
  number	= {4},
  urldate	= {2022-02-03},
  journal	= {Physical Review A},
  author	= {Malekakhlagh, Moein and Magesan, Easwar and McKay, David
		  C.},
  month		= oct,
  year		= {2020},
  pages		= {042605}
}

@Article{	  sheldon2016procedure,
  title		= {Procedure for systematically tuning up cross-talk in the
		  cross-resonance gate},
  volume	= {93},
  issn		= {2469-9926, 2469-9934},
  url		= {https://link.aps.org/doi/10.1103/PhysRevA.93.060302},
  doi		= {10.1103/PhysRevA.93.060302},
  number	= {6},
  urldate	= {2022-03-28},
  journal	= {Physical Review A},
  author	= {Sheldon, Sarah and Magesan, Easwar and Chow, Jerry M. and
		  Gambetta, Jay M.},
  month		= jun,
  year		= {2016},
  pages		= {060302}
}

@article{vinay_crosstalk_DD,
  title = {Suppression of Crosstalk in Superconducting Qubits Using Dynamical Decoupling},
  author = {Tripathi, Vinay and Chen, Huo and Khezri, Mostafa and Yip, Ka-Wa and Levenson-Falk, E.M. and Lidar, Daniel A.},
  journal = {Phys. Rev. Appl.},
  volume = {18},
  issue = {2},
  pages = {024068},
  numpages = {24},
  year = {2022},
  month = {Aug},
  publisher = {American Physical Society},
  doi = {10.1103/PhysRevApplied.18.024068},
  url = {https://link.aps.org/doi/10.1103/PhysRevApplied.18.024068}
}

@article{kono2020breaking,
  title={Breaking the trade-off between fast control and long lifetime of a superconducting qubit},
  author={Kono, Shingo and Koshino, Kazuki and Lachance-Quirion, Dany and Van Loo, Arjan F and Tabuchi, Yutaka and Noguchi, Atsushi and Nakamura, Yasunobu},
  journal={Nature communications},
  volume={11},
  number={1},
  pages={3683},
  year={2020},
  publisher={Nature Publishing Group UK London},
  doi={https://doi.org/10.1038/s41467-020-17511-y}
}

@article{qubit_frequency_design,
  title = {Mitigation of frequency collisions in superconducting quantum processors},
  author = {Osman, Amr and Fern\'andez-Pend\'as, Jorge and Warren, Christopher and Kosen, Sandoko and Scigliuzzo, Marco and Frisk Kockum, Anton and Tancredi, Giovanna and Fadavi Roudsari, Anita and Bylander, Jonas},
  journal = {Phys. Rev. Res.},
  volume = {5},
  issue = {4},
  pages = {043001},
  numpages = {17},
  year = {2023},
  month = {Oct},
  publisher = {American Physical Society},
  doi = {10.1103/PhysRevResearch.5.043001},
  url = {https://link.aps.org/doi/10.1103/PhysRevResearch.5.043001}
}

@Article{	  chi2022programmable,
  title		= {A programmable qudit-based quantum processor},
  volume	= {13},
  issn		= {2041-1723},
  url		= {https://www.nature.com/articles/s41467-022-28767-x},
  doi		= {10.1038/s41467-022-28767-x},
  number	= {1},
  urldate	= {2022-07-21},
  journal	= {Nature Communications},
  author	= {Chi, Yulin and Huang, Jieshan and Zhang, Zhanchuan and
		  Mao, Jun and Zhou, Zinan and Chen, Xiaojiong and Zhai,
		  Chonghao and Bao, Jueming and Dai, Tianxiang and Yuan,
		  Huihong and Zhang, Ming and Dai, Daoxin and Tang, Bo and
		  Yang, Yan and Li, Zhihua and Ding, Yunhong and Oxenløwe,
		  Leif K. and Thompson, Mark G. and O’Brien, Jeremy L. and
		  Li, Yan and Gong, Qihuang and Wang, Jianwei},
  month		= dec,
  year		= {2022},
  pages		= {1166}
}

@Article{	  deller2022quantum,
  title		= {Quantum approximate optimization algorithm for qudit
		  systems with long-range interactions},
  url		= {http://arxiv.org/abs/2204.00340},
  urldate	= {2022-08-04},
  publisher	= {arXiv},
  author	= {Deller, Yannick and Schmitt, Sebastian and Lewenstein,
		  Maciej and Lenk, Steve and Federer, Marika and
		  Jendrzejewski, Fred and Hauke, Philipp and Kasper,
		  Valentin},
  month		= apr,
  year		= {2022},
  journal	= {arXiv:2204.00340}
}

@Article{	  gonzalezcuadra2022hardware,
  title		= {Hardware efficient quantum simulation of non-abelian gauge
		  theories with qudits on {Rydberg} platforms},
  url		= {http://arxiv.org/abs/2203.15541},
  urldate	= {2022-07-21},
  publisher	= {arXiv},
  author	= {González-Cuadra, Daniel and Zache, Torsten V. and
		  Carrasco, Jose and Kraus, Barbara and Zoller, Peter},
  month		= mar,
  year		= {2022},
  journal	= {arXiv:2203.15541}
}

@Article{	  macdonell2021analog,
  title		= {Analog quantum simulation of chemical dynamics},
  volume	= {12},
  issn		= {2041-6520, 2041-6539},
  url		= {http://xlink.rsc.org/?DOI=D1SC02142G},
  doi		= {10.1039/D1SC02142G},
  number	= {28},
  urldate	= {2022-08-04},
  journal	= {Chemical Science},
  author	= {MacDonell, Ryan J. and Dickerson, Claire E. and Birch,
		  Clare J. T. and Kumar, Alok and Edmunds, Claire L. and
		  Biercuk, Michael J. and Hempel, Cornelius and Kassal,
		  Ivan},
  year		= {2021},
  pages		= {9794--9805}
}

@Article{	  hu2018beating,
  title		= {Beating the channel capacity limit for superdense coding
		  with entangled ququarts},
  volume	= {4},
  issn		= {2375-2548},
  url		= {https://www.science.org/doi/10.1126/sciadv.aat9304},
  doi		= {10.1126/sciadv.aat9304},
  number	= {7},
  urldate	= {2022-08-04},
  journal	= {Science Advances},
  author	= {Hu, Xiao-Min and Guo, Yu and Liu, Bi-Heng and Huang,
		  Yun-Feng and Li, Chuan-Feng and Guo, Guang-Can},
  month		= jul,
  year		= {2018},
  pages		= {eaat9304}
}

@Article{	  brennen2005efficient,
  title		= {Efficient {circuits} for {exact}-{universal}
		  {computations} with {qudits}},
  url		= {https://dl.acm.org/doi/abs/10.5555/2012086.2012095},
  urldate	= {2022-08-04},
  publisher	= {arXiv},
  author	= {Brennen, Gavin K. and Bullock, Stephen S. and O'Leary,
		  Dianne P.},
  month		= sep,
  year		= {2005},
  journal	= {Quantum Information \& Computation},
  volume	= {6},
  issue		= {4},
  pages		= {436 -- 454}
}

@Article{	  seifert2022timeefficient,
  title		= {Time-{efficient} {qudit} gates through {incremental}
		  {pulse} {re}-seeding},
  url		= {http://arxiv.org/abs/2206.14975},
  urldate	= {2022-08-05},
  publisher	= {arXiv},
  author	= {Seifert, Lennart Maximilian and Chadwick, Jason and
		  Litteken, Andrew and Chong, Frederic T. and Baker, Jonathan
		  M.},
  month		= jun,
  year		= {2022},
  journal	= { arXiv:2206.14975}
}

@Article{	  kasper2022universal,
  title		= {Universal quantum computation and quantum error correction
		  with ultracold atomic mixtures},
  volume	= {7},
  issn		= {2058-9565},
  url		= {https://iopscience.iop.org/article/10.1088/2058-9565/ac2d39},
  doi		= {10.1088/2058-9565/ac2d39},
  number	= {1},
  urldate	= {2022-08-05},
  pages		= {015008},
  journal	= {Quantum Science and Technology},
  author	= {Kasper, Valentin and González-Cuadra, Daniel and Hegde,
		  Apoorva and Xia, Andy and Dauphin, Alexandre and Huber,
		  Felix and Tiemann, Eberhard and Lewenstein, Maciej and
		  Jendrzejewski, Fred and Hauke, Philipp},
  month		= jan,
  year		= {2022}
}

@InProceedings{	  gokhale2019asymptotic,
  address	= {Phoenix Arizona},
  title		= {Asymptotic improvements to quantum circuits via qutrits},
  isbn		= {978-1-4503-6669-4},
  url		= {https://dl.acm.org/doi/10.1145/3307650.3322253},
  doi		= {10.1145/3307650.3322253},
  urldate	= {2022-07-15},
  booktitle	= {Proceedings of the 46th {International} {Symposium} on
		  {Computer} {Architecture}},
  publisher	= {ACM},
  author	= {Gokhale, Pranav and Baker, Jonathan M. and Duckering,
		  Casey and Brown, Natalie C. and Brown, Kenneth R. and
		  Chong, Frederic T.},
  month		= jun,
  year		= {2019},
  pages		= {554--566}
}

@article{gokhale_ON3_Measurement_2020,
  author={Gokhale, Pranav and Angiuli, Olivia and Ding, Yongshan and Gui, Kaiwen and Tomesh, Teague and Suchara, Martin and Martonosi, Margaret and Chong, Frederic T.},
  journal={IEEE Trans. Quantum Eng.},
  title={{$O(N^3)$ Measurement Cost for Variational Quantum Eigensolver on Molecular Hamiltonians}},
  year={2020},
  volume={1},
  number={},
  pages={1-24},
  doi={10.1109/TQE.2020.3035814},
  url = {https://doi.org/10.1109/TQE.2020.3035814}
 }

@article{patel2025quantum,
  title={Quantum Measurement for Quantum Chemistry on a Quantum Computer},
  author={Patel, Smik and Jayakumar, Praveen and Yen, Tzu-Ching and Izmaylov, Artur F},
  journal={arXiv preprint arXiv:2501.14968},
  year={2025},
  doi={https://doi.org/10.48550/arXiv.2501.14968}
}

@Article{	  kiktenko2015multilevel,
  title		= {Multilevel superconducting circuits as two-qubit systems:
		  {operations}, state preparation, and entropic
		  inequalities},
  volume	= {91},
  issn		= {1050-2947, 1094-1622},
  shorttitle	= {Multilevel superconducting circuits as two-qubit systems},
  url		= {https://link.aps.org/doi/10.1103/PhysRevA.91.042312},
  doi		= {10.1103/PhysRevA.91.042312},
  number	= {4},
  urldate	= {2022-07-21},
  journal	= {Physical Review A},
  author	= {Kiktenko, Evgeniy. O. and Fedorov, Aleksey. K. and Man'ko,
		  Olga V. and Man'ko, Vladimir I.},
  month		= apr,
  year		= {2015},
  pages		= {042312}
}

@Article{	  magesan2020effective,
  title		= {Effective {Hamiltonian} models of the cross-resonance
		  gate},
  volume	= {101},
  issn		= {2469-9926, 2469-9934},
  url		= {https://link.aps.org/doi/10.1103/PhysRevA.101.052308},
  doi		= {10.1103/PhysRevA.101.052308},
  number	= {5},
  urldate	= {2022-01-28},
  journal	= {Physical Review A},
  author	= {Magesan, Easwar and Gambetta, Jay M.},
  month		= may,
  year		= {2020},
  pages		= {052308}
}

@Article{	  sundaresan2020reducing,
  title		= {Reducing {unitary} and {spectator} {errors} in {cross}
		  {resonance} with {optimized} {rotary} {echoes}},
  volume	= {1},
  issn		= {2691-3399},
  url		= {https://link.aps.org/doi/10.1103/PRXQuantum.1.020318},
  doi		= {10.1103/PRXQuantum.1.020318},
  number	= {2},
  urldate	= {2022-04-08},
  journal	= {PRX Quantum},
  author	= {Sundaresan, Neereja and Lauer, Isaac and Pritchett, Emily
		  and Magesan, Easwar and Jurcevic, Petar and Gambetta, Jay
		  M.},
  month		= dec,
  year		= {2020},
  pages		= {020318}
}

@Article{	  di2013synthesis,
  title		= {Synthesis of multivalued quantum logic circuits by
		  elementary gates},
  volume	= {87},
  issn		= {1050-2947, 1094-1622},
  url		= {https://link.aps.org/doi/10.1103/PhysRevA.87.012325},
  doi		= {10.1103/PhysRevA.87.012325},
  number	= {1},
  urldate	= {2022-08-31},
  journal	= {Physical Review A},
  author	= {Di, Yao-Min and Wei, Hai-Rui},
  month		= jan,
  year		= {2013},
  pages		= {012325}
}

@Article{	  di2015optimal,
  title		= {Optimal synthesis of multivalued quantum circuits},
  volume	= {92},
  issn		= {1050-2947, 1094-1622},
  url		= {https://link.aps.org/doi/10.1103/PhysRevA.92.062317},
  doi		= {10.1103/PhysRevA.92.062317},
  number	= {6},
  urldate	= {2022-09-02},
  journal	= {Physical Review A},
  author	= {Di, Yao-Min and Wei, Hai-Rui},
  month		= dec,
  year		= {2015},
  pages		= {062317}
}

@Article{	  shende2006synthesis,
  author	= {Shende, V.V. and Bullock, S.S. and Markov, I.L.},
  journal	= {IEEE Transactions on Computer-Aided Design of Integrated
		  Circuits and Systems},
  title		= {Synthesis of quantum-logic circuits},
  year		= {2006},
  volume	= {25},
  number	= {6},
  pages		= {1000-1010},
  doi		= {10.1109/TCAD.2005.855930}
}

@Article{	  peterer2015coherence,
  title		= {Coherence and {decay} of {higher} {energy} {levels} of a
		  {superconducting} {transmon} {qubit}},
  volume	= {114},
  issn		= {0031-9007, 1079-7114},
  url		= {https://link.aps.org/doi/10.1103/PhysRevLett.114.010501},
  doi		= {10.1103/PhysRevLett.114.010501},
  number	= {1},
  urldate	= {2022-01-18},
  journal	= {Physical Review Letters},
  author	= {Peterer, Michael J. and Bader, Samuel J. and Jin, Xiaoyue
		  and Yan, Fei and Kamal, Archana and Gudmundsen, Theodore J.
		  and Leek, Peter J. and Orlando, Terry P. and Oliver,
		  William D. and Gustavsson, Simon},
  month		= jan,
  year		= {2015},
  pages		= {010501}
}

@Article{	  koch2007chargeinsensitive,
  title		= {Charge-insensitive qubit design derived from the {Cooper}
		  pair box},
  volume	= {76},
  issn		= {1050-2947, 1094-1622},
  url		= {https://link.aps.org/doi/10.1103/PhysRevA.76.042319},
  doi		= {10.1103/PhysRevA.76.042319},
  number	= {4},
  urldate	= {2022-01-18},
  journal	= {Physical Review A},
  author	= {Koch, Jens and Yu, Terri M. and Gambetta, Jay and Houck,
		  Andrew A. and Schuster, David I. and Majer, Johannes and
		  Blais, Alexandre and Devoret, Michel H. and Girvin, Steven
		  M. and Schoelkopf, Robert J.},
  month		= oct,
  year		= {2007},
  pages		= {042319}
}

@Article{	  kjaergaard2020superconducting,
  title		= {Superconducting {qubits}: {current} {state} of {play}},
  volume	= {11},
  issn		= {1947-5454, 1947-5462},
  shorttitle	= {Superconducting {Qubits}},
  url		= {https://www.annualreviews.org/doi/10.1146/annurev-conmatphys-031119-050605},
  number	= {1},
  urldate	= {2022-09-19},
  journal	= {Annual Review of Condensed Matter Physics},
  author	= {Kjaergaard, Morten and Schwartz, Mollie E. and
		  Braumüller, Jochen and Krantz, Philip and Wang, Joel I.-J.
		  and Gustavsson, Simon and Oliver, William D.},
  month		= mar,
  year		= {2020},
  pages		= {369--395}
}

@article{manhattan_data_01,
author = {Mooney, Gary J. and White, Gregory A. L. and Hill, Charles D. and Hollenberg, Lloyd C. L.},
title = {Whole-Device Entanglement in a 65-Qubit Superconducting Quantum Computer},
journal = {Advanced Quantum Technologies},
volume = {4},
number = {10},
pages = {2100061},
doi = {https://doi.org/10.1002/qute.202100061},
url = {https://advanced.onlinelibrary.wiley.com/doi/abs/10.1002/qute.202100061},
year = {2021}
}

@article{manhattan_data_02,
  title={High-performance superconducting quantum processors via laser annealing of transmon qubits},
  author={Zhang, Eric J and Srinivasan, Srikanth and Sundaresan, Neereja and Bogorin, Daniela F and Martin, Yves and Hertzberg, Jared B and Timmerwilke, John and Pritchett, Emily J and Yau, Jeng-Bang and Wang, Cindy and others},
  journal={Science Advances},
  volume={8},
  number={19},
  pages={eabi6690},
  year={2022},
  publisher={American Association for the Advancement of Science},
  doi={https://doi.org/10.1126/sciadv.abi6690}
}

@article{IBM_tunable_coupler,
  title = {Tunable Coupling Architecture for Fixed-Frequency Transmon Superconducting Qubits},
  author = {Stehlik, J. and Zajac, D. M. and Underwood, D. L. and Phung, T. and Blair, J. and Carnevale, S. and Klaus, D. and Keefe, G. A. and Carniol, A. and Kumph, M. and Steffen, Matthias and Dial, O. E.},
  journal = {Phys. Rev. Lett.},
  volume = {127},
  issue = {8},
  pages = {080505},
  numpages = {6},
  year = {2021},
  month = {Aug},
  publisher = {American Physical Society},
  doi = {10.1103/PhysRevLett.127.080505},
  url = {https://link.aps.org/doi/10.1103/PhysRevLett.127.080505}
}

@article{dicarlo2009demonstration,
  title={Demonstration of two-qubit algorithms with a superconducting quantum processor},
  author={DiCarlo, Leonardo and Chow, Jerry M and Gambetta, Jay M and Bishop, Lev S and Johnson, Blake R and Schuster, David I and Majer, J and Blais, Alexandre and Frunzio, Luigi and Girvin, SM and others},
  journal={Nature},
  volume={460},
  number={7252},
  pages={240--244},
  year={2009},
  publisher={Nature Publishing Group UK London},
  doi={https://doi.org/10.1038/nature08121}
}

@article{Google_2q_gates,
  title = {Demonstrating a Continuous Set of Two-Qubit Gates for Near-Term Quantum Algorithms},
  author = {Foxen, B. and Neill, C. and Dunsworth, A. and Roushan, P. and Chiaro, B. and Megrant, A. and Kelly, J. and Chen, Zijun and Satzinger, K. and Barends, R. and Arute, F. and Arya, K. and Babbush, R. and Bacon, D. and Bardin, J. C. and Boixo, S. and Buell, D. and Burkett, B. and Chen, Yu and Collins, R. and Farhi, E. and Fowler, A. and Gidney, C. and Giustina, M. and Graff, R. and Harrigan, M. and Huang, T. and Isakov, S. V. and Jeffrey, E. and Jiang, Z. and Kafri, D. and Kechedzhi, K. and Klimov, P. and Korotkov, A. and Kostritsa, F. and Landhuis, D. and Lucero, E. and McClean, J. and McEwen, M. and Mi, X. and Mohseni, M. and Mutus, J. Y. and Naaman, O. and Neeley, M. and Niu, M. and Petukhov, A. and Quintana, C. and Rubin, N. and Sank, D. and Smelyanskiy, V. and Vainsencher, A. and White, T. C. and Yao, Z. and Yeh, P. and Zalcman, A. and Neven, H. and Martinis, J. M.},
  collaboration = {Google AI Quantum},
  journal = {Phys. Rev. Lett.},
  volume = {125},
  issue = {12},
  pages = {120504},
  numpages = {6},
  year = {2020},
  month = {Sep},
  publisher = {American Physical Society},
  doi = {10.1103/PhysRevLett.125.120504},
  url = {https://link.aps.org/doi/10.1103/PhysRevLett.125.120504}
}

@article{PhysRevLett.129.010502,
  title = {Fluxonium: An Alternative Qubit Platform for High-Fidelity Operations},
  author = {Bao, Feng and Deng, Hao and Ding, Dawei and Gao, Ran and Gao, Xun and Huang, Cupjin and Jiang, Xun and Ku, Hsiang-Sheng and Li, Zhisheng and Ma, Xizheng and Ni, Xiaotong and Qin, Jin and Song, Zhijun and Sun, Hantao and Tang, Chengchun and Wang, Tenghui and Wu, Feng and Xia, Tian and Yu, Wenlong and Zhang, Fang and Zhang, Gengyan and Zhang, Xiaohang and Zhou, Jingwei and Zhu, Xing and Shi, Yaoyun and Chen, Jianxin and Zhao, Hui-Hai and Deng, Chunqing},
  journal = {Phys. Rev. Lett.},
  volume = {129},
  issue = {1},
  pages = {010502},
  numpages = {6},
  year = {2022},
  month = {Jun},
  publisher = {American Physical Society},
  doi = {10.1103/PhysRevLett.129.010502},
  url = {https://link.aps.org/doi/10.1103/PhysRevLett.129.010502}
}

@article{martinis2009superconducting,
  title={Superconducting phase qubits},
  author={Martinis, John M},
  journal={Quantum information processing},
  volume={8},
  number={2},
  pages={81--103},
  year={2009},
  publisher={Springer},
  doi={https://doi.org/10.1007/s11128-009-0105-1}
}

@article{chiorescu2003coherent,
  title={Coherent quantum dynamics of a superconducting flux qubit},
  author={Chiorescu, Irinel and Nakamura, Y and Harmans, CJP Ma and Mooij, JE},
  journal={Science},
  volume={299},
  number={5614},
  pages={1869--1871},
  year={2003},
  publisher={American Association for the Advancement of Science},
  doi={https://doi.org/10.1126/science.1081045}
}

@article{krantz_quantum_2019,
	title = {A {Quantum} {Engineer}'s {Guide} to {Superconducting} {Qubits}},
	volume = {6},
	issn = {1931-9401},
	url = {http://arxiv.org/abs/1904.06560},
	doi = {10.1063/1.5089550},
	number = {2},
	urldate = {2022-03-31},
	journal = {Applied Physics Reviews},
	author = {Krantz, Philip and Kjaergaard, Morten and Yan, Fei and Orlando, Terry P. and Gustavsson, Simon and Oliver, William D.},
	month = jun,
	year = {2019},
	note = {arXiv: 1904.06560},
	pages = {021318},
}

@Article{	  elder2020highfidelity,
  title		= {High-{fidelity} {measurement} of {qubits} {encoded} in
		  {multilevel} {superconducting} {circuits}},
  volume	= {10},
  issn		= {2160-3308},
  url		= {https://link.aps.org/doi/10.1103/PhysRevX.10.011001},
  doi		= {10.1103/PhysRevX.10.011001},
  number	= {1},
  urldate	= {2022-01-18},
  journal	= {Physical Review X},
  author	= {Elder, Salvatore S. and Wang, Christopher S. and Reinhold,
		  Philip and Hann, Connor T. and Chou, Kevin S. and Lester,
		  Brian J. and Rosenblum, Serge and Frunzio, Luigi and Jiang,
		  Liang and Schoelkopf, Robert J.},
  month		= jan,
  year		= {2020},
  pages		= {011001}
}

@Article{	  jurcevic2021demonstration,
  title		= {Demonstration of quantum volume 64 on a superconducting
		  quantum computing system},
  volume	= {6},
  issn		= {2058-9565},
  url		= {https://iopscience.iop.org/article/10.1088/2058-9565/abe519},
  doi		= {10.1088/2058-9565/abe519},
  number	= {2},
  urldate	= {2022-01-18},
  journal	= {Quantum Science and Technology},
  author	= {Jurcevic, Petar and Javadi-Abhari, Ali and Bishop, Lev S
		  and Lauer, Isaac and Bogorin, Daniela F and Brink, Markus
		  and others},
  month		= apr,
  year		= {2021},
  pages		= {025020}
}

@Article{	  mckay2017efficient,
  title		= {Efficient {Z} gates for quantum computing},
  volume	= {96},
  issn		= {2469-9926, 2469-9934},
  url		= {https://link.aps.org/doi/10.1103/PhysRevA.96.022330},
  doi		= {10.1103/PhysRevA.96.022330},
  number	= {2},
  urldate	= {2022-01-18},
  journal	= {Physical Review A},
  author	= {McKay, David C. and Wood, Christopher J. and Sheldon,
		  Sarah and Chow, Jerry M. and Gambetta, Jay M.},
  month		= aug,
  year		= {2017},
  pages		= {022330}
}

@Article{	  tripathi2019operation,
  title		= {Operation and intrinsic error budget of a two-qubit
		  cross-resonance gate},
  volume	= {100},
  issn		= {2469-9926, 2469-9934},
  url		= {http://arxiv.org/abs/1902.09054},
  doi		= {10.1103/PhysRevA.100.012301},
  number	= {1},
  urldate	= {2022-01-28},
  journal	= {Physical Review A},
  author	= {Tripathi, Vinay and Khezri, Mostafa and Korotkov,
		  Alexander N.},
  month		= jul,
  year		= {2019},
  pages		= {012301}
}

@Article{	  motzoi2009simple,
  title		= {Simple {pulses} for {elimination} of {leakage} in {weakly}
		  {nonlinear} {qubits}},
  volume	= {103},
  issn		= {0031-9007, 1079-7114},
  url		= {https://link.aps.org/doi/10.1103/PhysRevLett.103.110501},
  doi		= {10.1103/PhysRevLett.103.110501},
  number	= {11},
  urldate	= {2022-01-18},
  journal	= {Physical Review Letters},
  author	= {Motzoi, Felix and Gambetta, Jay M. and Rebentrost, Patrick
		  and Wilhelm, Frank K.},
  month		= sep,
  year		= {2009},
  pages		= {110501}
}

@Book{		  gambetta2013quantum,
  title		= {Quantum {information} {processing} - {lecture} {notes} of
		  the 44th {IFF} {spring} {school} 2013},
  isbn		= {978-3-89336-833-4},
  publisher	= {Forschungszentrum Jülich, Zentralbibliothek},
  author	= {Gambetta, Jay M.},
  editor	= {DiVincenzo, David P.},
  year		= {2013},
  url		= {https://juser.fz-juelich.de/record/153195}
}

@Article{	  johansson2012qutip,
  title		= {{QuTiP}: {an} open-source {Python} framework for the
		  dynamics of open quantum systems},
  volume	= {183},
  issn		= {00104655},
  shorttitle	= {{QuTiP}},
  url		= {https://linkinghub.elsevier.com/retrieve/pii/S0010465512000835},
  doi		= {10.1016/j.cpc.2012.02.021},
  number	= {8},
  urldate	= {2022-01-18},
  journal	= {Computer Physics Communications},
  author	= {Johansson, Robert and Nation, Paul D. and Nori, Franco},
  month		= aug,
  year		= {2012},
  pages		= {1760--1772}
}

@Article{	  nielsen2002simple,
  title		= {A simple formula for the average gate fidelity of a
		  quantum dynamical operation},
  volume	= {303},
  issn		= {03759601},
  url		= {https://linkinghub.elsevier.com/retrieve/pii/S0375960102012720},
  doi		= {10.1016/S0375-9601(02)01272-0},
  number	= {4},
  urldate	= {2022-01-18},
  journal	= {Physics Letters A},
  author	= {Nielsen, Michael A.},
  month		= oct,
  year		= {2002},
  pages		= {249--252}
}

@Article{	  chiesa2020,
  author	= {Alessandro Chiesa and Emilio Macaluso and Francesco
		  Petiziol and Sandro Wimberger and Paolo Santini and Stefano
		  Carretta},
  title		= {Molecular Nanomagnets as Qubits with Embedded
		  Quantum-Error Correction},
  journal	= {Journal of Physical Chemistry Letters},
  year		= {2020},
  volume	= {11},
  pages		= {8610-8615},
  doi		= {https://doi.org/10.1021/acs.jpclett.0c02213},
  url		= {https://pubs.acs.org/doi/abs/10.1021/acs.jpclett.0c02213}
}

@Article{	  prxgirvin,
  author	= {Marios H. Michael and Matti Silveri and Richard Brierley
		  and Victor V. Albert and Juha Salmilehto and Liang Jiang
		  and Steven M. Girvin},
  title		= {New Class of Quantum Error-Correcting Codes for a Bosonic
		  Mode},
  journal	= {Physical Review X},
  year		= {2016},
  volume	= {6},
  pages		= {031006},
  doi		= {https://doi.org/10.1103/PhysRevX.6.031006},
  url		= {https://journals.aps.org/prx/abstract/10.1103/PhysRevX.6.031006}
}

@Article{	  earnest2021pulse,
  title		= {Pulse-efficient circuit transpilation for quantum
		  applications on cross-resonance-based hardware},
  author	= {Earnest, Nathan and Tornow, Caroline and Egger, Daniel J},
  journal	= {Physical Review Research},
  volume	= {3},
  number	= {4},
  pages		= {043088},
  year		= {2021},
  publisher	= {APS},
  url		= {https://doi.org/10.1103/PhysRevResearch.3.043088}
}

@Article{	  werninghaus2021leakage,
  title		= {Leakage reduction in fast superconducting qubit gates via
		  optimal control},
  volume	= {7},
  issn		= {2056-6387},
  url		= {http://www.nature.com/articles/s41534-020-00346-2},
  doi		= {10.1038/s41534-020-00346-2},
  number	= {1},
  urldate	= {2022-01-19},
  journal	= {npj Quantum Information},
  author	= {Werninghaus, Max and Egger, Daniel J. and Roy, Federico
		  and Machnes, Shai and Wilhelm, Frank K. and Filipp,
		  Stefan},
  month		= dec,
  year		= {2021},
  pages		= {14}
}

@Article{	  rico2018nuclear,
  title		= {S{O}(3) “{nuclear} {physics}” with ultracold {gases}},
  volume	= {393},
  issn		= {00034916},
  url		= {https://linkinghub.elsevier.com/retrieve/pii/S0003491618300757},
  doi		= {10.1016/j.aop.2018.03.020},
  urldate	= {2022-10-10},
  journal	= {Annals of Physics},
  author	= {Rico, Enrique and Dalmonte, Marcello and Zoller, Peter and
		  Banerjee, Debarghya and B{\"o}gli, Michael and Stebler,
		  Pascal and Wiese, Uwe-Jens},
  month		= jun,
  year		= {2018},
  pages		= {466--483}
}

@Article{	  kraft2018characterizing,
  title		= {Characterizing {{genuine multilevel 1ntanglement}}},
  author	= {Kraft, Tristan and Ritz, Christina and Brunner, Nicolas
		  and Huber, Marcus and G{\"u}hne, Otfried},
  year		= {2018},
  month		= feb,
  journal	= {Physical Review Letters},
  volume	= {120},
  number	= {6},
  pages		= {060502},
  issn		= {0031-9007, 1079-7114},
  doi		= {10.1103/PhysRevLett.120.060502},
  url		= {https://link.aps.org/doi/10.1103/PhysRevLett.120.060502},
  urldate	= {2022-10-14},
  langid	= {english}
}

@Article{	  stricker2022experimental,
  title		= {Experimental {{single-setting quantum state tomography}}},
  author	= {Stricker, Roman and Meth, Michael and Postler, Lukas and
		  Edmunds, Claire and Ferrie, Chris and Blatt, Rainer and
		  Schindler, Philipp and Monz, Thomas and Kueng, Richard and
		  Ringbauer, Martin},
  year		= {2022},
  month		= oct,
  journal	= {PRX Quantum},
  volume	= {3},
  number	= {4},
  pages		= {040310},
  issn		= {2691-3399},
  doi		= {10.1103/PRXQuantum.3.040310},
  url		= {https://link.aps.org/doi/10.1103/PRXQuantum.3.040310},
  urldate	= {2022-11-09}
}

@incollection{brassard2002quantum,
  title={Quantum amplitude amplification and estimation},
  author={Brassard, Gilles and H{\o}yer, Peter and Mosca, Michele and Tapp, Alain},
  booktitle={Quantum computation and information (Washington, DC, 2000)},
  pages={53--74},
  year={2002},
  publisher={American Mathematical Society},
  doi={https://doi.org/10.1090/conm/305/05215}
}

@article{wiebe_bayesian_phase_estimation,
  title = {Efficient Bayesian Phase Estimation},
  author = {Wiebe, Nathan and Granade, Chris},
  journal = {Phys. Rev. Lett.},
  volume = {117},
  issue = {1},
  pages = {010503},
  numpages = {6},
  year = {2016},
  month = {Jun},
  publisher = {American Physical Society},
  doi = {10.1103/PhysRevLett.117.010503},
  url = {https://link.aps.org/doi/10.1103/PhysRevLett.117.010503}
}

@article{suzuki_amplitude_2020,
	title = {Amplitude estimation without phase estimation},
	volume = {19},
	issn = {1573-1332},
	url = {https://doi.org/10.1007/s11128-019-2565-2},
	doi = {10.1007/s11128-019-2565-2},
	number = {2},
	journal = {Quantum Information Processing},
	author = {Suzuki, Yohichi and Uno, Shumpei and Raymond, Rudy and Tanaka, Tomoki and Onodera, Tamiya and Yamamoto, Naoki},
	month = jan,
	year = {2020},
	pages = {75},
}

@article{VNCDR,
  title = {Unified approach to data-driven quantum error mitigation},
  author = {Lowe, Angus and Gordon, Max Hunter and Czarnik, Piotr and Arrasmith, Andrew and Coles, Patrick J. and Cincio, Lukasz},
  journal = {Phys. Rev. Res.},
  volume = {3},
  issue = {3},
  pages = {033098},
  numpages = {12},
  year = {2021},
  month = {Jul},
  publisher = {American Physical Society},
  doi = {10.1103/PhysRevResearch.3.033098},
  url = {https://link.aps.org/doi/10.1103/PhysRevResearch.3.033098}
}

@article{Bultrini2023unifying,
  doi = {10.22331/q-2023-06-06-1034},
  url = {https://doi.org/10.22331/q-2023-06-06-1034},
  title = {Unifying and benchmarking state-of-the-art quantum error mitigation techniques},
  author = {Bultrini, Daniel and Gordon, Max Hunter and Czarnik, Piotr and Arrasmith, Andrew and Cerezo, M. and Coles, Patrick J. and Cincio, Lukasz},
  journal = {{Quantum}},
  issn = {2521-327X},
  publisher = {{Verein zur F{\"{o}}rderung des Open Access Publizierens in den Quantenwissenschaften}},
  volume = {7},
  pages = {1034},
  month = jun,
  year = {2023}
}

@article{grinko2021iterative,
  title={Iterative quantum amplitude estimation},
  author={Grinko, Dmitry and Gacon, Julien and Zoufal, Christa and Woerner, Stefan},
  journal={npj Quantum Information},
  volume={7},
  number={1},
  pages={52},
  year={2021},
  publisher={Nature Publishing Group UK London},
  doi={https://doi.org/10.1038/s41534-021-00379-1}
}

@article{Wang_high_EJEC_transmon,
  title = {High-${E}_{J}/{E}_{C}$ transmon qudits with up to 12 levels},
  author = {Wang, Zihao and Parker, Rayleigh W. and Champion, Elizabeth and Blok, Machiel S.},
  journal = {Phys. Rev. Appl.},
  volume = {23},
  issue = {3},
  pages = {034046},
  numpages = {19},
  year = {2025},
  month = {Mar},
  publisher = {American Physical Society},
  doi = {10.1103/PhysRevApplied.23.034046},
  url = {https://link.aps.org/doi/10.1103/PhysRevApplied.23.034046}
}

@article{multi_qudit_tunable_couplers,
  title = {Experimental Realization of Two Qutrits Gate with Tunable Coupling in Superconducting Circuits},
  author = {Luo, Kai and Huang, Wenhui and Tao, Ziyu and Zhang, Libo and Zhou, Yuxuan and Chu, Ji and Liu, Wuxin and Wang, Biying and Cui, Jiangyu and Liu, Song and Yan, Fei and Yung, Man-Hong and Chen, Yuanzhen and Yan, Tongxing and Yu, Dapeng},
  journal = {Phys. Rev. Lett.},
  volume = {130},
  issue = {3},
  pages = {030603},
  numpages = {6},
  year = {2023},
  month = {Jan},
  publisher = {American Physical Society},
  doi = {10.1103/PhysRevLett.130.030603},
  url = {https://link.aps.org/doi/10.1103/PhysRevLett.130.030603}
}

@Article{	  shlyakhov2018quantum,
  title		= {Quantum Metrology with a Transmon Qutrit},
  author	= {Shlyakhov, A. R. and Zemlyanov, V. V. and Suslov, M. V.
		  and Lebedev, Andrey V. and Paraoanu, Gheorghe S. and
		  Lesovik, Gordey B. and Blatter, Gianni},
  year		= {2018},
  month		= feb,
  journal	= {Physical Review A},
  volume	= {97},
  number	= {2},
  pages		= {022115},
  issn		= {2469-9926, 2469-9934},
  doi		= {10.1103/PhysRevA.97.022115},
  url		= {https://link.aps.org/doi/10.1103/PhysRevA.97.022115},
  urldate	= {2022-10-14}
}

@Article{	  bruzewicz2019trappedion,
  title		= {Trapped-Ion Quantum Computing: {{Pprogress}} and
		  challenges},
  shorttitle	= {Trapped-Ion Quantum Computing},
  author	= {Bruzewicz, Colin D. and Chiaverini, John and McConnell,
		  Robert and Sage, Jeremy M.},
  year		= {2019},
  month		= jun,
  journal	= {Applied Physics Reviews},
  volume	= {6},
  number	= {2},
  pages		= {021314},
  issn		= {1931-9401},
  doi		= {10.1063/1.5088164},
  url		= {http://aip.scitation.org/doi/10.1063/1.5088164},
  urldate	= {2022-10-14},
  langid	= {english}
}

@Article{	  clarke2008superconducting,
  title		= {Superconducting Quantum Bits},
  author	= {Clarke, John and Wilhelm, Frank K.},
  year		= {2008},
  month		= jun,
  journal	= {Nature},
  volume	= {453},
  number	= {7198},
  pages		= {1031--1042},
  issn		= {0028-0836, 1476-4687},
  doi		= {10.1038/nature07128},
  url		= {http://www.nature.com/articles/nature07128},
  urldate	= {2022-10-14},
  langid	= {english}
}

@Article{shiquantumlogicentanglement2021,
  title		= {Quantum Logic and Entanglement by Neutral {{Rydberg}}
		  Atoms: Methods and Fidelity},
  shorttitle	= {Quantum Logic and Entanglement by Neutral {{Rydberg}}
		  Atoms},
  author	= {Shi, Xiaofeng},
  year		= {2021},
  volume	= {7},
  pages		= {023002},
  month		= jul,
  journal	= {Quantum Science and Technology},
  issn		= {2058-9565},
  doi		= {10.1088/2058-9565/ac18b8},
  url		= {https://iopscience.iop.org/article/10.1088/2058-9565/ac18b8},
  urldate	= {2022-01-26}
}

@Article{	  burkard2021semiconductor,
  title		= {Semiconductor {{spin qubits}}},
  author	= {Burkard, Guido and Ladd, Thaddeus D. and Nichol, John M.
		  and Pan, Andrew and Petta, Jason R.},
  year		= {2021},
  month		= dec,
  number	= {arXiv:2112.08863},
  journal	= {arXiv:2112.08863},
  url		= {http://arxiv.org/abs/2112.08863},
  urldate	= {2022-10-14},
  archiveprefix	= {arXiv}
}

@Article{	  blais2021circuit,
  title		= {Circuit Quantum Electrodynamics},
  author	= {Blais, Alexandre and Grimsmo, Arne L. and Girvin, Steven
		  M. and Wallraff, Andreas},
  year		= {2021},
  month		= may,
  journal	= {Reviews of Modern Physics},
  volume	= {93},
  number	= {2},
  pages		= {025005},
  issn		= {0034-6861, 1539-0756},
  doi		= {10.1103/RevModPhys.93.025005},
  url		= {https://link.aps.org/doi/10.1103/RevModPhys.93.025005},
  urldate	= {2022-10-25},
  langid	= {english}
}

@article{miao2022overcoming,
  title={Overcoming leakage in quantum error correction},
  author={Miao, Kevin C and McEwen, Matt and Atalaya, Juan and Kafri, Dvir and Pryadko, Leonid P and Bengtsson, Andreas and Opremcak, Alex and Satzinger, Kevin J and Chen, Zijun and Klimov, Paul V and others},
  journal={Nature Physics},
  volume={19},
  number={12},
  pages={1780--1786},
  year={2023},
  publisher={Nature Publishing Group UK London},
  doi={https://doi.org/10.1038/s41567-023-02226-w}
}

@article{kehrer_2024_qudit_readout,
  title = {Improving transmon qudit measurement on IBM Quantum hardware},
  author = {Kehrer, Tobias and Nadolny, Tobias and Bruder, Christoph},
  journal = {Phys. Rev. Res.},
  volume = {6},
  issue = {1},
  pages = {013050},
  numpages = {12},
  year = {2024},
  month = {Jan},
  publisher = {American Physical Society},
  doi = {10.1103/PhysRevResearch.6.013050},
  url = {https://link.aps.org/doi/10.1103/PhysRevResearch.6.013050}
}

@Article{	  roy2022realization,
  title		= {Realization of Two-Qutrit Quantum Algorithms on a
		  Programmable Superconducting Processor},
  author	= {Roy, Tanay and Li, Ziqian and Kapit, Eliot and Schuster,
		  David I.},
  year		= {2022},
  month		= nov,
  journal	= {arXiv:2211.06523},
  url		= {http://arxiv.org/abs/2211.06523},
  urldate	= {2022-11-15},
  archiveprefix	= {arXiv}
}

@Article{	  blok2021quantum,
  title		= {Quantum {information} {scrambling} on a {superconducting}
		  {qutrit} {processor}},
  volume	= {11},
  issn		= {2160-3308},
  url		= {https://link.aps.org/doi/10.1103/PhysRevX.11.021010},
  doi		= {10.1103/PhysRevX.11.021010},
  number	= {2},
  urldate	= {2022-01-18},
  journal	= {Physical Review X},
  author	= {Blok, Machiel S. and Ramasesh, Vinay V. and Schuster,
		  Thomas and O’Brien, Kevin and Kreikebaum, John-Mark and
		  Dahlen, Dar and Morvan, Alexis and Yoshida, Beni and Yao,
		  Norman Y. and Siddiqi, Irfan},
  year		= {2021},
  pages		= {021010}
}

@article{miller2024hardwaretailored,
  title = {Hardware-Tailored Diagonalization Circuits},
  author = {Miller, Daniel and Fischer, Laurin E. and Levi, Kyano and Kuehnke, Eric J. and Sokolov, Igor O. and Barkoutsos, Panagiotis Kl. and Eisert, Jens and Tavernelli, Ivano},
  year = {2024},
  month = nov,
  journal = {npj Quantum Information},
  volume = {10},
  number = {1},
  pages = {122},
  issn = {2056-6387},
  doi = {10.1038/s41534-024-00901-1},
  url = {https://www.nature.com/articles/s41534-024-00901-1},
  urldate = {2025-04-07},
  langid = {english}
}

@Article{	  miessen2021spin-boson,
  title		= {Quantum algorithms for quantum dynamics: A performance
		  study on the spin-boson model},
  author	= {Miessen, Alexander and Ollitrault, Pauline J. and
		  Tavernelli, Ivano},
  journal	= {Phys. Rev. Research},
  volume	= {3},
  issue		= {4},
  pages		= {043212},
  url		= {https://doi.org/10.1103/PhysRevResearch.3.043212},
  numpages	= {11},
  year		= {2021},
}

@Article{	  mazzolag2020,
  title		= {Nonadiabatic Molecular Quantum Dynamics with Quantum
		  Computers},
  author	= {Ollitrault, Pauline J. and Mazzola, Guglielmo and
		  Tavernelli, Ivano},
  journal	= {Phys. Rev. Lett.},
  volume	= {125},
  issue		= {26},
  pages		= {260511},
  url		= {https://doi.org/10.1103/PhysRevLett.125.260511},
  numpages	= {6},
  year		= {2020},
}

@Article{	  mathis2020,
  title		= {Toward scalable simulations of lattice gauge theories on
		  quantum computers},
  author	= {Mathis, Simon V. and Mazzola, Guglielmo and Tavernelli,
		  Ivano},
  journal	= {Phys. Rev. D},
  volume	= {102},
  issue		= {9},
  pages		= {094501},
  numpages	= {26},
  url		= {https://doi.org/10.1103/PhysRevD.102.094501},
  year		= {2020},
}

@Article{	  mazzola2021,
  title		= {Gauge-invariant quantum circuits for $U$(1) and Yang-Mills
		  lattice gauge theories},
  author	= {Mazzola, Giulia and Mathis, Simon V. and Mazzola,
		  Guglielmo and Tavernelli, Ivano},
  journal	= {Phys. Rev. Research},
  volume	= {3},
  issue		= {4},
  pages		= {043209},
  numpages	= {15},
  url		= {https://doi.org/10.1103/PhysRevResearch.3.043209},
  year		= {2021},
}

@Article{	  ollitrault_vib2020,
  author	= "Ollitrault, Pauline J. and Baiardi, Alberto and Reiher,
		  Markus and Tavernelli, Ivano",
  title		= "Hardware efficient quantum algorithms for vibrational
		  structure calculations",
  journal	= "Chem. Sci.",
  year		= "2020",
  url		= {https://doi.org/10.1039/D0SC01908A},
  volume	= "11",
  issue		= "26",
  pages		= "6842-6855"
}

@Book{		  nielsen_chuang_2010,
  place		= {Cambridge},
  title		= {Quantum Computation and Quantum Information: 10th
		  Anniversary Edition},
  doi		= {10.1017/CBO9780511976667},
  publisher	= {Cambridge University Press},
  author	= {Nielsen, Michael A. and Chuang, Isaac L.},
  year		= {2010},
  url		= {https://doi.org/10.1017/CBO9780511976667}
}

@Article{	  goss2022highfidelity,
  title		= {High-Fidelity qutrit entangling gates for superconducting
		  circuits},
  author	= {Goss, Noah and Morvan, Alexis and Marinelli, Brian and
		  Mitchell, Bradley K. and Nguyen, Long B. and Naik, Ravi K.
		  and Chen, Larry and J{\"u}nger, Christian and Kreikebaum,
		  John Mark and Santiago, David I. and Wallman, Joel J. and
		  Siddiqi, Irfan},
  year		= {2022},
  month		= dec,
  journal	= {Nature Communications},
  volume	= {13},
  number	= {1},
  pages		= {7481},
  issn		= {2041-1723},
  doi		= {10.1038/s41467-022-34851-z},
  url		= {https://www.nature.com/articles/s41467-022-34851-z},
  urldate	= {2022-12-08},
  langid	= {english}
}

@Article{	  gelfand1943imbedding,
  title		= {On the imbedding of normed rings into the ring of
		  operators in {Hilbert} space},
  volume	= {12},
  number	= {2},
  journal	= {Matematicheskii Sbornik},
  author	= {Gelfand, Israel and Neumark, Mark},
  year		= {1943},
  pages		= {197--217}
}

@Article{	  garcia2021learning,
  title		= {Learning to {measure}: {adaptive} {informationally}
		  {complete} {generalized} {measurements} for {quantum}
		  {algorithms}},
  volume	= {2},
  issn		= {2691-3399},
  url		= {https://link.aps.org/doi/10.1103/PRXQuantum.2.040342},
  doi		= {10.1103/PRXQuantum.2.040342},
  number	= {4},
  urldate	= {2022-01-18},
  journal	= {PRX Quantum},
  author	= {Garcia-Perez, Guillermo and Rossi, Matteo A.C. and
		  Sokolov, Boris and Tacchino, Francesco and Barkoutsos,
		  Panagiotis Kl. and Mazzola, Guglielmo and Tavernelli, Ivano
		  and Maniscalco, Sabrina},
  month		= nov,
  year		= {2021},
  pages		= {040342}
}

@Article{	  chen2007ancilla,
  title		= {Ancilla dimensions needed to carry out
		  positive-operator-valued measurement},
  volume	= {76},
  issn		= {1050-2947, 1094-1622},
  url		= {https://link.aps.org/doi/10.1103/PhysRevA.76.060303},
  doi		= {10.1103/PhysRevA.76.060303},
  number	= {6},
  urldate	= {2022-01-18},
  journal	= {Physical Review A},
  author	= {Chen, Ping-Xing and Bergou, János A. and Zhu, Shi-Yao and
		  Guo, Guang-Can},
  month		= dec,
  year		= {2007},
  pages		= {060303}
}

@Article{	  schirmer2002constructive,
  title		= {Constructive control of quantum systems using
		  factorization of unitary operators},
  volume	= {35},
  issn		= {0305-4470},
  url		= {https://iopscience.iop.org/article/10.1088/0305-4470/35/39/313},
  doi		= {10.1088/0305-4470/35/39/313},
  number	= {39},
  urldate	= {2022-01-18},
  journal	= {Journal of Physics A: Mathematical and General},
  author	= {Schirmer, Sophie G. and Greentree, Andrew D. and
		  Ramakrishna, Viswanath and Rabitz, Herschel},
  month		= oct,
  year		= {2002},
  pages		= {8315--8339}
}

@Book{		  golub1996matrix,
  address	= {Baltimore},
  edition	= {3rd ed},
  series	= {Johns {Hopkins} studies in the mathematical sciences},
  title		= {Matrix computations},
  publisher	= {Johns Hopkins University Press},
  author	= {Golub, Gene H. and Van Loan, Charles F.},
  year		= {1996}
}

@Article{	  wallraff2005approaching,
  title		= {Approaching {unit} {visibility} for {control} of a
		  {superconducting} {qubit} with {dispersive} {readout}},
  volume	= {95},
  issn		= {0031-9007, 1079-7114},
  url		= {https://link.aps.org/doi/10.1103/PhysRevLett.95.060501},
  doi		= {10.1103/PhysRevLett.95.060501},
  number	= {6},
  urldate	= {2022-01-18},
  journal	= {Physical Review Letters},
  author	= {Wallraff, Andreas and Schuster, David I. and Blais,
		  Alexandre and Frunzio, Luigi and Majer, Johannes and
		  Devoret, Michel H. and Girvin, Steven M. and Schoelkopf,
		  Robert J.},
  month		= aug,
  year		= {2005},
  pages		= {060501}
}

@article{purcell_filter_ref,
    author = {Reed, M. D. and Johnson, B. R. and Houck, A. A. and DiCarlo, L. and Chow, J. M. and Schuster, D. I. and Frunzio, L. and Schoelkopf, R. J.},
    title = {Fast reset and suppressing spontaneous emission of a superconducting qubit},
    journal = {Applied Physics Letters},
    volume = {96},
    number = {20},
    pages = {203110},
    year = {2010},
    month = {05},
    issn = {0003-6951},
    doi = {10.1063/1.3435463},
    url = {https://doi.org/10.1063/1.3435463},
}

@article{multiplexed_readout,
  title = {Rapid High-fidelity Multiplexed Readout of Superconducting Qubits},
  author = {Heinsoo, Johannes and Andersen, Christian Kraglund and Remm, Ants and Krinner, Sebastian and Walter, Theodore and Salath\'e, Yves and Gasparinetti, Simone and Besse, Jean-Claude and Poto\ifmmode \check{c}\else \v{c}\fi{}nik, Anton and Wallraff, Andreas and Eichler, Christopher},
  journal = {Phys. Rev. Appl.},
  volume = {10},
  issue = {3},
  pages = {034040},
  numpages = {14},
  year = {2018},
  month = {Sep},
  publisher = {American Physical Society},
  doi = {10.1103/PhysRevApplied.10.034040},
  url = {https://link.aps.org/doi/10.1103/PhysRevApplied.10.034040}
}

@article{macklin2015near,
  title={A near--quantum-limited Josephson traveling-wave parametric amplifier},
  author={Macklin, Chris and O’brien, K and Hover, D and Schwartz, ME and Bolkhovsky, V and Zhang, X and Oliver, WD and Siddiqi, I},
  journal={Science},
  volume={350},
  number={6258},
  pages={307--310},
  year={2015},
  publisher={American Association for the Advancement of Science},
  doi={https://doi.org/10.1126/science.aaa8525}
}

@Article{	  riste2013millisecond,
  title		= {Millisecond charge-parity fluctuations and induced
		  decoherence in a superconducting transmon qubit},
  volume	= {4},
  issn		= {2041-1723},
  url		= {http://www.nature.com/articles/ncomms2936},
  doi		= {10.1038/ncomms2936},
  number	= {1},
  urldate	= {2022-01-18},
  journal	= {Nature Communications},
  author	= {Riste, Diego and Bultink, Niels and Tiggelman, Marijn J.
		  and Schouten, Raymond N. and Lehnert, Konrad W. and
		  DiCarlo, Leonardo},
  month		= oct,
  year		= {2013},
  pages		= {1913}
}

@article{deutsch1992rapid,
  title={Rapid solution of problems by quantum computation},
  author={Deutsch, David and Jozsa, Richard},
  journal={Proceedings of the Royal Society of London. Series A: Mathematical and Physical Sciences},
  volume={439},
  number={1907},
  pages={553--558},
  year={1992},
  publisher={The Royal Society London}
}

@Article{	  gambetta2011analytic,
  title		= {Analytic control methods for high-fidelity unitary
		  operations in a weakly nonlinear oscillator},
  volume	= {83},
  issn		= {1050-2947, 1094-1622},
  url		= {https://link.aps.org/doi/10.1103/PhysRevA.83.012308},
  doi		= {10.1103/PhysRevA.83.012308},
  number	= {1},
  urldate	= {2022-01-18},
  journal	= {Physical Review A},
  author	= {Gambetta, Jay. M. and Motzoi, Felix and Merkel, Seth T.
		  and Wilhelm, Frank K.},
  month		= jan,
  year		= {2011},
  pages		= {012308}
}

@article{mcewen2021removing,
  title={Removing leakage-induced correlated errors in superconducting quantum error correction},
  author={McEwen, Matt and Kafri, Dvir and Chen, Z and Atalaya, Juan and Satzinger, KJ and Quintana, Chris and Klimov, Paul Victor and Sank, Daniel and Gidney, C and Fowler, AG and others},
  journal={Nature communications},
  volume={12},
  number={1},
  pages={1761},
  year={2021},
  publisher={Nature Publishing Group UK London},
  doi={https://doi.org/10.1038/s41467-021-21982-y}
}

@article{undonditional_Reset_ZRL,
  title = {Pulsed Reset Protocol for Fixed-Frequency Superconducting Qubits},
  author = {Egger, D.J. and Werninghaus, M. and Ganzhorn, M. and Salis, G. and Fuhrer, A. and M\"uller, P. and Filipp, S.},
  journal = {Phys. Rev. Appl.},
  volume = {10},
  issue = {4},
  pages = {044030},
  numpages = {7},
  year = {2018},
  month = {Oct},
  publisher = {American Physical Society},
  doi = {10.1103/PhysRevApplied.10.044030},
  url = {https://link.aps.org/doi/10.1103/PhysRevApplied.10.044030}
}

@article{gambetta2017building,
  title={Building logical qubits in a superconducting quantum computing system},
  author={Gambetta, Jay M and Chow, Jerry M and Steffen, Matthias},
  journal={npj quantum information},
  volume={3},
  number={1},
  pages={2},
  year={2017},
  publisher={Nature Publishing Group UK London},
  doi={https://doi.org/10.1038/s41534-016-0004-0}
}

@article{readout_induced_leakage,
  title = {Measurement-Induced State Transitions in a Superconducting Qubit: Beyond the Rotating Wave Approximation},
  author = {Sank, Daniel and Chen, Zijun and Khezri, Mostafa and Kelly, J. and Barends, R. and Campbell, B. and Chen, Y. and Chiaro, B. and Dunsworth, A. and Fowler, A. and Jeffrey, E. and Lucero, E. and Megrant, A. and Mutus, J. and Neeley, M. and Neill, C. and O'Malley, P. J. J. and Quintana, C. and Roushan, P. and Vainsencher, A. and White, T. and Wenner, J. and Korotkov, Alexander N. and Martinis, John M.},
  journal = {Phys. Rev. Lett.},
  volume = {117},
  issue = {19},
  pages = {190503},
  numpages = {6},
  year = {2016},
  month = {Nov},
  publisher = {American Physical Society},
  doi = {10.1103/PhysRevLett.117.190503},
  url = {https://link.aps.org/doi/10.1103/PhysRevLett.117.190503}
}

@Article{	  mckay2018qiskit,
  title		= {Qiskit {backend} {specifications} for {OpenQASM} and
		  {OpenPulse} {experiments}},
  url		= {http://arxiv.org/abs/1809.03452},
  urldate	= {2022-01-18},
  journal	= {arXiv:1809.03452 [quant-ph]},
  author	= {McKay, David C. and Alexander, Thomas and Bello, Luciano
		  and Biercuk, Michael J. and Bishop, Lev and Chen, Jiayin
		  and Chow, Jerry M. and Córcoles, Antonio D. and Egger,
		  Daniel and Filipp, Stefan and Gomez, Juan and Hush, Michael
		  and Javadi-Abhari, Ali and Moreda, Diego and Nation, Paul
		  and Paulovicks, Brent and Winston, Erick and Wood,
		  Christopher J. and Wootton, James and Gambetta, Jay M.},
  month		= sep,
  year		= {2018}
}

@Article{	  dariano2004quantum,
  title		= {Quantum {calibration} of {measurement} {instrumentation}},
  volume	= {93},
  issn		= {0031-9007, 1079-7114},
  url		= {https://link.aps.org/doi/10.1103/PhysRevLett.93.250407},
  doi		= {10.1103/PhysRevLett.93.250407},
  number	= {25},
  urldate	= {2022-01-18},
  journal	= {Physical Review Letters},
  author	= {D’Ariano, Giacomo Mauro and Maccone, Lorenzo and Presti,
		  Paoloplacido Lo},
  month		= dec,
  year		= {2004},
  pages		= {250407}
}

@Article{	  lundeen2009tomography,
  title		= {Tomography of quantum detectors},
  volume	= {5},
  issn		= {1745-2473, 1745-2481},
  url		= {http://www.nature.com/articles/nphys1133},
  doi		= {10.1038/nphys1133},
  number	= {1},
  urldate	= {2022-01-18},
  journal	= {Nature Physics},
  author	= {Lundeen, Jeff S. and Feito, Alvaro and Coldenstrodt-Ronge,
		  Hendrik and Pregnell, Kenny L. and Silberhorn, Christine
		  and Ralph, Timothy C. and Eisert, Jens and Plenio, Martin
		  B. and Walmsley, Ian A.},
  month		= jan,
  year		= {2009},
  pages		= {27--30}
}

@Article{	  fiurasek2001maximumlikelihood,
  title		= {Maximum-likelihood estimation of quantum measurement},
  volume	= {64},
  issn		= {1050-2947, 1094-1622},
  url		= {https://link.aps.org/doi/10.1103/PhysRevA.64.024102},
  doi		= {10.1103/PhysRevA.64.024102},
  number	= {2},
  urldate	= {2022-01-18},
  journal	= {Physical Review A},
  author	= {Fiurasek, Jaromir},
  month		= jul,
  year		= {2001},
  pages		= {024102}
}

@Article{maciejewski2020mitigation,
  title		= {Mitigation of readout noise in near-term quantum devices
		  by classical post-processing based on detector tomography},
  volume	= {4},
  issn		= {2521-327X},
  url		= {https://quantum-journal.org/papers/q-2020-04-24-257/},
  doi		= {10.22331/q-2020-04-24-257},
  urldate	= {2022-01-18},
  journal	= {Quantum},
  author	= {Maciejewski, Filip B. and Zimboras, Zoltan and Oszmaniec,
		  Michal},
  month		= apr,
  year		= {2020},
  pages		= {257}
}

@Article{	  puchala2018strategies,
  title		= {Strategies for optimal single-shot discrimination of
		  quantum measurements},
  volume	= {98},
  issn		= {2469-9926, 2469-9934},
  url		= {https://link.aps.org/doi/10.1103/PhysRevA.98.042103},
  doi		= {10.1103/PhysRevA.98.042103},
  number	= {4},
  urldate	= {2022-01-18},
  journal	= {Physical Review A},
  author	= {Puchala, Zbigniew and Pawela, Lukasz and Krawiec,
		  Aleksandra and Kukulski, Ryszard},
  month		= oct,
  year		= {2018},
  pages		= {042103}
}

@Article{	  jiang2020optimal,
  title		= {Optimal fermion-to-qubit mapping via ternary trees with
		  applications to reduced quantum states learning},
  volume	= {4},
  issn		= {2521-327X},
  url		= {https://quantum-journal.org/papers/q-2020-06-04-276/},
  doi		= {10.22331/q-2020-06-04-276},
  urldate	= {2022-01-18},
  journal	= {Quantum},
  author	= {Jiang, Zhang and Kalev, Amir and Mruczkiewicz, Wojciech
		  and Neven, Hartmut},
  month		= jun,
  year		= {2020},
  pages		= {276}
}

@Article{	  schreier2008suppressing,
  title		= {Suppressing charge noise decoherence in superconducting
		  charge qubits},
  volume	= {77},
  issn		= {1098-0121, 1550-235X},
  url		= {https://link.aps.org/doi/10.1103/PhysRevB.77.180502},
  doi		= {10.1103/PhysRevB.77.180502},
  number	= {18},
  urldate	= {2022-01-18},
  journal	= {Physical Review B},
  author	= {Schreier, Joseph A. and Houck, Andrew A. and Koch, Jens
		  and Schuster, David I. and Johnson, Bradley R. and Chow,
		  Jerry M. and Gambetta, Jay M. and Majer, Johannes and
		  Frunzio, Luigi and Devoret, Michel H. and Girvin, Steven M.
		  and Schoelkopf, Robert J.},
  month		= may,
  year		= {2008},
  pages		= {180502}
}

@Article{	  neugebauer2020neuralnetwork,
  title		= {Neural-network quantum state tomography in a two-qubit
		  experiment},
  volume	= {102},
  issn		= {2469-9926, 2469-9934},
  url		= {https://link.aps.org/doi/10.1103/PhysRevA.102.042604},
  doi		= {10.1103/PhysRevA.102.042604},
  number	= {4},
  urldate	= {2022-01-18},
  journal	= {Physical Review A},
  author	= {Neugebauer, Marcel and Fischer, Laurin and J{\"a}ger,
		  Alexander and Czischek, Stefanie and Jochim, Selim and
		  Weidemüller, Matthias and G{\"a}rttner, Martin},
  month		= oct,
  year		= {2020},
  pages		= {042604}
}

@InCollection{	  hradil2004maximumlikelihood,
  address	= {Berlin, Heidelberg},
  title		= {3 {maximum}-{likelihood} {methods in} {quantum}
		  {mechanics}},
  volume	= {649},
  isbn		= {978-3-540-22329-0 978-3-540-44481-7},
  url		= {http://link.springer.com/10.1007/978-3-540-44481-7_3},
  urldate	= {2022-01-18},
  booktitle	= {Quantum {State} {Estimation}},
  publisher	= {Springer Berlin Heidelberg},
  author	= {Hradil, Zdenek and Rehacek, Jaroslav and Fiurasek, Jaromir
		  and Jezek, Miroslav},
  editor	= {Paris, Matteo and Rehacek, Jaroslav},
  month		= aug,
  year		= {2004},
  doi		= {10.1007/978-3-540-44481-7_3},
  note		= {Series Title: Lecture Notes in Physics},
  pages		= {59--112}
}

@Article{	  bauer2020quantum,
  title		= {Quantum {algorithms} for {quantum} {chemistry} and
		  {quantum} {materials} {science}},
  volume	= {120},
  issn		= {0009-2665, 1520-6890},
  url		= {https://pubs.acs.org/doi/10.1021/acs.chemrev.9b00829},
  doi		= {10.1021/acs.chemrev.9b00829},
  number	= {22},
  urldate	= {2022-01-18},
  journal	= {Chemical Reviews},
  author	= {Bauer, Bela and Bravyi, Sergey and Motta, Mario and Chan,
		  Garnet Kin-Lic},
  month		= nov,
  year		= {2020},
  pages		= {12685--12717}
}

@Article{	  mcardle2020quantum,
  title		= {Quantum computational chemistry},
  volume	= {92},
  issn		= {0034-6861, 1539-0756},
  url		= {https://link.aps.org/doi/10.1103/RevModPhys.92.015003},
  doi		= {10.1103/RevModPhys.92.015003},
  number	= {1},
  urldate	= {2022-01-18},
  journal	= {Reviews of Modern Physics},
  author	= {McArdle, Sam and Endo, Suguru and Aspuru-Guzik, Alan and
		  Benjamin, Simon C. and Yuan, Xiao},
  month		= mar,
  year		= {2020},
  pages		= {015003}
}

@Article{	  motta2021emerging,
  author	= {Motta, Mario and Rice, Julia E.},
  title		= {Emerging quantum computing algorithms for quantum
		  chemistry},
  journal	= {WIREs Computational Molecular Science},
  volume	= {12},
  number	= {3},
  pages		= {e1580},
  doi		= {https://doi.org/10.1002/wcms.1580},
  url		= {https://wires.onlinelibrary.wiley.com/doi/abs/10.1002/wcms.1580},
  year		= {2022}
}

@Article{	  kandala2017hardwareefficient,
  title		= {Hardware-efficient variational quantum eigensolver for
		  small molecules and quantum magnets},
  volume	= {549},
  issn		= {0028-0836, 1476-4687},
  url		= {http://www.nature.com/articles/nature23879},
  doi		= {10.1038/nature23879},
  number	= {7671},
  urldate	= {2022-01-18},
  journal	= {Nature},
  author	= {Kandala, Abhinav and Mezzacapo, Antonio and Temme, Kristan
		  and Takita, Maika and Brink, Markus and Chow, Jerry M. and
		  Gambetta, Jay M.},
  year		= {2017},
  pages		= {242--246}
}

@Article{	  cerezo2021variational,
  title		= {Variational quantum algorithms},
  volume	= {3},
  issn		= {2522-5820},
  url		= {https://www.nature.com/articles/s42254-021-00348-9},
  doi		= {10.1038/s42254-021-00348-9},
  number	= {9},
  urldate	= {2022-01-18},
  journal	= {Nature Reviews Physics},
  author	= {Cerezo, Marco and Arrasmith, Andrew and Babbush, Ryan and
		  Benjamin, Simon C. and Endo, Suguru and Fujii, Keisuke and
		  McClean, Jarrod R. and Mitarai, Kosuke and Yuan, Xiao and
		  Cincio, Lukasz and Coles, Patrick J.},
  year		= {2021},
  pages		= {625--644}
}

@Article{	  peruzzo2014variational,
  title		= {A variational eigenvalue solver on a photonic quantum
		  processor},
  volume	= {5},
  issn		= {2041-1723},
  url		= {http://www.nature.com/articles/ncomms5213},
  doi		= {10.1038/ncomms5213},
  number	= {1},
  urldate	= {2022-01-18},
  journal	= {Nature Communications},
  author	= {Peruzzo, Alberto and McClean, Jarrod and Shadbolt, Peter
		  and Yung, Man-Hong and Zhou, Xiao-Qi and Love, Peter J. and
		  Aspuru-Guzik, Alan and O’Brien, Jeremy L.},
  year		= {2014},
  pages		= {4213}
}

@Article{	  wang2021noiseinduced,
  title		= {Noise-induced barren plateaus in variational quantum
		  algorithms},
  volume	= {12},
  issn		= {2041-1723},
  url		= {https://www.nature.com/articles/s41467-021-27045-6},
  doi		= {10.1038/s41467-021-27045-6},
  number	= {1},
  urldate	= {2022-01-18},
  journal	= {Nature Communications},
  author	= {Wang, Samson and Fontana, Enrico and Cerezo, Marco and
		  Sharma, Kunal and Sone, Akira and Cincio, Lukasz and Coles,
		  Patrick J.},
  year		= {2021},
  pages		= {6961}
}

@Article{	  crawford2021efficient,
  title		= {Efficient quantum measurement of {Pauli} operators in the
		  presence of finite sampling error},
  volume	= {5},
  issn		= {2521-327X},
  url		= {https://quantum-journal.org/papers/q-2021-01-20-385/},
  doi		= {10.22331/q-2021-01-20-385},
  urldate	= {2022-01-18},
  journal	= {Quantum},
  author	= {Crawford, Ophelia and Straaten, Barnaby van and Wang,
		  Daochen and Parks, Thomas and Campbell, Earl and Brierley,
		  Stephen},
  year		= {2021},
  pages		= {385}
}

@Article{	  huang2020predicting,
  title		= {Predicting many properties of a quantum system from very
		  few measurements},
  volume	= {16},
  issn		= {1745-2473, 1745-2481},
  url		= {http://www.nature.com/articles/s41567-020-0932-7},
  doi		= {10.1038/s41567-020-0932-7},
  number	= {10},
  urldate	= {2022-01-18},
  journal	= {Nature Physics},
  author	= {Huang, Hsin-Yuan and Kueng, Richard and Preskill, John},
  year		= {2020},
  pages		= {1050--1057}
}

@Article{	  brida2012ancillaassisted,
  title		= {Ancilla-Assisted Calibration of a Measuring Apparatus},
  volume	= {108},
  issn		= {0031-9007, 1079-7114},
  url		= {https://link.aps.org/doi/10.1103/PhysRevLett.108.253601},
  doi		= {10.1103/PhysRevLett.108.253601},
  number	= {25},
  urldate	= {2022-01-19},
  journal	= {Physical Review Letters},
  author	= {Brida, Giorgio and Ciavarella, Luigi and Degiovanni, Ivo
		  P. and Genovese, Marco and Migdall, Alan and Mingolla,
		  Maria G. and Paris, Matteo G. A. and Piacentini, Fabrizio
		  and Polyakov, Sergey V.},
  year		= {2012},
  pages		= {253601}
}

@Article{	  wecker2015,
  title		= {Progress towards practical quantum variational
		  algorithms},
  author	= {Wecker, Dave and Hastings, Matthew B. and Troyer,
		  Matthias},
  journal	= {Physical Review A},
  volume	= {92},
  issue		= {4},
  pages		= {042303},
  numpages	= {10},
  year		= {2015},
  month		= {Oct},
  publisher	= {American Physical Society},
  doi		= {10.1103/PhysRevA.92.042303},
  url		= {https://link.aps.org/doi/10.1103/PhysRevA.92.042303}
}

@Article{	  alexander2020qiskit,
  title		= {Qiskit pulse: programming quantum computers through the
		  cloud with pulses},
  volume	= {5},
  issn		= {2058-9565},
  shorttitle	= {Qiskit pulse},
  url		= {https://iopscience.iop.org/article/10.1088/2058-9565/aba404},
  doi		= {10.1088/2058-9565/aba404},
  number	= {4},
  urldate	= {2022-01-19},
  journal	= {Quantum Science and Technology},
  author	= {Alexander, Thomas and Kanazawa, Naoki and Egger, Daniel J
		  and Capelluto, Lauren and Wood, Christopher J and
		  Javadi-Abhari, Ali and C McKay, David},
  year		= {2020},
  pages		= {044006}
}

@Article{	  egger2018pulsed,
  title		= {Pulsed Reset Protocol for Fixed-Frequency Superconducting
		  Qubits},
  volume	= {10},
  issn		= {2331-7019},
  url		= {https://link.aps.org/doi/10.1103/PhysRevApplied.10.044030},
  doi		= {10.1103/PhysRevApplied.10.044030},
  number	= {4},
  urldate	= {2022-01-31},
  journal	= {Physical Review Applied},
  author	= {Egger, Daniel J. and Werninghaus, Max and Ganzhorn, Marc
		  and Salis, Gian and Fuhrer, Andreas and Müller, Peter and
		  Filipp, Stefan},
  year		= {2018},
  pages		= {044030}
}

@Article{	  egger2019entanglement,
  title		= {Entanglement Generation in Superconducting Qubits Using
		  Holonomic Operations},
  volume	= {11},
  issn		= {2331-7019},
  url		= {https://link.aps.org/doi/10.1103/PhysRevApplied.11.014017},
  doi		= {10.1103/PhysRevApplied.11.014017},
  number	= {1},
  urldate	= {2022-01-31},
  journal	= {Physical Review Applied},
  author	= {Egger, Daniel J. and Ganzhorn, Max and Salis, Gian and
		  Fuhrer, Andreas and Müller, Peter and Barkoutsos,
		  Panagiotis Kl. and Moll, Nikolaj and Tavernelli, Ivano and
		  Filipp, Stefan},
  year		= {2019},
  pages		= {014017}
}

@article{seskir_landscape_2022,
	title = {The landscape of the quantum start-up ecosystem},
	volume = {9},
	issn = {2196-0763},
	url = {https://doi.org/10.1140/epjqt/s40507-022-00146-x},
	doi = {10.1140/epjqt/s40507-022-00146-x},
	number = {1},
	journal = {EPJ Quantum Technology},
	author = {Seskir, Zeki Can and Korkmaz, Ramis and Aydinoglu, Arsev Umur},
	month = oct,
	year = {2022},
	pages = {27},
}

@article{bobier2024long,
  title={The long-term forecast for quantum computing still looks bright},
  author={Bobier, JF and Langione, M and Naudet-Baulieu, C and Cui, Z and Watanabe, E},
  journal={Boston Consulting Group},
  year={2024},
  url={https://www.bcg.com/publications/2024/long-term-forecast-for-quantum-computing-still-looks-bright}
}

@article{arute_quantum_2019,
	title = {Quantum supremacy using a programmable superconducting processor},
	volume = {574},
	issn = {1476-4687},
	url = {https://doi.org/10.1038/s41586-019-1666-5},
	doi = {10.1038/s41586-019-1666-5},
	number = {7779},
	journal = {Nature},
	author = {Arute, Frank and Arya, Kunal and Babbush, Ryan and Bacon, Dave and Bardin, Joseph C. and Barends, Rami and Biswas, Rupak and Boixo, Sergio and Brandao, Fernando G. S. L. and Buell, David A. and Burkett, Brian and Chen, Yu and Chen, Zijun and Chiaro, Ben and Collins, Roberto and Courtney, William and Dunsworth, Andrew and Farhi, Edward and Foxen, Brooks and Fowler, Austin and Gidney, Craig and Giustina, Marissa and Graff, Rob and Guerin, Keith and Habegger, Steve and Harrigan, Matthew P. and Hartmann, Michael J. and Ho, Alan and Hoffmann, Markus and Huang, Trent and Humble, Travis S. and Isakov, Sergei V. and Jeffrey, Evan and Jiang, Zhang and Kafri, Dvir and Kechedzhi, Kostyantyn and Kelly, Julian and Klimov, Paul V. and Knysh, Sergey and Korotkov, Alexander and Kostritsa, Fedor and Landhuis, David and Lindmark, Mike and Lucero, Erik and Lyakh, Dmitry and Mandrà, Salvatore and McClean, Jarrod R. and McEwen, Matthew and Megrant, Anthony and Mi, Xiao and Michielsen, Kristel and Mohseni, Masoud and Mutus, Josh and Naaman, Ofer and Neeley, Matthew and Neill, Charles and Niu, Murphy Yuezhen and Ostby, Eric and Petukhov, Andre and Platt, John C. and Quintana, Chris and Rieffel, Eleanor G. and Roushan, Pedram and Rubin, Nicholas C. and Sank, Daniel and Satzinger, Kevin J. and Smelyanskiy, Vadim and Sung, Kevin J. and Trevithick, Matthew D. and Vainsencher, Amit and Villalonga, Benjamin and White, Theodore and Yao, Z. Jamie and Yeh, Ping and Zalcman, Adam and Neven, Hartmut and Martinis, John M.},
	month = oct,
	year = {2019},
	pages = {505--510},
}

@article{zhong2020quantum,
  title={Quantum computational advantage using photons},
  author={Zhong, Han-Sen and Wang, Hui and Deng, Yu-Hao and Chen, Ming-Cheng and Peng, Li-Chao and Luo, Yi-Han and Qin, Jian and Wu, Dian and Ding, Xing and Hu, Yi and others},
  journal={Science},
  volume={370},
  number={6523},
  pages={1460--1463},
  year={2020},
  publisher={American Association for the Advancement of Science},
  doi={10.1126/science.abe8770}
}

@Article{	  sokolov2020quantum,
  title		= {Quantum orbital-optimized unitary coupled cluster methods
		  in the strongly correlated regime: {Can} quantum algorithms
		  outperform their classical equivalents?},
  volume	= {152},
  issn		= {0021-9606, 1089-7690},
  shorttitle	= {Quantum orbital-optimized unitary coupled cluster methods
		  in the strongly correlated regime},
  url		= {http://aip.scitation.org/doi/10.1063/1.5141835},
  doi		= {10.1063/1.5141835},
  number	= {12},
  urldate	= {2022-02-01},
  journal	= {The Journal of Chemical Physics},
  author	= {Sokolov, Igor O. and Barkoutsos, Panagiotis Kl. and
		  Ollitrault, Pauline J. and Greenberg, Donny and Rice, Julia
		  and Pistoia, Marco and Tavernelli, Ivano},
  month		= mar,
  year		= {2020},
  pages		= {124107}
}

@article{larose2024brief,
  title={A brief history of quantum vs classical computational advantage},
  author={LaRose, Ryan},
  journal={arXiv preprint arXiv:2412.14703},
  year={2024},
  url={https://arxiv.org/abs/2412.14703}
}

@Article{	  holmes2022connecting,
  title		= {Connecting Ansatz Expressibility to Gradient Magnitudes
		  and Barren Plateaus},
  volume	= {3},
  issn		= {2691-3399},
  url		= {https://link.aps.org/doi/10.1103/PRXQuantum.3.010313},
  doi		= {10.1103/PRXQuantum.3.010313},
  number	= {1},
  urldate	= {2022-02-01},
  journal	= {PRX Quantum},
  author	= {Holmes, Zoe and Sharma, Kunal and Cerezo, Marco and Coles,
		  Patrick J.},
  month		= jan,
  year		= {2022},
  pages		= {010313}
}

@article{haghshenas2025digital,
  title={Digital quantum magnetism at the frontier of classical simulations},
  author={Haghshenas, Reza and Chertkov, Eli and Mills, Michael and Kadow, Wilhelm and Lin, Sheng-Hsuan and Chen, Yi-Hsiang and Cade, Chris and Niesen, Ido and Begu{\v{s}}i{\'c}, Tomislav and Rudolph, Manuel S and others},
  journal={arXiv preprint arXiv:2503.20870},
  year={2025},
  url={https://arxiv.org/abs/2503.20870}
}

@Article{	  ollitrault2020quantum,
  title		= {Quantum equation of motion for computing molecular
		  excitation energies on a noisy quantum processor},
  volume	= {2},
  issn		= {2643-1564},
  url		= {https://link.aps.org/doi/10.1103/PhysRevResearch.2.043140},
  doi		= {10.1103/PhysRevResearch.2.043140},
  number	= {4},
  urldate	= {2022-02-21},
  journal	= {Physical Review Research},
  author	= {Ollitrault, Pauline J. and Kandala, Abhinav and Chen,
		  Chun-Fu and Barkoutsos, Panagiotis Kl. and Mezzacapo,
		  Antonio and Pistoia, Marco and Sheldon, Sarah and
		  W{\"o}rner, Stefan and Gambetta, Jay M. and Tavernelli,
		  Ivano},
  month		= oct,
  year		= {2020},
  pages		= {043140}
}

@Article{	  weidenfeller2022scaling,
  title		= {Scaling of the quantum approximate optimization algorithm
		  on superconducting qubit based hardware},
  url		= {http://arxiv.org/abs/2202.03459},
  urldate	= {2022-02-28},
  journal	= {arXiv:2202.03459 [quant-ph]},
  author	= {Weidenfeller, Johannes and Valor, Lucia C. and Gacon,
		  Julien and Tornow, Caroline and Bello, Luciano and
		  W{\"o}rner, Stefan and Egger, Daniel J.},
  month		= feb,
  year		= {2022}
}

@article{fischer2023universal,
  title = {Universal {{Qudit Gate Synthesis}} for {{Transmons}}},
  author = {Fischer, Laurin E. and Chiesa, Alessandro and Tacchino, Francesco and Egger, Daniel J. and Carretta, Stefano and Tavernelli, Ivano},
  year = {2023},
  month = aug,
  journal = {PRX Quantum},
  volume = {4},
  number = {3},
  pages = {030327},
  issn = {2691-3399},
  doi = {10.1103/PRXQuantum.4.030327},
  url = {https://link.aps.org/doi/10.1103/PRXQuantum.4.030327},
  urldate = {2023-08-29},
  langid = {english}
}

@article{hanisch2024soft,
  title={Soft information decoding with superconducting qubits},
  author={Hanisch, Maurice D and Het{\'e}nyi, Bence and Wootton, James R},
  journal={arXiv preprint arXiv:2411.16228},
  year={2024},
  doi={https://doi.org/10.48550/arXiv.2411.16228}
}

@article{dispersive_readout_review,
  title = {Enhancing Dispersive Readout of Superconducting Qubits through Dynamic Control of the Dispersive Shift: Experiment and Theory},
  author = {Swiadek, Fran\ifmmode \mbox{\c{c}}\else \c{c}\fi{}ois and Shillito, Ross and Magnard, Paul and Remm, Ants and Hellings, Christoph and Lacroix, Nathan and Ficheux, Quentin and Zanuz, Dante Colao and Norris, Graham J. and Blais, Alexandre and Krinner, Sebastian and Wallraff, Andreas},
  journal = {PRX Quantum},
  volume = {5},
  issue = {4},
  pages = {040326},
  numpages = {14},
  year = {2024},
  month = {Nov},
  publisher = {American Physical Society},
  doi = {10.1103/PRXQuantum.5.040326},
  url = {https://link.aps.org/doi/10.1103/PRXQuantum.5.040326}
}

@article{gacon2024scalable,
  title={Scalable Quantum Algorithms for Noisy Quantum Computers},
  author={Gacon, Julien},
  journal={arXiv preprint arXiv:2403.00940},
  year={2024},
  doi={https://doi.org/10.5075/epfl-thesis-11132}
}

@article{egger2023pulse,
  title = {Pulse Variational Quantum Eigensolver on Cross-Resonance-Based Hardware},
  author = {Egger, Daniel J. and Capecci, Chiara and Pokharel, Bibek and Barkoutsos, Panagiotis Kl. and Fischer, Laurin E. and Guidoni, Leonardo and Tavernelli, Ivano},
  year = {2023},
  journal = {Physical Review Research},
  volume = {5},
  number = {3},
  pages = {033159},
  issn = {2643-1564},
  doi = {10.1103/PhysRevResearch.5.033159},
  url = {https://link.aps.org/doi/10.1103/PhysRevResearch.5.033159},
  urldate = {2025-04-07},
  langid = {english},
  doi={https://doi.org/10.1103/PhysRevResearch.5.033159}
}

@Article{	  rossmannek2021quantum,
  title		= {Quantum HF/DFT-embedding algorithms for electronic
		  structure calculations: Scaling up to complex molecular
		  systems},
  author	= {Rossmannek, Max and Barkoutsos, Panagiotis Kl and
		  Ollitrault, Pauline J and Tavernelli, Ivano},
  journal	= {The Journal of Chemical Physics},
  volume	= {154},
  number	= {11},
  pages		= {114105},
  url		= {https://aip.scitation.org/doi/10.1063/5.0029536},
  year		= {2021},
  publisher	= {AIP Publishing LLC}
}

@Article{	  blunt2022perspective,
  title		= {Perspective on the {{Current State-of-the-Art}} of
		  {{Quantum Computing}} for {{Drug Discovery Applications}}},
  author	= {Blunt, Nick S. and Camps, Joan and Crawford, Ophelia and
		  Izs{\'a}k, R{\'o}bert and Leontica, Sebastian and Mirani,
		  Arjun and Moylett, Alexandra E. and Scivier, Sam A. and
		  S{\"u}nderhauf, Christoph and Schopf, Patrick and Taylor,
		  Jacob M. and Holzmann, Nicole},
  year		= {2022},
  month		= dec,
  journal	= {Journal of Chemical Theory and Computation},
  volume	= {18},
  number	= {12},
  pages		= {7001--7023},
  issn		= {1549-9618, 1549-9626},
  doi		= {10.1021/acs.jctc.2c00574},
  url		= {https://pubs.acs.org/doi/10.1021/acs.jctc.2c00574},
  urldate	= {2022-12-16},
  langid	= {english}
}

@article{manin1980computable,
  title={Computable and uncomputable},
  author={Manin, Yuri},
  journal={Sovetskoye Radio, Moscow},
  volume={128},
  pages={15},
  year={1980}
}

@Article{	  tilly2022variational,
  title		= {The {{Variational Quantum Eigensolver}}: {{A}} Review of
		  Methods and Best Practices},
  shorttitle	= {The {{Variational Quantum Eigensolver}}},
  author	= {Tilly, Jules and Chen, Hongxiang and Cao, Shuxiang and
		  Picozzi, Dario and Setia, Kanav and Li, Ying and Grant,
		  Edward and Wossnig, Leonard and Rungger, Ivan and Booth,
		  George H. and Tennyson, Jonathan},
  year		= {2022},
  month		= nov,
  journal	= {Physics Reports},
  volume	= {986},
  pages		= {1--128},
  issn		= {03701573},
  doi		= {10.1016/j.physrep.2022.08.003},
  url		= {https://linkinghub.elsevier.com/retrieve/pii/S0370157322003118},
  urldate	= {2022-12-16},
  langid	= {english}
}

@Article{	  ollitraultmolecularquantumdynamics2021,
  title		= {Molecular {{Quantum Dynamics}}: {{A Quantum Computing
		  Perspective}}},
  shorttitle	= {Molecular {{Quantum Dynamics}}},
  author	= {Ollitrault, Pauline J. and Miessen, Alexander and
		  Tavernelli, Ivano},
  year		= {2021},
  month		= dec,
  journal	= {Accounts of Chemical Research},
  volume	= {54},
  number	= {23},
  pages		= {4229--4238},
  issn		= {0001-4842, 1520-4898},
  doi		= {10.1021/acs.accounts.1c00514},
  url		= {https://pubs.acs.org/doi/10.1021/acs.accounts.1c00514},
  urldate	= {2022-02-02},
  langid	= {english}
}

@Article{	  seeley2012bravyikitaev,
  title		= {The {{Bravyi-Kitaev}} Transformation for Quantum
		  Computation of Electronic Structure},
  author	= {Seeley, Jacob T. and Richard, Martin J. and Love, Peter
		  J.},
  year		= {2012},
  month		= dec,
  journal	= {The Journal of Chemical Physics},
  volume	= {137},
  number	= {22},
  pages		= {224109},
  issn		= {0021-9606, 1089-7690},
  doi		= {10.1063/1.4768229},
  url		= {http://aip.scitation.org/doi/10.1063/1.4768229},
  urldate	= {2022-12-16},
  langid	= {english}
}

@Article{	  kirby2022secondquantized,
  title		= {Second-{{Quantized Fermionic Operators}} with
		  {{Polylogarithmic Qubit}} and {{Gate Complexity}}},
  author	= {Kirby, William and Fuller, Bryce and Hadfield, Charles and
		  Mezzacapo, Antonio},
  year		= {2022},
  month		= jun,
  journal	= {PRX Quantum},
  volume	= {3},
  number	= {2},
  pages		= {020351},
  issn		= {2691-3399},
  doi		= {10.1103/PRXQuantum.3.020351},
  url		= {https://link.aps.org/doi/10.1103/PRXQuantum.3.020351},
  urldate	= {2022-12-16},
  langid	= {english}
}

@Article{	  cleve1998quantum,
  title		= {Quantum algorithms revisited},
  author	= {Cleve, Richard and Ekert, Artur and Macchiavello, Chiara
		  and Mosca, Michele},
  journal	= {Proceedings of the Royal Society of London. Series A:
		  Mathematical, Physical and Engineering Sciences},
  volume	= {454},
  number	= {1969},
  pages		= {339--354},
  url		= {https://royalsocietypublishing.org/doi/10.1098/rspa.1998.0164},
  year		= {1998},
  publisher	= {The Royal Society}
}

@Article{	  shor1999polynomial,
  title		= {Polynomial-time algorithms for prime factorization and
		  discrete logarithms on a quantum computer},
  author	= {Shor, Peter W},
  journal	= {SIAM review},
  volume	= {41},
  number	= {2},
  url		= {https://epubs.siam.org/doi/10.1137/S0097539795293172},
  pages		= {303--332},
  year		= {1999},
  publisher	= {SIAM}
}

@Article{	  reiher2017elucidating,
  title		= {Elucidating reaction mechanisms on quantum computers},
  author	= {Reiher, Markus and Wiebe, Nathan and Svore, Krysta M and
		  Wecker, Dave and Troyer, Matthias},
  journal	= {Proceedings of the national academy of sciences},
  volume	= {114},
  number	= {29},
  pages		= {7555--7560},
  year		= {2017},
  url		= {https://www.pnas.org/doi/10.1073/pnas.1619152114},
  publisher	= {National Acad Sciences}
}

@Article{	  bittel2021training,
  title		= {Training variational quantum algorithms is np-hard},
  author	= {Bittel, Lennart and Kliesch, Martin},
  journal	= {Physical Review Letters},
  volume	= {127},
  number	= {12},
  pages		= {120502},
  url		= {https://journals.aps.org/prl/abstract/10.1103/PhysRevLett.127.120502},
  year		= {2021},
  publisher	= {APS}
}

@Article{	  lee2022there,
  title		= {Is there evidence for exponential quantum advantage in
		  quantum chemistry?},
  author	= {Lee, Seunghoon and Lee, Joonho and Zhai, Huanchen and
		  Tong, Yu and Dalzell, Alexander M and Kumar, Ashutosh and
		  Helms, Phillip and Gray, Johnnie and Cui, Zhi-Hao and Liu,
		  Wenyuan and others},
  journal	= {arXiv:2208.02199},
  url		= {https://arxiv.org/abs/2208.02199},
  year		= {2022}
}

@Article{	  nakanishi2019subspace,
  title		= {Subspace-search variational quantum eigensolver for
		  excited states},
  author	= {Nakanishi, Ken M and Mitarai, Kosuke and Fujii, Keisuke},
  journal	= {Physical Review Research},
  volume	= {1},
  url		= {https://journals.aps.org/prresearch/abstract/10.1103/PhysRevResearch.1.033062},
  number	= {3},
  pages		= {033062},
  year		= {2019},
  publisher	= {APS}
}

@Misc{		  obrien2022purificationbased,
  title		= {Purification-Based Quantum Error Mitigation of
		  Pair-Correlated Electron Simulations},
  author	= {O'Brien, T. E. and Anselmetti, G. and Gkritsis, F. and
		  Elfving, V. E. and Polla, S. and others},
  year		= {2022},
  month		= oct,
  number	= {arXiv:2210.10799},
  eprint	= {2210.10799},
  eprinttype	= {arxiv},
  publisher	= {{arXiv}},
  url		= {http://arxiv.org/abs/2210.10799},
  urldate	= {2022-11-10},
  archiveprefix	= {arXiv}
}

@article{schuster2024polynomial,
  title={A polynomial-time classical algorithm for noisy quantum circuits},
  author={Schuster, Thomas and Yin, Chao and Gao, Xun and Yao, Norman Y},
  journal={arXiv preprint arXiv:2407.12768},
  year={2024},
  url={schuster2024polynomial}
}

@article{angrisani2025simulating,
  title={Simulating quantum circuits with arbitrary local noise using Pauli Propagation},
  author={Angrisani, Armando and Mele, Antonio A and Rudolph, Manuel S and Cerezo, M and Holmes, Zoe},
  journal={arXiv preprint arXiv:2501.13101},
  year={2025},
  url={https://arxiv.org/abs/2501.13101}
}

@Article{	  ma2020quantum,
  title		= {Quantum simulations of materials on near-term quantum
		  computers},
  author	= {Ma, He and Govoni, Marco and Galli, Giulia},
  journal	= {npj Computational Materials},
  volume	= {6},
  number	= {1},
  url		= {https://www.nature.com/articles/s41524-020-00353-z},
  pages		= {1--8},
  year		= {2020},
  publisher	= {Nature Publishing Group}
}

@article{aharonov2025importance,
  title={On the importance of error mitigation for quantum computation},
  author={Aharonov, Dorit and Alberton, Ori and Arad, Itai and Atia, Yosi and Bairey, Eyal and Brakerski, Zvika and Cohen, Itsik and Golan, Omri and Gurwich, Ilya and Kenneth, Oded and others},
  journal={arXiv preprint arXiv:2503.17243},
  year={2025},
  url={https://arxiv.org/abs/2503.17243}
}

@article{cerezo2023does,
  title={Does provable absence of barren plateaus imply classical simulability? or, why we need to rethink variational quantum computing},
  author={Cerezo, Marco and Larocca, Martin and Garc{\'\i}a-Mart{\'\i}n, Diego and Diaz, Nelson L and Braccia, Paolo and Fontana, Enrico and Rudolph, Manuel S and Bermejo, Pablo and Ijaz, Aroosa and Thanasilp, Supanut and others},
  journal={arXiv preprint arXiv:2312.09121},
  year={2023},
  url={https://arxiv.org/abs/2312.09121}
}

@misc{reinholdt2025fundamental,
      title={Fundamental Limitations in Sample-Based Quantum Diagonalization Methods}, 
      author={Peter Reinholdt and Karl Michael Ziems and Erik Rosendahl Kjellgren and Sonia Coriani and Stephan P. A. Sauer and Jacob Kongsted},
      year={2025},
      url={https://arxiv.org/abs/2501.07231}, 
      journal={arXiv preprint arXiv:2501.07231},
      year={2025},
}

@Article{	  wallman2016noise,
  title		= {Noise tailoring for scalable quantum computation via
		  randomized compiling},
  author	= {Wallman, Joel J and Emerson, Joseph},
  journal	= {Physical Review A},
  volume	= {94},
  number	= {5},
  pages		= {052325},
  url		= {https://journals.aps.org/pra/abstract/10.1103/PhysRevA.94.052325},
  year		= {2016},
  publisher	= {APS}
}

@Article{	  crawford_efficient_quantum_2021,
  doi		= {10.22331/q-2021-01-20-385},
  url		= {https://doi.org/10.22331/q-2021-01-20-385},
  title		= {Efficient quantum measurement of {P}auli operators in the
		  presence of finite sampling error},
  author	= {Crawford, Ophelia and Straaten, Barnaby van and Wang,
		  Daochen and Parks, Thomas and Campbell, Earl and Brierley,
		  Stephen},
  journal	= {{Quantum}},
  issn		= {2521-327X},
  publisher	= {{Verein zur F{\"{o}}rderung des Open Access Publizierens
		  in den Quantenwissenschaften}},
  volume	= {5},
  pages		= {385},
  month		= jan,
  year		= {2021}
}

@Article{	  elben2023randomized,
  title		= {The Randomized Measurement Toolbox},
  author	= {Elben, Andreas and Flammia, Steven T. and Huang, Hsin-Yuan
		  and Kueng, Richard and Preskill, John and Vermersch,
		  Beno{\^i}t and Zoller, Peter},
  year		= {2023},
  month		= jan,
  journal	= {Nature Reviews Physics},
  volume	= {5},
  number	= {1},
  pages		= {9--24},
  issn		= {2522-5820},
  doi		= {10.1038/s42254-022-00535-2},
  url		= {https://doi.org/10.1038/s42254-022-00535-2}
}

@article{bertoni2024shallowshadows,
  title = {Shallow Shadows: Expectation Estimation Using Low-Depth Random Clifford Circuits},
  author = {Bertoni, Christian and Haferkamp, Jonas and Hinsche, Marcel and Ioannou, Marios and Eisert, Jens and Pashayan, Hakop},
  journal = {Phys. Rev. Lett.},
  volume = {133},
  issue = {2},
  pages = {020602},
  numpages = {7},
  year = {2024},
  month = {Jul},
  publisher = {American Physical Society},
  doi = {10.1103/PhysRevLett.133.020602},
  url = {https://link.aps.org/doi/10.1103/PhysRevLett.133.020602}
}

@article{childs2012hamiltonian,
  title={Hamiltonian simulation using linear combinations of unitary operations},
  author={Childs, Andrew M and Wiebe, Nathan},
  journal={Quantum Information \& Computation},
  volume={12},
  number={11-12},
  pages={901--924},
  year={2012},
  publisher={Rinton Press, Incorporated Paramus, NJ},
  doi={https://doi.org/10.26421/QIC12.11-12}
}

@article{quantum_signal_prcoessing,
  title = {Optimal Hamiltonian Simulation by Quantum Signal Processing},
  author = {Low, Guang Hao and Chuang, Isaac L.},
  journal = {Phys. Rev. Lett.},
  volume = {118},
  issue = {1},
  pages = {010501},
  numpages = {5},
  year = {2017},
  month = {Jan},
  publisher = {American Physical Society},
  doi = {10.1103/PhysRevLett.118.010501},
  url = {https://link.aps.org/doi/10.1103/PhysRevLett.118.010501}
}

@article{innocenti2023shadow,
  title = {Shadow Tomography on General Measurement Frames},
  author = {Innocenti, L. and Lorenzo, S. and Palmisano, I. and Albarelli, F. and Ferraro, A. and Paternostro, M. and Palma, G. M.},
  journal = {PRX Quantum},
  volume = {4},
  issue = {4},
  pages = {040328},
  numpages = {27},
  year = {2023},
  month = {Nov},
  publisher = {American Physical Society},
  doi = {10.1103/PRXQuantum.4.040328},
  url = {https://link.aps.org/doi/10.1103/PRXQuantum.4.040328}
}

@Article{	  nikolaeva2022decomposing,
  title		= {Decomposing the Generalized {{Toffoli}} Gate with
		  Qutrits},
  author	= {Nikolaeva, Anastasiia S. and Kiktenko, Evgeniy O. and
		  Fedorov, Aleksey K.},
  year		= {2022},
  month		= mar,
  journal	= {Physical Review A},
  volume	= {105},
  number	= {3},
  pages		= {032621},
  issn		= {2469-9926, 2469-9934},
  doi		= {10.1103/PhysRevA.105.032621},
  url		= {https://link.aps.org/doi/10.1103/PhysRevA.105.032621},
  urldate	= {2022-12-19},
  langid	= {english}
}

@inproceedings{bronn2017fast,
  title={Fast, high-fidelity readout of multiple qubits},
  author={Bronn, Nicholas T and Abdo, Baleegh and Inoue, Ken and Lekuch, Scott and C{\'o}rcoles, Antonio D and Hertzberg, Jared B and Takita, Maika and Bishop, Lev S and Gambetta, Jay M and Chow, Jerry M},
  booktitle={Journal of Physics: Conference Series},
  volume={834},
  number={1},
  pages={012003},
  year={2017},
  organization={IOP Publishing},
  doi={10.1088/1742-6596/834/1/012003}
}

@article{papivc2025near,
  title={Near-Term Fermionic Simulation with Subspace Noise Tailored Quantum Error Mitigation},
  author={Papi{\v{c}}, Miha and Algaba, Manuel G and Godinez-Ramirez, Emiliano and de Vega, In{\'e}s and Auer, Adrian and {\v{S}}imkovic IV, Fedor and Calzona, Alessio},
  journal={arXiv preprint arXiv:2503.11785},
  year={2025},
  doi={https://doi.org/10.48550/arXiv.2503.11785}
}

@book{engel2011density,
  title={Density functional theory},
  author={Engel, Eberhard},
  year={2011},
  publisher={Springer},
  doi={https://doi.org/10.1007/978-3-642-14090-7}
}

@article{bishop1991overview,
  title={An overview of coupled cluster theory and its applications in physics},
  author={Bishop, RF},
  journal={Theoretica chimica acta},
  volume={80},
  number={2},
  pages={95--148},
  year={1991},
  publisher={Springer},
  doi={https://doi.org/10.1007/BF01119617}
}

@article{CPMD,
author = {Hutter, Jürg},
title = {Car–Parrinello molecular dynamics},
journal = {WIREs Computational Molecular Science},
volume = {2},
number = {4},
pages = {604-612},
doi = {https://doi.org/10.1002/wcms.90},
url = {https://wires.onlinelibrary.wiley.com/doi/abs/10.1002/wcms.90},
year = {2012}
}

@book{schulman2012techniques,
  title={Techniques and applications of path integration},
  author={Schulman, Lawrence S},
  year={2012},
  publisher={Courier Corporation},
  doi={https://doi.org/10.1063/1.2914703}
}

@article{herman2023quantum,
  title={Quantum computing for finance},
  author={Herman, Dylan and Googin, Cody and Liu, Xiaoyuan and Sun, Yue and Galda, Alexey and Safro, Ilya and Pistoia, Marco and Alexeev, Yuri},
  journal={Nature Reviews Physics},
  volume={5},
  number={8},
  pages={450--465},
  year={2023},
  publisher={Nature Publishing Group UK London},
  doi={https://doi.org/10.1038/s42254-023-00603-1}
}

@book{schuld2021machine,
  title={Machine learning with quantum computers},
  author={Schuld, Maria and Petruccione, Francesco},
  volume={676},
  year={2021},
  publisher={Springer},
  doi={https://doi.org/10.1007/978-3-030-83098-4}
}

@Article{	  rigetti2010fully,
  title		= {Fully Microwave-Tunable Universal Gates in Superconducting
		  Qubits with Linear Couplings and Fixed Transition
		  Frequencies},
  author	= {Rigetti, Chad and Devoret, Michel},
  year		= {2010},
  month		= apr,
  journal	= {Physical Review B},
  volume	= {81},
  number	= {13},
  pages		= {134507},
  issn		= {1098-0121, 1550-235X},
  doi		= {10.1103/PhysRevB.81.134507},
  url		= {https://link.aps.org/doi/10.1103/PhysRevB.81.134507},
  urldate	= {2023-04-04},
  langid	= {english}
}

@Article{	  patterson2019calibration,
  title		= {Calibration of a {{Cross-Resonance Two-Qubit Gate Between
		  Directly Coupled Transmons}}},
  author	= {Patterson, A.D. and Rahamim, J. and Tsunoda, T. and
		  Spring, P.A. and Jebari, S. and Ratter, K. and
		  Mergenthaler, M. and Tancredi, G. and Vlastakis, B. and
		  Esposito, M. and Leek, P.J.},
  year		= {2019},
  month		= dec,
  journal	= {Physical Review Applied},
  volume	= {12},
  number	= {6},
  pages		= {064013},
  issn		= {2331-7019},
  doi		= {10.1103/PhysRevApplied.12.064013},
  url		= {https://link.aps.org/doi/10.1103/PhysRevApplied.12.064013},
  urldate	= {2023-04-04},
  langid	= {english}
}

@Article{	  iten2016quantum,
  title		= {Quantum Circuits for Isometries},
  author	= {Iten, Raban and Colbeck, Roger and Kukuljan, Ivan and
		  Home, Jonathan and Christandl, Matthias},
  year		= {2016},
  month		= mar,
  journal	= {Physical Review A},
  volume	= {93},
  number	= {3},
  pages		= {032318},
  issn		= {2469-9926, 2469-9934},
  doi		= {10.1103/PhysRevA.93.032318},
  url		= {https://link.aps.org/doi/10.1103/PhysRevA.93.032318},
  urldate	= {2023-04-25},
  langid	= {english}
}

@Article{	  chen2013qcompiler,
  title		= {Qcompiler: {{Quantum}} Compilation with the {{CSD}}
		  Method},
  shorttitle	= {Qcompiler},
  author	= {Chen, Y.G. and Wang, J.B.},
  year		= {2013},
  month		= mar,
  journal	= {Computer Physics Communications},
  volume	= {184},
  number	= {3},
  pages		= {853--865},
  issn		= {00104655},
  doi		= {10.1016/j.cpc.2012.10.019},
  url		= {https://linkinghub.elsevier.com/retrieve/pii/S0010465512003621},
  urldate	= {2023-04-25},
  langid	= {english}
}

@Article{	  cao2023emulatinga,
  title		= {Emulating Two Qubits with a Four-Level Transmon Qudit for
		  Variational Quantum Algorithms},
  author	= {Cao, Shuxiang and Bakr, Mustafa and Campanaro, Giulio and
		  Fasciati, Simone D. and Wills, James and Lall, Deep and
		  Shteynas, Boris and Chidambaram, Vivek and Rungger, Ivan
		  and Leek, Peter},
  year		= {2023},
  month		= mar,
  journal	= {arXiv:2303.04796},
  publisher	= {{arXiv}},
  url		= {http://arxiv.org/abs/2303.04796},
  urldate	= {2023-04-26},
  archiveprefix	= {arxiv}
}

@Article{	  seifert2023exploring,
  title		= {Exploring {{Ququart Computation}} on a {{Transmon}} Using
		  {{Optimal Control}}},
  author	= {Seifert, Lennart Maximilian and Li, Ziqian and Roy, Tanay
		  and Schuster, David I. and Chong, Frederic T. and Baker,
		  Jonathan M.},
  year		= {2023},
  month		= apr,
  journal	= {arXiv:2304.11159},
  publisher	= {{arXiv}},
  url		= {http://arxiv.org/abs/2304.11159},
  urldate	= {2023-04-26},
  archiveprefix	= {arxiv}
}

@Article{	  liu2023performing,
  title		= {Performing $\mathrm{SU}(d)$ Operations and Rudimentary
		  Algorithms in a Superconducting Transmon Qudit for $d=3$
		  and $d=4$},
  author	= {Liu, Pei and Wang, Ruixia and Zhang, Jing-Ning and Zhang,
		  Yingshan and Cai, Xiaoxia and Xu, Huikai and Li, Zhiyuan
		  and Han, Jiaxiu and Li, Xuegang and Xue, Guangming and Liu,
		  Weiyang and You, Li and Jin, Yirong and Yu, Haifeng},
  journal	= {Phys. Rev. X},
  volume	= {13},
  issue		= {2},
  pages		= {021028},
  numpages	= {21},
  year		= {2023},
  publisher	= {American Physical Society},
  doi		= {10.1103/PhysRevX.13.021028},
  url		= {https://link.aps.org/doi/10.1103/PhysRevX.13.021028}
}

@Article{	  simm2023two,
  title		= {Two qubits in one transmon--QEC without ancilla hardware},
  author	= {Simm, Alexander and Machnes, Shai and Wilhelm, Frank K},
  journal	= {arXiv:2302.14707},
  year		= {2023},
  url		= {https://arxiv.org/abs/2302.14707}
}

@Article{	  hadfield2022measurements,
  title		= {Measurements of {{Quantum Hamiltonians}} with
		  {{Locally-Biased Classical Shadows}}},
  author	= {Hadfield, Charles and Bravyi, Sergey and Raymond, Rudy and
		  Mezzacapo, Antonio},
  year		= {2022},
  month		= may,
  journal	= {Communications in Mathematical Physics},
  volume	= {391},
  number	= {3},
  pages		= {951--967},
  issn		= {0010-3616, 1432-0916},
  doi		= {10.1007/s00220-022-04343-8},
  url		= {https://link.springer.com/10.1007/s00220-022-04343-8},
  urldate	= {2022-07-04},
  langid	= {english}
}

@Article{	  li_general_1995,
  title		= {On General Frame Decompositions},
  author	= {Li, Shidong},
  year		= {1995},
  month		= jan,
  journal	= {Numerical Functional Analysis and Optimization},
  volume	= {16},
  number	= {9-10},
  pages		= {1181--1191},
  issn		= {0163-0563, 1532-2467},
  doi		= {10.1080/01630569508816668},
  url		= {http://www.tandfonline.com/doi/abs/10.1080/01630569508816668},
  urldate	= {2024-01-24},
  langid	= {english}
}

@Article{	  watanabe_information_1960,
  title		= {Information {{Theoretical Analysis}} of {{Multivariate
		  Correlation}}},
  author	= {Watanabe, Satosi},
  year		= {1960},
  month		= jan,
  journal	= {IBM Journal of Research and Development},
  volume	= {4},
  number	= {1},
  pages		= {66--82},
  issn		= {0018-8646, 0018-8646},
  doi		= {10.1147/rd.41.0066},
  url		= {http://ieeexplore.ieee.org/document/5392532/},
  urldate	= {2024-01-24}
}

@Article{	  wan2023matchgate,
  title		= {Matchgate {{Shadows}} for {{Fermionic Quantum
		  Simulation}}},
  author	= {Wan, Kianna and Huggins, William J. and Lee, Joonho and
		  Babbush, Ryan},
  year		= {2023},
  month		= dec,
  journal	= {Communications in Mathematical Physics},
  volume	= {404},
  number	= {2},
  pages		= {629--700},
  issn		= {1432-0916},
  doi		= {10.1007/s00220-023-04844-0},
  url		= {https://doi.org/10.1007/s00220-023-04844-0}
}

@Article{	  helsen2022thrifty,
  title		= {Thrifty Shadow Estimation: Reusing Quantum Circuits and
		  Bounding Tails},
  author	= {Helsen, Jonas and Walter, Michael},
  journal	= {Phys. Rev. Lett.},
  volume	= {131},
  issue		= {24},
  pages		= {240602},
  numpages	= {6},
  year		= {2023},
  publisher	= {American Physical Society},
  doi		= {10.1103/PhysRevLett.131.240602},
  url		= {https://link.aps.org/doi/10.1103/PhysRevLett.131.240602}
}

@article{malmi2024enhanced,
  title = {Enhanced observable estimation through classical optimization of informationally overcomplete measurement data: Beyond classical shadows},
  author = {Malmi, Joonas and Korhonen, Keijo and Cavalcanti, Daniel and Garc\'{\i}a-P\'erez, Guillermo},
  journal = {Phys. Rev. A},
  volume = {109},
  issue = {6},
  pages = {062412},
  numpages = {6},
  year = {2024},
  month = {Jun},
  publisher = {American Physical Society},
  doi = {10.1103/PhysRevA.109.062412},
  url = {https://link.aps.org/doi/10.1103/PhysRevA.109.062412}
}

@article{caprotti2024optimising,
  title = {Optimizing quantum tomography via shadow inversion},
  author = {Caprotti, Andrea and Morris, Joshua and Daki\ifmmode \acute{c}\else \'{c}\fi{}, Borivoje},
  journal = {Phys. Rev. Res.},
  volume = {6},
  issue = {3},
  pages = {033301},
  numpages = {7},
  year = {2024},
  month = {Sep},
  publisher = {American Physical Society},
  doi = {10.1103/PhysRevResearch.6.033301},
  url = {https://link.aps.org/doi/10.1103/PhysRevResearch.6.033301}
}

@article{fischer2024dual,
  title={Dual-frame optimization for informationally complete quantum measurements},
  author={Fischer, Laurin E and Dao, Timoth{\'e}e and Tavernelli, Ivano and Tacchino, Francesco},
  journal={Physical Review A},
  volume={109},
  number={6},
  pages={062415},
  year={2024},
  publisher={APS},
  doi={10.1103/PhysRevA.109.062415}
}

@article{fischer2022ancillafree,
  title = {Ancilla-Free Implementation of Generalized Measurements for Qubits Embedded in a Qudit Space},
  author = {Fischer, Laurin E. and Miller, Daniel and Tacchino, Francesco and Barkoutsos, Panagiotis Kl. and Egger, Daniel J. and Tavernelli, Ivano},
  year = {2022},
  month = jul,
  journal = {Physical Review Research},
  volume = {4},
  number = {3},
  pages = {033027},
  issn = {2643-1564},
  doi = {10.1103/PhysRevResearch.4.033027},
  url = {https://link.aps.org/doi/10.1103/PhysRevResearch.4.033027},
  urldate = {2022-08-04},
  langid = {english}
}

@Article{	  hu2023scrambled,
  title		= {Classical shadow tomography with locally scrambled quantum
		  dynamics},
  author	= {Hu, Hong-Ye and Choi, Soonwon and You, Yi-Zhuang},
  journal	= {Phys. Rev. Res.},
  volume	= {5},
  issue		= {2},
  pages		= {023027},
  numpages	= {21},
  year		= {2023},
  publisher	= {American Physical Society},
  doi		= {10.1103/PhysRevResearch.5.023027},
  url		= {https://link.aps.org/doi/10.1103/PhysRevResearch.5.023027}
}

@Article{	  seif2023shadow,
  title		= {Shadow Distillation: Quantum Error Mitigation with
		  Classical Shadows for Near-Term Quantum Processors},
  author	= {Seif, Alireza and Cian, Ze-Pei and Zhou, Sisi and Chen,
		  Senrui and Jiang, Liang},
  journal	= {PRX Quantum},
  volume	= {4},
  issue		= {1},
  pages		= {010303},
  numpages	= {14},
  year		= {2023},
  publisher	= {American Physical Society},
  doi		= {10.1103/PRXQuantum.4.010303},
  url		= {https://link.aps.org/doi/10.1103/PRXQuantum.4.010303}
}

@Article{	  levy2021classical,
  title		= {Classical shadows for quantum process tomography on
		  near-term quantum computers},
  author	= {Levy, Ryan and Luo, Di and Clark, Bryan K.},
  journal	= {Phys. Rev. Res.},
  volume	= {6},
  issue		= {1},
  pages		= {013029},
  numpages	= {18},
  year		= {2024},
  month		= {Jan},
  publisher	= {American Physical Society},
  doi		= {10.1103/PhysRevResearch.6.013029},
  url		= {https://link.aps.org/doi/10.1103/PhysRevResearch.6.013029}
}

@Article{	  gyurik2023limitations,
  title		= {Limitations of measure-first protocols in quantum machine
		  learning},
  author	= {Gyurik, Casper and Molteni, Riccardo and Dunjko, Vedran},
  journal	= { arXiv:2311.12618},
  year		= {2023},
  url		= {https://arxiv.org/abs/2311.12618}
}

@Article{	  jerbi2023shadows,
  title		= {Shadows of quantum machine learning},
  author	= {Jerbi, Sofiene and Gyurik, Casper and Marshall, Simon C
		  and Molteni, Riccardo and Dunjko, Vedran},
  journal	= { arXiv:2306.00061},
  year		= {2023},
  url		= {https://arxiv.org/abs/2306.00061}
}

@Article{	  huang2022provably,
  author	= {Hsin-Yuan Huang and Richard Kueng and Giacomo Torlai and
		  Victor V. Albert and John Preskill },
  title		= {Provably efficient machine learning for quantum many-body
		  problems},
  journal	= {Science},
  volume	= {377},
  number	= {6613},
  year		= {2022},
  url		= {https://www.science.org/doi/abs/10.1126/science.abk3333}
}

@Article{	  torlai2018neural,
  title		= {Neural-Network Quantum State Tomography},
  author	= {Torlai, Giacomo and Mazzola, Guglielmo and Carrasquilla,
		  Juan and Troyer, Matthias and Melko, Roger and Carleo,
		  Giuseppe},
  year		= {2018},
  month		= may,
  journal	= {Nature Physics},
  volume	= {14},
  number	= {5},
  pages		= {447--450},
  issn		= {1745-2473, 1745-2481},
  doi		= {10.1038/s41567-018-0048-5},
  url		= {https://www.nature.com/articles/s41567-018-0048-5},
  urldate	= {2024-01-24},
  langid	= {english}
}

@Article{	  garcia2021scrambling,
  title		= {Quantum scrambling with classical shadows},
  author	= {Garcia, Roy J. and Zhou, You and Jaffe, Arthur},
  journal	= {Phys. Rev. Res.},
  volume	= {3},
  issue		= {3},
  pages		= {033155},
  numpages	= {21},
  year		= {2021},
  month		= {Aug},
  publisher	= {American Physical Society},
  doi		= {10.1103/PhysRevResearch.3.033155},
  url		= {https://link.aps.org/doi/10.1103/PhysRevResearch.3.033155}
}

@Article{	  joshi2022probing,
  title		= {Probing Many-Body Quantum Chaos with Quantum Simulators},
  author	= {Joshi, Lata Kh and Elben, Andreas and Vikram, Amit and
		  Vermersch, Beno\^{\i}t and Galitski, Victor and Zoller,
		  Peter},
  journal	= {Phys. Rev. X},
  volume	= {12},
  issue		= {1},
  pages		= {011018},
  numpages	= {34},
  year		= {2022},
  month		= {Jan},
  publisher	= {American Physical Society},
  doi		= {10.1103/PhysRevX.12.011018},
  url		= {https://link.aps.org/doi/10.1103/PhysRevX.12.011018}
}

@Article{	  fano1961,
  title		= {Transmission of information: A statistical theory of
		  communications},
  author	= {Fano, Robert M and Hawkins, David},
  journal	= {American Journal of Physics},
  volume	= {29},
  number	= {11},
  pages		= {793--794},
  year		= {1961},
  publisher	= {AIP Publishing},
  url		= {https://doi.org/10.1119/1.1937609}
}

@Article{	  kim2023evidence,
  title		= {Evidence for the utility of quantum computing before fault
		  tolerance},
  author	= {Kim, Youngseok and Eddins, Andrew and Anand, Sajant and
		  Wei, Ken Xuan and Van Den Berg, Ewout and Rosenblatt, Sami
		  and Nayfeh, Hasan and Wu, Yantao and Zaletel, Michael and
		  Temme, Kristan and others},
  journal	= {Nature},
  volume	= {618},
  number	= {7965},
  pages		= {500--505},
  year		= {2023},
  publisher	= {Nature Publishing Group UK London},
  doi		= {10.1038/s41586-023-06096-3}
}

@Book{		  nielsen2010quantum,
  title		= {Quantum computation and quantum information},
  author	= {Nielsen, Michael A and Chuang, Isaac L},
  year		= {2010},
  publisher	= {Cambridge university press}
}

@article{anand2023classical,
  title={Classical benchmarking of zero noise extrapolation beyond the exactly-verifiable regime},
  author={Anand, Sajant and Temme, Kristan and Kandala, Abhinav and Zaletel, Michael},
  journal={arXiv preprint arXiv:2306.17839},
  year={2023},
  doi={https://doi.org/10.48550/arXiv.2306.17839}
}

@article{vandenBerg2024techniqueslearning,
  doi = {10.22331/q-2024-12-10-1556},
  url = {https://doi.org/10.22331/q-2024-12-10-1556},
  title = {Techniques for learning sparse {P}auli-{L}indblad noise models},
  author = {van den Berg, Ewout and Wocjan, Pawel},
  journal = {{Quantum}},
  issn = {2521-327X},
  publisher = {{Verein zur F{\"{o}}rderung des Open Access Publizierens in den Quantenwissenschaften}},
  volume = {8},
  pages = {1556},
  month = dec,
  year = {2024}
}

@article{abanin2025constructive,
  title={Constructive interference at the edge of quantum ergodic dynamics},
  author={Abanin, Dmitry A and Acharya, Rajeev and Aghababaie-Beni, Laleh and Aigeldinger, Georg and Ajoy, Ashok and Alcaraz, Ross and Aleiner, Igor and Andersen, Trond I and Ansmann, Markus and Arute, Frank and others},
  journal={arXiv preprint arXiv:2506.10191},
  year={2025},
  doi={https://doi.org/10.48550/arXiv.2506.10191}
}

@article{endo2018practical,
  title={Practical quantum error mitigation for near-future applications},
  author={Endo, Suguru and Benjamin, Simon C and Li, Ying},
  journal={Physical Review X},
  volume={8},
  number={3},
  pages={031027},
  year={2018},
  publisher={APS},
  url={https://doi.org/10.1103/PhysRevX.8.031027}
}

@Article{	  zhao2021fermionic,
  title		= {Fermionic Partial Tomography via Classical Shadows},
  author	= {Zhao, Andrew and Rubin, Nicholas C. and Miyake, Akimasa},
  journal	= {Phys. Rev. Lett.},
  volume	= {127},
  issue		= {11},
  pages		= {110504},
  numpages	= {8},
  year		= {2021},
  publisher	= {American Physical Society},
  doi		= {10.1103/PhysRevLett.127.110504},
  url		= {https://link.aps.org/doi/10.1103/PhysRevLett.127.110504}
}

@article{morales2025selection,
  title={Selection and improvement of product formulae for best performance of quantum simulation},
  author={Morales, Mauro ES and Costa, Pedro and Pantaleoni, Giacomo and Burgarth, Daniel K and Sanders, Yuval R and Berry, Dominic W},
  journal={Quantum Information \& Computation},
  volume={25},
  number={1},
  pages={1--35},
  year={2025},
  doi={https://doi.org/10.2478/qic-2025-0001}
}

@article{miessen2024benchmarking,
  title = {Benchmarking Digital Quantum Simulations Above Hundreds of Qubits Using Quantum Critical Dynamics},
  author = {Miessen, Alexander and Egger, Daniel J. and Tavernelli, Ivano and Mazzola, Guglielmo},
  journal = {PRX Quantum},
  volume = {5},
  issue = {4},
  pages = {040320},
  numpages = {19},
  year = {2024},
  month = {Nov},
  publisher = {American Physical Society},
  doi = {10.1103/PRXQuantum.5.040320},
  url = {https://link.aps.org/doi/10.1103/PRXQuantum.5.040320}
}

@article{hope2025quantum,
   author = "Zehr, Hope and Baiardi, Alberto and Tacchino, Francesco and Gandon, Anthony and Fischer, Laurin E. and Xu, Yue and DiFilippo, Frank P. and Guidoni, Leonardo and Haase, Pi A.B. and Talarico, Walter N. and Stella, Martina and Tarocco, Fabio and Nykänen, Anton and Fitzpatrick, Aaron and Miller, Aaron and Thiessen, Leander and Knecht, Stefan and Borrelli, Elsi-Mari and Maniscalco, Sabrina and Pavošević, Fabijan and Tavernelli, Ivano and Maytin, Edward and Krishna, Vijay",
   title = "Quantum Computing for Photosensitizer Design in Photodynamic Therapy",
   journal = "Annual Review of Biomedical Data Science",
   year = "2025",
   publisher = "Annual Reviews",
   url = "https://www.annualreviews.org/content/journals/10.1146/annurev-biodatasci-103123-095644",
   doi = "https://doi.org/10.1146/annurev-biodatasci-103123-095644"
}

@article{born1925quantenmechanik,
  title={Zur quantenmechanik},
  author={Born, Max and Jordan, Pascual},
  journal={Zeitschrift f{\"u}r Physik},
  volume={34},
  number={1},
  pages={858--888},
  year={1925},
  publisher={Springer}
}

@Article{	  huang_efficient_2021,
  title		= {Efficient Estimation of Pauli Observables by
		  Derandomization},
  volume	= {127},
  url		= {https://link.aps.org/doi/10.1103/PhysRevLett.127.030503},
  doi		= {10.1103/PhysRevLett.127.030503},
  pages		= {030503},
  number	= {3},
  journaltitle	= {Physical Review Letters},
  journal	= {Phys. Rev. Lett.},
  author	= {Huang, Hsin-Yuan and Kueng, Richard and Preskill, John},
  urldate	= {2023-03-06},
  date		= {2021-07-16},
  year		= {2021}
}

@Article{	  vermersch2023many,
  title		= {Many-body entropies and entanglement from
		  polynomially-many local measurements},
  author	= {Vermersch, Beno{\^\i}t and Ljubotina, Marko and Cirac, J
		  Ignacio and Zoller, Peter and Serbyn, Maksym and Piroli,
		  Lorenzo},
  journal	= { arXiv:2311.08108},
  url		= {http://arxiv.org/abs/2311.08108},
  year		= {2023}
}

@Article{	  filippov2023scalable,
  title		= {Scalable tensor-network error mitigation for near-term
		  quantum computing},
  author	= {Filippov, Sergei and Leahy, Matea and Rossi, Matteo AC and
		  Garc{\'\i}a-P{\'e}rez, Guillermo},
  journal	= { arXiv:2307.11740},
  url		= {http://arxiv.org/abs/2307.11740},
  year		= {2023}
}

@Article{	  hayashi_optimal_2006,
  title		= {Optimal Estimation of a Physical Observable's Expectation
		  Value for Pure States},
  author	= {Hayashi, A. and Horibe, M. and Hashimoto, T.},
  year		= {2006},
  month		= jun,
  journal	= {Physical Review A},
  volume	= {73},
  number	= {6},
  pages		= {062322},
  issn		= {1050-2947, 1094-1622},
  doi		= {10.1103/PhysRevA.73.062322},
  url		= {https://link.aps.org/doi/10.1103/PhysRevA.73.062322},
  urldate	= {2024-01-24},
  langid	= {english}
}

@Article{	  dariano_optimal_2006,
  title		= {Optimal Estimation of Quantum Observables},
  author	= {D'Ariano, Giacomo Mauro and Giovannetti, Vittorio and
		  Perinotti, Paolo},
  year		= {2006},
  month		= feb,
  journal	= {Journal of Mathematical Physics},
  volume	= {47},
  number	= {2},
  pages		= {022102},
  issn		= {0022-2488, 1089-7658},
  doi		= {10.1063/1.2168122},
  url		= {https://pubs.aip.org/jmp/article/47/2/022102/921200/Optimal-estimation-of-quantum-observables},
  urldate	= {2024-01-24},
  langid	= {english}
}

@Article{	  huang_predicting_2020,
  title		= {Predicting Many Properties of a Quantum System from Very
		  Few Measurements},
  author	= {Huang, Hsin-Yuan and Kueng, Richard and Preskill, John},
  year		= {2020},
  month		= oct,
  journal	= {Nature Physics},
  volume	= {16},
  number	= {10},
  pages		= {1050--1057},
  issn		= {1745-2473, 1745-2481},
  doi		= {10.1038/s41567-020-0932-7},
  url		= {https://www.nature.com/articles/s41567-020-0932-7},
  urldate	= {2023-07-27},
  langid	= {english}
}

@Article{	  d2004informationally,
  title		= {Informationally complete measurements and group
		  representation},
  author	= {d’Ariano, GM and Perinotti, P and Sacchi, MF},
  journal	= {Journal of Optics B: Quantum and Semiclassical Optics},
  volume	= {6},
  number	= {6},
  pages		= {S487},
  year		= {2004},
  publisher	= {IOP Publishing},
  url		= {https://iopscience.iop.org/article/10.1088/1464-4266/6/6/005}
}

@Article{	  krahmer_sparsity_2013,
  title		= {Sparsity and Spectral Properties of Dual Frames},
  author	= {Krahmer, Felix and Kutyniok, Gitta and Lemvig, Jakob},
  year		= {2013},
  month		= aug,
  journal	= {Linear Algebra and its Applications},
  volume	= {439},
  number	= {4},
  pages		= {982--998},
  issn		= {00243795},
  doi		= {10.1016/j.laa.2012.10.016},
  url		= {https://linkinghub.elsevier.com/retrieve/pii/S0024379512007264},
  urldate	= {2024-01-24},
  langid	= {english}
}

@Article{	  dariano_classical_2005,
  doi		= {10.1088/0305-4470/38/26/010},
  url		= {https://dx.doi.org/10.1088/0305-4470/38/26/010},
  year		= {2005},
  publisher	= {},
  volume	= {38},
  number	= {26},
  pages		= {5979},
  author	= {Giacomo Mauro D'Ariano and Paoloplacido Lo Presti and
		  Paolo Perinotti},
  title		= {Classical randomness in quantum measurements},
  journal	= {Journal of Physics A: Mathematical and General}
}

@Article{	  struchalin2021,
  title		= {Experimental Estimation of Quantum State Properties from
		  Classical Shadows},
  author	= {Struchalin, G.I. and Zagorovskii, Ya. A. and Kovlakov,
		  E.V. and Straupe, S.S. and Kulik, S.P.},
  journal	= {PRX Quantum},
  volume	= {2},
  issue		= {1},
  pages		= {010307},
  numpages	= {11},
  year		= {2021},
  month		= {Jan},
  publisher	= {American Physical Society},
  doi		= {10.1103/PRXQuantum.2.010307},
  url		= {https://link.aps.org/doi/10.1103/PRXQuantum.2.010307}
}

@Article{	  garcia-perez_learning_2021,
  title		= {Learning to measure: Adaptive informationally complete
		  generalized measurements for quantum algorithms},
  author	= {Garc{\'\i}a-P{\'e}rez, Guillermo and Rossi, Matteo AC and
		  Sokolov, Boris and Tacchino, Francesco and Barkoutsos,
		  Panagiotis Kl and Mazzola, Guglielmo and Tavernelli, Ivano
		  and Maniscalco, Sabrina},
  journal	= {PRX Quantum},
  url		= {https://link.aps.org/doi/10.1103/PRXQuantum.2.040342},
  volume	= {2},
  number	= {4},
  pages		= {040342},
  year		= {2021},
  publisher	= {APS}
}

@Article{	  gonthier2022measurements,
  title		= {Measurements as a roadblock to near-term practical quantum
		  advantage in chemistry: Resource analysis},
  author	= {Gonthier, J\'er\^ome F. and Radin, Maxwell D. and Buda,
		  Corneliu and Doskocil, Eric J. and Abuan, Clena M. and
		  Romero, Jhonathan},
  journal	= {Phys. Rev. Res.},
  volume	= {4},
  issue		= {3},
  pages		= {033154},
  numpages	= {14},
  year		= {2022},
  month		= {Aug},
  publisher	= {American Physical Society},
  doi		= {10.1103/PhysRevResearch.4.033154},
  url		= {https://link.aps.org/doi/10.1103/PhysRevResearch.4.033154}
}

@article{ivashkov2024high,
  title={High-fidelity, multiqubit generalized measurements with dynamic circuits},
  author={Ivashkov, Petr and Uchehara, Gideon and Jiang, Liang and Wang, Derek S and Seif, Alireza},
  journal={PRX Quantum},
  volume={5},
  number={3},
  pages={030315},
  year={2024},
  publisher={APS},
  doi={https://doi.org/10.1103/PRXQuantum.5.030315}
}

@Article{	  jnane2024error,
  title		= {Quantum Error Mitigated Classical Shadows},
  author	= {Jnane, Hamza and Steinberg, Jonathan and Cai, Zhenyu and
		  Nguyen, H. Chau and Koczor, B\'alint},
  journal	= {PRX Quantum},
  volume	= {5},
  issue		= {1},
  pages		= {010324},
  numpages	= {21},
  year		= {2024},
  month		= {Feb},
  publisher	= {American Physical Society},
  doi		= {10.1103/PRXQuantum.5.010324},
  url		= {https://link.aps.org/doi/10.1103/PRXQuantum.5.010324}
}

@Article{	  gopalakrishnan-2019,
  title		= {Unitary circuits of finite depth and infinite width from
		  quantum channels},
  author	= {Gopalakrishnan, Sarang and Lamacraft, Austen},
  journal	= {Phys. Rev. B},
  volume	= {100},
  issue		= {6},
  pages		= {064309},
  numpages	= {15},
  year		= {2019},
  month		= {Aug},
  publisher	= {American Physical Society},
  doi		= {10.1103/PhysRevB.100.064309},
  url		= {https://link.aps.org/doi/10.1103/PhysRevB.100.064309}
}

@thesis{zhu_quantum_2012,
	title = {Quantum State Estimation and Symmetric Informationally Complete {POMs}},
	url = {https://scholarbank.nus.edu.sg/handle/10635/35247},
	type = {Thesis},
	author = {Zhu, Huangjun},
	urldate = {2023-04-02},
	date = {2012-03-30},
	langid = {english},
    year={2012}
}

@Article{	  mi2022time,
  title		= {Time-crystalline eigenstate order on a quantum processor},
  author	= {Mi, Xiao and Ippoliti, Matteo and Quintana, Chris and
		  Greene, Ami and Chen, Zijun and Gross, Jonathan and Arute,
		  Frank and Arya, Kunal and Atalaya, Juan and Babbush, Ryan
		  and others},
  journal	= {Nature},
  volume	= {601},
  number	= {7894},
  pages		= {531--536},
  year		= {2022},
  publisher	= {Nature Publishing Group UK London},
  doi		= {https://doi.org/10.1038/s41586-021-04257-w}
}

@Article{	  miessen2023quantum,
  title		= {Quantum algorithms for quantum dynamics},
  author	= {Miessen, Alexander and Ollitrault, Pauline J and Tacchino,
		  Francesco and Tavernelli, Ivano},
  journal	= {Nature Computational Science},
  doi		= {10.1038/s43588-022-00374-2},
  volume	= {3},
  number	= {1},
  pages		= {25--37},
  year		= {2023},
  publisher	= {Nature Publishing Group US New York}
}

@Misc{		  robertson2024,
  title		= {Tensor Network enhanced Dynamic Multiproduct Formulas},
  author	= {Niall F. Robertson and Bibek Pokharel and Bryce Fuller and
		  Eric Switzer and Oles Shtanko and Mirko Amico and Adam
		  Byrne and Andrea D'Urbano and Salome Hayes-Shuptar and
		  Albert Akhriev and Nathan Keenan and Sergey Bravyi and
		  Sergiy Zhuk},
  year		= {2024},
  eprint	= {2407.17405},
  archiveprefix	= {arXiv},
  primaryclass	= {quant-ph},
  url		= {https://arxiv.org/abs/2407.17405}
}

@Article{ber-19a,
  author	= {Bertini, Bruno and Kos, Pavel and Prosen, Toma{\v z}},
  doi		= {10.1103/PhysRevLett.123.210601},
  issue		= {21},
  journal	= {Phys. Rev. Lett.},
  month		= {Nov},
  numpages	= {6},
  pages		= {210601},
  publisher	= {American Physical Society},
  title		= {Exact Correlation Functions for Dual-Unitary Lattice
		  Models in $1+1$ Dimensions},
  url		= {https://link.aps.org/doi/10.1103/PhysRevLett.123.210601},
  volume	= {123},
  year		= {2019},
}

@Article{	  temme2017error,
  title		= {Error Mitigation for Short-Depth Quantum Circuits},
  author	= {Temme, Kristan and Bravyi, Sergey and Gambetta, Jay M.},
  journal	= {Phys. Rev. Lett.},
  volume	= {119},
  issue		= {18},
  pages		= {180509},
  numpages	= {5},
  year		= {2017},
  month		= {Nov},
  publisher	= {American Physical Society},
  doi		= {10.1103/PhysRevLett.119.180509},
  url		= {https://link.aps.org/doi/10.1103/PhysRevLett.119.180509}
}

@Article{	  rajagopala2024,
  title		= {Hardware-Assisted Parameterized Circuit Execution},
  author	= {Rajagopala, Abhi D and Hashim, Akel and Fruitwala, Neelay
		  and Huang, Gang and Xu, Yilun and Hines, Jordan and
		  Siddiqi, Irfan and Klymko, Katherine and Nowrouzi, Kasra},
  journal	= {arXiv preprint arXiv:2409.03725},
  year		= {2024},
  url		= {https://arxiv.org/abs/2409.03725}
}

@Article{	  wack2021,
  title		= {Scale, Quality, and Speed: three key attributes to measure
		  the performance of near-term quantum computers},
  author	= {Wack, Andrew and Paik, Hanhee and Javadi-Abhari, Ali and
		  Jurcevic, Petar and Faro, Ismael and Gambetta, Jay M and
		  Johnson, Blake R},
  journal	= {arXiv preprint arXiv:2110.14108},
  year		= {2021},
  url		= {https://arxiv.org/abs/2110.14108}
}

@Article{	  eddins2024,
  title		= {Lightcone shading for classically accelerated quantum
		  error mitigation},
  author	= {Eddins, Andrew and Tran, Minh C and Rall, Patrick},
  journal	= {arXiv preprint arXiv:2409.04401},
  year		= {2024},
  url		= {https://arxiv.org/abs/2409.04401}
}

@article{robledomoreno2025,
author = {Javier Robledo-Moreno  and Mario Motta  and Holger Haas  and Ali Javadi-Abhari  and Petar Jurcevic  and William Kirby  and Simon Martiel  and Kunal Sharma  and Sandeep Sharma  and Tomonori Shirakawa  and Iskandar Sitdikov  and Rong-Yang Sun  and Kevin J. Sung  and Maika Takita  and Minh C. Tran  and Seiji Yunoki  and Antonio Mezzacapo },
title = {Chemistry beyond the scale of exact diagonalization on a quantum-centric supercomputer},
journal = {Science Advances},
volume = {11},
number = {25},
pages = {eadu9991},
year = {2025},
doi = {10.1126/sciadv.adu9991},
URL = {https://www.science.org/doi/abs/10.1126/sciadv.adu9991},
}

@article{SQD_2,
author = {Liepuoniute, Ieva and Doney, Kirstin D. and Robledo Moreno, Javier and Job, Joshua A. and Friend, William S. and Jones, Gavin O.},
title = {Quantum-Centric Computational Study of Methylene Singlet and Triplet States},
journal = {Journal of Chemical Theory and Computation},
volume = {21},
number = {10},
pages = {5062-5070},
year = {2025},
doi = {10.1021/acs.jctc.5c00075},
note ={PMID: 40357738},
URL = { https://doi.org/10.1021/acs.jctc.5c00075},
}

@article{yu2025sample,
  title={Sample-based Krylov Quantum Diagonalization},
  author={Yu, Jeffery and Robledo Moreno, Javier and Iosue, Joseph and Bertels, Luke and Claudino, Daniel and Fuller, Bryce and Groszkowski, Peter and Humble, Travis S and Jurcevic, Petar and Kirby, William and others},
  journal={arXiv e-prints},
  pages={arXiv--2501},
  year={2025},
  doi={https://doi.org/10.48550/arXiv.2501.09702}
}

@book{cullum2002lanczos,
  title={Lanczos algorithms for large symmetric eigenvalue computations: Vol. I: Theory},
  author={Cullum, Jane K and Willoughby, Ralph A},
  year={2002},
  publisher={SIAM}
}

@Article{	  ber-19b,
  author	= {Bertini, Bruno and Kos, Pavel and Prosen, Toma{\v z}},
  doi		= {10.1103/PhysRevX.9.021033},
  issue		= {2},
  journal	= {Phys. Rev. X},
  month		= {May},
  numpages	= {27},
  pages		= {021033},
  publisher	= {American Physical Society},
  title		= {Entanglement Spreading in a Minimal Model of Maximal
		  Many-Body Quantum Chaos},
  url		= {https://link.aps.org/doi/10.1103/PhysRevX.9.021033},
  volume	= {9},
  year		= {2019}
}

@Article{	  ber-20a,
  author	= {Bertini, Bruno and Piroli, Lorenzo},
  doi		= {10.1103/PhysRevB.102.064305},
  issue		= {6},
  journal	= {Phys. Rev. B},
  month		= {Aug},
  numpages	= {25},
  pages		= {064305},
  publisher	= {American Physical Society},
  title		= {Scrambling in random unitary circuits: Exact results},
  url		= {https://link.aps.org/doi/10.1103/PhysRevB.102.064305},
  volume	= {102},
  year		= {2020}
}

@Article{	  ber-20b,
  title		= {Operator entanglement in local quantum circuits I: Chaotic
		  dual-unitary circuits},
  author	= {Bertini, Bruno and Kos, Pavel and Prosen, Toma{\v{z}}},
  journal	= {SciPost Physics},
  volume	= {8},
  number	= {4},
  pages		= {067},
  year		= {2020},
  url		= {https://doi.org/10.21468/SciPostPhys.8.4.067}
}

@Article{	  ber-21a,
  author	= {Bertini, Bruno and Kos, Pavel and Prosen, Toma{\v z}},
  doi		= {10.1007/s00220-021-04139-2},
  id		= {Bertini2021},
  isbn		= {1432-0916},
  journal	= {Commun. Math. Phys.},
  number	= {1},
  pages		= {597--620},
  title		= {Random Matrix Spectral Form Factor of Dual-Unitary Quantum
		  Circuits},
  url		= {https://doi.org/10.1007/s00220-021-04139-2},
  volume	= {387},
  year		= {2021}
}

@article{rudolph2025simulating,
  title={Simulating and Sampling from Quantum Circuits with 2D Tensor Networks},
  author={Rudolph, Manuel S and Tindall, Joseph},
  journal={arXiv preprint arXiv:2507.11424},
  year={2025},
  doi={https://doi.org/10.48550/arXiv.2507.11424}
}

@misc{povm_toolbox,
  author       = {qiskit-community},
  title        = {povm toolbox},
  year         = {2025},
  howpublished = {\url{https://github.com/qiskit-community/povm-toolbox}},
  note         = {GitHub repository}
}

@Article{	  cla-20a,
  author	= {Claeys, Pieter W. and Lamacraft, Austen},
  doi		= {10.1103/PhysRevResearch.2.033032},
  issue		= {3},
  journal	= {Phys. Rev. Res.},
  month		= {Jul},
  numpages	= {20},
  pages		= {033032},
  publisher	= {American Physical Society},
  title		= {Maximum velocity quantum circuits},
  url		= {https://link.aps.org/doi/10.1103/PhysRevResearch.2.033032},
  volume	= {2},
  year		= {2020}
}

@Article{	  ipp-21a,
  title		= {Postselection-Free Entanglement Dynamics via Spacetime
		  Duality},
  author	= {Ippoliti, Matteo and Khemani, Vedika},
  journal	= {Phys. Rev. Lett.},
  volume	= {126},
  issue		= {6},
  pages		= {060501},
  numpages	= {7},
  year		= {2021},
  month		= {Feb},
  publisher	= {American Physical Society},
  doi		= {10.1103/PhysRevLett.126.060501},
  url		= {https://link.aps.org/doi/10.1103/PhysRevLett.126.060501}
}

@Article{	  pir-20a,
  author	= {Piroli, Lorenzo and Bertini, Bruno and Cirac, J. Ignacio
		  and Prosen, Toma{\v z}},
  doi		= {10.1103/PhysRevB.101.094304},
  issue		= {9},
  journal	= {Phys. Rev. B},
  month		= {Mar},
  numpages	= {16},
  pages		= {094304},
  publisher	= {American Physical Society},
  title		= {Exact dynamics in dual-unitary quantum circuits},
  url		= {https://link.aps.org/doi/10.1103/PhysRevB.101.094304},
  volume	= {101},
  year		= {2020}
}

@Article{	  zho-22a,
  author	= {Zhou, Tianci and Harrow, Aram W.},
  doi		= {10.1103/PhysRevB.106.L201104},
  issue		= {20},
  journal	= {Phys. Rev. B},
  month		= {Nov},
  numpages	= {6},
  pages		= {L201104},
  publisher	= {American Physical Society},
  title		= {Maximal entanglement velocity implies dual unitarity},
  url		= {https://link.aps.org/doi/10.1103/PhysRevB.106.L201104},
  volume	= {106},
  year		= {2022}
}

@article{lanes2025framework,
  title={A Framework for Quantum Advantage},
  author={Lanes, Olivia and Beji, Mourad and Corcoles, Antonio D and Dalyac, Constantin and Gambetta, Jay M and Henriet, Loic and Javadi-Abhari, Ali and Kandala, Abhinav and Mezzacapo, Antonio and Porter, Christopher and others},
  journal={arXiv preprint arXiv:2506.20658},
  year={2025},
  doi={https://doi.org/10.48550/arXiv.2506.20658}
}

@Article{	  ber-18a,
  author	= {Bertini, Bruno and Kos, Pavel and Prosen, Toma\v{z}},
  doi		= {10.1103/PhysRevLett.121.264101},
  issue		= {26},
  journal	= {Phys. Rev. Lett.},
  month		= {Dec},
  numpages	= {6},
  pages		= {264101},
  publisher	= {American Physical Society},
  title		= {Exact Spectral Form Factor in a Minimal Model of Many-Body
		  Quantum Chaos},
  url		= {https://link.aps.org/doi/10.1103/PhysRevLett.121.264101},
  volume	= {121},
  year		= {2018}
}

@article{abughanem2025ibm,
  title={IBM quantum computers: Evolution, performance, and future directions},
  author={AbuGhanem, Muhammad},
  journal={The Journal of Supercomputing},
  volume={81},
  number={5},
  pages={687},
  year={2025},
  publisher={Springer},
  doi={https://doi.org/10.1007/s11227-025-07047-7}
}

@Article{	  suz-22,
  title		= {Computational power of one-and two-dimensional
		  dual-unitary quantum circuits},
  author	= {Suzuki, Ryotaro and Mitarai, Kosuke and Fujii, Keisuke},
  journal	= {Quantum},
  volume	= {6},
  pages		= {631},
  year		= {2022},
  publisher	= {Verein zur F{\"o}rderung des Open Access Publizierens in
		  den Quantenwissenschaften},
  url		= {https://doi.org/10.22331/q-2022-01-24-631}
}

@Article{	  fisher2023random,
  title		= {Random quantum circuits},
  author	= {Fisher, Matthew PA and Khemani, Vedika and Nahum, Adam and
		  Vijay, Sagar},
  journal	= {Annual Review of Condensed Matter Physics},
  volume	= {14},
  pages		= {335--379},
  year		= {2023},
  publisher	= {Annual Reviews},
  url		= {https://doi.org/10.1146/annurev-conmatphys-031720-030658}
}

@Article{	  hangleiter2023computational,
  title		= {Computational advantage of quantum random sampling},
  author	= {Hangleiter, Dominik and Eisert, Jens},
  journal	= {Reviews of Modern Physics},
  volume	= {95},
  number	= {3},
  pages		= {035001},
  year		= {2023},
  publisher	= {APS},
  url		= {https://link.aps.org/doi/10.1103/RevModPhys.95.035001}
}

@Article{	  fisher18,
  title		= {Quantum Zeno effect and the many-body entanglement
		  transition},
  author	= {Li, Yaodong and Chen, Xiao and Fisher, Matthew P. A.},
  journal	= {Phys. Rev. B},
  volume	= {98},
  issue		= {20},
  pages		= {205136},
  numpages	= {9},
  year		= {2018},
  month		= {Nov},
  publisher	= {American Physical Society},
  doi		= {10.1103/PhysRevB.98.205136},
  url		= {https://link.aps.org/doi/10.1103/PhysRevB.98.205136}
}

@Article{	  skinner19,
  title		= {Measurement-Induced Phase Transitions in the Dynamics of
		  Entanglement},
  author	= {Skinner, Brian and Ruhman, Jonathan and Nahum, Adam},
  journal	= {Phys. Rev. X},
  volume	= {9},
  issue		= {3},
  pages		= {031009},
  numpages	= {21},
  year		= {2019},
  month		= {Jul},
  publisher	= {American Physical Society},
  doi		= {10.1103/PhysRevX.9.031009},
  url		= {https://link.aps.org/doi/10.1103/PhysRevX.9.031009}
}

@Article{	  chan18,
  title		= {Solution of a Minimal Model for Many-Body Quantum Chaos},
  author	= {Chan, Amos and De Luca, Andrea and Chalker, J. T.},
  journal	= {Phys. Rev. X},
  volume	= {8},
  issue		= {4},
  pages		= {041019},
  numpages	= {17},
  year		= {2018},
  month		= {Nov},
  publisher	= {American Physical Society},
  doi		= {10.1103/PhysRevX.8.041019},
  url		= {https://link.aps.org/doi/10.1103/PhysRevX.8.041019}
}
\addcontentsline{toc}{chapter}{Bibliography}

\end{document}